\documentclass[12pt,draft]{cernrep}

\input epsf

\usepackage{amssymb}
\usepackage{mathrsfs}

\usepackage{feynmf}

\def\stackrel#1#2{\mathrel{\mathop{#2}\limits^{#1}}}

\long\def\symbolfootnote[#1]#2{\begingroup%
   \def\thefootnote{\fnsymbol{footnote}}\footnote[#1]{#2}\endgroup} 

\def\dj{\hbox{d\kern-0.347em \vrule width 0.3em height 1.252ex depth
-1.21ex \kern 0.051em}}

\def\dslash{\partial\!\!\!/} 
\def\Dslash{D\!\!\!\!/\,\,}
\def\fslash#1{\slash\!\!\!#1}
\def\Fslash#1{\slash\!\!\!\!#1}

\numberwithin{equation}{section}

\title{Introductory Lectures on Quantum Field Theory\thanks{Based on 
lectures delivered by 
L.A.-G. at the 3rd. CERN-CLAF School of High-Energy Physics, Malarg\"ue 
(Argentina), February 27-March 12, 2005; at the 5th CERN-CLAF School
of High-Energy Physics, Medell\'{\i}n (Colombia), March 15-28, 2009;
at the 6th CERN-CLAF School of High Energy Physics, Natal (Brazil), March 23-April 5, 2011;
and at the 2012 Asia-Europe-Pacific School of High Energy Physics, Fukuoka (Japan), October 14-27, 2012.}}

\author{Luis \'Alvarez-Gaum\'e$^{\,a,\,}$\thanks{Luis.Alvarez-Gaume@cern.ch}{} 
{\,\, and} 
Miguel A. V\'azquez-Mozo$^{\,b,\,c,\,}$\thanks{Miguel.Vazquez-Mozo@cern.ch,
vazquez@usal.es}}

\institute{$^{a}\,$Physics Department, Theory Division, CERN, CH-1211
Geneva 23, Switzerland \\
$^{b}\,$Departamento de F\'{\i}sica Fundamental, Universidad de Salamanca, 
Plaza de la Merced s/n, E-37008 Salamanca, Spain \\
$^{c}\,$Instituto Universitario de F\'{\i}sica Fundamental y
Matem\'aticas (IUFFyM), Universidad de Salamanca, Salamanca, Spain}

\begin{document}
\maketitle

\begin{abstract}
In these lectures we present a few topics in Quantum Field Theory in detail.
Some of them are conceptual and some more practical. They have been
selected because they appear frequently in current applications to Particle
Physics and String Theory.
\end{abstract}

\newpage

\tableofcontents 

\newpage

\section{Introduction}

These notes summarize lectures presented at the 2005 CERN-CLAF school in Malarg{\"u}e,
Argentina, the 2009 CERN-CLAF school in Medell\'{\i}n, Colombia, the 2011 CERN-CLAF school in 
Natal (Brazil), and the 2012 Asia-Europe-Pacific School of High Energy Physics in Fukuoka (Japan).  
The audience in all occasions was composed to a
large extent by students in experimental High Energy Physics with an
important minority of theorists. In nearly ten hours it is quite
difficult to give a reasonable introduction to a subject as vast as
Quantum Field Theory.  For this reason the lectures were intended to
provide a review of those parts of the subject to be used later by
other lecturers. Although a cursory acquaitance with th subject of
Quantum Field Theory is helpful, the only requirement to follow the
lectures it is a working knowledge of Quantum Mechanics and Special
Relativity.

The guiding principle in choosing the topics presented (apart to serve
as introductions to later courses) was to present some basic aspects
of the theory that present conceptual subtleties. Those topics one
often is uncomfortable with after a first introduction to the subject.
Among them we have selected:

\begin{itemize}
\item[-]
The need to introduce quantum fields, with the great complexity this
implies.

\item[-]
Quantization of gauge theories and the r\^ole of topology in quantum
phenomena. We have included a brief study of the Aharonov-Bohm effect
and Dirac's explanation of the quantization of the electric charge in
terms of magnetic monopoles. 

\item[-]
Quantum aspects of global and gauge symmetries and their breaking.

\item[-]
Anomalies.

\item[-]
The physical idea behind the process of renormalization of quantum
field theories.

\item[-]
Some more specialized topics, like the creation of particle by classical
fields and the very basics of supersymmetry.

\end{itemize}

These notes have been written following closely the original
presentation, with numerous clarifications. Sometimes the treatment
given to some subjects has been extended, in particular the
discussion of the Casimir effect and particle creation by classical
backgrounds.  Since no group theory was assumed, we have included an
Appendix with a review of the basics concepts. 

By lack of space and purpose, few proofs have been included. Instead,
very often we illustrate a concept or property by describing a
physical situation where it arises. A very much expanded version of these lectures,
following the same philosophy but including many other topics,
has appeared in book form in \cite{ours}.
For full details and proofs we refer the reader to 
the many textbooks in the subject, and in particular in the
ones provided in the bibliography
\cite{bjorken,itzykson,ramond,peskin,weinberg,
deligne,zee,dewitt,nair,banks}.  Specially modern presentations, very much
in the spirit of these lectures, can be found in references
\cite{peskin,weinberg,nair,banks}.  We should nevertheless warn the reader
that we have been a bit cavalier about references. Our aim has been to
provide mostly a (not exhaustive) list of reference for further
reading. We apologize to those authors who feel misrepresented.

{\bf Acknowlegments.}
L.A.-G would like to thank the organizers of the 2012 Asia-Europe-Pacific School of High Energy Physics
for their kindness and hospitality, and in particular
Nick Ellis, Hellen Haller, Masami Yokoyama, Lydia Roos and Francesco Riva for plenty of interesting discussion.
The work of M.A.V.-M. has been partially supported by Spanish Science 
Ministry Grants FPA2009-10612, FPA2012-34456, and FIS2012-30926, 
Basque Government Grant IT-357-07, and Spanish Consolider-Ingenio 2010 Programme CPAN (CSD2007-00042).

\subsection{A note about notation}

Before starting it is convenient to review the notation used. Through
these notes we will be using the metric $\eta_{\mu\nu}={\rm
diag\,}(1,-1,-1,-1)$. Derivatives with respect to the four-vector
$x^{\mu}=(ct,\vec{x})$ will be denoted by the shorthand
\begin{eqnarray}
\partial_{\mu}\equiv {\partial\over \partial x^{\mu}}
=\left({1\over c}{\partial\over\partial t},\vec{\nabla}\right).
\end{eqnarray}
As usual space-time indices will be labelled by Greek letters ($\mu,\nu,\ldots
=0,1,2,3$) while Latin indices will be used for spatial directions ($i,j,\ldots
=1,2,3$). In many expressions we will use the notation
$\sigma^{\mu}=(\mathbf{1},\sigma^{i})$ where $\sigma^{i}$ are the
Pauli matrices
\begin{eqnarray}
\sigma^{1}=\left(
\begin{array}{rr}
0 & 1 \\
1 & 0
\end{array}
\right), \hspace*{1cm}
\sigma^{2}=\left(
\begin{array}{rr}
0 & -i \\
i & 0
\end{array}
\right), \hspace*{1cm}
\sigma^{3}=\left(
\begin{array}{rr}
1 & 0 \\
0 & -1
\end{array}
\right).
\end{eqnarray}
Sometimes we use of the Feynman's slash notation  
$\fslash{a} =
\gamma^{\mu}a_{\mu}$. Finally, unless stated otherwise, 
we work in natural units $\hbar=c=1$.

\section{Why do we need Quantum Field Theory after all?}
\label{sec:why}

In spite of the impressive success of Quantum Mechanics in describing
atomic physics, it was immediately clear after its formulation that
its relativistic extension was not free of difficulties. These
problems were clear already to Schr\"odinger, whose first guess for a
wave equation of a free relativistic particle was the Klein-Gordon
equation
\begin{eqnarray}
\left({\partial^2\over \partial t^2}-{\nabla}^2+m^2\right)
\psi(t,\vec{x})=0.
\label{KG}
\end{eqnarray}
This equation follows directly from the relativistic ``mass-shell'' identity 
$E^2=\vec{p}^{\,2}+m^2$  using the correspondence principle
\begin{eqnarray}
E&\rightarrow& {i}{\partial\over\partial t}, \nonumber \\
\vec{p} &\rightarrow& -i \vec{\nabla}.
\end{eqnarray}

Plane wave solutions to the wave equation (\ref{KG}) are readily
obtained
\begin{eqnarray}
\psi(t,\vec{x})=e^{-ip_{\mu}x^{\mu}}=e^{-iE t+i\vec{p}\cdot\vec{x}}
\hspace*{1cm} \mbox{with} \hspace*{1cm} E=\pm \omega_{p}
\equiv \pm\sqrt{\vec{p}^{\,2}+m^2}.
\end{eqnarray}
In order to have a complete basis of functions, one must include plane
wave with both $E>0$ and $E<0$. This implies that given the
conserved current
\begin{eqnarray}
j_{\mu}={i\over 2}\Big(\psi^{*}\partial_{\mu}\psi-
\partial_{\mu}\psi^{*}\,\psi\Big),
\end{eqnarray}
its time-component is $j^{0}=E$ and therefore does 
not define a positive-definite probability density.

A complete, properly normalized, continuous basis of solutions of the
Klein-Gordon equation (\ref{KG}) labelled by the momentum $\vec{p}$
can be defined as
\begin{eqnarray}
f_{p}(t,\vec{x})&=&{1\over
(2\pi)^{3\over 2}\sqrt{2\omega_{p}}}\,e^{-i\omega_{p}t+i\vec{p}\cdot\vec{x}},
\nonumber \\ 
f_{-p}(t,\vec{x})&=&{1\over
(2\pi)^{3\over 2}\sqrt{2\omega_{p}}}\,e^{i\omega_{p}t-i\vec{p}
\cdot\vec{x}}.
\label{basis}
\end{eqnarray}
Given the inner product
\begin{eqnarray}
\langle\psi_{1}|\psi_{2}\rangle = i\int d^{3}x 
\Big(\psi_{1}^{*}\partial_{0}\psi_{2}-\partial_{0}\psi_{1}^{*}\,\psi_{2}
\Big) \nonumber 
\end{eqnarray}
the states (\ref{basis}) form an orthonormal basis
\begin{eqnarray}
\langle f_{p}|f_{p'}\rangle &=&  \delta(\vec{p}-\vec{p}\,'), \nonumber \\
\langle f_{-p}|f_{-p'}\rangle &=&  -\delta(\vec{p}-\vec{p}\,'), \\
\langle f_{p}|f_{-p'}\rangle &=& 0.
\end{eqnarray}

\begin{figure}
\centerline{\epsfxsize=0.5truein\epsfbox{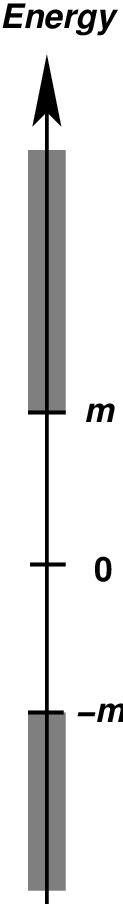}}
\caption[]{Spectrum of the Klein-Gordon wave equation}
\label{fig:spectrum}
\end{figure}

The wave functions $f_{p}(t,x)$ describes states with momentum $\vec{p}$ and
energy given by $\omega_{p}=\sqrt{\vec{p}^{\,\,2}+m^2}$. 
On the other hand, the states $|f_{-p}\rangle$ not only have a negative scalar 
product but they actually correspond to negative energy states
\begin{eqnarray}
i\partial_{0}f_{-p}(t,\vec{x})= -\sqrt{\vec{p}^{\,\,2}+m^2}\,
f_{-p}(t,\vec{x}).
\end{eqnarray}
Therefore the energy spectrum of the theory satisfies $|E|>m$ and is
unbounded from below (see Fig. \ref{fig:spectrum}). Although in a case
of a free theory the absence of a ground state is not necessarily a
fatal problem, once the theory is coupled to the electromagnetic
field this is the source of all kinds of disasters, since 
nothing can prevent the decay of any state by emission
of electromagnetic radiation. 

The problem of the instability of the ``first-quantized'' relativistic
wave equation can be heuristically tackled in the case of
spin-${1\over 2}$ particles, described by the Dirac equation
\begin{eqnarray}
\left(-i\mathbf{\beta}{\partial\over \partial t}
+\vec{\mathbf{\alpha}}\cdot\vec{\nabla}
-m\right)\psi(t,\vec{x})=0,
\label{eq:dirac}
\end{eqnarray}
where $\vec{\alpha}$ and ${\beta}$ are $4\times 4$ matrices
\begin{eqnarray}
\mathbf{\alpha}^{i}=\left(
\begin{array}{cc}
0 & i\sigma^{i} \\
-i\sigma^{i} & 0 
\end{array}
\right), \hspace*{1cm}
\mathbf{\beta}=\left(
\begin{array}{cc}
0 & \mathbf{1} \\
\mathbf{1} & 0 
\end{array}
\right),
\end{eqnarray}
with $\sigma^{i}$ the Pauli matrices, and the wave function
$\psi(t,\vec{x})$ has four components. The wave equation
(\ref{eq:dirac}) can be thought of as a kind of ``square root'' of the
Klein-Gordon equation (\ref{KG}), since the latter can be obtained as
\begin{eqnarray}
\left(-i\mathbf{\beta}{\partial\over \partial t}
+\vec{\mathbf{\alpha}}\cdot\vec{\nabla}
-m\right)^{\dagger}\left(-i\mathbf{\beta}{\partial\over \partial t}
+\vec{\mathbf{\alpha}}\cdot\vec{\nabla}
-m\right)\psi(t,\vec{x})=
\left({\partial^2\over \partial t^2}-{\nabla}^2+m^2\right)
\psi(t,\vec{x}).
\end{eqnarray}

An analysis of Eq. (\ref{eq:dirac}) along the lines of the one presented above
for the Klein-Gordon equation leads again to the existence of negative energy
states and a spectrum unbounded from below as in Fig. \ref{fig:spectrum}. 
Dirac, however, solved the instability problem by pointing out that now
the particles are fermions and therefore they are subject to Pauli's 
exclusion principle. Hence, each state in the spectrum can be occupied
by at most one particle, so the states with $E=m$ can be made stable
if we assume that {\em all} the negative energy states are filled.

\begin{figure}
\centerline{\epsfxsize=5.0truein\epsfbox{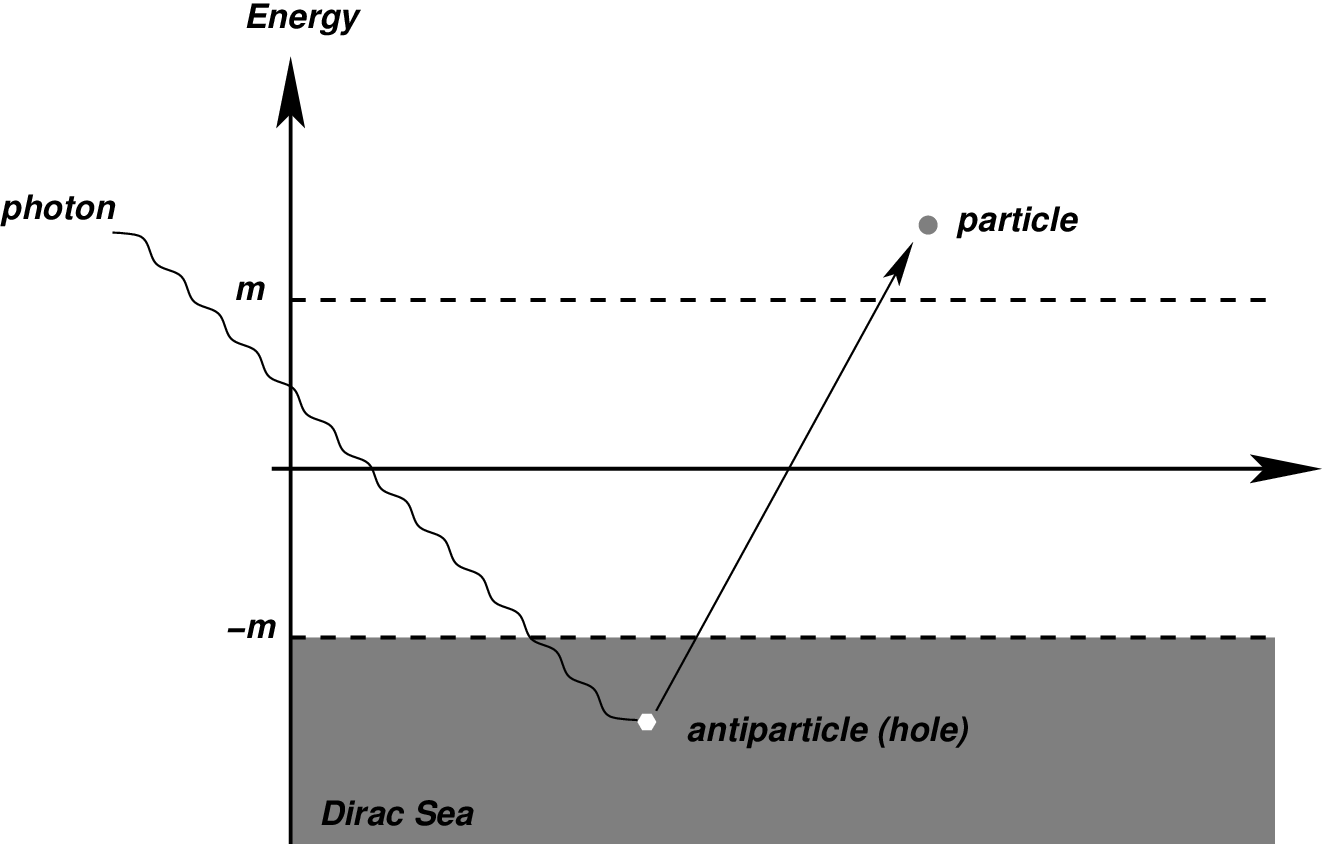}}
\caption[]{Creation of a particle-antiparticle pair in the Dirac see
picture}
\label{fig:dirac_sea}
\end{figure}

If Dirac's idea restores the stability of the spectrum by introducing
a stable vacuum where all negative energy states are occupied, the
so-called Dirac sea, it also leads directly to the conclusion that a
single-particle interpretation of the Dirac equation is
not possible. Indeed, a photon with enough energy ($E>2m$) can excite
one of the electrons filling the negative energy states, leaving
behind a ``hole'' in the Dirac see (see Fig. \ref{fig:dirac_sea}).
This hole behaves as a particle with equal mass and opposite charge
that is interpreted as a positron,
so there is no escape to the conclusion that interactions will produce
pairs particle-antiparticle out of the vacuum.

In spite of the success of the heuristic interpretation of negative
energy states in the Dirac equation this is not the end of the story.
In 1929 Oskar Klein stumbled into an apparent paradox when trying to
describe the scattering of a relativistic electron by a square
potential using Dirac's wave equation \cite{klein} (for pedagogical
reviews see
\cite{klein-review1,klein-review2}). In order to capture 
the essence of the problem without entering into unnecessary
complication we will study Klein's paradox in the context of the
Klein-Gordon equation.

\begin{figure}
\centerline{\epsfxsize=5.0truein\epsfbox{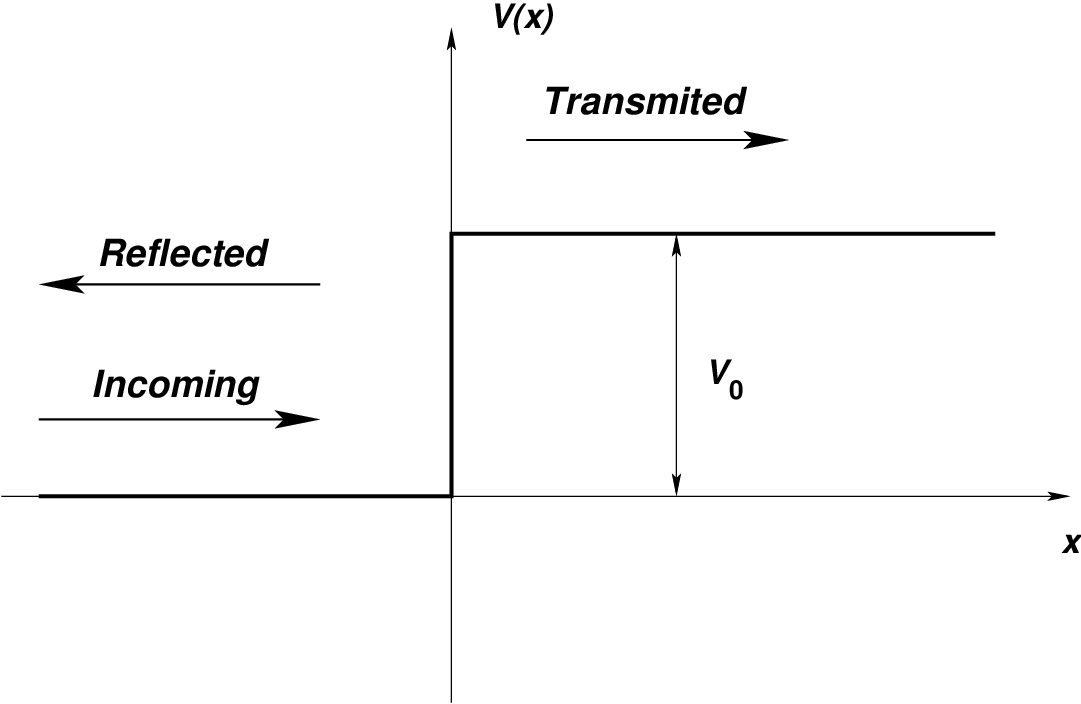}}
\caption[]{Illustration of the Klein paradox.}
\label{fig:potential}
\end{figure}

Let us consider a square potential with height $V_{0}>0$ of the type
showed in Fig. \ref{fig:potential}. A solution to the wave equation in 
regions I and II is given by
\begin{eqnarray}
\psi_{I}(t,x)&=& e^{-iEt+ip_{1}x}+R e^{-iEt-ip_{1}x}, \nonumber \\
\psi_{II}(t,x)&=& T e^{-iEt+p_{2}x},
\end{eqnarray}
where the mass-shell condition implies that 
\begin{eqnarray}
p_{1}=\sqrt{E^2-m^2}, \hspace*{1cm}
p_{2}=\sqrt{(E-V_{0})^2-m^2}.
\end{eqnarray}
The constants $R$ and $T$ are computed
by matching the two solutions across the boundary $x=0$. The
conditions $\psi_{I}(t,0)=
\psi_{II}(t,0)$ and $\partial_{x}\psi_{I}(t,0)=\partial_{x}\psi_{II}(t,0)$
imply that
\begin{eqnarray}
T={2p_{1}\over p_{1}+p_{2}}, \hspace*{1cm}
R={p_{1}-p_{2}\over p_{1}+p_{2}}.
\end{eqnarray}

At first sight one would expect a behavior similar to the one encountered
in the nonrelativistic case. If the kinetic energy is bigger than $V_{0}$
both a transmitted and reflected wave are expected, whereas when the 
kinetic energy is smaller than $V_{0}$ one only expect to find a reflected
wave, the transmitted wave being exponentially damped within a distance
of a Compton wavelength inside the barrier.

Indeed this is what happens if $E-m>V_{0}$. In this case both $p_{1}$ and
$p_{2}$ are real and we have a partly reflected, and a partly transmitted wave.
In the same way, if $V_{0}-2m<E-m<V_{0}$ then $p_{2}$ is
imaginary and there is total reflection. 

However, in the case when $V_{0}>2m$ and the energy is in the range
$0<E-m<V_{0}-2m$ a completely different situation arises. In this
case one finds that both $p_{1}$ and $p_{2}$ are real and therefore
the incoming wave function is partially reflected and partially
transmitted across the barrier. This is a shocking result, since it
implies that there is a nonvanishing probability of finding the
particle at any point across the barrier with negative kinetic energy
($E-m-V_{0}<0$)! This weird result is known as Klein's paradox.

As with the negative energy states, the Klein paradox results from our
insistence in giving a single-particle interpretation to the relativistic
wave function. Actually, a multiparticle analysis of the paradox 
\cite{klein-review1} shows that what happens when $0<E-m<V_{0}-2m$ is that
the reflection of the incoming particle by the barrier is accompanied
by the creation of pairs particle-antiparticle out of the energy of
the barrier (notice that for this to happen it is required that
$V_{0}>2m$, the threshold for the creation of a particle-antiparticle
pair). 

Actually, this particle creation can be understood by noticing that the sudden
potential step in Fig. \ref{fig:potential} localizes the incoming
particle with mass $m$ in distances smaller than its Compton
wavelength $\lambda={1\over m}$. This can be seen by replacing the square
potential by another one where the potential varies smoothly from $0$ to
$V_{0}>2m$ in distances scales larger than $1/m$. This case was
worked out by Sauter shortly after Klein pointed out the paradox
\cite{Sauter}.  He considered a situation where the regions with
$V=0$ and $V=V_{0}$ are connected by a region of length $d$ with a
linear potential $V(x)={V_{0}x\over d}$. When $d>{1\over m}$ he found
that the transmission coefficient is exponentially small\footnote{In
section (\ref{sec:schwinger}) we will see how, in the case of the
Dirac field, this exponential behavior can be associated with the
creation of electron-positron pairs due to a constant electric field
(Schwinger effect).}.

The creation of particles is impossible to avoid whenever one tries to
locate a particle of mass $m$ within its Compton wavelength. Indeed,
from Heisenberg uncertainty relation we find that if $\Delta x\sim
{1\over m}$, the fluctuations in the momentum will be of order $\Delta
p\sim m$ and fluctuations in the energy of order
\begin{eqnarray}
\Delta E\sim m
\end{eqnarray}
can be expected.
Therefore, in a relativistic theory, the fluctuations of the energy
are enough to allow the creation of particles out of
the vacuum. In the case of a spin-$1\over 2$ particle, the Dirac sea
picture shows clearly how, when the energy fluctuations are of order
$m$, electrons from the Dirac sea can be excited to positive energy
states, thus creating electron-positron pairs.

\begin{figure}
\centerline{\epsfxsize=3.5truein\epsfbox{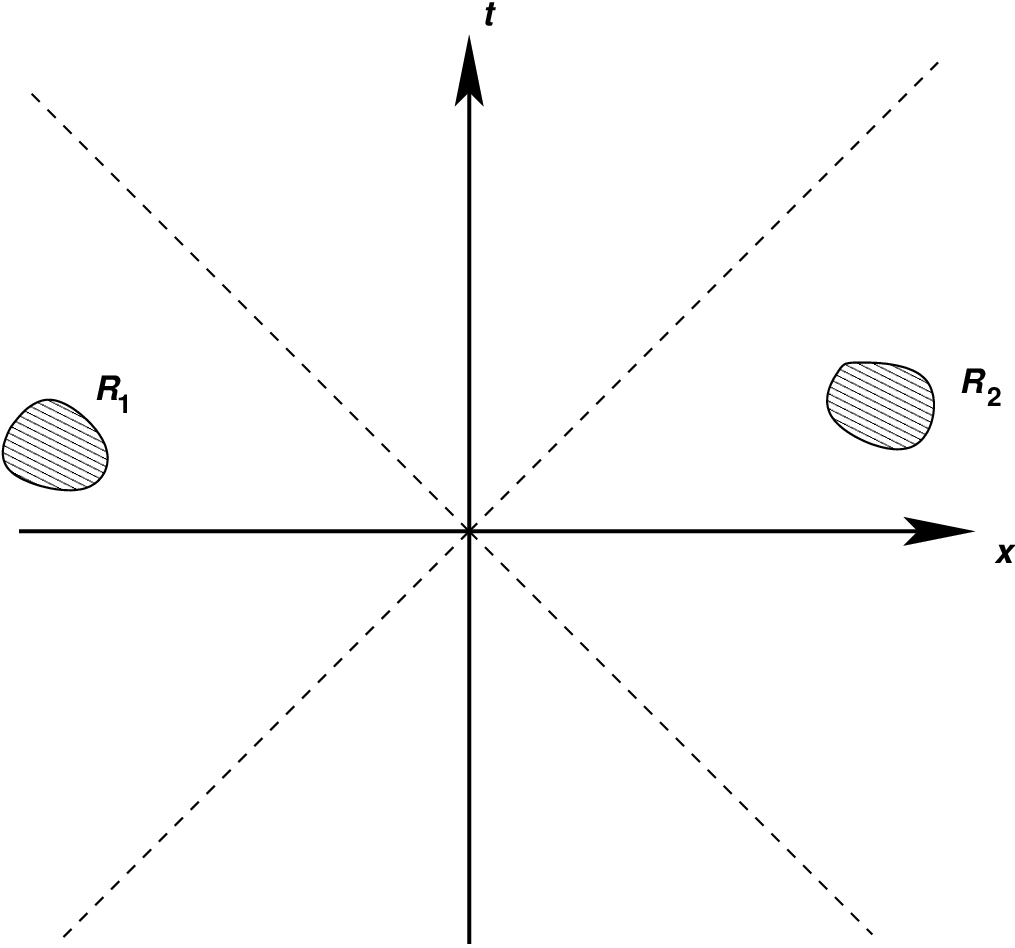}}
\caption[]{Two regions $R_{1}$, $R_{2}$ that are causally disconnected.}
\label{fig:causal}
\end{figure}

It is possible to see how the multiparticle interpretation is forced
upon us by relativistic invariance. In non-relativistic Quantum
Mechanics observables are represented by self-adjoint operator that in
the Heisenberg picture depend on time. Therefore measurements are
localized in time but are global in space.  The situation is radically
different in the relativistic case.  Because no signal can propagate
faster than the speed of light, measurements have to be localized both
in time and space. Causality demands then that two measurements
carried out in causally-disconnected regions of space-time cannot
interfere with each other. In mathematical terms this means that if
$\mathcal{O}_{R_{1}}$ and $\mathcal{O}_{R_{2}}$ are the observables
associated with two measurements localized in two causally-disconnected
regions $R_{1}$, $R_{2}$ (see Fig. \ref{fig:causal}), they satisfy
\begin{eqnarray}
[\mathcal{O}_{R_{1}},\mathcal{O}_{R_{2}}]=0, \hspace*{1cm}
\mbox{if $(x_{1}-x_{2})^2<0$, for all $x_1\in R_{1}$, $x_{2}\in R_{2}$}.
\label{commutator}
\end{eqnarray}

Hence, in a relativistic theory, the basic operators in the Heisenberg 
picture must depend on the space-time position $x^{\mu}$. Unlike the 
case in non-relativistic quantum mechanics, here the position $\vec{x}$ 
{\em is not} an observable, but just a label, similarly to the case of
time in ordinary quantum mechanics. Causality is then imposed microscopically
by requiring
\begin{eqnarray}
[\mathcal{O}(x),\mathcal{O}(y)]=0, \hspace*{1cm} \mbox{if $(x-y)^2<0$}.
\label{microcausality}
\end{eqnarray}
A smeared operator $\mathcal{O}_{R}$ over a space-time
region $R$ can then be defined as
\begin{eqnarray}
\mathcal{O}_{R}=\int d^{4}x\, \mathcal{O}(x)\,f_{R}(x)
\end{eqnarray}
where $f_{R}(x)$ is the characteristic function associated with  
$R$,
\begin{eqnarray}
f_{R}(x)=\left\{
\begin{array}{lll}
1 & & x\in R \\
0 & & x\notin R
\end{array}
\right. .
\end{eqnarray}
Eq. (\ref{commutator}) follows now from the microcausality condition
(\ref{microcausality}).

\begin{figure}
\centerline{\epsfxsize=4.0truein\epsfbox{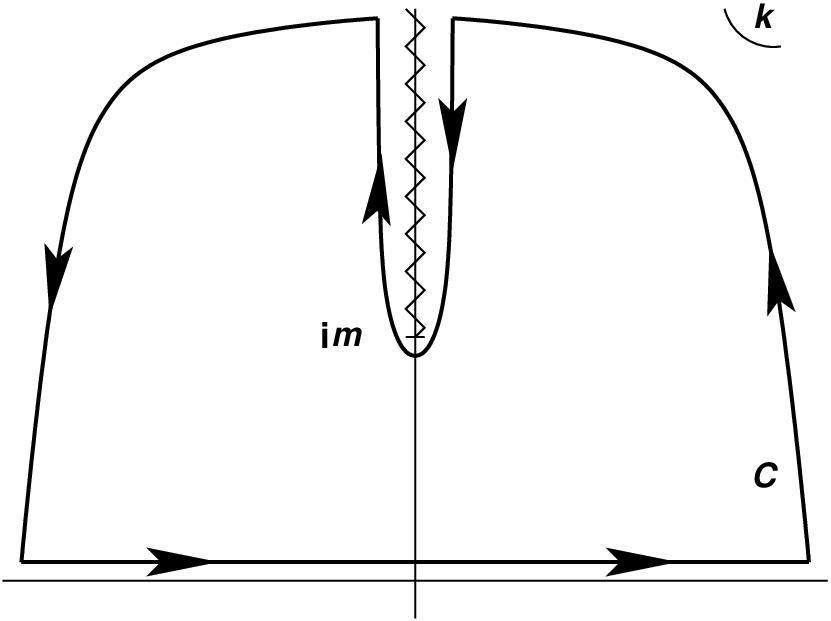}}
\caption[]{Complex contour $C$ for the computation of the integral in Eq.
(\ref{integral_wf}).}
\label{fig:contour}
\end{figure}

Therefore, relativistic invariance forces the introduction of quantum 
fields. It is only when we insist in keeping a single-particle interpretation
that we crash against causality violations. To illustrate the point, let us
consider a single particle wave function $\psi(t,\vec{x})$ that initially 
is localized in the position $\vec{x}=0$
\begin{eqnarray}
\psi(0,\vec{x})=\delta(\vec{x}).
\end{eqnarray}
Evolving this wave function using the Hamiltonian $H=\sqrt{-\nabla^2+m^2}$
we find that the wave function can be written as
\begin{eqnarray}
\psi(t,\vec{x})=e^{-it\sqrt{-\nabla^2+m^2}}\delta(\vec{x})=
\int {d^{3}k\over (2\pi)^{3}}e^{i\vec{k}\cdot\vec{x}-it\sqrt{k^2+m^2}}.
\end{eqnarray}
Integrating over the angular variables, the wave function can be
recast in the form
\begin{eqnarray}
\psi(t,\vec{x})={1\over 2\pi^2|\vec{x}|}\int_{-\infty}^{\infty}
k\,dk\,e^{ik|\vec{x}|}\,e^{-it\sqrt{k^2+m^2}}.
\label{integral_wf}
\end{eqnarray}
The resulting integral can be evaluated using the complex integration
contour $C$ shown in Fig. \ref{fig:contour}. The result is that, for any $t>0$,
one finds that $\psi(t,\vec{x})\neq 0$ for any $\vec{x}$. If we insist in 
interpreting the wave function $\psi(t,\vec{x})$ as the probability 
density of finding the particle at the location $\vec{x}$ in the time $t$
we find that the probability leaks out of the light cone, thus violating 
causality.

\section{From classical to quantum fields}
\label{sec:class2quant}

We have learned how the consistency of quantum mechanics with 
special relativity forces us to abandon the single-particle interpretation
of the wave function. Instead we have to consider quantum fields 
whose elementary excitations are associated with particle states, as we will
see below. 

In any scattering experiment, the only information available to us is
the set of quantum number associated with the set of free particles in
the initial and final states. Ignoring for the moment other quantum
numbers like spin and flavor, one-particle states are labelled by
the three-momentum $\vec{p}$ and span the single-particle Hilbert
space $\mathcal{H}_{1}$
\begin{eqnarray}
|\vec{p}\rangle \in \mathcal{H}_{1}, \hspace*{1cm} 
\langle\vec{p}|\vec{p}\,'\rangle=\delta(\vec{p}-\vec{p}\,')\,.
\label{eq:nonrelativistic_states}
\end{eqnarray}
The states $\{|\vec{p}\rangle\}$ form a basis of $\mathcal{H}_{1}$
and therefore satisfy the closure relation
\begin{eqnarray}
\int d^{3}p\,|\vec{p}\rangle\langle\vec{p}|
=\mathbf{1}
\end{eqnarray}
The group of spatial rotations acts unitarily on the states $|\vec{p}\rangle$.
This means that for every rotation $R\in {\rm SO}(3)$ there is a unitary
operator $\mathcal{U}(R)$ such that
\begin{eqnarray}
\mathcal{U}(R)|\vec{p}\rangle=|R\vec{p}\rangle
\label{eq:transformation}
\end{eqnarray}
where $R\vec{p}$ represents the action of the rotation on the vector
$\vec{k}$, $(R\vec{p})^{i}=R^{i}_{\,\,\,j}k^{j}$. Using a spectral 
decomposition, the momentum operator
$\widehat{{P}}^{i}$ can be written as
\begin{eqnarray}
\widehat{{P}}^{i}=
\int d^{3}p\,|\vec{p}\rangle\,{p}^{i}\,\langle\vec{p}|
\label{eq:closure_NR}
\end{eqnarray}
With the help of Eq. (\ref{eq:transformation}) it is straightforward
to check that the momentum operator transforms as a vector under
rotations:
\begin{eqnarray}
\mathcal{U}(R)^{-1}\,\widehat{{P}}^{i}\,\mathcal{U}(R)=
\int d^{3}p\,|R^{-1}\vec{p}\rangle\,\,{p}^{i}\,\,
\langle R^{-1}\vec{p}|
=R^{i}_{\,\,\,j}\widehat{{P}}^{j},
\end{eqnarray}
where we have used that the integration measure is invariant under SO$(3)$.

Since, as we argued above, we are forced to deal with multiparticle states,
it is convenient to introduce creation-annihilation operators associated
with a single-particle state of momentum $\vec{p}$
\begin{eqnarray}
[a(\vec{p}),a^{\dagger}(\vec{p}\,')]=\delta(\vec{p}-\vec{p}\,'),
\hspace*{1cm} [a(\vec{p}),a(\vec{p}\,')]=
[a^{\dagger}(\vec{p}),a^{\dagger}(\vec{p}\,')]=0,
\end{eqnarray}
such that the state $|\vec{p}\rangle$ is created out of the Fock space
vacuum $|0\rangle$ (normalized such that $\langle 0|0\rangle=1$) by
the action of a creation operator $a^{\dagger}(\vec{p})$
\begin{eqnarray}
|\vec{p}\rangle= a^{\dagger}(\vec{p})|0\rangle, \hspace*{1cm}
a(\vec{p})|0\rangle=0\,\,\,\,\forall \vec{p}.
\end{eqnarray}

Covariance under spatial rotations is all we need if we are interested
in a nonrelativistic theory. However in a relativistic quantum field theory
we must preserve more that SO$(3)$, actually we need the expressions to be
covariant under the full Poincar\'e group ISO$(1,3)$ consisting in 
spatial rotations, boosts and space-time translations. Therefore, in order
to build the Fock space of the theory we need two key ingredients: 
first an invariant normalization for the states, since we want 
a normalized state in one reference frame to be normalized  
in any other inertial frame. And secondly a relativistic invariant 
integration measure in momentum space, so the spectral decomposition of 
operators is covariant under the full Poincar\'e group.

Let us begin with the invariant measure. Given an invariant function
$f(p)$ of the four-mo\-men\-tum $p^{\mu}$ of a particle of mass $m$ with
positive energy $p^{0}>0$, there is an integration measure which is
invariant under proper Lorentz transformations\footnote{The factors
of $2\pi$ are introduced for later convenience.}
\begin{eqnarray}
\int {d^{4}p\over (2\pi)^{4}}\,(2\pi)\delta(p^2-m^2)\,\theta(p^{0})\,f(p), 
\end{eqnarray}
where $\theta(x)$ represent the Heaviside step function. 
The integration over $p^{0}$ can be easily done using the $\delta$-function
identity
\begin{eqnarray}
\delta[f(x)]=\sum_{x_i={\rm zeros\,\,of\,\,}f}{1\over |f'(x_{i})|}
\delta(x-x_{i}),
\label{eq:delta_identity}
\end{eqnarray}
which in our case implies that
\begin{eqnarray}
\delta(p^2-m^2)={1\over 2p^{0}}\,
\delta\left(p^{0}-\sqrt{\vec{p}^{\,2}+m^{2}}\right)
+{1\over 2p^{0}}\,
\delta\left(p^{0}+\sqrt{\vec{p}^{\,2}+m^{2}}\right).
\end{eqnarray}
The second term in the previous expression correspond to states with 
negative energy and therefore does not contribute to the integral. 
We can write then
\begin{eqnarray}
\int {d^{4}p\over (2\pi)^{4}}\,(2\pi)\delta(p^2-m^2)\,\theta(p^{0})\,f(p)=
\int {d^{3}p\over (2\pi)^{3}}{1\over 2\sqrt{\vec{p}^{\,2}+m^2}}
\,f\left(\sqrt{\vec{p}^{\,2}+m^2},\vec{p}\right).
\end{eqnarray}
Hence, the relativistic invariant measure is given by 
\begin{eqnarray}
\int {d^{3}p\over (2\pi)^{3}}{1\over 2\omega_{p}} \hspace*{1cm}
\mbox{with} \hspace*{1cm}
\omega_{p}\equiv \sqrt{\vec{p}^{\,2}+m^2}.
\end{eqnarray}

Once we have an invariant measure the next step is to find an
invariant normalization for the states. We work with a basis
$\{|p\rangle\}$ of eigenstates of the four-momentum operator
$\widehat{P}^{\mu}$
\begin{eqnarray}
\widehat{P}^{0}|p\rangle = \omega_{p}|p\rangle, \hspace*{2cm}
\widehat{{P}}^{i}|p\rangle = {p}^{\,\,i}|p\rangle .
\end{eqnarray}
Since the states $|p\rangle$ are eigenstates of the three-momentum operator
we can express them in terms of the non-relativistic states $|\vec{p}\rangle$
that we introduced in Eq. (\ref{eq:nonrelativistic_states})
\begin{eqnarray}
|p\rangle = N(\vec{p})|\vec{p}\rangle
\label{eq:normalization}
\end{eqnarray}
with $N(\vec{p})$ a normalization to be determined now. The states 
$\{|p\rangle \}$ form a complete basis, so they should satisfy the 
Lorentz invariant closure relation
\begin{eqnarray}
\int {d^{4}p\over (2\pi)^{4}}\,(2\pi)\delta(p^2-m^2)\,\theta(p^{0})\,
|p\rangle\,\langle p|=\mathbf{1}
\label{eq:closure_R}
\end{eqnarray}
At the same time, this closure relation can be expressed, using Eq. 
(\ref{eq:normalization}), in terms of the nonrelativistic basis of 
states $\{|\vec{p}\rangle\}$ as
\begin{eqnarray}
\int {d^{4}p\over (2\pi)^{4}}\,(2\pi)\delta(p^2-m^2)\,\theta(p^{0})\,
|p\rangle\,\langle p|=\int {d^{3}p\over (2\pi)^{3}}{1\over 2\omega_{p}}
|N(p)|^{2}\,|\vec{p}\rangle\,\langle\vec{p}|.
\end{eqnarray}
Using now Eq. (\ref{eq:closure_NR}) for the nonrelativistic states,
expression (\ref{eq:closure_R}) follows provided
\begin{eqnarray}
|N(\vec{p})|^{2}=(2\pi)^{3}\,(2\omega_{p}).
\end{eqnarray}
Taking the overall phase in Eq. (\ref{eq:normalization}) so that
$N(p)$ is real, we define the Lorentz invariant states $|p\rangle$ as
\begin{eqnarray}
|p\rangle = (2\pi)^{3\over 2}\,\sqrt{2\omega_{p}}\,|\vec{p}\rangle,
\end{eqnarray}
and given the normalization of $|\vec{p}\rangle$ we find the normalization 
of the relativistic states to be
\begin{eqnarray}
\langle p|p'\rangle=(2\pi)^{3}(2\omega_{p})\delta(\vec{p}-\vec{p}\, ').
\label{eq:rel_normalization}
\end{eqnarray}

Although not obvious at first sight, the previous normalization is 
Lorentz invariant. Although it is not difficult to show this in general, here
we  consider the simpler case of 1+1 dimensions where the two 
components $(p^{0},p^{1})$ of the on-shell momentum can be parametrized in 
terms of a single hyperbolic angle $\lambda$ as
\begin{eqnarray}
p^{0}=m \cosh\lambda, \hspace*{2cm} p^{1}=m\sinh\lambda.
\end{eqnarray}
Now, the combination $2\omega_{p}\delta(p^{1}-p^{1}{}')$ can be written as
\begin{eqnarray}
2\omega_{p}\delta(p^{1}-p^{1}{}')=2m\cosh\lambda\,\delta(m\sinh\lambda-
m\sinh\lambda')=2\delta(\lambda-\lambda'),
\label{eq:1+1}
\end{eqnarray}
where we have made use of the property (\ref{eq:delta_identity}) of
the $\delta$-function.  Lorentz transformations in $1+1$ dimensions
are labelled by a parameter $\xi\in\mathbb{R}$ and act on the momentum
by shifting the hyperbolic angle $\lambda\rightarrow
\lambda+\xi$. However, Eq. (\ref{eq:1+1}) is invariant under a common
shift of $\lambda$ and $\lambda'$, so the whole expression is
obviously invariant under Lorentz transformations.

To summarize what we did so far, we have succeed in constructing a
Lorentz covariant basis of states for the one-particle Hilbert space 
$\mathcal{H}_{1}$. The generators of the Poincar\'e group act on
the states $|p\rangle$ of the basis as
\begin{eqnarray}
\widehat{P}^{\mu}|p\rangle=p^{\mu}|p\rangle, \hspace*{2cm}
\mathcal{U}(\Lambda)|p\rangle=|\Lambda^{\mu}_{\,\,\,\nu}\,p^{\nu}\rangle 
\equiv |\Lambda p\rangle \hspace*{0.5cm}
\mbox{with} \hspace*{0.5cm} \Lambda\in {\rm SO}(1,3).
\end{eqnarray}
This is compatible with the Lorentz invariance of the normalization that 
we have checked above
\begin{eqnarray}
\langle p|p'\rangle = \langle p|\mathcal{U}(\Lambda)^{-1}
\mathcal{U}(\Lambda)|p'\rangle=
\langle \Lambda p|\Lambda p'\rangle.
\label{eq:transform_R}
\end{eqnarray}
On $\mathcal{H}_{1}$ the operator $\widehat{P}^{\mu}$ admits
the following spectral representation
\begin{eqnarray}
\widehat{P}^{\mu}=\int {d^{3}p\over (2\pi)^{3}}{1\over 2\omega_{p}}
|p\rangle\,p^{\mu}\,\langle p|\,.
\end{eqnarray}
Using (\ref{eq:transform_R}) and the fact that the measure is invariant
under Lorentz transformation, one can easily show that $\widehat{P}^{\mu}$ 
transform covariantly under SO$(1,3)$
\begin{eqnarray}
\mathcal{U}(\Lambda)^{-1}\widehat{P}^{\mu}\mathcal{U}(\Lambda)
=\int {d^{3}p\over (2\pi)^{3}}
{1\over 2\omega_{p}}|\Lambda^{-1}p\rangle\,p^{\mu}\,\langle \Lambda^{-1}p|
=\Lambda^{\mu}_{\,\,\,\nu}\widehat{P}^{\nu}.
\end{eqnarray}

A set of covariant creation-annihilation operators can be constructed
now in terms of the operators $a(\vec{p})$, $a^{\dagger}(\vec{p})$
introduced above
\begin{eqnarray}
\alpha(\vec{p})\equiv (2\pi)^{3\over 2}\sqrt{2\omega_{p}}a(\vec{p}), 
\hspace*{2cm}
\alpha^{\dagger}(\vec{p})\equiv (2\pi)^{3\over 2}\sqrt{2\omega_{p}}
a^{\dagger}(\vec{p})
\end{eqnarray}
with the Lorentz invariant commutation relations
\begin{eqnarray}
{}[\alpha(\vec{p}),\alpha^{\dagger}(\vec{p}\,')]&=&(2\pi)^{3}
(2\omega_{p})\delta(\vec{p}-\vec{p}\,'), \nonumber \\
{}[\alpha(\vec{p}),\alpha(\vec{p}\,')]
&=&[\alpha^{\dagger}(\vec{p}),\alpha^{\dagger}(\vec{p}\,')]=0.
\label{eq:Lorentz_inv_cr}
\end{eqnarray}
Particle states are created by acting with any number of
creation operators $\alpha(\vec{p})$ on the Poincar\'e invariant vacuum state
$|0\rangle$ satisfying
\begin{eqnarray}
\langle 0|0\rangle=1, \hspace*{1cm} \widehat{P}^{\mu}|0\rangle=0, 
\hspace*{1cm} 
\mathcal{U}(\Lambda)|0\rangle=|0\rangle, \hspace*{0.5cm} \forall \Lambda
\in {\rm SO}(1,3).
\end{eqnarray}
A general one-particle state $|f\rangle\in\mathcal{H}_{1}$ can be then written
as
\begin{eqnarray}
|f\rangle=\int {d^{3}p\over (2\pi)^{3}}{1\over 2\omega_{p}}f(\vec{p})
\alpha^{\dagger}(\vec{p})|0\rangle,
\end{eqnarray}
while a $n$-particle state $|f\rangle\in\mathcal{H}_{1}^{\otimes\,n}$ can
be expressed as
\begin{eqnarray}
|f\rangle=\int \prod_{i=1}^{n}{d^{3}p_{i}\over (2\pi)^{3}}
{1\over 2\omega_{p_{i}}}f(\vec{p}_{1},\ldots,\vec{p}_{n})
\alpha^{\dagger}(\vec{p}_{1})\ldots\alpha^{\dagger}(\vec{p}_{n})|0\rangle.
\end{eqnarray}
That this states are Lorentz invariant can be checked by noticing that
from the definition of the creation-annihilation operators follows the
transformation
\begin{eqnarray}
\mathcal{U}(\Lambda)\alpha(\vec{p})\mathcal{U}(\Lambda)^{\dagger}
=\alpha(\Lambda\vec{p})
\label{eq:transform_alphas} 
\end{eqnarray}
and the corresponding one for creation operators.

As we have argued above, the very fact that measurements have to be localized
implies the necessity of introducing quantum fields. Here we will consider the
simplest case of a scalar quantum field $\phi(x)$ satisfying the following
properties:

\begin{itemize}

\item[-]
{\bf Hermiticity.}
\begin{eqnarray}
\phi^{\dagger}(x)=\phi(x).
\end{eqnarray}

\item[-]
{\bf Microcausality.} Since measurements cannot interfere with each other
when performed in causally disconnected points of space-time, the commutator
of two fields have to vanish outside the relative ligth-cone
\begin{eqnarray}
[\phi(x),\phi(y)]=0, \hspace*{2cm} (x-y)^{2}<0.
\end{eqnarray}

\item[-]
{\bf Translation invariance.}
\begin{eqnarray}
e^{i\widehat{P}\cdot a}\phi(x)e^{-i\widehat{P}\cdot a}=
\phi(x-a).
\end{eqnarray}

\item[-]
{\bf Lorentz invariance.}
\begin{eqnarray}
\mathcal{U}(\Lambda)^{\dagger}\phi(x)\mathcal{U}(\Lambda)=\phi(\Lambda^{-1}x).
\end{eqnarray}

\item[-]
{\bf Linearity.}
To simplify matters we will also assume that  $\phi(x)$ is linear
in the creation-annihilation operators $\alpha(\vec{p})$, 
$\alpha^{\dagger}(\vec{p})$
\begin{eqnarray}
\phi(x)=\int {d^{3}p\over (2\pi)^{3}}{1\over 2\omega_{p}}
\left[f(\vec{p},x)\alpha(\vec{p})+
g(\vec{p},x)\alpha^{\dagger}(\vec{p})\right].
\end{eqnarray}
Since $\phi(x)$ should be hermitian we are forced to take
$f(\vec{p},x)^{*} =g(\vec{p},x)$. Moreover, $\phi(x)$ satisfies the
equations of motion of a free scalar field,
$(\partial_{\mu}\partial^{\mu}+m^{2})\phi(x)=0$, only if
$f(\vec{p},x)$ is a complete basis of solutions of the Klein-Gordon
equation. These considerations leads to the expansion
\begin{eqnarray}
\phi(x)=\int {d^{3}p\over (2\pi)^{3}}{1\over 2\omega_{p}}
\left[e^{-i\omega_{p}t+i\vec{p}\cdot\vec{x}}\alpha(\vec{p})+
e^{i\omega_{p}t-i\vec{p}\cdot\vec{x}}
\alpha^{\dagger}(\vec{p})\right].
\end{eqnarray}

\end{itemize}

Given the expansion of the scalar field in terms of the creation-annihilation
operators it can be checked that $\phi(x)$ and $\partial_{t}\phi(x)$
satisfy the equal-time canonical commutation relations
\begin{eqnarray}
[\phi(t,\vec{x}),\partial_{t}\phi(t,\vec{y})]=i\delta(\vec{x}-\vec{y})
\end{eqnarray}
The general commutator $[\phi(x),\phi(y)]$ can be also computed to be
\begin{eqnarray}
[\phi(x),\phi(x')]=i\Delta(x-x').
\end{eqnarray}
The function $\Delta(x-y)$ is given by
\begin{eqnarray}
i\Delta(x-y)&=&-{\rm Im\,\,}\int {d^{3}p\over (2\pi)^{3}}{1\over 2\omega_{p}}
e^{-i\omega_{p}(t-t')+i\vec{p}\cdot (\vec{x}-\vec{x}\,')}\nonumber \\
&=&\int {d^{4}p\over (2\pi)^{4}}(2\pi)\delta(p^{2}-m^2)
\varepsilon(p^{0})e^{-ip\cdot (x-x')},
\label{eq:propagator}
\end{eqnarray}
where $\varepsilon(x)$ is defined as
\begin{eqnarray}
\varepsilon(x)\equiv \theta(x)-\theta(-x)=\left\{
\begin{array}{rr}
1 & x>0 \\
-1 & x<0
\end{array}
\right. .
\end{eqnarray}

Using the last expression in Eq. (\ref{eq:propagator}) it is easy to 
show that $i\Delta(x-x')$ vanishes when $x$ and $x'$ are space-like 
separated. Indeed, if $(x-x')^{2}<0$ there is always a reference frame
in which both events are simultaneous, and since $i\Delta(x-x')$ is 
Lorentz invariant we can compute it in this reference frame. In this case
$t=t'$ and the exponential in the second line of (\ref{eq:propagator})
does not depend on $p^{0}$. Therefore, the integration over $k^{0}$
gives
\begin{eqnarray}
\int_{-\infty}^{\infty}dp^{0}\varepsilon(p^{0})\delta(p^{2}-m^{2})&=&
\int_{-\infty}^{\infty}dp^{0}\left[{1\over 2\omega_{p}}\varepsilon(p^{0})
\delta(p^{0}-\omega_{p})+{1\over 2\omega_{p}}\varepsilon(p^{0})
\delta(p^{0}+\omega_{p})\right] \nonumber \\
&=&{1\over 2\omega_{p}}-{1\over 2\omega_{p}}=0.
\end{eqnarray}
So we have concluded that $i\Delta(x-x')=0$ if $(x-x')^{2}<0$, as
required by microcausality.  Notice that the situation is completely
different when $(x-x')^{2}\geq 0$, since in this case the exponential
depends on $p^{0}$ and the integration over this component of the
momentum does not vanish.

\subsection{Canonical quantization}

So far we have contented ourselves with requiring a number of
properties to the quantum scalar field: existence of asymptotic
states, locality, microcausality and relativistic invariance. With
these only ingredients we have managed to go quite far. The previous
can also be obtained using canonical
quantization.  One starts with a classical free scalar field
theory in Hamiltonian formalism and obtains the quantum theory by
replacing Poisson brackets by commutators.  Since this quantization
procedure is based on the use of the canonical formalism, which gives
time a privileged r\^ole, it is important to check at the end of the
calculation that the resulting quantum theory is Lorentz invariant. In
the following we will briefly overview the canonical quantization of
the Klein-Gordon scalar field.

The starting point is the action functional $S[\phi(x)]$ which, in the
case of a free real scalar field of mass $m$ is given by
\begin{eqnarray}
S[\phi(x)]\equiv \int d^{4}x \,\mathcal{L}(\phi,\partial_{\mu}\phi)=
{1\over 2}\int d^{4}x \,\left(\partial_{\mu}\phi\partial^{\mu}\phi-
{m^{2}}\phi^2\right).
\end{eqnarray}
The equations of motion are obtained, as usual, from the Euler-Lagrange
equations
\begin{eqnarray}
\partial_{\mu}\left[\partial\mathcal{L}\over \partial(\partial_{\mu}\phi)
\right]-{\partial\mathcal{L}\over \partial\phi}=0 \hspace*{1cm}
\Longrightarrow \hspace*{1cm} (\partial_{\mu}\partial^{\mu}+m^{2})\phi=0.
\label{eq:eomKG}
\end{eqnarray}

The momentum canonically conjugated to the field $\phi(x)$ is given by
\begin{eqnarray}
\pi(x)\equiv {\partial\mathcal{L}\over \partial(\partial_{0}\phi)}
={\partial\phi\over\partial t}.
\end{eqnarray}
In the Hamiltonian formalism the physical system is described not in terms
of the generalized coordinates and their time derivatives but in terms
of the generalized coordinates and their canonically conjugated momenta. 
This is achieved by a Legendre transformation after which the dynamics of
the system is determined by the Hamiltonian function
\begin{eqnarray}
H\equiv \int d^{3}x \left(\pi{\partial\phi\over\partial t}-\mathcal{L}\right) 
= {1\over 2}\int d^{3}x\left[
\pi^2+(\vec{\nabla}\phi)^{2}+m^{2}\right].
\end{eqnarray}

The equations of motion can be written in terms of the Poisson
rackets. Given two functional $A[\phi,\pi]$, $B[\phi,\pi]$ of the
canonical variables
\begin{eqnarray}
A[\phi,\pi]=\int d^{3}x \mathcal{A}(\phi,\pi), \hspace*{1cm}
B[\phi,\pi]=\int d^{3}x \mathcal{B}(\phi,\pi).
\end{eqnarray}
Their Poisson bracket is defined by
\begin{eqnarray}
\{A,B\}\equiv \int d^{3}x\left[{\delta {A}\over \delta \phi}
{\delta{B}\over \delta\pi}-
{\delta{A}\over \delta\pi}{\delta{B}\over \delta\phi}
\right],
\end{eqnarray}
where ${\delta\over \delta \phi}$ denotes the functional derivative 
defined as
\begin{eqnarray}
{\delta A\over \delta\phi}\equiv {\partial\mathcal{A}\over 
\partial\phi}-\partial_{\mu}\left[{\partial\mathcal{A}
\over \partial(\partial_{\mu}\phi)}\right]
\end{eqnarray}
Then, the canonically conjugated fields satisfy the
following equal time Poisson brackets
\begin{eqnarray}
\{\phi(t,\vec{x}),\phi(t,\vec{x}\,')\}&=&\{\pi(t,\vec{x}),
\pi(t,\vec{x}\,')\}=0,\nonumber \\
\{\phi(t,\vec{x}),\pi(t,\vec{x}\,')\}&=&\delta(\vec{x}-\vec{x}\,').
\label{eq:etccr}
\end{eqnarray}

Canonical quantization proceeds now by replacing classical fields with
operators and Poisson brackets with commutators according to the rule
\begin{eqnarray}
i\{\cdot,\cdot\} \longrightarrow [\cdot,\cdot].
\end{eqnarray}
In the case of the scalar field, a general solution of the field equations
(\ref{eq:eomKG}) can be obtained by working with the Fourier
transform
\begin{eqnarray}
(\partial_{\mu}\partial^{\mu}+m^{2})\phi(x)=0 \hspace*{1cm}
\Longrightarrow \hspace*{1cm} 
(-p^{2}+m^{2})\widetilde{\phi}(p)=0,
\end{eqnarray}
whose general solution can be written as\footnote{In momentum space,
the general solution to this equation is
$\widetilde{\phi}(p)=f(p)\delta(p^{2}-m^{2})$, with $f(p)$ a completely 
general function of $p^{\mu}$. The solution in position space is obtained 
by inverse Fourier transform.}
\begin{eqnarray}
\phi(x)&=&\int {d^{4}p\over (2\pi)^{4}}(2\pi)\delta(p^{2}-m^{2})\theta(p^{0})
\left[\alpha(p)e^{-ip\cdot x}+\alpha(p)^{*}e^{ip\cdot x}\right] \nonumber \\
&=& \int {d^{3}p\over (2\pi)^{3}}{1\over 2\omega_{p}}
\left[\alpha(\vec{p}\,)e^{-i\omega_{p}t + \vec{p}\cdot \vec{x}}
+\alpha(\vec{p}\,)^{*}e^{i\omega_{p}t-\vec{p}\cdot \vec{x}}\right]
\label{eq:general_sol_phi}
\end{eqnarray}
and we have required $\phi(x)$ to be real. The conjugate momentum is
\begin{eqnarray}
\pi(x)= -{i\over 2}\int {d^{3}p\over (2\pi)^{3}}
\left[\alpha(\vec{p}\,)e^{-i\omega_{p}t + \vec{p}\cdot \vec{x}}
+\alpha(\vec{p}\,)^{*}e^{i\omega_{p}t-\vec{p}\cdot \vec{x}}\right].
\end{eqnarray}

Now $\phi(x)$ and $\pi(x)$ are promoted to operators by replacing the 
functions $\alpha(\vec{p})$, $\alpha(\vec{p})^{*}$ by the corresponding
operators
\begin{eqnarray}
\alpha(\vec{p}\,)\longrightarrow \widehat{\alpha}(\vec{p}\,), \hspace*{2cm}
\alpha(\vec{p}\,)^{*}\longrightarrow \widehat{\alpha}^{\dagger}(\vec{p}\,).
\end{eqnarray}
Moreover, demanding $[\phi(t,\vec{x}),\pi(t,\vec{x}\,')]=
i\delta(\vec{x}-\vec{x}\,')$ forces the operators
$\widehat{\alpha}(\vec{p})$, $\widehat{\alpha}(\vec{p})^{\dagger}$ to
have the commutation relations found in
Eq. (\ref{eq:Lorentz_inv_cr}). Therefore they are identified as a set
of creation-annihilation operators creating states with well-defined
momentum $\vec{p}$ out of the vacuum $|0\rangle$. In the
canonical quantization formalism the concept of particle appears as a
result of the quantization of a classical field.

Knowing the expressions of $\widehat{\phi}$ and $\widehat{\pi}$ 
in terms of the creation-annihilation operators we can proceed to evaluate
the Hamiltonian operator. After a simple calculation one arrives to the  
expression 
\begin{eqnarray} 
\widehat{H}=\int d^{3}p\, \left[ 
\omega_{p}\widehat{\alpha}^{\dagger}(\vec{p})\widehat{\alpha}(\vec{p}) 
+{1\over 2}\omega_{p}\,\delta(\vec{0})\right] .
\label{eq:hamiltonianKG}  
\end{eqnarray}
The first term has a simple physical interpretation since
$\widehat{\alpha}^{\dagger}(\vec{p})\widehat{\alpha}(\vec{p})$ is the
number operator of particles with momentum $\vec{p}$. The second
divergent term can be eliminated if we defined the
normal-ordered Hamiltonian $:\!\!\widehat{H}\!\!:$ with the vacuum 
energy subtracted
\begin{eqnarray}
:\!\!\widehat{H}\!\!:\equiv \widehat{H}-\langle 0|\widehat{H}|0\rangle =\int
d^{3}p\,\omega_{p}\,
\widehat{\alpha}^{\dagger}(\vec{p}\,)\,\widehat{\alpha}(\vec{p}\,)
\end{eqnarray}

It is interesting to try to make sense of the divergent
term in Eq.  (\ref{eq:hamiltonianKG}). This term have two sources of
divergence.  One is associated with the delta function evaluated at
zero coming from the fact that we are working in a
infinite volume. It can be regularized for large but finite volume by
replacing $\delta(\vec{0})\sim V$. Hence, it is of infrared origin. 
The second one comes from the integration of $\omega_{p}$ at large values of 
the momentum and it is then an ultraviolet divergence. The infrared 
divergence can be regularized by considering the scalar field to be living
in a box of finite volume $V$. In this case the vacuum energy is 
\begin{eqnarray}
E_{\rm vac}\equiv \langle 0|\widehat{H}|0\rangle = \sum_{\vec{p}}
{1\over 2}\omega_{p}.
\end{eqnarray}
Written in this way the interpretation of the vacuum energy is
straightforward. A free scalar quantum field can be seen as a infinite
collection of harmonic oscillators per unit volume, each one labelled
by $\vec{p}$.  Even if those oscillators are not excited, they
contribute to the vacuum energy with their zero-point energy, given by
${1\over 2}\omega_{p}$.  This vacuum contribution to the energy add up
to infinity even if we work at finite volume, since even then there
are modes with arbitrary high momentum contributing to the sum,
$p_{i}={n_{i}\pi \over L_{i}}$, with $L_{i}$ the sides of the box of
volume $V$ and $n_{i}$ an integer. Hence, this divergence is of
ultraviolet origin.

\subsection{The Casimir effect}

\begin{figure}
\centerline{\epsfxsize=4.5truein\epsfbox{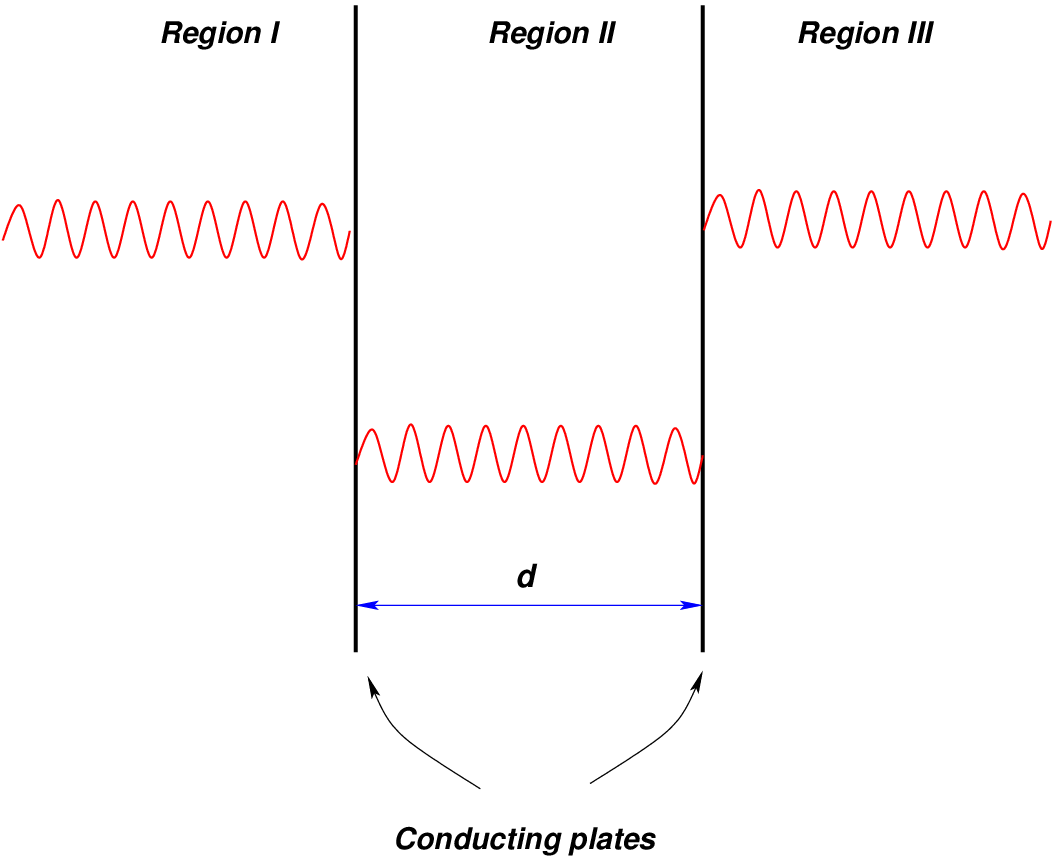}}
\caption[]{Illustration of the Casimir effect. In regions I and II 
the spetrum of modes of the momentum $p_{\perp}$ is continuous, while
in the space between the plates (region II) it is quantized in units
of $\pi\over d$.}
\label{fig:casimir}
\end{figure}

The presence of a vacuum energy is not characteristic of the scalar
field.  It is also present in other cases, in particular in quantum
electrodynamics.  Although one might be tempted to discarding this
infinite contribution to the energy of the vacuum as unphysical, it
has observable consequences. In 1948 Hendrik Casimir pointed out
\cite{casimir} that although a formally divergent vacuum energy would
not be observable, any variation in this energy would be (see
\cite{casimir-reviews} for comprehensive reviews).

To show this he devised the following experiment. Consider a couple of
infinite, perfectly conducting plates placed parallel to each other at
a distance $d$ (see Fig. \ref{fig:casimir}). Because the conducting
plates fix the boundary condition of the vacuum modes of the
electromagnetic field these are discrete in between the plates (region
II), while outside there is a continuous spectrum of modes (regions I
and III). In order to calculate the force between the plates we can
take the vacuum energy of the electromagnetic field as given by the
contribution of two scalar fields corresponding to the two
polarizations of the photon. Therefore we can use the formulas derived
above.

A naive calculation of the vacuum energy in this system gives a
divergent result. This infinity can be removed, however, by
substracting the vacuum energy corresponding to the situation where
the plates are removed
\begin{eqnarray}
E(d)_{\rm reg}=E(d)_{\rm vac}-E(\infty)_{\rm vac}
\end{eqnarray}
This substraction cancels the contribution of the modes outside the 
plates. Because of the boundary conditions imposed by the plates
the momentum of the modes perpendicular to the plates are quantized
according to $p_{\perp}={n\pi\over d}$, with $n$ a non-negative integer.
If we consider that the size of the plates is much larger than their
separation $d$ we can take the momenta parallel to the plates
$\vec{p}_{\parallel}$ as continuous. For $n>0$ we have two polarizations
for each vacuum mode of the electromagnetic field, each contributing like
${1\over 2}\sqrt{\vec{p}_{\parallel}^{\,2}+p_{\perp}^{2}}$ to the 
vacuum energy. On the other hand, when $p_{\perp}=0$ the corresponding 
modes of the field are effectively (2+1)-dimensional and therefore
there is only one polarization. Keeping this in mind, we can write
\begin{eqnarray}
E(d)_{\rm reg}&=&S\int {d^{2}p_{\parallel}\over (2\pi)^{2}}
{1\over 2}
|\vec{p}_{\parallel}|+
2S\int {d^{2}p_{\parallel}\over (2\pi)^{2}}\sum_{n=1}^{\infty}
{1\over 2}
\sqrt{\vec{p}_{\parallel}^{\,2}+\left({n\pi\over d}\right)^{2}}\nonumber \\
&-&
2Sd \int {d^{3}p\over (2\pi)^{3}}{1\over 2}|\vec{p}\,|
\end{eqnarray}
where $S$ is the area of the plates. The factors of 2 take into account
the two propagating degrees of freedom of the electromagnetic field, as
discussed above. In order to ensure the convergence of
integrals and infinite sums we
can introduce an exponential damping factor\footnote{Actually, one could
introduce any cutoff function $f(p_{\perp}^{2}+p_{\parallel}^{2})$ 
going to zero fast enough as $p_{\perp}$, $p_{\parallel}\rightarrow \infty$.
The result is independent of the particular function used in the
calculation.}
\begin{eqnarray}
E(d)_{\rm reg}&=&
{1\over 2}S
\int {d^{2}p_{\perp}\over (2\pi)^{2}}e^{-{1\over \Lambda}
|\vec{p}_{\parallel}\,|}|\vec{p}_{\parallel}\,|+
S\sum_{n=1}^{\infty}
\int {d^{2}p_{\parallel}\over (2\pi)^{2}}e^{-{1\over \Lambda}
\sqrt{\vec{p}_{\parallel}^{\,2}+\left({n\pi\over d}\right)^{2}}}
\sqrt{\vec{p}_{\parallel}^{\,2}+\left({n\pi\over d}\right)^{2}}
\nonumber \\
&-& Sd\int_{-\infty}^{\infty} {dp_{\perp}\over
2\pi}\int {d^{2}p_{\parallel}\over
(2\pi)^{2}}e^{-{1\over \Lambda}\sqrt{\vec{p}_{\parallel}^{\,2}+p_{\perp}^{2}}}
\sqrt{\vec{p}_{\parallel}^{\,2}+p_{\perp}^{2}}
\end{eqnarray}
where $\Lambda$ is an ultraviolet cutoff.
It is now straightforward to see that if we define the function
\begin{eqnarray}
F(x)={1\over 2\pi}\int_{0}^{\infty}y\, dy\, e^{-{1\over \Lambda}
\sqrt{y^{2}+\left({x\pi\over d}\right)^{2}}}
\sqrt{y^{2}+\left({x\pi\over d}\right)^{2}}={1\over 4\pi}
\int_{\left({x\pi\over d}\right)^{2}}^{\infty}
dz\,e^{-{\sqrt{z}\over \Lambda}}\sqrt{z}
\end{eqnarray}
the regularized vacuum energy can be written as
\begin{eqnarray}
E(d)_{\rm reg}=S\left[{1\over 2}F(0)+\sum_{n=1}^{\infty}F(n)-
\int_{0}^{\infty}dx\,F(x)\right]
\end{eqnarray}
This expression can be evaluated using the Euler-MacLaurin formula
\cite{abramowitz}
\begin{eqnarray}
\sum_{n=1}^{\infty}F(n)-\int_{0}^{\infty} dx\,F(x)&=&
-{1\over 2}\left[F(0)+F(\infty)\right]+{1\over 12}\left[
F'(\infty)-F'(0)\right]\nonumber \\
&-&{1\over 720}\left[F'''(\infty)-F'''(0)\right]+\ldots
\end{eqnarray}
Since for our function $F(\infty)=F'(\infty)=F'''(\infty)=0$ and
$F'(0)=0$, the value of $E(d)_{\rm reg}$ is determined by $F'''(0)$.
Computing this term and removing the ultraviolet cutoff, $\Lambda
\rightarrow\infty$ we find the result
\begin{eqnarray}
E(d)_{\rm reg}={S\over 720}F'''(0)=-{\pi^{2}S\over 720 d^{3}}.
\end{eqnarray}
Then, the force per unit area between the plates is given by
\begin{eqnarray}
P_{\rm Casimir}=-{\pi^{2}\over 240}{1\over d^{4}}.
\end{eqnarray}
The minus sign shows that the force between the plates is
attractive. This is the so-called Casimir effect. It was
experimentally measured in 1958 by Sparnaay \cite{sparnaay} and 
since then the Casimir effect has been checked with better and
better precission in a variety of situations \cite{casimir-reviews}.

\section{Theories and Lagrangians}
\label{sec:theories&lagrangians}

Up to this point we have used a scalar field to illustrate our discussion
of the quantization procedure. However, nature is richer than
that and it is necessary to consider other fields with more complicated
behavior under Lorentz transformations.  Before considering other fields
we pause and study the properties of the Lorentz group.

\subsection{Representations of the Lorentz group}
\label{sec:lorentz}

In four dimensions the Lorentz group has six generators. Three of them 
correspond to the generators of the group of rotations in 
three dimensions SO(3). In terms of the generators $J_{i}$ of the
group a finite 
rotation of angle $\varphi$ with respect to an axis determined by 
a unitary vector $\vec{e}$ can be written as
\begin{eqnarray}
R(\vec{e},\varphi)=e^{-i\varphi\,\vec{e}\cdot\vec{J}}, 
\hspace*{1cm} \vec{J}=\left(
\begin{array}{r}
J_{1} \\
J_{2} \\
J_{3}
\end{array}\right).
\end{eqnarray}
The other three generators of the Lorentz group are associated with
boosts $M_{i}$ along the three spatial directions. A boost
with rapidity $\lambda$ along a direction $\vec{u}$ is given by
\begin{eqnarray}
B(\vec{u},\lambda)=e^{-i\lambda\,\vec{u}\cdot\vec{M}}, 
\hspace*{1cm} \vec{M}=\left(
\begin{array}{r}
M_{1} \\
M_{2} \\
M_{3}
\end{array}\right).
\end{eqnarray}
These six generators satisfy the algebra
\begin{eqnarray}
[J_{i},J_{j}] &=& i\epsilon_{ijk}J_{k}, \nonumber \\
{}[J_{i},M_{k}] &=& i\epsilon_{ijk}M_{k}, 
\label{eq:lorentz}\\
{}[M_{i},M_{j}] &=& -i\epsilon_{ijk}J_{k}, \nonumber 
\end{eqnarray}
The first line corresponds to the commutation relations of SO(3), while 
the second one implies that the generators of the boosts transform like
a vector under rotations. 

At first sight, to find representations of the algebra (\ref{eq:lorentz}) 
might seem difficult. The problem is greatly simplified
if we consider the following combination of the generators
\begin{eqnarray}
J_{k}^{\pm}={1\over 2}(J_{k}\pm iM_{k}).
\label{eq:jpm}
\end{eqnarray}
Using (\ref{eq:lorentz}) it is easy to prove that the new generators
$J_{k}^{\pm}$ satisfy the algebra
\begin{eqnarray}
[J_{i}^{\pm},J_{j}^{\pm}]&=&i\epsilon_{ijk}J_{k}^{\pm}, \nonumber \\
{}[J_{i}^{+},J_{j}^{-}] &=& 0.
\end{eqnarray}
Then the Lorentz algebra (\ref{eq:lorentz}) is actually equivalent to
two copies of the algebra of ${\rm SU(2)}\approx {\rm
SO}(3)$. Therefore the irreducible representations of the Lorentz
group can be obtained from the well-known representations of
SU(2). Since the latter ones are labelled by the spin
$\mathbf{s}=k+{1\over 2},k$ (with $k\in \mathbb{N}$), any representation of
the Lorentz algebra can be identified by specifying
$\mathbf{(s_{+},s_{-})}$, the spins of the representations of the two
copies of SU(2) that made up the algebra (\ref{eq:lorentz}). 

To get familiar with this way of labelling the representations of the
Lorentz group we study some particular examples. Let us start
with the simplest one $\mathbf{(s_{+},s_{-})=(0,0)}$. This state is
a singlet under $J^{\pm}_{i}$ and therefore also under rotations and boosts.
Therefore we have a scalar.

The next interesting cases are $\mathbf{({1\over 2},0)}$ and 
$\mathbf{(0,{1\over 2})}$. They correspond respectively to a right-handed 
and a left-handed Weyl spinor. Their properties will be studied in more
detail below. In the case of $\mathbf{({1\over 2},{1\over 2})}$, since
from Eq. (\ref{eq:jpm}) we see that $J_{i}=J_{i}^{+}+J_{i}^{-}$ the
rules of addition of angular momentum tell us that there
are two states, one of them transforming as a vector and another one as a
scalar under three-dimensional rotations. Actually, a more detailed
analysis shows that the singlet state corresponds to the time component of
a vector and the states combine to form a vector under the Lorentz group.

There are also more ``exotic'' representations. For example we can
consider the $\mathbf{(1,0)}$ and $\mathbf{(0,1)}$ representations
corresponding respectively to a selfdual and an anti-selfdual rank-two
antisymmetric tensor. In Table \ref{table:1} we summarize the previous
discussion.

\begin{table}
\begin{center}
\begin{tabular}{|c|l|}
\hline 
{\bf Representation} & {\bf Type of field} \\
\hline
 & \\
$\mathbf{(0,0)}$ & {\rm Scalar} \\
 & \\
$\mathbf{({1\over 2},0)}$ & Right-handed spinor \\
 & \\
$\mathbf{(0,{1\over 2})}$ & Left-handed spinor \\
 & \\
$\mathbf{({1\over 2},{1\over 2})}$ & Vector \\
 & \\
$\mathbf{(1,0)}$ & Selfdual antisymmetric 2-tensor \\
 & \\
$\mathbf{(0,1)}$ & Anti-selfdual antisymmetric 2-tensor \\
 & \\
\hline
\end{tabular}
\caption{Representations of the Lorentz group}
\label{table:1}
\end{center}
\end{table}

To conclude our discussion of the representations of the Lorentz group
we notice that under a parity transformation the generators of SO(1,3)
transform as
\begin{eqnarray}
P:J_{i}\longrightarrow J_{i}, \hspace*{1cm} 
P:M_{i}\longrightarrow -M_{i}
\end{eqnarray} 
this means that $P:J_{i}^{\pm}\longrightarrow J^{\mp}_{i}$ and therefore
a representation $\mathbf{(s_{1},s_{2})}$ is transformed into
$\mathbf{(s_{2},s_{1})}$. This means that, for example, a vector 
$\mathbf{({1\over 2},{1\over 2})}$ is invariant under parity, whereas 
a left-handed Weyl spinor $\mathbf{({1\over 2},0)}$ transforms into a
right-handed one $\mathbf{(0,{1\over 2})}$ and vice versa.

\subsection{Spinors}

{\bf Weyl spinors.}
Let us go back to the two spinor representations of the Lorentz group, namely
$\mathbf{({1\over 2},0)}$ and $\mathbf{(0,{1\over 2})}$. These representations
can be explicitly constructed using the Pauli matrices as
\begin{eqnarray}
J^{+}_{i}&=&{1\over 2}\sigma^{i}, 
\hspace*{1.1cm} J^{-}_{i}=0 \hspace*{1.9cm} 
\mbox{for  \hspace*{0.5cm} $\mathbf{({1\over 2},0)}$},
\nonumber 
\\
J^{+}_{i}&=& 0 , 
\hspace*{1.5cm} J^{-}_{i}={1\over 2}\sigma^{i} \hspace*{1.5cm} 
\mbox{for \hspace*{0.5cm} $\mathbf{(0,{1\over 2})}$}.
\end{eqnarray}
We denote by $u_{\pm}$ a complex two-component object that
transforms in the representation $\mathbf{s}_{\pm}={1\over 2}$ of
$J_{\pm}^{i}$.  If we define
$\sigma^{\mu}_{\pm}=(\mathbf{1},\pm\sigma^{i})$ we can construct the
following vector quantities
\begin{eqnarray}
u^{\dagger}_{+}\sigma^{\mu}_{+}u_{+}, \hspace*{2cm}
u^{\dagger}_{-}\sigma^{\mu}_{-}u_{-}.
\label{eq:u's}
\end{eqnarray}
Notice that since $(J_{i}^{\pm})^{\dagger}=J_{i}^{\mp}$ the hermitian
conjugated fields $u_{\pm}^{\dagger}$ are in the $\mathbf{(0,{1\over
2})}$ and $\mathbf{({1\over 2},0)}$ respectively. 

To construct a free Lagrangian for the fields $u_{\pm}$ we have
to look for quadratic combinations of the fields that are Lorentz scalars.
If we also demand invariance under global phase rotations
\begin{eqnarray}
u_{\pm}\longrightarrow e^{i\theta}u_{\pm}
\end{eqnarray}
we are left with just one possibility up to a sign
\begin{eqnarray}
\mathcal{L}^{\pm}_{\rm Weyl}=iu^{\dagger}_{\pm}\left(\partial_{t}\pm 
\vec{\sigma}\cdot\vec{\nabla}\right)u_{\pm} 
= iu^{\dagger}_{\pm}\sigma_{\pm}^{\mu}\partial_{\mu} u_{\pm}.
\label{eq:weyl_lag}
\end{eqnarray}
This is the Weyl Lagrangian. In order to grasp the physical meaning of
the spinors $u_{\pm}$ we write the equations of motion
\begin{eqnarray}
\left(\partial_{0}\pm\vec{\sigma}\cdot\vec{\nabla}\right)u_{\pm}=0.
\label{eq:weyleom}
\end{eqnarray}
Multiplying this equation on the left by $\left(\partial_{0}\mp
\vec{\sigma}\cdot\vec{\nabla}\right)$ and applying the algebraic properties
of the Pauli matrices we conclude that $u_{\pm}$ satisfies the massless
Klein-Gordon equation
\begin{eqnarray}
\partial_{\mu}\partial^{\mu}\,u_{\pm}=0,
\end{eqnarray}
whose solutions are:
\begin{eqnarray}
u_{\pm}(x)=u_{\pm}(k)e^{-ik\cdot x}, \hspace*{1cm} \mbox{with}
\hspace*{0.5cm} k^{0}=|\vec{k}|.
\end{eqnarray}
Plugging these solutions back into the equations of motion (\ref{eq:weyleom})
we find
\begin{eqnarray}
\left(|\vec{k}|\mp\vec{k}\cdot\vec{\sigma}\right)u_{\pm}=0,
\end{eqnarray}
which implies
\begin{eqnarray}
u_{+}: & \hspace*{1cm} & {\vec{\sigma}\cdot\vec{k}\over |\vec{k}|}=1,  
\nonumber \\
u_{-}: &  \hspace*{1cm}& {\vec{\sigma}\cdot\vec{k}\over |\vec{k}|}=-1. 
\label{eq:helicity}
\end{eqnarray}
Since the spin operator is defined as $\vec{s}={1\over
2}\vec{\sigma}$, the previous expressions give the chirality of the
states with wave function $u_{\pm}$, i.e. the projection of spin along
the momentum of the particle. Therefore we conclude that $u_{+}$ is a
Weyl spinor of positive helicity $\lambda={1\over 2}$, while $u_{-}$
has negative helicity $\lambda=-{1\over 2}$. This agrees with our
assertion that the representation $\mathbf{({1\over 2},0)}$
corresponds to a right-handed Weyl fermion (positive chirality)
whereas $\mathbf{(0,{1\over 2})}$ is a left-handed Weyl fermion
(negative chirality). For example, in the Standard Model neutrinos are
left-handed Weyl spinors and therefore transform in the representation
$\mathbf{(0,{1\over 2})}$ of the Lorentz group.

Nevertheless, it is possible that we were too restrictive in
constructing the Weyl Lagrangian (\ref{eq:weyl_lag}). There we
constructed the invariants from the vector bilinears (\ref{eq:u's})
corresponding to the product representations
\begin{eqnarray}
\mbox{$\mathbf{({1\over 2},{1\over 2})}
=\mathbf{({1\over 2},0)}\otimes\mathbf{(0,{1\over 2})}$}
\hspace*{0.5cm} \mbox{and} \hspace*{0.5cm}
\mbox{$({1\over 2},{1\over 2})
=\mathbf{(0,{1\over 2})}\otimes\mathbf{({1\over 2},0)}$}.
\end{eqnarray}
In particular our insistence in demanding the Lagrangian to be
invariant under the global symmetry $u_{\pm}\rightarrow
e^{i\theta}u_{\pm}$ rules out the scalar term that appears in the
product representations
\begin{eqnarray}
\mbox{$\mathbf{({1\over 2},0)}\otimes\mathbf{({1\over 2},0)}=\mathbf{(1,0)}
\oplus\mathbf{(0,0)}$}, \hspace*{2cm}
\mbox{$\mathbf{(0,{1\over 2})}\otimes\mathbf{(0,{1\over 2})}=\mathbf{(0,1)}
\oplus\mathbf{(0,0)}$}.
\end{eqnarray}
The singlet representations corresponds to the antisymmetric combinations
\begin{eqnarray}
\epsilon_{ab}u_{\pm}^{a}u_{\pm}^{b},
\label{eq:majorana1}
\end{eqnarray}
where $\epsilon_{ab}$ is the antisymmetric symbol $\epsilon_{12}=
-\epsilon_{21}=1$. 

At first sight it might seem that the term (\ref{eq:majorana1}) vanishes
identically because of the antisymmetry of the $\epsilon$-symbol. 
However we should keep in mind that the spin-statistic theorem (more on 
this later) demands that fields with half-integer spin have to satisfy
the Fermi-Dirac statistics and therefore satisfy anticommutation relations,
whereas fields of integer spin follow the statistic of Bose-Einstein and,
as a consequence, quantization replaces Poisson brackets by commutators.
This implies that the components of the Weyl fermions $u_{\pm}$ 
are anticommuting Grassmann fields
\begin{eqnarray}
u_{\pm}^{a}u_{\pm}^{b}+u_{\pm}^{b}u_{\pm}^{a}=0.
\end{eqnarray}
It is important to realize that, strictly speaking, fermions (i.e., 
objects that satisfy the Fermi-Dirac statistics)
do not exist classically. The reason is that they satisfy the Pauli 
exclusion principle and therefore each quantum state can be occupied,
at most, by one fermion. Therefore the na\"\i ve definition of 
the classical limit as a limit of large occupation numbers cannot be 
applied.  Fermion field  do not really make sense classically.

Since the combination (\ref{eq:majorana1}) does not vanish and we can
construct a new Lagrangian
\begin{eqnarray}
\mathcal{L}^{\pm}_{\rm Weyl}=
iu^{\dagger}_{\pm}\sigma_{\pm}^{\mu}\partial_{\mu} u_{\pm}
-{m\over 2}\epsilon_{ab}u_{\pm}^{a}u_{\pm}^{b}+\mbox{h.c.}
\end{eqnarray}
This mass term, called of Majorana type, is allowed if we do not
worry about breaking the global U(1) symmetry
$u_{\pm}\rightarrow e^{i\theta}u_{\pm}$. This is not the case, for
example, of charged chiral fermions, since the Majorana mass violates
the conservation of electric charge or any other gauge U(1) charge. In the
Standard Model, however, there is no such a problem if we introduce
Majorana masses for right-handed neutrinos, since they are singlet under
all standard model gauge groups. Such a term will break, however, 
the global U(1) lepton number charge because the operator $\epsilon_{ab}
\nu_{R}^{a}\nu_{R}^{b}$ changes the lepton number by two units

{\bf Dirac spinors.}
We have seen that parity interchanges the representations $\mathbf{({1\over
2},0)}$ and $\mathbf{(0,{1\over 2})}$, i.e. it changes right-handed
with left-handed fermions
\begin{eqnarray}
P:u_{\pm}\longrightarrow u_{\mp}.
\end{eqnarray}
An obvious way to build a parity invariant theory is to introduce a pair
or Weyl fermions $u_{+}$ and $u_{+}$. Actually, these two fields can be 
combined in a single four-component spinor
\begin{eqnarray}
\psi=\left(
\begin{array}{c}
u_{+} \\
u_{-}
\end{array}\right)
\end{eqnarray}
transforming in the reducible representation $\mathbf{({1\over 2},0)}
\oplus \mathbf{(0,{1\over 2})}$. 

Since now we have both $u_{+}$ and $u_{-}$ simultaneously at our disposal
the equations of motion for $u_{\pm}$, $i\sigma_{\pm}^{\mu}\partial_{\mu}
u_{\pm}=0$ can be modified, while keeping them linear, to
\begin{eqnarray}
\left.
\begin{array}{l}
i\sigma_{+}^{\mu}\partial_{\mu}u_{+}=mu_{-} \\
\\
i\sigma_{-}^{\mu}\partial_{\mu}u_{-}=mu_{+}
\end{array}
\right\} \hspace*{0.5cm} \Longrightarrow \hspace*{0.5cm} i\left(
\begin{array}{cc}
\sigma_{+}^{\mu} & 0 \\
0 & \sigma_{-}^{\mu}
\end{array}
\right)\partial_{\mu}\psi=m\left(
\begin{array}{cc}
0 & \mathbf{1} \\
\mathbf{1} & 0
\end{array}
\right)\psi.
\end{eqnarray}
These equations of motion can be derived from the Lagrangian 
density
\begin{eqnarray}
\mathcal{L}_{\rm Dirac}=i\psi^{\dagger}\left(
\begin{array}{cc}
\sigma_{+}^{\mu} & 0 \\
0 & \sigma_{-}^{\mu}
\end{array}
\right)\partial_{\mu}\psi-m\psi^{\dagger}
\left(
\begin{array}{cc}
0 & \mathbf{1} \\
\mathbf{1} & 0
\end{array}
\right)
\psi.
\label{eq:dirac0}
\end{eqnarray}
To simplify the notation it is useful to define the Dirac $\gamma$-matrices as
\begin{eqnarray}
\gamma^{\mu}=\left(
\begin{array}{cc}
0 & \sigma_{-}^{\mu}  \\
\sigma_{+}^{\mu} & 0
\end{array}
\right)
\label{eq:gamma_matrices}
\end{eqnarray}
and the Dirac conjugate spinor $\overline{\psi}$
\begin{eqnarray}
\overline{\psi}\equiv \psi^{\dagger}\gamma^{0}=\psi^{\dagger}
\left(
\begin{array}{cc}
0 & \mathbf{1} \\
\mathbf{1} & 0
\end{array}
\right).
\end{eqnarray}
Now the Lagrangian (\ref{eq:dirac0}) can be written in the more compact form
\begin{eqnarray}
\mathcal{L}_{\rm Dirac}=\overline{\psi}\left(i\gamma^{\mu}\partial_{\mu}
-m\right)\psi.
\label{eq:dirac_eq}
\end{eqnarray}
The associated equations of motion give the Dirac equation (\ref{eq:dirac})
with the identifications
\begin{eqnarray}
\gamma^{0}=\beta, \hspace*{1cm} \gamma^{i}=i\alpha^{i}.
\end{eqnarray}

In addition, the $\gamma$-matrices defined in
(\ref{eq:gamma_matrices}) satisfy the Clifford algebra
\begin{eqnarray}
\{\gamma^{\mu},\gamma^{\nu}\}=2\eta^{\mu\nu}.
\label{eq:gamma}
\end{eqnarray}
In $D$ dimensions this algebra admits representations of dimension
$2^{[{D\over 2}]}$. When $D$ is even the Dirac fermions $\psi$
transform in a reducible representation of the Lorentz group.  In the
case of interest, $D=4$ this is easy to prove by
defining the matrix
\begin{eqnarray}
\gamma^{5}=-i\gamma^{0}\gamma^{1}\gamma^{2}\gamma^{3}=
\left(
\begin{array}{cc}
\mathbf{1} & 0 \\
0 & -\mathbf{1}
\end{array}
\right).
\label{eq:gamma5}
\end{eqnarray}
We see that 
$\gamma^{5}$ anticommutes with all other $\gamma$-matrices. This implies
that 
\begin{eqnarray}
[\gamma^{5},\sigma^{\mu\nu}]=0, \hspace*{1cm} \mbox{with} \hspace*{1cm}
\sigma^{\mu\nu}=-{i\over 4}[\gamma^{\mu},\gamma^{\nu}].
\end{eqnarray}
Because of Schur's lemma (see Appendix) this implies 
that the representation of the
Lorentz group provided by $\sigma^{\mu\nu}$ is reducible into
subspaces spanned by the eigenvectors of $\gamma^{5}$ with the same
eigenvalue. If we define the projectors $P_{\pm}={1\over
2}(1\pm\gamma^{5})$ these subspaces correspond to
\begin{eqnarray}
P_{+}\psi=\left(
\begin{array}{c}
u_{+} \\
0
\end{array}
\right), 
\hspace*{1cm}
P_{-}\psi=\left(
\begin{array}{c}
0 \\
u_{-}
\end{array}
\right), 
\hspace*{1cm}
\end{eqnarray}
which are precisely the Weyl spinors introduced before.

Our next task is to quantize the Dirac Lagrangian. This will be done 
along the lines used for the Klein-Gordon field, starting with a general
solution to the Dirac equation and introducing the corresponding set of
creation-annihilation operators. Therefore we start by 
looking for a complete basis of solutions to the Dirac equation.
In the case of the
scalar field the elements of the basis were labelled by their four-momentum
$k^{\mu}$. Now, however, we have more degrees of freedom since we are
dealing with a spinor which means that we have to add extra labels. 
Looking back at Eq. (\ref{eq:helicity}) we can define the helicity operator
for a Dirac spinor as
\begin{eqnarray}
\lambda={1\over 2}\vec{\sigma}\cdot{\vec{k}\over|\vec{k}|}\left(
\begin{array}{cc}
\mathbf{1} & 0 \\
0 & \mathbf{1}
\end{array}
\right).
\end{eqnarray}
Hence, each element of the basis of functions is labelled by its four-momentum
$k^{\mu}$ and the corresponding eigenvalue $s$ of the helicity operator. For 
positive energy solutions we then propose the ansatz
\begin{eqnarray}
u(k,s)e^{-ik\cdot x}, \hspace*{1cm} s=\pm{1\over 2},
\end{eqnarray}
where $u_{\alpha}(k,s)$ ($\alpha=1,\ldots,4$) is a four-component
spinor. Substituting in the Dirac equation we obtain
\begin{eqnarray}
(\fslash{k}-m)u(k,s)=0.
\label{eq:positive}
\end{eqnarray}
In the same way, for negative energy solutions we have
\begin{eqnarray}
v(k,s)e^{ik\cdot x}, \hspace*{1cm} s=\pm{1\over 2},
\end{eqnarray}
where $v(k,s)$ has to satisfy
\begin{eqnarray}
(\fslash{k}+m)v(k,s)=0.
\label{eq:negative}
\end{eqnarray}
Multiplying Eqs. (\ref{eq:positive}) and (\ref{eq:negative}) on the 
left respectively by $(\fslash{k}\mp m)$ we find that the momentum is on
the mass shell, $k^{2}=m^{2}$. Because of this, the wave function 
for both positive- and negative-energy solutions can be labeled as well using
the three-momentum $\vec{k}$ of the particle, $u(\vec{k},s)$, 
$v(\vec{k},s)$.

A detailed analysis shows that the functions $u(\vec{k},s)$, $v(\vec{k},s)$
satisfy the properties
\begin{eqnarray}
\overline{u}(\vec{k},s)u(\vec{k},s)=2m, \hspace*{1.3cm}  & \hspace*{1cm} & 
\hspace*{1.5cm} \overline{v}(\vec{k},s)v(\vec{k},s)=-2m,  \nonumber \\
\overline{u}(\vec{k},s)\gamma^{\mu} u(\vec{k},s) = 2k^{\mu},\hspace*{1.2cm} & \hspace*{1cm} &
\hspace*{1.2cm}\overline{v}(\vec{k},s)\gamma^{\mu} v(\vec{k},s)=2k^{\mu},
\label{eq:varrelspindir}\\
\sum_{s=\pm{1\over 2}}u_{\alpha}(\vec{k},s)\overline{u}_{\beta}(\vec{k},s)
=(\fslash{k}+m)_{\alpha\beta}, &\hspace*{0.8cm} &
\hspace*{0.3cm} 
{}\sum_{s=\pm{1\over 2}}v_{\alpha}(\vec{k},s)\overline{v}_{\beta}(\vec{k},s)
=(\fslash{k}-m)_{\alpha\beta}, \nonumber 
\end{eqnarray}
with $k^{0}=\omega_{k}=\sqrt{\vec{k}^{\,2}+m^{2}}$. 
Then, a general solution to the Dirac equation including creation and
annihilation operators can be written as:
\begin{eqnarray}
\widehat{\psi}(t,\vec{x})=\int {d^{3}k\over (2\pi)^{3}}{1\over 2\omega_{k}}
\sum_{s=\pm{1\over 2}}
\left[u(\vec{k},s)\,
\widehat{b}(\vec{k},s)e^{-i\omega_{k}t+i\vec{k}\cdot\vec{x}}
+v(\vec{k},s)\,\widehat{d}^{\dagger}(\vec{k},s)
e^{i\omega_{k}t-i\vec{k}\cdot\vec{x}}\right].
\end{eqnarray}

The operators $\widehat{b}^{\dagger}(\vec{k},s)$, 
$\widehat{b}(\vec{k})$
respectively create and annihilate a spin-${1\over 2}$ particle (for 
example, an electron) out of the vacuum with momentum $\vec{k}$ and helicity 
$s$. Because we are dealing with half-integer spin fields, the spin-statistics
theorem forces canonical anticommutation relations for $\widehat{\psi}$
which means that the creation-annihilation operators satisfy the 
algebra\footnote{To simplify notation, and since there is no risk of 
confusion, we drop from now on the hat to indicate operators.}
\begin{eqnarray}
\{b(\vec{k},s),b^{\dagger}(\vec{k}\,',s')\}&=&
\delta(\vec{k}-\vec{k}\,\,')\delta_{ss'}, \nonumber \\
\{b(\vec{k},s),b(\vec{k}\,',s')\}&=&
\{b^{\dagger}(\vec{k},s),b^{\dagger}(\vec{k}\,',s')\}=0.
\end{eqnarray}

In the case of 
$d(\vec{k},s)$, $d^{\dagger}(\vec{k},s)$ we have a 
set of creation-annihilation operators for the corresponding antiparticles
(for example positrons).
This is clear if we notice that $d^{\dagger}(\vec{k},s)$
can be seen as the annihilation operator of a negative energy state of
the Dirac equation with wave function $v_{\alpha}(\vec{k},s)$. As we saw, in 
the Dirac sea picture this corresponds to the creation of an antiparticle
out of the vacuum (see Fig. \ref{fig:dirac_sea}). The creation-annihilation
operators for antiparticles also satisfy the fermionic algebra
\begin{eqnarray}
\{d(\vec{k},s),d^{\dagger}(\vec{k}\,',s')\}&=&
\delta(\vec{k}-\vec{k}\,\,')\delta_{ss'}, \nonumber \\
\{d(\vec{k},s),d(\vec{k}\,',s')\}&=&
\{d^{\dagger}(\vec{k},s),d^{\dagger}(\vec{k}\,',s')\}=0.
\label{eq:ds}
\end{eqnarray}
All other anticommutators between $b(\vec{k},s)$, 
$b^{\dagger}(\vec{k},s)$ and $d(\vec{k},s)$,
$d^{\dagger}(\vec{k},s)$ vanish.

The Hamiltonian
operator for the Dirac field is
\begin{eqnarray}
\widehat{H}={1\over 2}\sum_{s=\pm{1\over 2}}\int {d^{3}k\over (2\pi)^{3}}\,\left[
b^{\dagger}(\vec{k},s)b(\vec{k},s)-
d(\vec{k},s)d^{\dagger}(\vec{k},s)\right].
\label{eq:hamilton_dirac1}
\end{eqnarray}
At this point we realize again of the necessity of quantizing the
theory using anticommutators instead of commutators. Had we use canonical
commutation relations, the second term inside the integral in 
(\ref{eq:hamilton_dirac1}) would give the number operator 
$d^{\dagger}(\vec{k},s)d(\vec{k},s)$ with a minus sign
in front. As a consequence the Hamiltonian would be unbounded from below and
we would be facing again the instability of the theory 
already noticed in the 
context of relativistic quantum mechanics. However, because of the 
{\em anticommutation} relations (\ref{eq:ds}), the Hamiltonian 
(\ref{eq:hamilton_dirac1}) takes the form
\begin{eqnarray}
\widehat{H}=\sum_{s=\pm{1\over 2}}\int {d^{3}k\over (2\pi)^{3}}
{1\over 2\omega_{k}}\,\left[\omega_{k}
b^{\dagger}(\vec{k},s)b(\vec{k},s)+\omega_{k}
d^{\dagger}(\vec{k},s)d(\vec{k},s)\right]-
2\int d^{3}k\,\omega_{k}\delta(\vec{0}).
\label{eq:hamilton_dirac2}
\end{eqnarray}
As with the scalar field, we find a divergent vacuum energy
contribution due to the zero-point energy of the infinite number of
harmonic oscillators. Unlike
the Klein-Gordon field, the vacuum energy is negative. In
section \ref{sec:supersymmetry} we will see that in certain type of
theories called supersymmetric, where the number of bosonic and
fermionic degrees of freedom is the same, there is a cancellation of
the vacuum energy.  The divergent contribution can be
removed by the normal order prescription
\begin{eqnarray}
:\!\!\widehat{H}\!\!:=\sum_{s=\pm{1\over 2}}\int {d^{3}k\over (2\pi)^{3}}
{1\over 2\omega_{k}}\,\left[\omega_{k}
b^{\dagger}(\vec{k},s)b(\vec{k},s)+\omega_{k}
d^{\dagger}(\vec{k},s)d(\vec{k},s)\right].
\label{eq:fermion_hamiltonian}
\end{eqnarray}

Finally, let us mention that using the Dirac equation it is easy to prove
that there is a conserved four-current given by
\begin{eqnarray}
j^{\mu}=\overline{\psi}\gamma^{\mu}\psi, \hspace*{1cm}
\partial_{\mu}j^{\mu}=0.
\end{eqnarray}
As we will explain further in sec. \ref{sec:symmetries} this current is
associated to the invariance of the Dirac Lagrangian under the global phase
shift $\psi\rightarrow e^{i\theta}\psi$. In electrodynamics the associated
conserved charge
\begin{eqnarray}
Q=e\int d^{3}x\,j^{0}
\end{eqnarray}
is identified with the electric charge.

\subsection{Gauge fields}
\label{sec:gauge_fields}

In classical electrodynamics the basic quantities are the electric and
magnetic fields $\vec{E}$, $\vec{B}$. These can be expressed
in terms of the scalar and vector potential $(\varphi,
\vec{A})$
\begin{eqnarray}
\vec{E}&=&-\vec{\nabla}\varphi -{\partial\vec{A}\over \partial t}, 
\nonumber \\
\vec{B}&=& \vec{\nabla}\times\vec{A}.
\end{eqnarray}
From these equations it follows that there is an ambiguity in the definition
of the potentials given by the gauge transformations
\begin{eqnarray}
\varphi(t,\vec{x})\rightarrow 
\varphi(t,\vec{x})+{\partial\over\partial t}\epsilon(t,\vec{x}),
\hspace*{1cm} 
\vec{A}(t,\vec{x}) \rightarrow \vec{A}(t,\vec{x})
-\vec{\nabla}\epsilon(t,\vec{x}).
\label{eq:gauge_trans}
\end{eqnarray}
Classically $(\varphi,\vec{A})$ are seen as only a  convenient way to solve 
the Maxwell equations, but without physical relevance.

The equations of electrodynamics can be recast in a manifestly Lorentz
invariant form using the four-vector gauge potential 
$A^{\mu}=(\varphi,\vec{A})$ and the antisymmetric rank-two 
tensor:  $F_{\mu\nu}=\partial_{\mu}A_{\nu}-\partial_{\nu}
A_{\mu}$.  Maxwell's equations become
\begin{eqnarray}
\partial_{\mu}F^{\mu\nu} &=& j^{\mu}, \nonumber \\
\epsilon^{\mu\nu\sigma\eta}\partial_{\nu}F_{\sigma\eta} &=& 0,
\label{eq:maxwell_covariant}
\end{eqnarray}
where the four-current $j^{\mu}=(\rho,\vec{\jmath})$ contains the
charge density and the electric current. The field strength tensor
$F_{\mu\nu}$ and the Maxwell equations are invariant under gauge
transformations (\ref{eq:gauge_trans}), which in covariant form read
\begin{eqnarray}
A_{\mu} \longrightarrow A_{\mu}+\partial_{\mu}\epsilon.
\end{eqnarray}
Finally, the equations of motion of charged particles are given, in
covariant form, by
\begin{eqnarray}
m{du^{\mu}\over d\tau}=eF^{\mu\nu}u_{\nu},
\end{eqnarray}
where $e$ is the charge of the particle and $u^{\mu}(\tau)$ its four-velocity
as a function of the proper time.

The physical r\^ole of the vector potential becomes manifest only in
Quantum Mechanics. Using the prescription of minimal substitution
$\vec{p}\rightarrow
\vec{p}-e\vec{A}$, the Schr\"odinger equation describing a
particle with charge $e$ moving in an electromagnetic field is
\begin{eqnarray}
i\partial_{t}\Psi=\left[-{1\over 2m}\left(\vec{\nabla}-ie\vec{A}\right)^{2}
+e\varphi\right]\Psi.
\label{eq:schr_em}
\end{eqnarray}
Because of the explicit dependence on the electromagnetic potentials
$\varphi$ and $\vec{A}$, this equation seems to change under the gauge
transformations (\ref{eq:gauge_trans}).  This is physically acceptable
only if the ambiguity does not affect the probability density given by
$|\Psi(t,\vec{x})|^{2}$. Therefore, a gauge transformation of the
electromagnetic potential should amount to a change in the
(unobservable) phase of the wave function. This is indeed what
happens: the Schr\"odinger equation (\ref{eq:schr_em}) is invariant
under the gauge transformations (\ref{eq:gauge_trans}) provided the
phase of the wave function is transformed at the same time according
to
\begin{eqnarray}
\Psi(t,\vec{x}) \longrightarrow e^{-ie\,\epsilon(t,\vec{x})}\Psi(t,\vec{x}).
\end{eqnarray}

\begin{figure}
\centerline{\epsfxsize=3.5truein\epsfbox{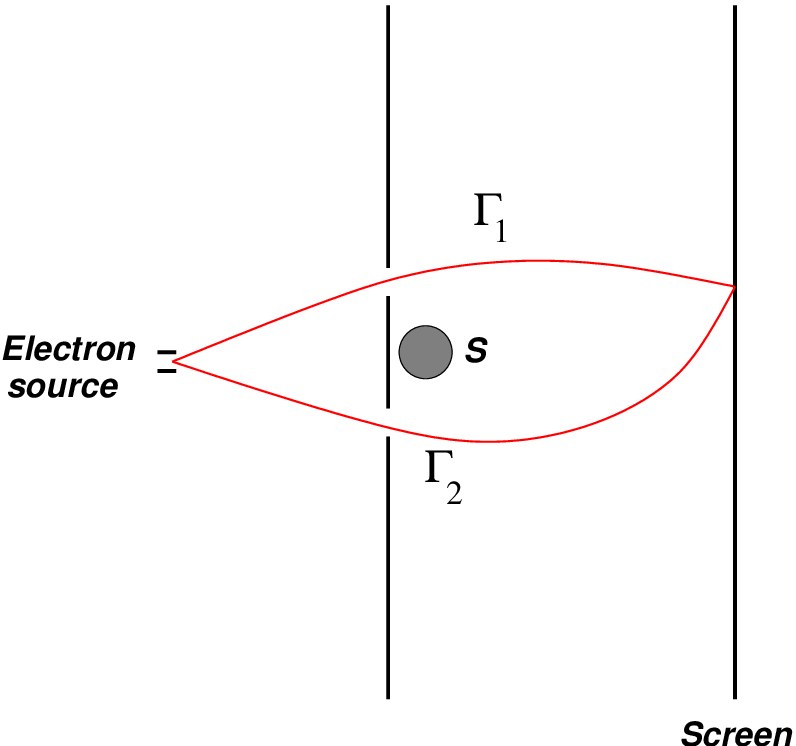}}
\caption[]{Illustration of an interference experiment to show the
Aharonov-Bohm effect. $S$ represent the solenoid in whose interior 
the magnetic field is confined.}
\label{fig:aharonov_bohm}
\end{figure}

{\bf Aharonov-Bohm effect.}
This interplay between gauge transformations and the phase of the wave
function give rise to surprising phenomena. The first evidence of the
r\^ole played by the electromagnetic potentials at the quantum level 
was pointed out by Yakir Aharonov and David Bohm \cite{aharonov_bohm}.
Let us consider a double slit experiment as shown in Fig. 
\ref{fig:aharonov_bohm}, where we have placed a shielded solenoid just 
behind the first screen. Although the magnetic field is confined to
the interior of the solenoid, the vector potential is nonvanishing
also outside. Of course the value of $\vec{A}$ outside the solenoid is
a pure gauge, i.e. $\vec{\nabla}\times\vec{A}=\vec{0}$, however
because the region outside the solenoid is not simply connected
the vector potential cannot be gauged to zero everywhere. If we denote
by $\Psi_{1}^{(0)}$ and $\Psi_{2}^{(0)}$ the wave functions for each
of the two electron beams in the absence of the solenoid, the total
wave function once the magnetic field is switched on can be written as
\begin{eqnarray}
\Psi&=&e^{ie\int_{\Gamma_{1}}\vec{A}\cdot d\vec{x}}\Psi_{1}^{(0)}+
e^{ie\int_{\Gamma_{2}}\vec{A}\cdot d\vec{x}}\Psi_{2}^{(0)} \nonumber \\ 
&=&e^{ie\int_{\Gamma_{1}}\vec{A}\cdot d\vec{x}}\left[\Psi_{1}^{(0)}
+e^{ie\oint_{\Gamma}\vec{A}\cdot d\vec{x}}\Psi_{2}^{(0)}\right],
\label{eq:extra_phase}
\end{eqnarray}
where $\Gamma_{1}$ and $\Gamma_{2}$ are two curves surrounding the solenoid 
from different sides, and $\Gamma$ is any closed loop surrounding it. 
Therefore the relative phase between the two beams gets an extra term 
depending on the 
value of the vector potential outside the solenoid as 
\begin{eqnarray} 
U=\exp\left[ie\oint_{\Gamma}\vec{A}\cdot d\vec{x}\right]. 
\label{eq:wilson} 
\end{eqnarray} 
Because of the change in the relative phase of the electron wave 
functions, the presence of the vector potential becomes observable
even if the electrons do not feel the magnetic field. If we perform 
the double-slit experiment when the magnetic field inside the solenoid 
is switched off we will observe the usual interference pattern on the 
second screen.  However if now the magnetic field is switched on, 
because of the phase (\ref{eq:extra_phase}), a change in the interference 
pattern will appear. This is the Aharonov-Bohm effect. 
                                        
The first question that comes up is what happens with gauge
invariance.  Since we said that $\vec{A}$ can be changed by a gauge
transformation it seems that the resulting interference patters might
depend on the gauge used. Actually, the phase $U$ in (\ref{eq:wilson})
is independent of the gauge although, unlike other gauge-invariant
quantities like $\vec{E}$ and $\vec{B}$, is nonlocal. Notice that,
since $\vec{\nabla}\times\vec{A}=\vec{0}$ outside the solenoid, the
value of $U$ does not change under continuous deformations of the
closed curve $\Gamma$, so long as it does not cross the solenoid.
  
{\bf The Dirac monopole.}  It is very easy to check that the vacuum
Maxwell equations remain invariant under the transformation
\begin{eqnarray}
\vec{E}-i\vec{B} \longrightarrow e^{i\theta}(\vec{E}-i\vec{B}), \hspace*{1cm}
\theta\in[0,2\pi]
\label{eq:duality}
\end{eqnarray}
which, in particular, for $\theta={\pi\over 2}$ interchanges the electric 
and the magnetic fields: $\vec{E}\rightarrow\vec{B}$, $\vec{B}\rightarrow
-\vec{E}$. This duality symmetry is however broken in the presence of electric
sources. Nevertheless the  Maxwell equations can be ``completed'' by 
introducing sources for the magnetic field $(\rho_{m},\vec{\jmath}_{m})$ 
in such a way that the duality (\ref{eq:duality}) is restored when
supplemented by the transformation
\begin{eqnarray}
\rho-i\rho_{m}\longrightarrow e^{i\theta}(\rho-i\rho_{m}),  \hspace*{1cm}
\vec{\jmath}-i\vec{\jmath}_{m} \longrightarrow e^{i\theta}(
\vec{\jmath}-i\vec{\jmath}_{m}).
\end{eqnarray}
Again for $\theta=\pi/2$ the electric and magnetic  sources get interchanged.
\begin{figure}
\centerline{\epsfxsize=3.2truein\epsfbox{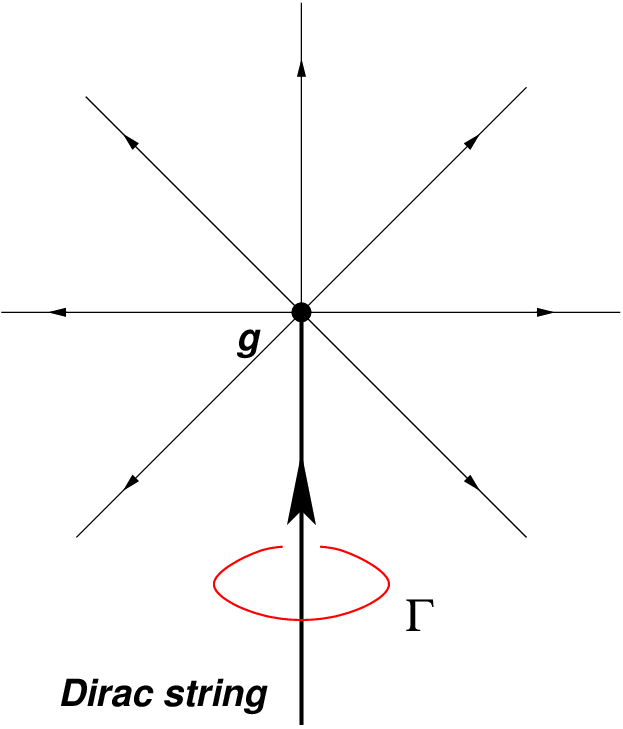}}
\caption[]{The Dirac monopole.}
\label{fig:dirac_monopole}
\end{figure}

In 1931 Dirac \cite{dirac-monopole} studied the possibility of finding
solutions of the completed Maxwell equation with a magnetic
monopoles of charge $g$, i.e. solutions to 
\begin{eqnarray}
\vec{\nabla}\cdot\vec{B}=g\,\delta(\vec{x}).
\label{eq:monopole1}
\end{eqnarray}
Away from the position of the monopole $\vec{\nabla}\cdot\vec{B}=0$
and the magnetic field can be still derived locally from a vector potential
$\vec{A}$ according to $\vec{B}=\vec{\nabla}\times\vec{A}$. However, 
the vector potential cannot be 
regular everywhere since otherwise Gauss law would imply that the
magnetic flux threading a closed surface around the monopole should vanish,
contradicting (\ref{eq:monopole1}). 

We look now for solutions to Eq. (\ref{eq:monopole1}). Working in 
spherical coordinates we find 
\begin{eqnarray}
{B}_{r}={g\over |\vec{x}|^{2}}, \hspace*{1cm}
B_{\varphi}=B_{\theta}=0.
\end{eqnarray}
Away from the position of the monopole ($\vec{x}\neq\vec{0}$) the
magnetic field can be derived from the vector potential
\begin{eqnarray}
{A}_{\varphi}={g\over |\vec{x}|}\tan{\theta\over 2}, 
\hspace*{1cm} A_{r}=A_{\theta}=0.
\label{eq:dirac_monopole}
\end{eqnarray}
As expected we find that this vector potential is actually singular
around the half-line $\theta=\pi$ (see
Fig. \ref{fig:dirac_monopole}).  This singular line starting at the
position of the mo\-no\-pole is called the Dirac string and its position
changes with a change of gauge but cannot be eliminated by any gauge
transformation. Physically we can see it as an infinitely thin
solenoid confining a magnetic flux entering into the magnetic mo\-no\-pole
from infinity that equals the outgoing magnetic flux from the mo\-no\-pole.

Since the position of the Dirac string depends on the gauge chosen it
seems that the presence of mono\-poles introduces an ambiguity. This
would be rather strange, since Maxwell equations are gauge invariant
also in the presence of magnetic sources. The solution to this
apparent riddle lies in the fact that the Dirac string does not pose
any consistency problem as far as it does not produce any physical
effect, i.e. if its presence turns out to be undetectable.  From our
discussion of the Aharonov-Bohm effect we know that the wave function
of charged particles pick up a phase (\ref{eq:wilson}) when
surrounding a region where magnetic flux is confined (for example the
solenoid in the Aharonov-Bohm experiment). As explained above, the
Dirac string associated with the monopole can be seen as a infinitely 
thin solenoid. Therefore the Dirac string will be unobservable if 
the phase picked up by the wave function of a charged particle is 
equal to one. A simple calculation shows that this happens if
\begin{eqnarray}
e^{i\,e\,g}=1 \hspace*{1cm}\Longrightarrow\hspace*{1cm} e\,g=2\pi n
\,\,\,\mbox{with}\,\,\, n\in\mathbb{Z}.
\end{eqnarray}
Interestingly, this discussion leads to the conclusion that the
presence of a single magnetic mono\-poles somewhere in the Universe
implies for consistency the quantization of the electric charge in
units of ${2\pi\over g}$, where $g$ the magnetic charge of the monopole.

{\bf Quantization of the electromagnetic field.}
We now proceed to the quantization of the electromagnetic field
in the absence of sources $\rho=0$, $\vec{\jmath}=\vec{0}$. In this
case the Maxwell equations (\ref{eq:maxwell_covariant}) can be derived
from the Lagrangian density
\begin{eqnarray}
\mathcal{L}_{\rm Maxwell}=-{1\over 4}F_{\mu\nu}F^{\mu\nu}=
{1\over 2}\left(\vec{E}^{\,2}-\vec{B}^{\,2}\right).
\end{eqnarray}
Although in general the procedure to quantize the Maxwell Lagrangian is
not very different from the one used for the Klein-Gordon or the Dirac field,
here we need to deal with a new ingredient: gauge 
invariance. Unlike the cases studied so far, here the photon field
$A_{\mu}$ is not unambiguously defined because the action and the equations
of motion are insensitive to the gauge transformations $A_{\mu}
\rightarrow A_{\mu}+\partial_{\mu}\varepsilon$. A first consequence of this
symmetry is that the theory has less physical degrees of freedom than
one would expect from the fact that we are dealing with a vector field.

The way to tackle the problem of gauge invariance is to fix the freedom
in choosing the electromagnetic potential before quantization. This can
be done in several ways, for example by imposing the 
Lorentz gauge fixing condition
\begin{eqnarray}
\partial_{\mu}A^{\mu}=0.
\label{eq:lorentz-condition}
\end{eqnarray}
Notice that this condition does not fix completely the gauge freedom
since Eq. (\ref{eq:lorentz-condition}) is left invariant by gauge
transformations satisfying
$\partial_{\mu}\partial^{\mu}\varepsilon=0$. One of the advantages,
however, of the Lorentz gauge is that it is covariant and therefore
does not pose any danger to the Lorentz invariance of the quantum
theory. Besides, applying it to the Maxwell equation
$\partial_{\mu}F^{\mu\nu}=0$ one finds
\begin{eqnarray}
0=\partial_{\mu}\partial^{\mu}A^{\nu}-\partial_{\nu}\left(
\partial_{\mu}A^{\mu}\right)=\partial_{\mu}\partial^{\mu}A^{\nu},
\label{eq:KG-Amu}
\end{eqnarray}
which means that since $A_{\mu}$ satisfies the massless Klein-Gordon
equation the photon, the quantum of the electromagnetic field, has
zero mass.

Once gauge invariance is fixed $A_{\mu}$ is expanded in a complete basis
of solutions to (\ref{eq:KG-Amu}) and the canonical commutation relations 
are imposed
\begin{eqnarray}
\widehat{A}_{\mu}(t,\vec{x})=\sum_{\lambda=\pm 1}\int
{d^{3}k\over (2\pi)^{3}}{1\over 2|\vec{k}|}\left[
\epsilon_{\mu}(\vec{k},\lambda)\widehat{a}
(\vec{k},\lambda)e^{-i|\vec{k}|t+i\vec{k}\cdot\vec{x}}
+\epsilon_{\mu}(\vec{k},\lambda)^{*}\,\widehat{a}^{\dagger}
(\vec{k},\lambda)e^{i|\vec{k}|t-i\vec{k}\cdot\vec{x}}
\right]
\end{eqnarray}
where $\lambda=\pm 1$ represent the helicity of the photon, and
$\epsilon_{\mu}(\vec{k},\lambda)$ are solutions to the equations of
motion with well defined momentum an helicity.  Because of
(\ref{eq:lorentz-condition}) the polarization vectors have to be
orthogonal to $k_{\mu}$
\begin{eqnarray}
k^{\mu}\epsilon_{\mu}(\vec{k},\lambda)=k^{\mu}
\epsilon_{\mu}(\vec{k},\lambda)^{*}=0.
\label{eq:transvcondpolvec}
\end{eqnarray}
The canonical commutation relations imply that 
\begin{eqnarray}
[\widehat{a}(\vec{k},\lambda),\widehat{a}^{\dagger}
(\vec{k}\,',\lambda')]&=&(2\pi)^{3}(2|\vec{k}|)\delta(\vec{k}-\vec{k}\,')\delta_{\lambda\lambda'}
\nonumber \\
{}[\widehat{a}(\vec{k},\lambda),\widehat{a}(\vec{k}\,',\lambda')]&=&
{}[\widehat{a}^{\dagger}(\vec{k},\lambda),\widehat{a}^{\dagger}
(\vec{k}\,',\lambda')]=0.
\end{eqnarray}
Therefore  $\widehat{a}(\vec{k},\lambda)$, $\widehat{a}^{\dagger}
(\vec{k},\lambda)$ form a set of creation-annihilation operators
for photons with momentum $\vec{k}$ and helicity $\lambda$.  

Behind the simple construction presented above there are a number of
subleties related with gauge invariance. In particular the gauge 
freedom seem to introduce states in the Hilbert space with negative
probability. A careful analysis shows that when gauge invariance
if properly handled these spurious states decouple from physical states
and can be eliminated. The details can be found in standard textbooks
\cite{ours}-\cite{banks}.

{\bf Coupling gauge fields to matter.}  Once we know how to quantize
the electromagnetic field we consider theories containing electrically
charged particles, for example electrons. To couple the Dirac
Lagrangian to electromagnetism we use as guiding principle what we
learned about the Schr\"odinger equation for a charged particle. There
we saw that the gauge ambiguity of the electromagnetic potential is
compensated with a U(1) phase shift in the wave function. In the case
of the Dirac equation we know that the Lagrangian is invariant under
$\psi\rightarrow e^{ie\varepsilon}\psi$, with $\varepsilon$ a
constant. However this invariance is broken as soon as one identifies
$\varepsilon$ with the gauge transformation parameter of the
electromagnetic field which depends on the position.

Looking at the Dirac Lagrangian (\ref{eq:dirac_eq}) it is easy to see
that in order to promote the global U(1) symmetry into a local one,
$\psi\rightarrow e^{-ie\varepsilon(x)}\psi$, it suffices to replace
the ordinary derivative $\partial_{\mu}$ by a covariant one $D_{\mu}$
satisfying
\begin{eqnarray}
D_{\mu}\left[e^{-ie\varepsilon(x)}\psi\right]=e^{-ie\varepsilon(x)}
D_{\mu}\psi.
\end{eqnarray}
This covariant derivative can be constructed in terms of the
gauge potential $A_{\mu}$ as
\begin{eqnarray}
D_{\mu}=\partial_{\mu}+ieA_{\mu}.
\label{eq:covariant_deriv}
\end{eqnarray}
The Lagrangian of a spin-${1\over 2}$ field coupled
to electromagnetism is written as
\begin{eqnarray}
\mathcal{L}_{\rm QED}=-{1\over 4}F_{\mu\nu}F^{\mu\nu}
+\overline{\psi}(i\Fslash{D}-m)\psi,
\label{eq:qed_lagrangian}
\end{eqnarray}
invariant under the gauge transformations
\begin{eqnarray}
\psi\longrightarrow e^{-ie\varepsilon(x)}\psi, \hspace*{1cm}
A_{\mu}\longrightarrow A_{\mu}+\partial_{\mu}\varepsilon(x).
\end{eqnarray}

Unlike the theories we have seen so far, the Lagrangian 
(\ref{eq:qed_lagrangian}) describe an interacting theory. By plugging
(\ref{eq:covariant_deriv}) into the Lagrangian we find that the
interaction between fermions and photons to be
\begin{eqnarray}
\mathcal{L}_{\rm QED}^{\rm (int)}=
-eA_{\mu}\,\overline{\psi}\gamma^{\mu}\psi.
\label{eq:interac_QED}
\end{eqnarray}
As advertised above, in the Dirac theory the electric current four-vector is
given by $j^{\mu}=e\overline{\psi}\gamma^{\mu}\psi$. 

The quantization of interacting field theories poses new problems that
we did not meet in the case of the free theories.
In particular in most cases it is not possible to solve the
theory exactly.  When this happens the physical observables have to be
computed in perturbation theory in powers of the coupling constant. An
added problem appears when computing quantum corrections to the
classical result, since in that case the computation of observables
are plagued with infinities that should be taken care of. We will go
back to this problem in section \ref{sec:renormalization}.

{\bf Nonabelian gauge theories.}
Quantum electrodynamics (QED) is the simplest example of a gauge
theory coupled to matter based in the abelian gauge symmetry of local
U(1) phase rotations. However, it is possible also to construct gauge
theories based on nonabelian groups. Actually, our knowledge of 
the strong and weak interactions is based on the use of such nonabelian
generalizations of QED.

Let us consider a gauge group ${G}$ with generators
$T^{a}$, $a=1,\ldots,{\rm dim\,}{G}$ satisfying the Lie 
algebra\footnote{Some basics facts about Lie groups have been summarized
in Appendix A.}
\begin{eqnarray}
[T^{a},T^{b}]=if^{abc}T^{c}.
\end{eqnarray}
A gauge field taking values on the Lie algebra of $\mathcal{G}$ can
be introduced $A_{\mu}\equiv A_{\mu}^{a}T^{a}$ which transforms under
a gauge transformations as
\begin{eqnarray}
A_{\mu}\longrightarrow -{1\over ig}U\partial_{\mu}U^{-1}
+UA_{\mu}U^{-1}, \hspace*{1cm} U=e^{i\chi^{a}(x)T^{a}},
\end{eqnarray}
where $g$ is the coupling constant. The associated field strength is
defined as
\begin{eqnarray}
F^{a}_{\mu\nu}=\partial_{\mu}A_{\nu}^{a}-\partial_{\nu}A_{\mu}^{a}
+gf^{abc}A^{b}_{\mu}A^{c}_{\nu}.
\label{eq:fmunuNA}
\end{eqnarray}
Notice that this definition of the $F^{a}_{\mu\nu}$ reduces to the one
used in QED in the abelian case when $f^{abc}=0$. In general, however, 
unlike the case of QED the field strength is not gauge invariant. 
In terms of $F_{\mu\nu}=F_{\mu\nu}^{a}T^{a}$ it transforms as
\begin{eqnarray}
F_{\mu\nu}\longrightarrow UF_{\mu\nu}U^{-1}.
\end{eqnarray}

The coupling of matter to a nonabelian gauge field is done by introducing 
again a covariant derivative. For a field in a
representation of $\mathcal{G}$
\begin{eqnarray}
\Phi \longrightarrow U \Phi
\end{eqnarray}
the covariant derivative is given by
\begin{eqnarray}
D_{\mu}\Phi=\partial_{\mu}\Phi-ig A_{\mu}^{a}T^{a}\Phi.
\label{eq:covariant_derivative}
\end{eqnarray}
With the help of this we can write a generic
Lagrangian for a nonabelian gauge field coupled to scalars $\phi$ 
and spinors $\psi$ as
\begin{eqnarray}
\mathcal{L}=-{1\over 4}F_{\mu\nu}^{a}F^{\mu\nu\,a}+
i\overline{\psi}\Fslash{D}\psi+ \overline{D_{\mu}\phi}D^{\mu}\phi
-\overline{\psi}\left[M_{1}(\phi)+i\gamma_{5}M_{2}(\phi)\right]\psi
-V(\phi).
\label{eq:nonabelian}
\end{eqnarray}
In order to keep the theory renormalizable we have to restrict
$M_{1}(\phi)$ and $M_{2}(\phi)$ to be at most linear in $\phi$ whereas
$V(\phi)$ have to be at most of quartic order. The Lagrangian of the
Standard Model is of the form (\ref{eq:nonabelian}).

\subsection{Understanding gauge symmetry}
\label{sec:gauge_symmetry}

In classical mechanics the use of the Hamiltonian formalism starts
with the replacement of generalized velocities by momenta
\begin{eqnarray}
p_{i}\equiv {\partial L\over \partial \dot{q}_{i}} \hspace*{1cm}
\Longrightarrow \hspace*{1cm} \dot{q}_{i}= \dot{q}_{i}(q,p).
\end{eqnarray}
Most of the times there is no problem in inverting the relations 
$p_{i}=p_{i}(q,\dot{q})$. However in some systems these relations might
not be invertible and result in a number of constraints of the type
\begin{eqnarray}
f_{a}(q,p)=0,  \hspace*{1cm} a=1,\ldots,N_{1}.
\label{eq:constraints_dirac}
\end{eqnarray}
These systems are called degenerate or constrained \cite{dirac_book,
henneaux-teitelboim}. 

The presence of constraints of the type (\ref{eq:constraints_dirac})
makes the formulation of the Hamiltonian formalism more involved. The
first problem is related to the ambiguity in defining
the Hamiltonian, since the addition of any linear combination of the
constraints do not modify its value.  Secondly, one has to make sure
that the constraints are consistent with the time evolution in the
system. In the language of Poisson brackets this means that
further constraints have to be imposed in the form
\begin{eqnarray}
\{f_{a},H\}\approx 0.
\label{eq:second}
\end{eqnarray}
Following \cite{dirac_book} we use the symbol $\approx$ to indicate
a ``weak'' equality that holds when the constraints $f_{a}(q,p)=0$ are
satisfied.  Notice however that since the computation of the Poisson
brackets involves derivatives, the constraints can be used only after
the bracket is computed.  In principle the conditions
(\ref{eq:second}) can give rise to a new set of constraints
$g_{b}(q,p)=0$, $b=1,\ldots,N_{2}$. Again these constraints have to be
consistent with time evolution and we have to repeat the procedure. 
Eventually this finishes when a set of constraints is found that do not
require any further constraint to be preserved by the 
time evolution\footnote{In
principle it is also possible that the procedure finishes because some
kind of inconsistent identity is found. In this case the system itself is 
inconsistent as it is the case with the Lagrangian $L(q,\dot{q})=q$.}.

Once we find all the constraints of a degenerate system we consider
the so-called first class constraints $\phi_{a}(q,p)=0$,
$a=1,\ldots,M$, which are those whose Poisson bracket vanishes weakly
\begin{eqnarray}
\{\phi_{a},\phi_{b}\}=c_{abc}\phi_{c}\approx 0.
\end{eqnarray}
The constraints that do not satisfy this condition, called second class 
constraints, can be eliminated by modifying the Poisson bracket 
\cite{dirac_book}. Then the total Hamiltonian of the theory is 
defined by
\begin{eqnarray}
H_{T}=p_{i}q_{i}-L+\sum_{a=1}^{M}\lambda(t)\phi_{a}.
\end{eqnarray}

What has all this to do with gauge invariance? The interesting answer
is that for a singular system the first class constraints $\phi_{a}$ 
generate gauge transformations. Indeed, because $\{\phi_{a},\phi_{b}\}
\approx 0 \approx \{\phi_{a},H\}$ the transformations
\begin{eqnarray}
q_{i} &\longrightarrow& q_{i}+\sum_{a}^{M}\varepsilon_{a}(t)\{q_{i},\phi_{a}\}, 
\nonumber \\ 
p_{i} &\longrightarrow& p_{i}+\sum_{a}^{M}\varepsilon_{a}(t)\{p_{i},\phi_{a}\} 
\end{eqnarray}
leave invariant the state of the system. This ambiguity in the description
of the system in terms of the generalized coordinates and momenta 
can be traced back 
to the equations of motion in Lagrangian language. Writing them in 
the form
\begin{eqnarray}
{\partial^{2}L\over\partial \dot{q}_{i}\partial\dot{q}_{j}}\ddot{q}_{j}=
-{\partial^{2}L\over\partial\dot{q}_{i}\partial q_{j}}\dot{q}_{j}+
{\partial L\over \partial q_{i}},
\end{eqnarray}
we find that order to determine the accelerations in terms of the
positions and velocities the matrix ${{\partial^{2}L\over\partial
\dot{q}_{i}\partial\dot{q}_{j}}}$ has to be invertible. However, the
existence of constraints (\ref{eq:constraints_dirac}) precisely
implies that the determinant of this matrix vanishes and therefore the
time evolution is not uniquely determined in terms of the initial
conditions.

Let us apply this to Maxwell electrodynamics described by the Lagrangian
\begin{eqnarray}
L=-{1\over 4}\int d^{3}\,F_{\mu\nu}F^{\mu\nu}.
\end{eqnarray} 
The generalized momentum
conjugate to $A_{\mu}$ is given by
\begin{eqnarray}
\pi^{\mu}={\delta L\over \delta (\partial_{0}A_{\mu})}=F^{0\mu}.
\end{eqnarray}
In particular for the time component we find the constraint $\pi^{0}=0$.
The Hamiltonian is given by
\begin{eqnarray}
H=\int d^{3}x\left[\pi^{\mu}\partial_{0}A_{\mu}-\mathcal{L}\right]
=\int d^{3}x\left[{1\over 2}\left(\vec{E}^{\,2}+\vec{B}^{\,2}\right)
+\pi^{0}\partial_{0}A_{0}+A_{0}\vec{\nabla}\cdot\vec{E}\right].
\end{eqnarray}
Requiring the consistency of the constraint $\pi^{0}=0$ we find a 
second constraint
\begin{eqnarray}
\{\pi^{0},H\}\approx\partial_{0}\pi^{0}+\vec{\nabla}\cdot\vec{E}=0.
\end{eqnarray}
Together with the first constraint $\pi^{0}=0$ this one implies Gauss'
law $\vec{\nabla}\cdot\vec{E}=0$. These two constrains have vanishing
Poisson bracket and therefore they are first class. Therefore the
total Hamiltonian is given by
\begin{eqnarray}
H_{T}=H+\int d^{3}x\left[\lambda_{1}(x)\pi^{0}+\lambda_{2}(x)
\vec{\nabla}\cdot\vec{E}\right],
\end{eqnarray}
where we have absorbed $A_{0}$ in the definition of the arbitrary
functions $\lambda_{1}(x)$ and $\lambda_{2}(x)$. Actually, we can
fix part of the ambiguity taking $\lambda_{1}=0$. Notice that, because
$A_{0}$ has been included in the multipliers, fixing $\lambda_{1}$ 
amounts to fixing the value of $A_{0}$ and therefore it is equivalent to
taking a temporal gauge. In this case the Hamiltonian is
\begin{eqnarray}
H_{T}=\int d^{3}x\left[{1\over 2}\left(\vec{E}^{\,2}+\vec{B}^{\,2}\right)
+\varepsilon(x)\vec{\nabla}\cdot\vec{E}\right]
\end{eqnarray}
and we are left just with Gauss' law as the only constraint. Using the
canonical commutation relations
\begin{eqnarray}
\{A_{i}(t,\vec{x}),E_{j}(t,\vec{x}\,')\}=\delta_{ij}\delta(\vec{x}-
\vec{x}\,')
\end{eqnarray}
we find that the remaining gauge transformations are generated by
Gauss' law
\begin{eqnarray}
\delta A_{i}=\{A_{i},\int d^{3} x' \,
\varepsilon\,\vec{\nabla}\cdot\vec{E}\}=\partial_{i}\varepsilon,
\end{eqnarray}
while leaving $A_{0}$ invariant, so for consistency with the general
gauge transformations the function $\varepsilon(x)$ should be
independent of time. Notice that the constraint
$\vec{\nabla}\cdot\vec{E}=0$ can be implemented by demanding
$\vec{\nabla}\cdot
\vec{A}=0$ which reduces the three degrees of freedom of $\vec{A}$ to 
the two physical degrees of freedom of the photon.

So much for the classical analysis. In the quantum theory the constraint 
$\vec{\nabla}\cdot\vec{E}=0$ has to be imposed on the physical states
$|{\rm phys}\rangle$. This is done by defining the following unitary operator
on the Hilbert space
\begin{eqnarray}
\mathcal{U}(\varepsilon)\equiv \exp\left(i\int d^{3}x\, \varepsilon(\vec{x})\,
\vec{\nabla}\cdot\vec{E}\right).
\end{eqnarray}
By definition, physical states should not change when a gauge transformations
is performed. This is implemented by requiring that the operator 
$\mathcal{U}(\varepsilon)$ acts trivially on a physical state
\begin{eqnarray}
\mathcal{U}(\varepsilon)|{\rm phys}\rangle = |{\rm phys}\rangle
\hspace*{1cm} \Longrightarrow \hspace*{1cm} (\vec{\nabla}\cdot\vec{E})
|{\rm phys}\rangle =0.
\end{eqnarray}
In the presence of charge density $\rho$, the condition that physical states
are annihilated by Gauss' law changes to $(\vec{\nabla}\cdot\vec{E}
-\rho)|{\rm phys}\rangle =0$.

The role of gauge transformations in the quantum theory is very
illuminating in understanding the real r\^ole of gauge invariance
\cite{jackiw_rmp}. As we have learned, the existence of a gauge
symmetry in a theory reflects a degree of redundancy in the
description of physical states in terms of the degrees of freedom
appearing in the Lagrangian. In Classical Mechanics, for example, the
state of a system is usually determined by the value of the canonical
coordinates $(q_{i}, p_{i})$. We know, however, that this is not the
case for constrained Hamiltonian systems where the transformations
generated by the first class constraints change the value of $q_{i}$
and $p_{i}$ withoug changing the physical state. In the case of
Maxwell theory for every physical configuration determined by the
gauge invariant quantities $\vec{E}$, $\vec{B}$ there is an infinite
number of possible values of the vector potential that are related by
gauge transformations $\delta A_{\mu} =\partial_{\mu}\varepsilon$. 

In the quantum theory this means that the Hilbert space of physical
states is defined as the result of identifying all states related by
the operator $\mathcal{U}(\varepsilon)$ with any gauge function
$\varepsilon(x)$ into a single physical state $|{\rm phys}\rangle$.
In other words, each physical state corresponds to a whole orbit of
states that are transformed among themselves by gauge transformations.

This explains the necessity of gauge fixing. In order to avoid the
redundancy in the states a further condition can be given that selects
one single state on each orbit. In the case
of Maxwell electrodynamics the conditions $A_{0}=0$,
$\vec{\nabla}\cdot\vec{A}=0$ selects a value of the gauge potential
among all possible ones giving the same value for the electric and
magnetic fields.

\begin{figure}
\centerline{\epsfxsize=4.5truein\epsfbox{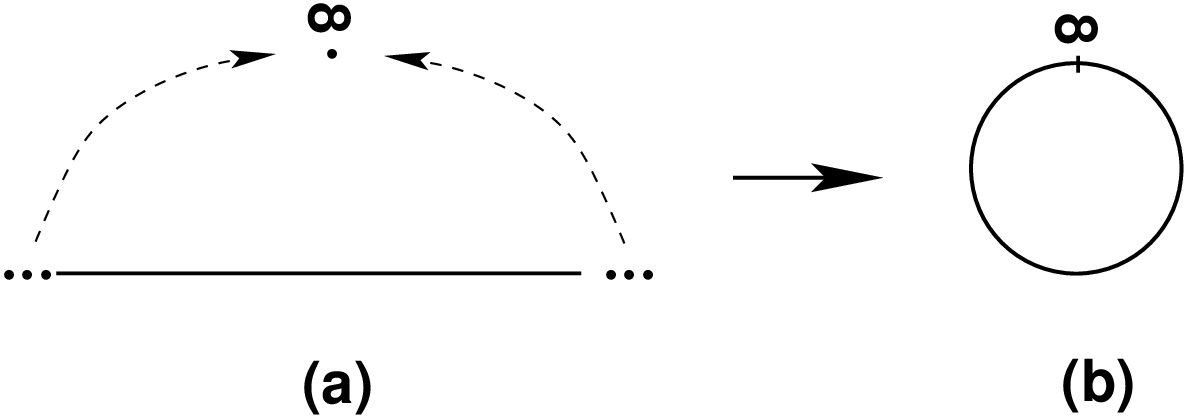}}
\caption[]{Compactification of the real line (a) into the 
circumference $S^{1}$ (b) by adding the point at infinity.}
\label{fig:circle}
\end{figure}

Since states have to be identified by gauge transformations the
topology of the gauge group plays an important physical r\^ole. To
illustrate the point let us first deal with a toy model of a U(1)
gauge theory in 1+1 dimensions.  Later we will be more general.  In
the Hamiltonian formalism gauge transformations $g(\vec{x})$ are
functions defined on $\mathbb{R}$ with values on the gauge group U(1)
\begin{eqnarray}
g:\mathbb{R}\longrightarrow U(1).
\end{eqnarray}
We assume that $g(x)$ is regular at infinity. In this case we can 
add to the real line $\mathbb{R}$ the point at infinity to compactify it
into the circumference $S^{1}$ (see Fig. \ref{fig:circle}). Once this is
done $g(x)$ are functions defined on $S^{1}$ with values on $U(1)=S^{1}$
that can be parametrized as
\begin{eqnarray}
g:S^{1}\longrightarrow U(1), \hspace*{1cm} g(x)=e^{i\alpha(x)},
\end{eqnarray} 
with $x\in [0,2\pi]$. 

Because $S^{1}$ does have a nontrivial topology, $g(x)$ can be divided into
topological sectors. These sectors are labelled by an
integer number $n\in\mathbb{Z}$ and are defined by
\begin{eqnarray}
\alpha(2\pi)=\alpha(0)+2\pi\,n\,\,.
\end{eqnarray}
Geometrically $n$ gives the number of times that the spatial $S^{1}$ winds
around the $S^{1}$ defining the gauge group U(1). This winding number can
be written in a more sophisticated way as
\begin{eqnarray}
\oint_{S^{1}}g(x)^{-1}dg(x)=2\pi n\,\,,
\end{eqnarray}
where the integral is along the spatial $S^{1}$. 

In $\mathbb{R}^{3}$ a similar situation happens with the gauge
group\footnote{Although we present for simplicity only the case of
SU(2), similar arguments apply to any simple group.}  SU(2).  If we
demand $g(\vec{x})\in \mbox{SU(2)}$ to be regular at infinity
$|\vec{x}|\rightarrow\infty$ we can compactify $\mathbb{R}^{3}$ into a
three-dimensional sphere $S^{3}$, exactly as we did in 1+1
dimensions. On the other hand, the function $g(\vec{x})$ can be
written as
\begin{eqnarray}
g(\vec{x})=a^{0}(x)\mathbf{1}+\vec{a}(x)\cdot\vec{\sigma}
\end{eqnarray}
and the conditions $g(x)^{\dagger}g(x)=\mathbf{1}$, $\det{g}=1$
implies that $(a^{0})^{2}+\vec{a}^{\,2}=1$. Therefore SU(2) is a
three-dimensional sphere and $g(x)$ defines a function
\begin{eqnarray}
g:S^{3}\longrightarrow S^{3}.
\end{eqnarray}
As it was the case in 1+1 dimensions here the gauge transformations 
$g(x)$ are also divided into topological sectors labelled this time by
the winding number
\begin{eqnarray}
n={1\over 24\pi^{2}}\int_{S^{3}}d^{3}x\,\epsilon_{ijk}{\rm Tr\,}\left[
\left(g^{-1}\partial_{i}g\right)\left(g^{-1}\partial_{i}g\right) 
\left(g^{-1}\partial_{i}g\right)\right]\in \mathbb{Z}.
\end{eqnarray}

In the two cases analyzed we find that due to the
nontrivial topology of the gauge group manifold the gauge
transformations are divided into different sectors labelled by an
integer $n$. Gauge transformations with different values of $n$ cannot
be smoothly deformed into each other.  The sector with $n=0$
corresponds to those gauge transformations that can be connected with
the identity. 

Now we can be a bit more formal. Let us consider a gauge theory in 3+1
dimensions with gauge group ${G}$ and let us denote by $\mathcal{G}$ the
set of all gauge transformations $\mathcal{G}=\{g:S^{3}\rightarrow G\}$.
At the same time we define $\mathcal{G}_{0}$ as the set of transformations
in $\mathcal{G}$ that can be smoothly deformed into the identity.
Our theory will have topological sectors if
\begin{eqnarray}
\mathcal{G}/\mathcal{G}_{0}\neq \mathbf{1}.
\end{eqnarray}
In the case of the electromagnetism we have seen that Gauss' law 
annihilates physical states. For a nonabelian theory the analysis
is similar and leads to the condition
\begin{eqnarray}
\mathcal{U}(g_{0})|{\rm phys}\rangle\equiv \exp\left[i\int d^{3}x\,
\chi^{a}(\vec{x})\vec{\nabla}\cdot\vec{E}^{a}\right]|{\rm phys}\rangle= 
|{\rm phys}\rangle,
\label{eq:u_cc}
\end{eqnarray}
where $g_{0}(\vec{x})=e^{i\chi^{a}(\vec{x})T^{a}}$ is in the connected
component of the identity $\mathcal{G}_{0}$. The important point to
realize here is that only the elements of $\mathcal{G}_{0}$ can be
written as exponentials of the infinitesimal generators. Since this
generators annihilate the physical states this implies that
$\mathcal{U}(g_{0})|{\rm phys}\rangle =|{\rm phys}\rangle$ only when
$g_{0}\in\mathcal{G}_{0}$.

What happens then with the other topological sectors? If $g\in
\mathcal{G}/\mathcal{G}_{0}$ there is still a unitary operator 
$\mathcal{U}(g)$ that realizes gauge transformations on the Hilbert
space of the theory. However since $g$ is not in the connected
component of the identity, 
it cannot be written as the exponential of Gauss' law. Still
gauge invariance is preserved if $\mathcal{U}(g)$ only changes the
overall global phase of the physical states. For example, if $g_{1}$
is a gauge transformation with winding number $n=1$
\begin{eqnarray}
\mathcal{U}(g_{1})|{\rm phys}\rangle = e^{i\theta}|{\rm phys}\rangle.
\label{eq:theta}
\end{eqnarray}
It is easy to convince oneself that all transformations with winding
number $n=1$ have the same value of $\theta$ modulo $2\pi$. This can
be shown by noticing that if $g(\vec{x})$ has winding number $n=1$
then $g(\vec{x})^{-1}$ has opposite winding number $n=-1$. Since the 
winding number is additive,
given two transformations $g_{1}$, $g_{2}$ with winding number 1, 
$g_{1}^{-1}g_{2}$ has winding number $n=0$. This implies that 
\begin{eqnarray}
|{\rm phys}\rangle=\mathcal{U}(g_{1}^{-1}g_{2})|{\rm phys}\rangle= 
\mathcal{U}(g_{1})^{\dagger}\mathcal{U}(g_{2})
|{\rm phys}\rangle=e^{i(\theta_{2}-\theta_{1})}|{\rm phys}\rangle
\end{eqnarray}
and we conclude that $\theta_{1}=\theta_{2} \mbox{ mod }2\pi$.
Once we know this it is straightforward to conclude that a gauge
transformation $g_{n}(\vec{x})$ with winding number $n$ has the following
action on physical states
\begin{eqnarray}
\mathcal{U}(g_{n})|{\rm phys}\rangle = e^{in\theta}|{\rm phys}\rangle,
\hspace*{1cm} n\in\mathbb{Z}.
\end{eqnarray}

To find a physical interpretation of this result we are going to look
for similar things in other physical situations. One of then is
borrowed from condensed matter physics and refers to the quantum
states of electrons in the periodic potential produced by the ion
lattice in a solid.  For simplicity we discuss the one-dimensional
case where the minima of the potential are separated by a distance
$a$. When the barrier between consecutive degenerate vacua is high
enough we can neglect tunneling between different vacua and consider
the ground state $|na\rangle$ of the potential near the minimum
located at $x=na$ ($n\in\mathbb{Z}$) as possible vacua of the
theory. This vacuum state is, however, not invariant under lattice
translations
\begin{eqnarray}
e^{ia\widehat{P}}|na\rangle = |(n+1)a\rangle.
\end{eqnarray}
However, it is possible to define a new vacuum state
\begin{eqnarray}
|k\rangle = \sum_{n\in\mathbb{Z}}e^{-ikna}|na\rangle,
\end{eqnarray}
which under $e^{ia\widehat{P}}$ transforms by a global phase
\begin{eqnarray}
e^{ia\widehat{P}}|k\rangle = \sum_{n\in\mathbb{Z}}e^{-ikna}|(n+1)a\rangle
=e^{ika}|k\rangle.
\end{eqnarray}
This ground state is labelled by the momentum $k$ and corresponds to the
Bloch wave function. 

This looks very much the same as what we found for nonabelian gauge
theories.  The vacuum state labelled by $\theta$ plays a
r\^ole similar to the Bloch wave function for the periodic potential with
the identification of $\theta$ with the momentum $k$.  To make this
analogy more precise let us write the Hamiltonian for nonabelian gauge
theories
\begin{eqnarray}
H={1\over 2}\int d^{3}x\,\left(\vec{\pi}_{a}\cdot\vec{\pi}_{a}+
\vec{B}_{a}\cdot\vec{B}_{a}\right)
={1\over 2}\int d^{3}x\,\left(\vec{E}_{a}\cdot\vec{E}_{a}+
\vec{B}_{a}\cdot\vec{B}_{a}\right),
\end{eqnarray}
where we have used the expression of the canonical momenta $\pi^{i}_{a}$
and we assume that the Gauss' law constraint is satisfied. Looking at
this Hamiltonian we can interpret the first term within the brackets as
the kinetic energy $T={1\over 2}\vec{\pi}_{a}\cdot\vec{\pi}_{a}$ and the
second term as the potential energy $V={1\over 2}\vec{B}_{a}\cdot\vec{B}_{a}$.
Since $V\geq 0$ we can identify the vacua of the theory as those $\vec{A}$
for which $V=0$, modulo gauge transformations. This happens wherever 
$\vec{A}$ is a pure gauge. However, since we know that the gauge 
transformations are labelled by the winding number we can have an 
infinite number of vacua which cannot be continuously connected with one
another using trivial gauge transformations. 
Taking a representative gauge transformation $g_{n}(\vec{x})$ in
the sector with winding number $n$, these vacua will be associated with 
the gauge potentials
\begin{eqnarray}
\vec{A}=-{1\over ig}g_{n}(\vec{x})\vec{\nabla}g_{n}(\vec{x})^{-1},
\end{eqnarray}
modulo topologically trivial gauge transformations. Therefore the
theory is characterized by an infinite number of vacua $|n\rangle$
labelled by the winding number. These vacua are not gauge invariant.
Indeed, a gauge transformation with $n=1$ will change the winding number
of the vacua in one unit
\begin{eqnarray}
\mathcal{U}(g_{1})|n\rangle = |n+1\rangle.
\end{eqnarray}
Nevertheless a gauge invariant vacuum can be defined as
\begin{eqnarray}
|\theta\rangle = \sum_{n\in\mathbb{Z}}e^{-in\theta}|n\rangle, \hspace*{1cm}
\mbox{with $\theta\in\mathbb{R}$}
\end{eqnarray}
satisfying
\begin{eqnarray}
\mathcal{U}(g_{1})|\theta\rangle = e^{i\theta}|\theta\rangle.
\end{eqnarray}

We have concluded that the nontrivial topology of the gauge group have
very important physical consequences for the quantum theory. In
particular it implies an ambiguity in the definition of the
vacuum. Actually, this can also be seen in a Lagrangian analysis. In
constructing the Lagrangian for the nonabelian version of Maxwell
theory we only consider the term $F_{\mu\nu}^{a}F^{\mu\nu\,a}$.
However this is not the only Lorentz and gauge invariant term that
contains just two derivatives. We can write the more general
Lagrangian
\begin{eqnarray}
\mathcal{L}&=&-{1\over 4}F_{\mu\nu}^{a}F^{\mu\nu\,a}
-{\theta g^{2}\over 32\pi^{2}}F^{a}_{\mu\nu}\widetilde{F}^{\mu\nu\,a},
\label{eq:theta-term}
\end{eqnarray}
where $\widetilde{F}^{a}_{\mu\nu}$ is the dual of the field strength 
defined by
\begin{eqnarray}
\widetilde{F}_{\mu\nu}^{a}={1\over 2}\epsilon_{\mu\nu\sigma\lambda}
F^{\sigma\lambda}.
\end{eqnarray}
The extra term in (\ref{eq:theta-term}),  proportional to
$\vec{E}^{\,a}\cdot\vec{B}^{\,a}$, is actually a total derivative and
does not change the equations of motion or the quantum perturbation
theory. Nevertheless it has several important physical consequences.
One of them is that it violates both parity $P$ and the combination of 
charge conjugation and parity $CP$. This means that since strong interactions
are described by a nonabelian gauge theory with group SU(3) there is an
extra source of $CP$ violation which puts a strong bound on the value of 
$\theta$. One of the consequences of a term like (\ref{eq:theta-term}) in the
QCD Lagrangian is a nonvanishing electric dipole moment for the neutron
\cite{beyond}. The fact that this is not observed impose a very strong
bound on the value of the $\theta$-parameter
\begin{eqnarray}
|\theta|<10^{-9}
\end{eqnarray}
From a theoretical point of view it is still to be fully understood why 
$\theta$ either vanishes or has a very small value.

Finally, the $\theta$-vacuum structure of gauge theories that we found in
the Hamiltonian formalism can be also obtained using path integral techniques
form the Lagrangian (\ref{eq:theta-term}). The second term in
Eq. (\ref{eq:theta-term}) gives then a contribution that depends on the winding
number of the corresponding gauge configuration.

\section{Towards computational rules: Feynman diagrams}
\label{sec:feyndiag}

\begin{fmffile}{anomalies}

As the basic tool to describe the physics of elementary particles, the final aim of 
Quantum Field Theory is the calculation of observables. Most of the information we have
about the physics of subatomic particles comes from scattering experiments. Typically, these experiments
consist of arranging two or more particles to collide with a certain energy and to setup an array of 
detectors, sufficiently far away from the region where the collision takes place,  that register the 
outgoing products of the collision and their momenta (together with other relevant quantum numbers).

Next we discuss how these cross sections can be computed from quantum mechanical amplitudes
and how these amplitudes themselves can be evaluated in perturbative Quantum Field Theory.
We keep our discussion rather heuristic and avoid technical details that can be found in standard texts 
\cite{bjorken}-\cite{banks}. The techniques 
described will be illustrated with the calculation of the cross section for Compton scattering at
low energies.

\subsection{Cross sections and S-matrix amplitudes}
\label{sec:smatrix}

\begin{figure}
\centerline{\epsfxsize=3.2truein\epsfbox{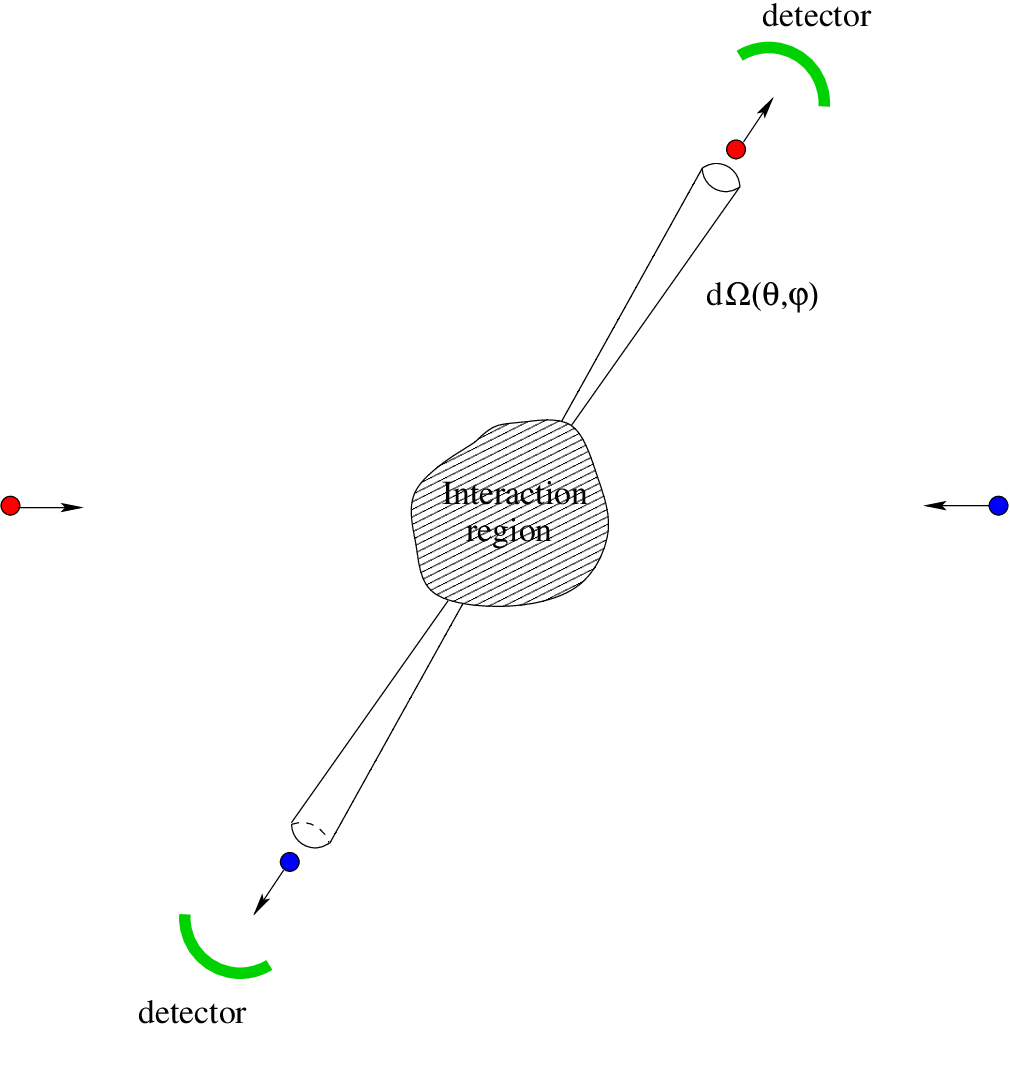}}
\caption[]{Schematic setup of a two-to-two-particles single scattering event in the center of mass reference frame.}
\label{fig:scattering}
\end{figure}

In order to fix ideas let us consider the simplest case of a collision experiment where two 
particles collide to produce again two particles in the final state. The aim of such an 
experiments is a direct measurement of the number of particles per unit time
${dN\over dt}(\theta,\varphi)$ registered by the detector flying within a solid angle $d\Omega$ in the direction 
specified by the polar angles $\theta$, $\varphi$ (see Fig. \ref{fig:scattering}). 
On general grounds we know that this quantity has to be proportional to the flux of incoming 
particles\footnote{This is defined as the number of particles that enter the interaction region per unit time and per unit
area perpendicular to the direction of the beam.}, 
$f_{\rm in}$. 
The proportionality constant defines the differential cross section
\begin{eqnarray}
{dN\over dt}(\theta,\varphi)=f_{\rm in}{d\sigma\over d\Omega}(\theta,\varphi).
\end{eqnarray}
In natural units $f_{\rm in}$ has dimensions of (length)$^{-3}$, and then the differential cross section has dimensions of 
(length)$^{2}$. It depends, apart from the direction $(\theta,\varphi)$,  on the parameters
of the collision (energy, impact parameter, etc.) as well as on the masses and spins of the incoming particles.

Differential cross sections measure the angular distribution of the products of the collision. It is also 
physically interesting to quantify how effective the interaction between the particles is
to produce a nontrivial dispersion. This is measured by the total cross section, which is obtained by
integrating the differential cross section over all directions
\begin{eqnarray}
\sigma=\int_{-1}^{1}d(\cos\theta)\int_{0}^{2\pi}d\varphi \,{d\sigma\over d\Omega}(\theta,\varphi).
\end{eqnarray}
To get some physical intuition of the meaning of the total cross section we can think of the classical 
scattering of a point particle off a sphere of radius $R$.  The particle undergoes a collision only when 
the impact parameter is smaller than the radius of the sphere and a calculation of the total cross section
yields $\sigma=\pi R^{2}$. This is precisely the cross area that the sphere presents to incoming particles.
 
In Quantum Mechanics in general and in Quantum Field Theory in particular the starting point for
the calculation of cross sections is the probability amplitude for the corresponding process. In a
scattering experiment one prepares a system with a given number of particles with definite momenta 
$\vec{p}_{1},\ldots,\vec{p}_{n}$. In the Heisenberg picture this is described
by a time independent state labelled by the incoming momenta of the particles (to keep things simple we
consider spinless particles) that we denote by 
\begin{eqnarray}
|\vec{p}_{1},\ldots,\vec{p}_{n};{\rm in}\rangle.
\label{eq:instate}
\end{eqnarray}
On the other hand, as a result of the scattering experiment a number $k$ of particles with momenta 
$\vec{p}_{1}{}',\ldots,\vec{p}_{k}{}'$
are detected. Thus, the system is now in the ``out'' Heisenberg picture state 
\begin{eqnarray}
|\vec{p}_{1}{}',\ldots,\vec{p}_{k}{}';{\rm out}\rangle
\label{eq:outstate}
\end{eqnarray}
labelled by the momenta of the particles detected at late times. The probability amplitude of detecting $k$ particles in the 
final state with momenta  $\vec{p}_{1}{}',\ldots,\vec{p}_{k}{}'$ in the collision of $n$ particles with initial momenta  
$\vec{p}_{1},\ldots,\vec{p}_{n}$ defines the $S$-matrix amplitude
\begin{eqnarray}
S({\rm in}\rightarrow {\rm out})=
\langle \vec{p}_{1}{}',\ldots,\vec{p}_{k}{}';{\rm out}|\vec{p}_{1},\ldots,\vec{p}_{n};{\rm in}\rangle.
\label{eq:smatrixdef}
\end{eqnarray}

It is very important to keep in mind that both the (\ref{eq:instate}) and (\ref{eq:outstate}) are time-independent states in the Hilbert space of a
very complicated interacting theory. However, since both at early and late times the incoming and outgoing particles
are well apart from each other, the ``in'' and ``out'' states can be thought as two states $|\vec{p}_{1},\ldots,\vec{p}_{n}\rangle$
and $|\vec{p}_{1}{}',\ldots,\vec{p}_{k}{}'\rangle $ of the Fock space of the corresponding free theory in which the coupling constants
are zero. Then, the overlaps (\ref{eq:smatrixdef})
can be written in terms of the matrix elements of an $S$-matrix operator $\widehat{S}$ acting on the free Fock space
\begin{eqnarray}
\langle \vec{p}_{1}{}',\ldots,\vec{p}_{k}{}';{\rm out}|\vec{p}_{1},\ldots,\vec{p}_{n};{\rm in}\rangle=
\langle \vec{p}_{1}{}',\ldots,\vec{p}_{k}{}'|\widehat{S}|\vec{p}_{1},\ldots,\vec{p}_{n}\rangle.
\end{eqnarray}
The operator $\widehat{S}$ is unitary, $\widehat{S}^{\dagger}=\widehat{S}^{-1}$, and its matrix elements are analytic in the 
external momenta.

In any scattering experiment there is the possibility that the particles do not interact at all and the system is left in the 
same initial state. Then it is useful to write the
 $S$-matrix operator as
\begin{eqnarray}
\widehat{S}=\mathbf{1}+i\widehat{T},
\end{eqnarray} 
where $\mathbf{1}$ represents the identity operator. In this way, 
all nontrivial interactions are encoded in the matrix elements of the $T$-operator 
$\langle \vec{p}_{1}{}',\ldots,\vec{p}_{k}{}'|i\widehat{T}|\vec{p}_{1},\ldots,\vec{p}_{n}\rangle$. Since momentum has to be conserved,
a global delta function can be factored out from these matrix elements to define the invariant scattering amplitude $i\mathcal{M}$
\begin{eqnarray}
\langle \vec{p}_{1}{}',\ldots,\vec{p}_{k}{}'|i\widehat{T}|\vec{p}_{1},\ldots,\vec{p}_{n}\rangle=
(2\pi)^{4}\delta^{(4)}\Bigg(\sum_{\rm initial}p_{i}-\sum_{\rm final}p'_{f}\Bigg)i\mathcal{M}(\vec{p}_{1},\ldots,\vec{p}_{n};
\vec{p}_{1}{}',\ldots,\vec{p}_{k}{}')
\label{eq:invampdef}
\end{eqnarray}

Total and differential cross sections can be now computed from the invariant amplitudes.
Here we consider the most common situation in which two particles with 
momenta $\vec{p}_{1}$ and $\vec{p}_{2}$ collide to produce  a number of particles in the final
state with momenta $\vec{p}_{i}{}'$. In this case the total cross section is given by
\begin{eqnarray}
\sigma={1\over (2\omega_{p_{1}}) (2\omega_{p_{2}}) |\vec{v}_{12}|}\begin{Large}\int\end{Large}
\Bigg[\prod\limits_{
\begin{array}{c}
  \\[-0.7cm]
\mbox{\tiny final} \\[-0.3cm]
\mbox{\tiny states}
\end{array}
}{d^{3}p_{i}'\over (2\pi)^{3}}{1\over 2\omega_{p_{i}'}}
\Bigg]\Big|\mathcal{M}_{i\rightarrow f}\Big|^{2}(2\pi)^{4}\delta^{(4)}\Bigg(p_{1}+p_{2}-\sum\limits_{
\begin{array}{c}
  \\[-0.7cm]
\mbox{\tiny final} \\[-0.3cm]
\mbox{\tiny states}
\end{array}
}p_{i}'\Bigg),
\end{eqnarray}
where $\vec{v}_{12}$ is the relative velocity of the two scattering particles. The corresponding differential 
cross section can be computed by dropping the integration over the directions of the final momenta. We will use
this expression later in Section \ref{sec:compton} to evaluate the cross section of Compton scattering.

We seen how particle cross sections are determined by the invariant amplitude for the corresponding proccess, 
i.e. $S$-matrix amplitudes. In general, in Quantum Field Theory it is not possible to compute exactly these amplitudes.
However, in many physical situations it can be argued that interactions are weak enough to allow for a perturbative 
evaluation. In what follows we will describe how $S$-matrix elements can be computed
in perturbation theory using Feynman diagrams and rules. These are very convenient bookkeeping techniques 
allowing both to keep track of all contributions to a process at a given order in perturbation theory, and computing 
the different contributions.

\subsection{Feynman rules}
\label{sec:subsecfr}

The basic quantities to be computed in Quantum Field Theory are vacuum expectation values of products of the 
operators of the theory.  Particularly useful are time-ordered Green functions,
\begin{eqnarray}
\langle \Omega|T\Big[\mathcal{O}_{1}(x_{1})\ldots\mathcal{O}_{n}(x_{n})\Big]|\Omega\rangle,
\label{eq:basic_correlation}
\end{eqnarray}
where $|\Omega\rangle$ is the the ground state of the theory and the time ordered product is defined 
\begin{eqnarray}
T\Big[\mathcal{O}_{i}(x)\mathcal{O}_{j}(y)\Big]=\theta(x^{0}-y^{0})\mathcal{O}_{i}(x)\mathcal{O}_{j}(y)
+\theta(y^{0}-x^{0})\mathcal{O}_{j}(y)\mathcal{O}_{i}(x).
\end{eqnarray}
The generalization to products with more than two operators is straightforward: operators are always multiplied
in time order, those evaluated at earlier times always to the right. The interest of these kind of correlation functions lies
in the fact that they can be related to $S$-matrix amplitudes through the so-called reduction formula. To keep our 
discussion as simple as possible we will not derived it or even write it down in full detail. Its form for different theories
can be found in any textbook. Here it suffices to say that the reduction formula simply states that 
any $S$-matrix amplitude can be written in terms of the Fourier transform of a time-ordered correlation function. Morally speaking
\begin{eqnarray}
&\langle \vec{p}_{1}{}\!',\ldots,\vec{p}_{m}{}\!';{\rm out}|\vec{p}_{1},\ldots,\vec{p}_{n};{\rm in}\rangle &\nonumber \\[0.2cm]
&\Downarrow & \\[0.2cm]
&\displaystyle{\int}  d^{4}x_{1}\ldots\displaystyle{\int} d^{4}y_{n}\langle\Omega|T\Big[\phi(x_{1})^{\dagger}\ldots\phi(x_{m})^{\dagger}
\phi(y_{1})\ldots \phi(y_{n})
\Big]|\Omega\rangle\, e^{ip_{1}{}\!'\cdot x_{1}}\ldots e^{-ip_{n}\cdot y_{n}},&
\nonumber
\end{eqnarray}
where $\phi(x)$ is the field whose elementary excitations are the particles involved in the scattering.

The reduction formula reduces the problem of computing $S$-matrix amplitudes to that of 
evaluating time-ordered correlation functions of field operators. These quantities are easy to compute exactly 
in the free theory. For an interacting theory the situation is more complicated, however. Using path 
integrals, the vacuum expectation value of the time-ordered product of a number of operators can be expressed
as
\begin{eqnarray}
\langle\Omega|T\Big[\mathcal{O}_{1}(x_{1})\ldots\mathcal{O}_{n}(x_{n})\Big]|\Omega\rangle=
{\displaystyle{\int}\mathscr{D}\phi\mathscr{D}\phi^{\dagger}\,
\mathcal{O}_{1}(x_{1})\ldots\mathcal{O}_{n}(x_{n})\,e^{iS[\phi,\phi^{\dagger}]}\over
\displaystyle{\int}\mathscr{D}\phi\mathscr{D}\phi^{\dagger}\,e^{iS[\phi,\phi^{\dagger}]}
}.
\label{eq:gmlpi}
\end{eqnarray} 
For an theory with interactions, neither the path integral in the numerator or in the denominator is Gaussian and 
they cannot be calculated exactly. However, Eq. (\ref{eq:gmlpi}) is still very useful. The action $S[\phi,\phi^{\dagger}]$ can
be split into the free (quadratic) piece and the interaction part
\begin{eqnarray}
S[\phi,\phi^{\dagger}]=S_{0}[\phi,\phi^{\dagger}]+S_{\rm int}[\phi,\phi^{\dagger}].
\end{eqnarray}
All dependence in the coupling constants of the theory comes from the second piece. Expanding now
$\exp[iS_{\rm int}]$ in power series of the coupling constant we find that each term in the series expansion of both the numerator 
and the denominator has the structure
\begin{eqnarray}
\int\mathscr{D}\phi\mathscr{D}\phi^{\dagger}\Big[\ldots\Big]e^{iS_{0}[\phi,\phi^{\dagger}]},
\label{eq:generictermpi}
\end{eqnarray}
where  ``$\ldots$'' denotes certain monomial of fields. The important point is that now the integration measure only involves the free action, 
and the path integral in (\ref{eq:generictermpi}) is Gaussian and therefore can be computed exactly. The same conclusion can be reached
using the operator formalism. In this case the correlation function (\ref{eq:basic_correlation}) can be expressed in terms of 
correlation functions of operators in the interaction picture. The advantage of using this picture is that the fields satisfy the free
equations of motion and therefore can be expanded in creation-annihilation operators. The correlations functions are then easily 
computed using Wick's theorem.

Putting together all the previous ingredients we can calculate $S$-matrix amplitudes in a perturbative series in the
coupling constants of the field theory. This can be done using Feynman diagrams and rules, a very economical way to  
compute each term in the perturbative expansion of the $S$-matrix amplitude
for a given process. We will not detail the the construction of Feynman rules but just present them heuristically. 

For the sake of concreteness we focus on the case of QED first. Going back to Eq. (\ref{eq:qed_lagrangian}) we 
expand the covariant derivative to write the action
\begin{eqnarray}
S_{\rm QED}=\int d^{4}x\left[-{1\over 4}F_{\mu\nu}F^{\mu\nu}
+\overline{\psi}(i\fslash{\partial}-m)\psi+e \overline{\psi}\gamma^{\mu}\psi A_{\mu}\right].
\end{eqnarray}
The action contains two types of particles, photons and fermions, that we represent by straight and wavy lines respectively
\begin{eqnarray*}
\parbox{42mm}{
\begin{fmfgraph*}(90,60)
\fmfleft{i1}
\fmfright{o1}
\fmf{fermion}{i1,o1}
\end{fmfgraph*} 
}\hspace*{1cm}
\parbox{42mm}{
\begin{fmfgraph*}(90,60)
\fmfleft{i1}
\fmfright{o1}
\fmf{photon}{i1,o1}
\end{fmfgraph*}
}
\end{eqnarray*}
The arrow in the fermion line does not represent the direction of the momentum but the flux of (negative) charge. This distinguishes particles
form antiparticles: if the fermion propagates from left to right (i.e. in the direction of the charge flux) it represents a particle, whereas when 
it does from right to left it corresponds to an antiparticle. Photons are not charged and therefore wavy lines do not have orientation.

Next we turn to the interaction part of the action containing a photon field, a spinor and its conjugate. In a Feynman diagram this 
corresponds to the vertex
\begin{eqnarray*}
\parbox{42mm}{
\begin{fmfgraph*}(90,60)
\fmfleft{i1,o1}
\fmfright{i2}
\fmf{fermion}{i1,v1,o1}
\fmf{photon}{i2,v1}
\end{fmfgraph*}
}
\end{eqnarray*}
Now, in order to compute an $S$-matrix amplitude to a given order in the coupling constant $e$ for a process with
certain number of incoming and outgoing asymptotic states one only has to draw all possible diagrams with as
many vertices as the order in perturbation theory, 
and the corresponding number and type of external legs. It is very important to keep in mind that in joining the fermion lines
among the different building blocks of the diagram one has to respect their orientation. This reflects the conservation of the
electric charge. In addition one should only consider diagrams that are topologically non-equivalent, i.e. that they cannot be 
smoothly deformed into one another keeping the external legs fixed\footnote{From the point of view of the operator formalism, 
the requirement of 
considering only diagrams that are topologically nonequivalent comes from the fact that each diagram represents
a certain Wick contraction in the correlation function of interaction-picture operators.}. 

To show in a practical way how Feynman diagrams are drawn, we consider Bhabha scattering, i.e. the 
elastic dispersion of an electron and a positron:
\begin{eqnarray*}
e^{+}+e^{-} \longrightarrow e^{+}+e^{-}.
\end{eqnarray*}
Our problem is to compute the $S$-matrix amplitude to the leading order in the electric charge. Because 
the QED vertex contains a photon line and our process does not have photons either in the initial or the final
states we find that drawing a Feynman diagram requires at least two vertices. In fact, the leading 
contribution is of order $e^{2}$ and comes from the following two diagrams, each containing two vertices:

\begin{eqnarray*}
\parbox{42mm}{
\begin{fmfgraph*}(90,60)
\fmfleft{i1,i2}
\fmfright{o1,o2}
\fmflabel{$e^{-}$}{i1}
\fmflabel{$e^{+}$}{i2}
\fmflabel{$e^{-}$}{o1}
\fmflabel{$e^{+}$}{o2}
\fmf{fermion}{i1,v1,i2}
\fmf{photon}{v1,v2}
\fmf{fermion}{o2,v2,o1}
\end{fmfgraph*}
} +(-1)\times \hspace*{0.7cm}
\parbox{42mm}{
\begin{fmfgraph*}(90,60)
\fmfleft{i1,i2}
\fmfright{o1,o2}
\fmflabel{$e^{-}$}{i1}
\fmflabel{$e^{+}$}{i2}
\fmflabel{$e^{-}$}{o1}
\fmflabel{$e^{+}$}{o2}
\fmf{fermion}{i1,v1,o1}
\fmf{photon}{v1,v2}
\fmf{fermion}{o2,v2,i2}
\end{fmfgraph*}
} 
 \\
\end{eqnarray*}
Incoming and outgoing particles appear respectively on the left and the right of this diagram. Notice how the identification of
electrons and positrons is done comparing the direction of the charge flux with the direction of propagation. 
For electrons the flux of charges goes in the direction
of propagation, whereas for positrons the two directions are opposite. These are the only two diagrams that can be drawn at this
order in perturbation theory. It is important to include a relative minus sign between the two contributions. 
To understand the origin of this sign we have to remember that in the operator formalism Feynman diagrams are just a way 
to encode a particular Wick contraction of field operators in the interaction picture. The factor of $-1$ reflects the relative sign 
in Wick contractions represented by the two diagrams, due to the fermionic character of the Dirac field.

We have learned how to draw Feynman diagrams in QED. Now one needs to compute the contribution of each one to the 
corresponding amplitude using the so-called Feynman rules. The idea is simple: given a diagram, each of its building blocks
(vertices as well as external and internal lines) has an associated contribution that allows the calculation of the corresponding
diagram. In the case of QED in the Feynman gauge, we have the following correspondence for vertices and internal propagators:

\begin{eqnarray*}
\parbox{42mm}{
\begin{fmfgraph*}(90,60)
\fmfleft{i1}
\fmfright{o1}
\fmflabel{$\alpha$}{i1}
\fmflabel{$\beta$}{o1}
\fmf{fermion}{i1,o1}
\end{fmfgraph*}
} &\Longrightarrow \hspace*{0.5cm} & \left({i\over \fslash{p}-m+i\varepsilon}\right)_{\beta\alpha} \\
\parbox{42mm}{
\begin{fmfgraph*}(90,60)
\fmfleft{i1}
\fmfright{o1}
\fmflabel{$\mu$}{i1}
\fmflabel{$\nu$}{o1}
\fmf{photon}{i1,o1}
\end{fmfgraph*}
}&\Longrightarrow \hspace*{0.5cm} & {-i\eta_{\mu\nu}\over p^{2}+i\varepsilon} \\
\parbox{42mm}{
\begin{fmfgraph*}(90,60)
\fmfleft{i1,o1}
\fmfright{i2}
\fmflabel{$\mu$}{i2}
\fmflabel{$\alpha$}{i1}
\fmflabel{$\beta$}{o1}
\fmf{fermion}{i1,v1,o1}
\fmf{photon}{i2,v1}
\end{fmfgraph*}
}&\Longrightarrow \hspace*{0.5cm} & -ie\gamma^{\mu}_{\beta\alpha} (2\pi)^{4}\delta^{(4)}(p_{1}+p_{2}+p_{3}).\\[0.5cm]
\end{eqnarray*}
A change in the gauge would reflect in an extra piece in the photon propagator.
The delta function implementing conservation of momenta is written using the convention that all momenta
are entering the vertex. In addition, one has to perform an integration over all momenta running in internal lines with the measure
\begin{eqnarray}
\int {d^{d}p\over (2\pi)^{4}},
\end{eqnarray}
and introduce a factor of $-1$ for each fermion loop in the diagram\footnote{The contribution of each diagram comes also multiplied
by a degeneracy factor that takes into account in how many ways a given Wick contraction can be done. In QED, however, these
factors are equal to 1 for many diagrams.}.

In fact, some of the integrations over internal momenta can actually be done using the 
delta function at the vertices, leaving just a global delta function implementing the total momentum 
conservation in the diagram [cf. Eq. (\ref{eq:invampdef})]. It is even possible that all integrations can be eliminated in this way. This is the
case when we have tree level diagrams, i.e. those without closed loops. In the case of diagrams with loops there will be as many 
remaining integrations as the number of independent loops in the diagram. 

The need to perform integrations over internal momenta in loop diagrams has important consequences in Quantum Field 
Theory. The reason is that in many cases the resulting integrals are ill-defined, i.e. are divergent either at small or large
values of the loop momenta. In the first case one speaks of {\em infrared divergences} and usually they cancel once all contributions
to a given process are added together. More profound, however, 
are the divergences appearing at large internal momenta. These {\em ultraviolet divergences} cannot be cancelled and have to be
dealt through the renormalization procedure. We will discuss this problem in some detail in Section \ref{sec:renormalization}.

Were we computing time-ordered (amputated) correlation function of operators, this would be all. However, in the case of 
$S$-matrix amplitudes this is not the whole story. In addition to the previous rules here one needs to attach contributions
also to the external legs in the diagram. These are the wave functions of the corresponding asymptotic states
containing information about the spin and momenta of the incoming and outgoing particles. In
the case of QED these contributions are:

\begin{eqnarray*}
\mbox{Incoming fermion:} \hspace*{1cm} 
\parbox{42mm}{
\begin{fmfgraph*}(90,60)
\fmfleft{i1}
\fmfright{o1}
\fmflabel{$\alpha$}{i1}
\fmf{fermion}{i1,v1}
\fmfblob{.25w}{v1}
\fmf{phantom}{v1,o1}
\end{fmfgraph*}
} &\Longrightarrow \hspace*{0.5cm} & u_{\alpha}(\vec{p},s)
\end{eqnarray*}
\begin{eqnarray*}
\mbox{Incoming antifermion:} \hspace*{1cm} 
\parbox{42mm}{
\begin{fmfgraph*}(90,60)
\fmfleft{i1}
\fmfright{o1}
\fmflabel{$\alpha$}{i1}
\fmf{fermion}{v1,i1}
\fmfblob{.25w}{v1}
\fmf{phantom}{v1,o1}
\end{fmfgraph*}
} &\Longrightarrow \hspace*{0.5cm} & \overline{v}_{\alpha}(\vec{p},s) 
\hspace*{0.5cm}
\end{eqnarray*}
\begin{eqnarray*}
\mbox{Outgoing fermion:} \hspace*{1cm} 
\parbox{42mm}{
\begin{fmfgraph*}(90,60)
\fmfleft{i1}
\fmfright{o1}
\fmflabel{$\alpha$}{o1}
\fmf{fermion}{v1,o1}
\fmfblob{.25w}{v1}
\fmf{phantom}{i1,v1}
\end{fmfgraph*}
} &\Longrightarrow \hspace*{0.5cm} & \overline{u}_{\alpha}(\vec{p},s)  
\end{eqnarray*}
\begin{eqnarray*}
\mbox{Outgoing antifermion:} \hspace*{1cm} 
\parbox{42mm}{
\begin{fmfgraph*}(90,60)
\fmfleft{i1}
\fmfright{o1}
\fmflabel{$\alpha$}{o1}
\fmf{fermion}{o1,v1}
\fmfblob{.25w}{v1}
\fmf{phantom}{i1,v1}
\end{fmfgraph*}
} &\Longrightarrow \hspace*{0.5cm} & v_{\alpha}(p,s)  
\hspace*{1.5cm}
\end{eqnarray*}
\begin{eqnarray*}
\mbox{Incoming photon:} \hspace*{1cm} 
\parbox{42mm}{
\begin{fmfgraph*}(90,60)
\fmfleft{i1}
\fmfright{o1}
\fmflabel{$\mu$}{i1}
\fmf{photon}{i1,v1}
\fmfblob{.25w}{v1}
\fmf{phantom}{v1,o1}
\end{fmfgraph*}
} &\Longrightarrow \hspace*{0.5cm} &  \epsilon_{\mu}(\vec{k},\lambda)
\hspace*{1cm}
\end{eqnarray*}
\begin{eqnarray*}
\mbox{Outgoing photon:}  
\parbox{42mm}{
\begin{fmfgraph*}(90,60)
\fmfleft{i1}
\fmfright{o1}
\fmflabel{$\mu$}{o1}
\fmf{photon}{v1,o1}
\fmfblob{.25w}{v1}
\fmf{phantom}{i1,v1}
\end{fmfgraph*}
} &\Longrightarrow \hspace*{0.5cm} & \epsilon_{\mu}(\vec{k},\lambda)^{*}
\end{eqnarray*}
Here we have assumed that the momenta for incoming (resp. outgoing) particles are entering (resp. leaving)
the diagram. It is important also to keep in mind that in the computation of $S$-matrix amplitudes all external states are on-shell.
In Section \ref{sec:compton} we illustrate the 
use of the Feynman rules for QED with the case of the Compton scattering. 

The application of Feynman diagrams to carry out computations in perturbation theory is extremely convenient.
It provides a very useful bookkeeping technique  to account for all contributions to a process at a given 
order in the coupling constant. This does not mean that the calculation of Feynman diagrams is an 
easy task. The number of diagrams contributing to the process grows very fast with the order in 
perturbation theory and the integrals that appear in calculating loop diagrams also get 
very complicated. This means that, generically, the calculation of Feynman diagrams beyond the 
first few orders very often requires the use of computers. 

Above we have illustrated the Feynman rules with the case of QED. Similar rules can be computed for other 
interacting quantum field theories with scalar, vector or spinor fields. In the case of the nonabelian gauge theories 
introduced in Section \ref{sec:gauge_fields} we have:

\begin{eqnarray*}
\parbox{42mm}{
\begin{fmfgraph*}(90,60)
\fmfleft{i1}
\fmfright{o1}
\fmflabel{$\alpha,i$}{i1}
\fmflabel{$\beta,j$}{o1}
\fmf{fermion}{i1,o1}
\end{fmfgraph*}
} &\Longrightarrow \hspace*{0.5cm} & \left({i\over \fslash{p}-m+i\varepsilon}\right)_{\beta\alpha}\delta_{ij} 
\end{eqnarray*}
\begin{eqnarray*}
\parbox{42mm}{
\begin{fmfgraph*}(90,60)
\fmfleft{i1}
\fmfright{o1}
\fmflabel{$\mu,a$}{i1}
\fmflabel{$\nu,b$}{o1}
\fmf{gluon}{i1,o1}
\end{fmfgraph*}
}&\Longrightarrow \hspace*{0.5cm} & {-i\eta_{\mu\nu}\over p^{2}+i\varepsilon} \delta^{ab}
\hspace*{1.7cm}
\end{eqnarray*}
\begin{eqnarray*}
\parbox{42mm}{
\begin{fmfgraph*}(90,60)
\fmfleft{i1,o1}
\fmfright{i2}
\fmflabel{$\mu,a$}{i2}
\fmflabel{$\alpha,i$}{i1}
\fmflabel{$\beta,j$}{o1}
\fmf{fermion}{i1,v1,o1}
\fmf{gluon}{i2,v1}
\end{fmfgraph*}
}&\Longrightarrow \hspace*{0.5cm} & -ig\gamma^{\mu}_{\beta\alpha} t^{a}_{ij} 
\hspace*{6.7cm} \\[0.3cm]
\end{eqnarray*}
\begin{eqnarray*}
\parbox{42mm}{
\begin{fmfgraph*}(90,60)
\fmfleft{i1,o1}
\fmfright{i2}
\fmflabel{$\mu,a$}{i2}
\fmflabel{$\nu,b$}{i1}
\fmflabel{$\sigma,c$}{o1}
\fmf{gluon}{i1,v1,o1}
\fmf{gluon}{i2,v1}
\end{fmfgraph*}
}&\Longrightarrow \hspace*{0.5cm} & g\,f^{abc}\Big[\eta^{\mu\nu}(p^{\sigma}_{1}-p^{\sigma}_{2})+
\mbox{permutations}\Big]
\hspace*{2.2cm}\\[0.3cm]
\end{eqnarray*}
\begin{eqnarray*}
\parbox{42mm}{
\begin{fmfgraph*}(90,60)
\fmfleft{i1,o1}
\fmfright{i2,o2}
\fmflabel{$\mu,a$}{i1}
\fmflabel{$\nu,b$}{i2}
\fmflabel{$\sigma,c$}{o1}
\fmflabel{$\lambda,d$}{o2}
\fmf{gluon}{i1,v1,o1}
\fmf{gluon}{i2,v1,o2}
\end{fmfgraph*}
}&\Longrightarrow \hspace*{0.5cm} & -ig^{2}\Big[f^{abe}f^{cde}\Big(\eta^{\mu\sigma}\eta^{\nu\lambda}-
\eta^{\mu\lambda}\eta^{\nu\sigma}\Big)+\mbox{permutations}\Big]
\\[0.5cm]
\end{eqnarray*}

It is not our aim here to give a full and detailed description of the Feynman rules for nonabelian gauge theories. It suffices
to point out that, unlike the case of QED, here the gauge fields can interact among themselves. Indeed, the 
three and four gauge field vertices are a consequence of the cubic and quartic terms in the action 
\begin{eqnarray}
S=-{1\over 4}\int d^{4}x\,F_{\mu\nu}^{a}F^{\mu\nu\,a},
\end{eqnarray}
where the nonabelian gauge field strength $F^{a}_{\mu\nu}$ is given in Eq. (\ref{eq:fmunuNA}). 
The self-interaction of the nonabelian gauge fields has crucial dynamical consequences and its at the very heart of
its success in describing the physics of elementary particles. 

\subsection{An example: Compton scattering}
\label{sec:compton}

To illustrate the use of Feynman diagrams and Feynman rules we compute the cross section for the dispersion of photons
by free electrons, the so-called Compton scattering:
\begin{eqnarray*}
\gamma(k,\lambda)+e^{-}(p,s)\longrightarrow \gamma(k',\lambda')+e^{-}(p',s').
\end{eqnarray*}
In brackets we have indicated the momenta for the different particles, as well as the polarizations and spins
of the incoming and outgoing photon and electrons respectively. The first step is to identify all the diagrams contributing to the process
at leading order. Taking into account that the vertex of QED contains two fermion and one photon leg, 
it is straightforward to realize that any diagram contributing to the process at hand must contain 
at least two vertices. Hence the leading contribution is of
order $e^{2}$. A first diagram we can draw is:

\begin{eqnarray*}
\parbox{42mm}{
\begin{fmfgraph*}(90,60)
\fmfleft{i1,i2}
\fmfright{o1,o2}
\fmflabel{$p,s$}{i2}
\fmflabel{$k,\lambda$}{i1}
\fmflabel{$p',s'$}{o2}
\fmflabel{$k',\lambda'$}{o1}
\fmf{photon}{i1,v1}
\fmf{fermion}{i2,v1}
\fmf{fermion}{v1,v2}
\fmf{photon}{v2,o1}
\fmf{fermion}{v2,o2}
\end{fmfgraph*}
} \\
\end{eqnarray*}
This is, however, not the only possibility. Indeed, there is a second possible diagram:

\begin{eqnarray*}
\parbox{40mm}{
\begin{fmfgraph*}(90,60)
\fmfleft{i1,i2}
\fmfright{o1,o2}
\fmflabel{$p,s$}{i2}
\fmflabel{$k,\lambda$}{i1}
\fmflabel{$p',s'$}{o1}
\fmflabel{$k',\lambda'$}{o2}
\fmf{photon}{i1,v1}
\fmf{fermion}{i2,v2,v1,o1}
\fmf{photon}{v2,o2}
\end{fmfgraph*}
} \\[0.3cm]
\end{eqnarray*}
It is important to stress that these two diagrams are topologically nonequivalent, since deforming one
into the other would require changing the label of the external legs. Therefore the leading $\mathcal{O}(e^{2})$ 
amplitude has to be computed adding the contributions from both of them. 

Using the Feynman rules of QED we find
\begin{eqnarray}
\parbox{20mm}{
\begin{fmfgraph*}(45,30)
\fmfleft{i1,i2}
\fmfright{o1,o2}
\fmf{photon}{i1,v1}
\fmf{fermion}{i2,v1}
\fmf{fermion}{v1,v2}
\fmf{photon}{v2,o1}
\fmf{fermion}{v2,o2}
\end{fmfgraph*}
}\!\!\!+
\parbox{20mm}{
\begin{fmfgraph*}(45,30)
\fmfleft{i1,i2}
\fmfright{o1,o2}
\fmf{photon}{i1,v1}
\fmf{fermion}{i2,v2,v1,o1}
\fmf{photon}{v2,o2}
\end{fmfgraph*}
}=\,\,\,(ie)^{2}\overline{u}(\vec{p}\,',s')\fslash{\epsilon}\,'(\vec{k}\,',\lambda')^{*}{\fslash{p}+\fslash{k}+m_{e}
\over (p+k)^{2}-m^{2}_{e}}\fslash{\epsilon}(\vec{k},\lambda)u(\vec{p},s) \nonumber \\
+\,\,\,(ie)^{2}\overline{u}(\vec{p}\,',s')\fslash{\epsilon}(\vec{k},\lambda){\fslash{p}-\fslash{k}'+m_{e}
\over (p-k')^{2}-m^{2}_{e}}\fslash{\epsilon}\,'(\vec{k}\,',\lambda')^{*}u(\vec{p},s) .
\label{eq:compton1}
\end{eqnarray}
Because the leading order contributions only involve tree-level diagrams, there is no integration over
internal momenta and therefore we are left with a purely algebraic expression for the amplitude. To 
get an explicit expression we begin by simplifying the numerators. The following simple identity 
turns out to be very useful for this task
\begin{eqnarray}
\fslash{a}\fslash{b}=-\fslash{b}\fslash{a}+2(a\cdot b) \mathbf{1}.
\label{eq:identity_slash}
\end{eqnarray}
Indeed, looking at the first term in Eq. (\ref{eq:compton1}) we have 
\begin{eqnarray}
(\fslash{p}+\fslash{k}+m_{e})\fslash{\epsilon}(\vec{k},\lambda)u(\vec{p},s)&=&
-\fslash{\epsilon}(\vec{k},\lambda)(\fslash{p}-m_{e})u(\vec{p},s)+
\fslash{k}\fslash{\epsilon}(\vec{k},\lambda)u(\vec{p},s) \nonumber \\
&+&
2p\cdot {\epsilon}(\vec{k},\lambda)u(\vec{p},s),
\end{eqnarray}
where we have applied the identity (\ref{eq:identity_slash}) on the first term inside the parenthesis. The first
term on the right-hand side of this equation vanishes identically because of Eq. (\ref{eq:positive}). The expression can be
further simplified if we restrict our attention to the Compton scattering at low energy when electrons are 
nonrelativistic. This means that all spatial momenta are much smaller than the electron mass
\begin{eqnarray}
|\vec{p}|,|\vec{k}|,|\vec{p}\,'|, |\vec{k}\,'|\ll m_{e}.
\label{eq:lowenergyapproxCs}
\end{eqnarray}
In this approximation we have that $p^{\mu}, p'{}^{\mu}\approx(m_{e},\vec{0})$ and therefore
\begin{eqnarray}
p\cdot {\epsilon}(\vec{k},\lambda)=0.
\label{eq:transversality_elecmomentum}
\end{eqnarray}
This follows from the absence of temporal photon polarization. Then we conclude that at low energies
\begin{eqnarray}
(\fslash{p}+\fslash{k}+m_{e})\fslash{\epsilon}(\vec{k},\lambda)u(\vec{p},s)=
\fslash{k}\fslash{\epsilon}(\vec{k},\lambda)u(\vec{p},s)
\end{eqnarray}
and similarly for the second term in Eq. (\ref{eq:compton1})
\begin{eqnarray}
(\fslash{p}-\fslash{k}'+m_{e})\fslash{\epsilon}\,'(\vec{k}',\lambda')^{*}u(\vec{p},s)=
-\fslash{k}'\fslash{\epsilon}\,'(\vec{k}',\lambda')^{*}u(\vec{p},s).
\end{eqnarray}

Next, we turn to the denominators in Eq. (\ref{eq:compton1}). 
As it was explained in Section \ref{sec:subsecfr}, in computing scattering amplitudes incoming and 
outgoing particles should 
have on-shell momenta,
\begin{eqnarray}
p^{2}=m^{2}_{e}=p'{}^{2} \hspace*{0.9cm} \mbox{and} \hspace*{0.9cm} k^{2}=0=k'{}^{2}.
\end{eqnarray}
Then, the two denominator in Eq. (\ref{eq:compton1}) simplify respectively to
\begin{eqnarray}
(p+k)^{2}-m^{2}_{e}=p^{2}+k^{2}+2p\cdot k-m^{2}_{e}=2p\cdot k=2\omega_{p}|\vec{k}|-2\vec{p}\cdot
\vec{k}
\end{eqnarray}
and
\begin{eqnarray}
(p-k')^{2}-m^{2}_{e}=p^{2}+k'{}^{2}+2p\cdot k'-m^{2}_{e}=-2p\cdot k'
=-2\omega_{p}|\vec{k}\,'|+2\vec{p}\cdot\vec{k}\,'.
\end{eqnarray}
Working again in the low energy approximation (\ref{eq:lowenergyapproxCs}) these two expressions simplify to
\begin{eqnarray}
(p+k)^{2}-m^{2}_{e}\approx
2m_{e}|\vec{k}|, \hspace*{1cm}
(p-k')^{2}-m^{2}_{e}\approx
-2m_{e}|\vec{k}\,'|.
\end{eqnarray}
Putting together all these expressions we find that at low energies
\begin{eqnarray}
\parbox{20mm}{
\begin{fmfgraph*}(45,30)
\fmfleft{i1,i2}
\fmfright{o1,o2}
\fmf{photon}{i1,v1}
\fmf{fermion}{i2,v1}
\fmf{fermion}{v1,v2}
\fmf{photon}{v2,o1}
\fmf{fermion}{v2,o2}
\end{fmfgraph*}
}\!\!\!+
\parbox{20mm}{
\begin{fmfgraph*}(45,30)
\fmfleft{i1,i2}
\fmfright{o1,o2}
\fmf{photon}{i1,v1}
\fmf{fermion}{i2,v2,v1,o1}
\fmf{photon}{v2,o2}
\end{fmfgraph*}
} \hspace*{8cm} \nonumber \\
\approx\,\,\, {(ie)^{2}\over 2m_{e}}\overline{u}(\vec{p}\,',s')\left[\fslash{\epsilon}\,'(\vec{k}\,'\lambda')^{*}
{\fslash{k}\over |\vec{k}|}\epsilon(\vec{k},\lambda)
+\epsilon(\vec{k},\lambda)
{\fslash{k}'\over |\vec{k}\,'|}\fslash{\epsilon}\,'(\vec{k}\,'\lambda')^{*}
\right]u(\vec{p},s).
\end{eqnarray}
Using now again the identity (\ref{eq:identity_slash}) a number of times as well as the transversality condition of the
polarization vectors (\ref{eq:transvcondpolvec}) we end up with a handier equation
\begin{eqnarray}
\parbox{20mm}{
\begin{fmfgraph*}(45,30)
\fmfleft{i1,i2}
\fmfright{o1,o2}
\fmf{photon}{i1,v1}
\fmf{fermion}{i2,v1}
\fmf{fermion}{v1,v2}
\fmf{photon}{v2,o1}
\fmf{fermion}{v2,o2}
\end{fmfgraph*}
}\!\!\!+
\parbox{20mm}{
\begin{fmfgraph*}(45,30)
\fmfleft{i1,i2}
\fmfright{o1,o2}
\fmf{photon}{i1,v1}
\fmf{fermion}{i2,v2,v1,o1}
\fmf{photon}{v2,o2}
\end{fmfgraph*}
} 
&\approx & {e^{2}\over m_{e}}\Big[\epsilon(\vec{k},\lambda)\cdot\epsilon'(\vec{k}\,',\lambda')^{*}\Big]
\overline{u}(\vec{p}\,',s'){\fslash{k}\over |\vec{k}|}u(\vec{p},s) \nonumber \\
&+&{e^{2}\over 2m_{e}}\overline{u}(\vec{p}\,',s')\fslash{\epsilon}(\vec{k},\lambda)
\fslash{\epsilon}\,'(\vec{k}\,',\lambda')^{*}\left({\fslash{k}\over |\vec{k}|}-{\fslash{k}'\over |\vec{k}\,'|}\right)
u(\vec{p},s). \hspace*{1cm}
\end{eqnarray}
With a little bit of effort we can show that the second term on the right-hand side vanishes. First we notice that
in the low energy limit $|\vec{k}|\approx |\vec{k}\,'|$. If in addition we make use the conservation of momentum
${k}-{k}\,'={p}\,'-{p}$ and the identity (\ref{eq:positive})
\begin{eqnarray}
\overline{u}(\vec{p}\,',s')\fslash{\epsilon}(\vec{k},\lambda)
\fslash{\epsilon}\,'(\vec{k}\,',\lambda')^{*}\left({\fslash{k}\over |\vec{k}|}-{\fslash{k}'\over |\vec{k}\,'|}\right)
u(\vec{p},s) \hspace*{3cm} \nonumber \\
\approx\,\,\, {1\over |\vec{k}|}
\overline{u}(\vec{p}\,',s')\fslash{\epsilon}(\vec{k},\lambda)
\fslash{\epsilon}\,'(\vec{k}\,',\lambda')^{*}(\fslash{p}'-m_{e})u(\vec{p},s).
\end{eqnarray}
Next we use the identity (\ref{eq:identity_slash}) to take the term $(\fslash{p}'-m_{e})$ to the right. Taking into 
account that in the low energy limit the electron four-momenta are orthogonal to the photon polarization vectors
[see Eq. (\ref{eq:transversality_elecmomentum})] we conclude that
\begin{eqnarray}
\overline{u}(\vec{p}\,',s')\fslash{\epsilon}(\vec{k},\lambda)
\fslash{\epsilon}\,'(\vec{k}\,',\lambda')^{*}(\fslash{p}'-m_{e})u(\vec{p},s) \hspace*{3cm} \nonumber \\
=
\overline{u}(\vec{p}\,',s')(\fslash{p}'-m_{e})\fslash{\epsilon}(\vec{k},\lambda)
\fslash{\epsilon}\,'(\vec{k}\,',\lambda')^{*}u(\vec{p},s)=0
\end{eqnarray}
where the last identity follows from the equation satisfied by the conjugate positive-energy spinor, 
$\overline{u}(\vec{p}\,',s')(\fslash{p}'-m_{e})=0$. 

After all these lengthy manipulations we have finally arrived at the expression of the invariant amplitude for 
the Compton scattering at low energies
\begin{eqnarray}
i\mathcal{M}={e^{2}\over m_{e}}
\Big[\epsilon(\vec{k},\lambda)\cdot\epsilon'(\vec{k}\,',\lambda')^{*}\Big]
\overline{u}(\vec{p}\,',s'){\fslash{k}\over |\vec{k}|}u(\vec{p},s).
\end{eqnarray}
The calculation of the cross section involves computing the modulus squared of this quantity. For many physical applications,
however, one is interested in the dispersion of photons with a given polarization by electrons that are not polarized, i.e. whose
spins are randomly distributed. In addition in many situations either we are not interested, or there is no way to measure the final
polarization of the outgoing electron. This is for example the situation in cosmology, where we do not have any information about
the polarization of the free electrons in the primordial plasma before or after the scattering with photons (although we have ways
to measure the polarization of the scattered photons). 

To describe this physical situations we have to average over initial electron polarization (since we do not know them) and sum over
all possible final electron polarization (because our detector is blind to this quantum number),
\begin{eqnarray}
\overline{|i\mathcal{M}|^{2}}={1\over 2}\left({e^{2}\over m_{e}|\vec{k}|}\right)^{2}
\Big|\epsilon(\vec{k},\lambda)\cdot\epsilon'(\vec{k}\,',\lambda')^{*}\Big|^{2}
\sum_{s=\pm{1\over 2}}\sum_{s'=\pm{1\over 2}}\Big|
\overline{u}(\vec{p}\,',s'){\fslash{k}}u(\vec{p},s)
\Big|^{2}.
\end{eqnarray}
The factor of ${1\over 2}$ comes from averaging over the two possible polarizations of the incoming electrons. 
The sums in this expression can be calculated without much difficulty. Expanding the absolute value explicitly 
\begin{eqnarray}
\sum_{s=\pm{1\over 2}}\sum_{s'=\pm{1\over 2}}\Big|
\overline{u}(\vec{p}\,',s'){\fslash{k}}u(\vec{p},s)
\Big|^{2}
=\sum_{s=\pm{1\over 2}}\sum_{s'=\pm{1\over 2}}
\Big[u(\vec{p},s)^{\dagger}\fslash{k}^{\dagger}\overline{u}(\vec{p}\,',s')^{\dagger}\Big]\Big[
\overline{u}(\vec{p}\,',s'){\fslash{k}}u(\vec{p},s)
\Big],
\end{eqnarray}
using that $\gamma^{\mu}{}^\dagger=\gamma^{0}\gamma^{\mu}\gamma^{0}$ and after some
manipulation one finds that
\begin{eqnarray}
\sum_{s=\pm{1\over 2}}\sum_{s'=\pm{1\over 2}}\Big|
\overline{u}(\vec{p}\,',s'){\fslash{k}}u(\vec{p},s)
\Big|^{2}&=&
\left[\sum_{s=\pm{1\over 2}}u_{\alpha}(\vec{p},s)\overline{u}_{\beta}(\vec{p},s)\right](\fslash{k})_{\beta\sigma}
\left[\sum_{s'=\pm{1\over 2}}u_{\sigma}(\vec{p}\,',s')\overline{u}_{\rho}(\vec{p}\,',s')\right](\fslash{k})_{\rho\alpha} \nonumber \\
&=&{\rm Tr\,}\Big[(\fslash{p}+m_{e})\fslash{k}(\fslash{p}\,'+m_{e})\fslash{k}\Big],
\end{eqnarray}
where the final expression has been computed using the completeness relations in Eq. (\ref{eq:varrelspindir}). The final evaluation
of the trace can be done using the standard Dirac matrices identities. 
Here we compute it applying again the relation (\ref{eq:identity_slash}) to commute $\fslash{p}'$ and 
$\fslash{k}$. Using that $k^{2}=0$ and that we are working in the low energy limit we have\footnote{We use also the fact that the
trace of the product of an odd number of Dirac matrices is always zero.}
\begin{eqnarray}
{\rm Tr\,}\Big[(\fslash{p}+m_{e})\fslash{k}(\fslash{p}\,'+m_{e})\fslash{k}\Big]=2(p\cdot k)(p'\cdot k){\rm Tr\,}\mathbf{1}
\approx 8m_{e}^{2}|\vec{k}|^{2}.
\end{eqnarray}
This gives the following value for the invariant amplitude
\begin{eqnarray}
\overline{|i\mathcal{M}|^{2}}=4e^{4}\Big|\epsilon(\vec{k},\lambda)\cdot\epsilon'(\vec{k}\,',\lambda')^{*}\Big|^{2}
\end{eqnarray}

Plugging $\overline{|i\mathcal{M}|^{2}}$ into the formula for the differential cross section we get 
\begin{eqnarray}
{d\sigma\over d\Omega}={1\over 64\pi^{2}m_{e}^{2}}\overline{|i\mathcal{M}|^{2}}=\left({e^{2}\over 4\pi m_{e}}\right)^{2}
\Big|\epsilon(\vec{k},\lambda)\cdot\epsilon'(\vec{k}\,',\lambda')^{*}\Big|^{2}.
\end{eqnarray}
The prefactor of the last equation is precisely the square of the so-called classical electron radius $r_{\rm cl}$. 
In fact, the previous differential cross section 
can be rewritten as
\begin{eqnarray}
{d\sigma\over d\Omega}={3\over 8\pi}\sigma_{T}\Big|\epsilon(\vec{k},\lambda)\cdot\epsilon'(\vec{k}\,',\lambda')^{*}\Big|^{2},
\label{eq:compton_lowenergy_cs}
\end{eqnarray}
where $\sigma_{T}$ is the total Thomson cross section
\begin{eqnarray}
\sigma_{T}={e^{4}\over 6\pi m_{e}^{2}}={8\pi\over 3}r_{\rm cl}^{2}.
\end{eqnarray}

The result (\ref{eq:compton_lowenergy_cs}) is relevant in many areas of Physics, but its importance is paramount in the
study of the cosmological microwave background (CMB). Just before recombination the universe is filled by a plasma of electrons
interacting with photons via Compton scattering, with temperatures of the order of 1 keV. Electrons are then nonrelativistic 
($m_{e}\sim 0.5$ MeV) and the approximations leading to Eq. (\ref{eq:compton_lowenergy_cs}) are fully valid. Because we do
not know the polarization state of the photons before being scattered by electrons we have to consider the cross section 
averaged over incoming photon polarizations. From Eq. (\ref{eq:compton_lowenergy_cs}) we see that this is proportional
to
\begin{eqnarray}
{1\over 2}\sum_{\lambda=1,2}\Big|\epsilon(\vec{k},\lambda)\cdot\epsilon'(\vec{k}\,',\lambda')^{*}\Big|^{2}=
\left[{1\over 2}\sum_{\lambda=1,2}\epsilon_{i}(\vec{k},\lambda)\epsilon_{j}(\vec{k},\lambda)^{*}\right]
\epsilon_{j}(\vec{k}\,',\lambda')\epsilon_{i}(\vec{k}\,',\lambda')^{*}.
\end{eqnarray}
The sum inside the brackets can be computed using the normalization of the polarization vectors, $|\vec{\epsilon}\,(\vec{k},
\lambda)|^{2}=1$, and the transversality condition $\vec{k}\cdot \vec{\epsilon}(\vec{k},\lambda)=0$
 \begin{eqnarray}
{1\over 2}\sum_{\lambda=1,2}\Big|\epsilon(\vec{k},\lambda)\cdot\epsilon'(\vec{k}\,',\lambda')^{*}\Big|^{2}&=&
{1\over 2}\left(\delta_{ij}-{k_{i}k_{j}\over |\vec{k}|^{2}}\right)
\epsilon_{j}'(\vec{k}\,',\lambda')\epsilon_{i}'(\vec{k}\,',\lambda')^{*} \nonumber \\
&=& {1\over 2}\Big[1-|\vec{\ell}\cdot\vec{\epsilon}\,'(\vec{k}\,',\lambda')|^{2}
\Big],
\end{eqnarray}
where $\vec{\ell}={\vec{k}\over |\vec{k}|}$ is the unit vector in the direction of the incoming photon.

From the last equation we conclude that Thomson scattering suppresses all polarizations parallel to the direction of the incoming 
photon $\vec{\ell}$, whereas the differential cross section reaches the maximum in the plane normal to $\vec{\ell}$. 
If photons would collide with the electrons in the plasma with the same intensity from all directions, the result would 
be an unpolarized CMB radiation. The fact that polarization is actually measured in the CMB carries crucial information about the
physics of the plasma before recombination and, as a consequence, about the very early universe (see for example \cite{dodelson} for
a throughout discussion).

\section{Symmetries}
\label{sec:symmetries}

\subsection{Noether's theorem}

In Classical Mechanics and Classical Field Theory there is a basic
result that relates symmetries and conserved charges. This is called Noether's
theorem and states that for each continuous symmetry of the system
there is conserved current. In its simplest version in Classical
Mechanics it can be easily proved. Let us consider a Lagrangian
$L(q_{i},\dot{q}_{i})$ which is invariant under a transformation
$q_{i}(t)\rightarrow q'_{i}(t,\epsilon)$ labelled by a parameter
$\epsilon$.  This means that $L(q',\dot{q}')=L(q,\dot{q})$ without
using the equations of motion\footnote{The following result can 
be also derived a more general situations where the Lagrangian changes
by a total time derivative.}. If $\epsilon\ll 1$ we can consider an
infinitesimal variation of the coordinates
$\delta_{\epsilon}q_{i}(t)$ and the invariance of
the Lagrangian implies
\begin{eqnarray}
0=\delta_{\epsilon}L(q_{i},\dot{q}_{i})=
{\partial L\over \partial q_{i}}\delta_{\epsilon}q_{i}+
{\partial L\over \partial \dot{q}_{i}}{\delta_{\epsilon}\dot{q}_{i}}=
\left[{\partial L\over \partial q_{i}}-{d\over dt}
{\partial L\over \partial \dot{q}_{i}}\right]\delta_{\epsilon}q_{i}
+{d\over dt}
\left({\partial L\over \partial\dot{q}_{i}}\delta_{\epsilon}q_{i}\right).
\label{eq:noether1}
\end{eqnarray}
When $\delta_{\epsilon}q_{i}$ is applied on a solution to the
equations of motion the term inside the square brackets vanishes and
we conclude that there is a conserved quantity
\begin{eqnarray}
\dot{Q}=0 \hspace*{0.5cm} \mbox{with} \hspace*{0.5cm}
Q\equiv {\partial L\over \partial \dot{q}_{i}}\delta_{\epsilon}q_{i}.
\end{eqnarray}
Notice that in this derivation it is crucial that the symmetry depends on
a continuous parameter since otherwise the infinitesimal variation of 
the Lagrangian in Eq. (\ref{eq:noether1}) does not make sense.

In Classical Field Theory a similar result holds. Let us consider for
simplicity a theory of a single field $\phi(x)$. We say that the 
variations $\delta_{\epsilon}\phi$ depending on a continuous parameter 
$\epsilon$ are a symmetry of the theory if, without using the equations
of motion, the Lagrangian density changes by 
\begin{eqnarray}
\delta_{\epsilon}\mathcal{L}=\partial_{\mu}K^{\mu}.
\end{eqnarray}
If this happens then the action remains invariant and so do the equations
of motion. Working out now the variation of $\mathcal{L}$ under 
$\delta_{\epsilon}\phi$ we find
\begin{eqnarray}
\partial_{\mu}K^{\mu}={\partial\mathcal{L}\over \partial(\partial_{\mu}\phi)}
\partial_{\mu}\delta_{\epsilon}\phi+{\partial\mathcal{L}\over\partial\phi}
\delta_{\epsilon}\phi=
\partial_{\mu}\left({\partial\mathcal{L}\over \partial(\partial_{\mu}\phi)}
\delta_{\epsilon}\phi\right)+\left[{\partial\mathcal{L}\over\partial\phi}-
\partial_{\mu}\left({\partial\mathcal{L}\over \partial(\partial_{\mu}\phi)}
\right)\right]\delta_{\epsilon}\phi.
\end{eqnarray}
If $\phi(x)$ is a solution to the equations of motion the last terms 
disappears, and we find that there is a conserved current
\begin{eqnarray}
\partial_{\mu}J^{\mu}=0 \hspace*{0.5cm} \mbox{with} \hspace*{0.5cm}
J^{\mu}={\partial\mathcal{L}\over \partial(\partial_{\mu}\phi)}
\delta_{\epsilon}\phi-K^{\mu}.
\end{eqnarray}

Actually a conserved current implies the existence of a  charge
\begin{eqnarray}
Q\equiv \int d^{3}x\,J^{0}(t,\vec{x})
\label{eq:canonical}
\end{eqnarray}
which is conserved
\begin{eqnarray}
{dQ\over dt}=\int d^{3}x\,{\partial_{0}}J^{0}(t,\vec{x})=-
\int d^{3}x\,{\partial_{i}}J^{i}(t,\vec{x})=0,
\end{eqnarray}
provided the fields vanish at infinity fast enough. Moreover, the
conserved charge $Q$ is a Lorentz scalar. After canonical quantization
the charge $Q$ defined by Eq.  (\ref{eq:canonical}) is
promoted to an operator that generates the symmetry on the fields
\begin{eqnarray}
\delta\phi=i[\phi,Q].
\label{eq:generators}
\end{eqnarray}

As an example we can consider a scalar field $\phi(x)$ which under a
coordinate transformation 
$x\rightarrow x'$ changes as $\phi'(x')=\phi(x)$. In particular performing
a space-time translation $x^{\mu\,'}=x^{\mu}+a^{\mu}$ we have
\begin{eqnarray}
\phi'(x)-\phi(x)=-a^{\mu}\partial_{\mu} \phi+\mathcal{O}(a^2)
\hspace*{0.5cm} \Longrightarrow \hspace*{0.5cm} \delta\phi=-a^{\mu}
\partial_{\mu}\phi.
\end{eqnarray}
Since the Lagrangian density is also a scalar quantity, it transforms under
translations as
\begin{eqnarray}
\delta\mathcal{L}=-a^{\mu}\partial_{\mu}\mathcal{L}.
\end{eqnarray}
Therefore the corresponding conserved charge is
\begin{eqnarray}
J^{\mu}=-{\partial\mathcal{L}\over\partial(\partial_{\mu}\phi)}
a^{\nu}\partial_{\nu}\phi+a^{\mu}\mathcal{L}\equiv -a_{\nu}T^{\mu\nu},
\end{eqnarray}
where we introduced the energy-momentum tensor
\begin{eqnarray}
T^{\mu\nu}={\partial\mathcal{L}\over\partial(\partial_{\mu}\phi)}
\partial^{\nu}\phi-\eta^{\mu\nu}\mathcal{L}.
\end{eqnarray}
We find that associated with the invariance of the theory with 
respect to space-time translations there are four conserved currents
defined by $T^{\mu\nu}$ with $\nu=0,\ldots,3$, each one associated with the
translation along a space-time direction. These four currents 
form a rank-two tensor under Lorentz transformations satisfying 
\begin{eqnarray}
\partial_{\mu}T^{\mu\nu}=0.
\end{eqnarray}
The associated conserved charges are given by
\begin{eqnarray}
P^{\nu}=\int d^{3}x\,T^{0\nu}
\end{eqnarray}
and correspond to the total energy-momentum content of the field configuration. 
Therefore
the energy density of the field is given by $T^{00}$ while $T^{0i}$ is the 
momentum density. In the quantum theory the $P^{\mu}$ are the generators of
space-time translations.

Another example of a symmetry related with a physically relevant
conserved charge is the global phase invariance of the Dirac
Lagrangian (\ref{eq:dirac_eq}), $\psi\rightarrow e^{i\theta}\psi$. For
small $\theta$ this corresponds to variations
$\delta_{\theta}\psi=i\theta\psi$,
$\delta_{\theta}\overline{\psi}=-i\theta\overline{\psi}$ which by
Noether's theorem result in the conserved charge
\begin{eqnarray}
j^{\mu}=\overline{\psi}\gamma^{\mu}\psi, \hspace*{1cm} \partial_{\mu}j^{\mu}=0.
\end{eqnarray}
Thus implying the existence of a conserved charge 
\begin{eqnarray}
Q=\int d^{3}x\overline{\psi}\gamma^{0}\psi=
\int d^{3}x\psi^{\dagger}\psi.
\end{eqnarray}
In physics there are several instances of global U(1) symmetries that
act as phase shifts on spinors. This is the case, for example, of the
baryon and lepton number conservation in the Standard Model. A more
familiar case is the U(1) local symmetry associated with
electromagnetism. Notice that although in this case we are dealing
with a local symmetry, $\theta\rightarrow e\alpha(x)$, the invariance
of the Lagrangian holds in particular for global transformations and
therefore there is a conserved current
$j^{\mu}=e\overline{\psi}\gamma^{\mu}\psi$. In
Eq. (\ref{eq:interac_QED}) we saw that the spinor is coupled to the
photon field precisely through this current. Its time component is
the electric charge density $\rho$, while the spatial
components are the current density vector $\vec{\jmath}$.

This analysis can be carried over also to nonabelian unitary global symmetries
acting as
\begin{eqnarray}
\psi_{i}\longrightarrow U_{ij}\psi_{j}, \hspace*{1cm} U^{\dagger}U=\mathbf{1}
\end{eqnarray}
and leaving invariant the Dirac Lagrangian when we have
several fermions.  If we write the matrix
$U$ in terms of the hermitian group generators $T^{a}$ as
\begin{eqnarray}
U=\exp\left(i\alpha_{a}T^{a}\right), \hspace*{1cm} (T^{a})^{\dagger}=T^{a},
\end{eqnarray}
we find the conserved current
\begin{eqnarray}
j^{\mu\,a}=\overline{\psi}_{i}T^{a}_{ij}\gamma^{\mu}\psi_{j}, 
\hspace*{1cm} \partial_{\mu}j^{\mu}=0.
\end{eqnarray}
This is the case, for example of the approximate flavor symmetries in
hadron physics.  The simplest example is the isospin symmetry that
mixes the quarks $u$ and $d$
\begin{eqnarray}
\left(
\begin{array}{c}
u \\
d
\end{array}
\right) \longrightarrow 
M\left(
\begin{array}{c}
u \\
d
\end{array}
\right), \hspace*{1cm} M\in {\rm SU(2)}.
\end{eqnarray}
Since the proton is a bound state of two quarks $u$ and one quark $d$ while
the neutron is made out of one quark $u$ and two quarks $d$, this isospin
symmetry reduces at low energies to the well known isospin transformations
of nuclear physics that mixes protons and neutrons.

\subsection{Symmetries in the quantum theory}

We have seen that in canonical quantization the conserved charges $Q^{a}$
associated
to symmetries by Noether's theorem are operators implementing the symmetry
at the quantum level. Since the charges are conserved they must commute with
the Hamiltonian
\begin{eqnarray}
[Q^{a},H]=0.
\end{eqnarray}
There are several possibilities in the quantum mechanical
realization of a symmetry:

{\bf Wigner-Weyl realization.}
In this case the ground state of the theory $|0\rangle$ is invariant
under the symmetry. Since the symmetry is generated by $Q^{a}$ this means that
\begin{eqnarray}
\mathcal{U}(\alpha)|0\rangle\equiv e^{i\alpha_{a}Q^{a}}|0\rangle = |0\rangle
\hspace*{0.5cm}
\Longrightarrow \hspace*{0.5cm} Q^{a}|0\rangle =0.
\label{eq:invariance}
\end{eqnarray}
At the same time the fields of the theory have to transform according to some
irreducible representation of the group generated by the $Q^{a}$. From Eq.
(\ref{eq:generators}) it is easy to prove that
\begin{eqnarray}
\mathcal{U}(\alpha)\phi_{i}\mathcal{U}(\alpha)^{-1}=
U_{ij}(\alpha)\phi_{j},
\end{eqnarray}
where $U_{ij}(\alpha)$ is an element of the representation in which
the field $\phi_{i}$ transforms. If we consider now the quantum state
associated with the operator $\phi_{i}$ 
\begin{eqnarray}
|i\rangle = \phi_{i}|0\rangle
\end{eqnarray}
we find that because of the invariance of the vacuum (\ref{eq:invariance}) 
the states $|i\rangle$ transform in the same representation as $\phi_{i}$
\begin{eqnarray}
\mathcal{U}(\alpha)|i\rangle=\mathcal{U}(\alpha)\phi_{i}
\mathcal{U}(\alpha)^{-1}\mathcal{U}(\alpha)|0\rangle =
U_{ij}(\alpha)\phi_{j}|0\rangle=U_{ij}(\alpha)|j\rangle.
\end{eqnarray}
Therefore the spectrum of the theory is classified in multiplets of the
symmetry group. In addition, since $[H,\mathcal{U}(\alpha)]=0$ all states
in the same multiplet have the same energy. If we consider one-particle
states, then going to the rest frame we conclude that all states in the 
same multiplet have exactly the same mass.

{\bf Nambu-Goldstone realization.}
In our previous discussion the result that the spectrum of the theory 
is classified according to multiplets of the symmetry group depended crucially
on the invariance of the ground state. However this condition is not mandatory
and one can relax it to consider theories where the vacuum state is not
left invariant by the symmetry
\begin{eqnarray}
e^{i\alpha_{a}Q^{a}}|0\rangle \neq |0\rangle
\hspace*{0.5cm}
\Longrightarrow \hspace*{0.5cm} Q^{a}|0\rangle \neq 0.
\label{eq:nambu_goldstone}
\end{eqnarray}
In this case it is also said that the symmetry is spontaneously broken by
the vacuum.

To illustrate the consequences of (\ref{eq:nambu_goldstone}) we consider
the example of a number scalar fields $\varphi^{i}$ ($i=1,\ldots,N$) 
whose dynamics is governed by the Lagrangian
\begin{eqnarray}
\mathcal{L}={1\over 2}\partial_{\mu}\varphi^{i}\partial^{\mu}\varphi^{i}
-V(\varphi),
\end{eqnarray}
where we assume that $V(\phi)$ is bounded from below.
This theory is globally invariant under the transformations
\begin{eqnarray}
\delta\varphi^{i}=\epsilon^{a}(T^{a})^{i}_{j}\varphi^{j},
\end{eqnarray}
with $T^{a}$, $a=1,\ldots,{1\over 2}N(N-1)$  the generators of
the group SO$(N)$.

To analyze the structure of vacua of the theory we construct the Hamiltonian
\begin{eqnarray}
H=\int d^{3}x\left[{1\over 2}\pi^{i}\pi^{i}+{1\over 2}\vec{\nabla}\varphi^{i}
\cdot\vec{\nabla}\varphi^{i}+V(\varphi)\right]
\end{eqnarray}
and look for the minimum of 
\begin{eqnarray}
\mathcal{V}(\varphi)=\int d^{3}x\left[{1\over 2}\vec{\nabla}\varphi^{i}
\cdot\vec{\nabla}\varphi^{i}+V(\varphi)\right].
\end{eqnarray}
Since we are interested in finding constant field configurations, $\vec{\nabla}
\varphi=\vec{0}$ to preserve translational  invariance, 
the vacua of the potential $\mathcal{V}(\varphi)$ 
coincides with the vacua of $V(\varphi)$. Therefore the minima of the
potential correspond to the vacuum expectation 
values\footnote{For simplicity we consider that the minima of $V(\phi)$
occur at zero potential.}
\begin{eqnarray}
\langle\varphi^{i}\rangle: \hspace*{1cm} V(\langle\varphi^{i}\rangle)=0, 
\hspace*{0.5cm} \left.{\partial V\over\partial\varphi^{i}}\right|_{\varphi^{i}
=\langle\varphi^{i}\rangle}=0.
\end{eqnarray}

We divide the generators $T^{a}$ of SO($N$) into two groups: Those denoted by
$H^{\alpha}$ ($\alpha=1,\ldots,h$) that satisfy
\begin{eqnarray}
(H^{\alpha})^{i}_{j}\langle\varphi^{j}\rangle=0.
\end{eqnarray}
This means that the vacuum configuration $\langle\varphi^{i}\rangle$
is left invariant by the transformation generated by $H^{\alpha}$. For
this reason we call them {\em unbroken generators}. Notice that the
commutator of two unbroken generators also annihilates the vacuum 
expectation value,
$[H^{\alpha},H^{\beta}]_{ij}\langle\varphi^{j}\rangle=0$. Therefore
the generators $\{H^{\alpha}\}$ form a subalgebra of the algebra of the
generators of SO($N$). The subgroup of the symmetry group generated by
them is realized \`a la Wigner-Weyl.

The remaining generators $K^{A}$, with $A=1,\ldots,{1\over 2}N(N-1)-h$, by 
definition do not preserve the vacuum expectation value of the field
\begin{eqnarray}
(K^{A})^{i}_{j}\langle\varphi^{j}\rangle\neq 0.
\end{eqnarray}
These will be called the {\em broken generators}. Next we prove a very important
result concerning the broken generators known as the Goldstone theorem: 
for each generator broken by the vacuum expectation value there is a massless
excitation.

The mass matrix of the excitations around the vacuum
$\langle\varphi^{i}\rangle$ is determined by the quadratic part of the
potential. Since we assumed that $V(\langle\varphi\rangle)=0$ and we
are expanding around a minimum, the first term in the expansion of the
potential $V(\varphi)$ around the vacuum expectation values is given
by
\begin{eqnarray}
V(\varphi)=\left.{\partial^{2}V\over \partial\varphi^{i}\partial\varphi^{j}}
\right|_{\varphi=\langle\varphi\rangle}(\varphi^{i}-\langle\varphi^{i}\rangle)
(\varphi^{j}-\langle\varphi^{j}\rangle)+\mathcal{O}
\left[(\varphi-\langle\varphi\rangle)^3\right]
\end{eqnarray}
and the mass matrix is:
\begin{eqnarray}
M_{ij}^{2}\equiv 
\left.{\partial^{2}V\over \partial\varphi^{i}\partial\varphi^{j}}
\right|_{\varphi=\langle\varphi\rangle}.
\end{eqnarray}
In order to avoid a cumbersome notation we do not show explicitly the
dependence of the mass matrix on the vacuum expectation values
$\langle\varphi^{i}\rangle$.

To extract some information about the possible zero modes of the
mass matrix, we write down the conditions that follow from the invariance
of the potential under $\delta\varphi^{i}=\epsilon^{a}(T^{a})^{i}_{j}
\varphi^{j}$. At first order in $\epsilon^{a}$
\begin{eqnarray}
\delta V(\varphi)=
\epsilon^{a}{\partial V\over \partial\varphi^{i}}(T^{a})^{i}_{j}
\varphi^{j}=0.
\end{eqnarray}
Differentiating this expression with respect to $\varphi^{k}$ we arrive
at
\begin{eqnarray}
{\partial^{2} V\over \partial\varphi^{i}\partial\varphi^{k}}(T^{a})^{i}_{j}
\varphi^{j}
+{\partial V\over \partial\varphi^{i}}(T^{a})^{i}_{k}=0.
\end{eqnarray}
Now we evaluate this expression in the vacuum $\varphi^{i}=\langle\varphi^{i}
\rangle$. Then the derivative in the second term cancels while the 
second derivative in the first one gives the mass matrix. Hence we find
\begin{eqnarray}
M^{2}_{ik}(T^{a})^{i}_{j}\langle\varphi^{j}\rangle=0.
\end{eqnarray}
Now we can write this expression for both broken and unbroken
generators.  For the unbroken ones, since
$(H^{\alpha})^{i}_{j}\langle\varphi^{j}\rangle=0$, we find a trivial
identity $0=0$. On the other hand for the broken generators we have
\begin{eqnarray}
M^{2}_{ik}(K^{A})^{i}_{j}\langle\varphi^{j}\rangle=0.
\end{eqnarray}
Since $(K^{A})^{i}_{j}\langle\varphi^{j}\rangle\neq 0$ this equation
implies that the mass matrix has as many zero modes as broken
generators. Therefore we have proven Goldstone's theorem: associated
with each broken symmetry there is a massless mode in the theory.
Here we have presented a classical proof of the theorem. In the quantum 
theory the proof follows the same lines as the one presented here
but one has to consider the effective action containing the effects
of the quantum corrections to the classical Lagrangian.

As an example to illustrate this theorem,  we consider a SO(3) invariant
scalar field theory with a ``mexican hat'' potential 
\begin{eqnarray}
V(\vec{\varphi})={\lambda\over 4}\left(\vec{\varphi}^{\,\,2}-a^2\right)^2.
\end{eqnarray}
The vacua of the theory correspond to the configurations satisfying
$\langle\vec{\varphi}\rangle^{\,\,2}=a^2$. In field space this
equation describes a two-dimensional sphere and each solution is just
a point in that sphere.  Geometrically it is easy to visualize that a
given vacuum field configuration, i.e. a point in the sphere, is
preserved by SO(2) rotations around the axis of the sphere that passes
through that point. Hence the vacuum expectation value of the scalar
field breaks the symmetry according to
\begin{eqnarray}
\langle\vec{\varphi}\rangle:\hspace*{0.5cm}
{\rm SO(3)}\longrightarrow {\rm SO(2)}.
\end{eqnarray}
Since SO(3) has three generators and SO(2) only one we see that two
generators are broken and therefore there are two massless Goldstone bosons.
Physically this massless modes can be thought of as corresponding to 
excitations along the surface of the sphere 
$\langle\vec{\varphi}\rangle^{\,\,2}=a^2$.

Once a minimum of the potential has been chosen we can proceed to
quantize the excitations around it. Since the vacuum only leaves
invariant a SO(2) subgroup of the original SO(3) symmetry group it
seems that the fact that we are expanding around a particular vacuum
expectation value of the scalar field has resulted in a lost of
symmetry. This is however not the case. The full quantum theory is
symmetric under the whole symmetry group SO(3). This is reflected in
the fact that the physical properties of the theory do not depend on
the particular point of the sphere $\langle\vec{\varphi}
\rangle^{\,\,2}=a^2$ that we
have chosen. Different vacua are related by the
full SO(3) symmetry and therefore should give the same physics.

It is very important to realize that given a theory with a vacuum
determined by $\langle\vec{\varphi}\rangle$ all other possible vacua
of the theory are unaccessible in the infinite volume limit. This means
that two vacuum states $|0_{1}\rangle$, $|0_{2}\rangle$ corresponding
to different vacuum expectation values of the scalar field are
orthogonal $\langle 0_{1}|0_{2}\rangle=0$ and cannot be connected by any
local observable $\Phi(x)$, $\langle 0_{1}|\Phi(x)|0_{2}\rangle=0$.
Heuristically this can be understood by noticing that in the infinite
volume limit switching from one vacuum into another one requires changing
the vacuum expectation value of the field everywhere in space at the same
time, something that cannot be done by any local operator.  Notice  that this
is  radically different to our expectations based on the Quantum Mechanics
of a system with a finite number of  degrees of  freedom.

In High Energy Physics the typical example of a Goldstone boson is the
pion, associated with the spontaneous breaking of the global chiral
isospin ${\rm SU(2)}_{L}\times{\rm SU(2)}_{R}$ symmetry. This symmetry
acts independently in the left- and right-handed spinors as
\begin{eqnarray}
\left(
\begin{array}{c}
u_{L,R} \\
d_{L,R}
\end{array}
\right) \longrightarrow 
M_{L,R}\left(
\begin{array}{c}
u_{L,R} \\
d_{L,R}
\end{array}
\right), \hspace*{1cm} M_{L,R}\in {\rm SU}(2)_{L,R}
\end{eqnarray}
Presumably since the quarks are confined at low energies this symmetry is
spontaneously broken down to the diagonal SU(2) acting in the same way
on the left- and right-handed components of the spinors. Associated
with this symmetry breaking there is a Goldstone mode which is
identified as the pion. Notice, nevertheless, that the
SU(2)$_{L}\times$SU(2)$_{R}$ would be an exact global symmetry of the
QCD Lagrangian only in the limit when the masses of the quarks are
zero $m_{u},m_{d}\rightarrow 0$.  Since these quarks have nonzero
masses the chiral symmetry is only approximate and as a consequence
the corresponding Goldstone boson is not massless. That is why 
pions have  masses, although they are the lightest particle among the hadrons.

Symmetry breaking appears also in many places in condensed matter. For
example, when a solid crystallizes from a liquid the translational
invariance that is present in the liquid phase is broken to a discrete
group of translations that represent the crystal lattice. This symmetry
breaking has Goldstone bosons associated which are identified with
phonons which are the quantum excitation modes of the vibrational
degrees of freedom of the lattice.  

{\bf The Higgs mechanism.}
Gauge symmetry seems to prevent a vector field from having a mass. This is
obvious once we realize that a term in the Lagrangian like
$m^{2}A_{\mu}A^{\mu}$ is incompatible with gauge invariance.  

However certain physical situations seem to require massive vector fields.
This happened for example during the 1960s in the study of weak interactions.
The Glashow model gave a common description of both electromagnetic and
weak interactions based on a gauge theory with group SU(2)$\times$U(1)
but, in order to reproduce Fermi's four-fermion theory of the $\beta$-decay
it was necessary that two of the vector fields involved would be massive.
Also in condensed matter physics massive vector fields are required to
describe certain systems, most notably in superconductivity.

The way out to this situation is found in the concept of spontaneous 
symmetry breaking discussed previously. The consistency of the quantum theory
requires gauge invariance, but this invariance can be realized \`a la
Nambu-Goldstone. When this is the case the full gauge symmetry is not
explicitly present in the effective action constructed around the particular
vacuum chosen by the theory. This makes possible the existence of
mass terms for gauge fields without jeopardizing the consistency of the
full theory, which is still invariant under the whole gauge group.

To illustrate the Higgs mechanism we study the simplest example,
the Abelian Higgs model: a U(1) gauge field coupled to a
self-interacting charged complex scalar field $\Phi$ with Lagrangian
\begin{eqnarray}
\mathcal{L}=-{1\over 4}F_{\mu\nu}F^{\mu\nu}+\overline{D_{\mu}\Phi}
D^{\mu}\Phi-{\lambda\over 4}\left(\overline{\Phi}\Phi-\mu^{2}\right)^{2},
\end{eqnarray}
where the covariant derivative is given by
Eq. (\ref{eq:covariant_deriv}).  This theory is invariant under the
gauge transformations 
\begin{eqnarray}
\Phi\rightarrow e^{i\alpha(x)}\Phi, \hspace*{1cm}
A_{\mu}\rightarrow A_{\mu}+\partial_{\mu}\alpha(x).
\label{eq:gauge_ahm}
\end{eqnarray} 
The minimum of the potential is defined by the equation $|\Phi|=\mu$. 
We have a continuum of different vacua labelled by the phase of
the scalar field. None of these vacua, however, is
invariant under the gauge symmetry
\begin{eqnarray}
\langle\Phi\rangle = \mu e^{i\vartheta_{0}}\rightarrow 
\mu e^{i\vartheta_{0}+i\alpha(x)}
\label{eq:gauge_vacua}
\end{eqnarray}
and therefore the symmetry is spontaneously broken
Let us study now the theory around one of these vacua, for example 
$\langle\Phi\rangle=\mu$, by writing the field $\Phi$ in terms of 
the excitations around this particular vacuum
\begin{eqnarray}
\Phi(x)=\left[\mu+{1\over \sqrt{2}}\sigma(x)\right]e^{i\vartheta(x)}.
\label{eq:ansatz_ahm}
\end{eqnarray}
Independently of whether we are expanding around a particular vacuum
for the scalar field we should keep in mind that the whole Lagrangian
is still gauge invariant under (\ref{eq:gauge_ahm}).
This means that performing a gauge transformation with parameter 
$\alpha(x)=-\vartheta(x)$ we can get rid of the phase in Eq. 
(\ref{eq:ansatz_ahm}). Substituting then $\Phi(x)=\mu+{1\over\sqrt{2}}
\sigma(x)$ in the Lagrangian we find
\begin{eqnarray}
\mathcal{L}&=&-{1\over 4}F_{\mu\nu}F^{\mu\nu}+e^{2}\mu^{2}A_{\mu}A^{\mu}+
{1\over 2}\partial_{\mu}\sigma\partial^{\mu}\sigma-{1\over 2}{\lambda\mu^{2}}
\sigma^{2} \nonumber \\
&-&{\lambda\mu}\sigma^{3}-{\lambda\over 4}\sigma^{4} 
+ e^{2}\mu A_{\mu}A^{\mu}\sigma+e^{2}A_{\mu}A^{\mu}\sigma^{2} .
\label{eq:abelianHiggs}
\end{eqnarray}
What are the excitation of the theory around the vacuum
$\langle\Phi\rangle=\mu$?  First we find a massive real scalar field
$\sigma(x)$.  The important point however is that the vector field
$A_{\mu}$ now has a mass given by
\begin{eqnarray}
m^{2}_{\gamma}=2e^{2}\mu^{2}.
\label{eq:mass_photon}
\end{eqnarray}
The remarkable thing about this way of giving a mass to the photon is
that at no point we have given up gauge invariance. The symmetry is
only hidden. Therefore in quantizing the theory we can
still enjoy all the advantages of having a gauge theory but at the same
time we have managed to generate 
a mass for the gauge field.

It is surprising, however, that in the Lagrangian
(\ref{eq:abelianHiggs}) we did not found any massless mode. Since the
vacuum chosen by the scalar field breaks the $U(1)$  generator of U(1) we
would have expected one
masless particle from Goldstone's theorem. To understand the
fate of the missing Goldstone boson we have to revisit the calculation
leading to Eq. (\ref{eq:abelianHiggs}). Were we dealing with a global
U(1) theory, the Goldstone boson would correspond to excitation of the
scalar field along the valley of the potential and the phase
$\vartheta(x)$ would be the massless Goldstone boson. However we have
to keep in mind that in computing the Lagrangian we managed to get rid
of $\vartheta(x)$ by shifting it into $A_{\mu}$ using a gauge
transformation. Actually by identifying the gauge parameter with the
Goldstone excitation we have completely fixed the gauge and the
Lagrangian (\ref{eq:abelianHiggs}) does not have any gauge symmetry
left.

A massive vector field has three polarizations: two transverse ones
$\vec{k}\cdot\vec{\epsilon}\,(\vec{k},\pm 1)=0$ plus a longitudinal
one $\vec{\epsilon}_{L}(\vec{k})\sim \vec{k}$.  In gauging
away the massless Goldstone boson $\vartheta(x)$ we have transformed
it into the longitudinal polarization of the massive vector field. In
the literature this is usually expressed saying that the Goldstone
mode is ``eaten up'' by the longitudinal component of the gauge field.
It is important to realize that in spite of the fact that the
Lagrangian (\ref{eq:abelianHiggs}) looks pretty different from the one
we started with we have not lost any degrees of freedom. We
started with the two polarizations of the photon plus the two degrees
of freedom associated with the real and imaginary components of the
complex scalar field. After symmetry breaking we end up with the three
polarizations of the massive vector field and the degree of freedom of
the real scalar field $\sigma(x)$.

We can also understand the Higgs mechanism in the light of our
discussion of gauge symmetry in section \ref{sec:gauge_symmetry}.  In
the Higgs mechanism the invariance of the theory under infinitesimal
gauge transformations is not explicitly broken, and this implies that
Gauss' law is satisfied quantum mechanically,
$\vec{\nabla}\cdot\vec{E}_{a}|{\rm phys}\rangle=0$.  The theory
remains invariant under gauge transformations in the connected
component of the identity $\mathcal{G}_{0}$, the ones generated by
Gauss' law. This does not pose any restriction on the possible
breaking of the invariance of the theory with respect to
transformations that cannot be continuously deformed to the
identity. Hence in the Higgs mechanism the invariance under gauge
transformation that are not in the connected component of the
identity, $\mathcal{G}/\mathcal{G}_{0}$, can be broken.
Let us try to put it in more precise terms.  As we learned in section
\ref{sec:gauge_symmetry}, in the Hamiltonian formulation of the
theory finite energy gauge field configurations tend to a pure gauge
at spatial infinity
\begin{eqnarray}
\vec{A}_{\mu}(\vec{x}) {\longrightarrow} -{1\over ig}g(\vec{x})
\vec{\nabla}g(\vec{x})^{-1}, \hspace*{1cm} |\vec{x}|\rightarrow \infty
\end{eqnarray}
The set transformations $g_{0}(\vec{x})\in \mathcal{G}_{0}$ that tend
to the identity at infinity are the ones generated by Gauss'
law. However, one can also consider in general gauge transformations
$g(\vec{x})$ which, as $|\vec{x}|\rightarrow\infty$, approach any
other element $g\in G$. The quotient $\mathcal{G}_{\infty}\equiv
\mathcal{G}/\mathcal{G}_{0}$ gives a copy of the gauge group at
infinity. There is no reason, however, why this group should not be
broken, and in general it is if the gauge symmetry is spontaneously
broken. Notice that this is not a threat to the consistency of the
theory. Properties like the decoupling of unphysical states are
guaranteed by the fact that Gauss' law is satisfied quantum
mechanically and are not affected by the breaking of $\mathcal{G}_{\infty}$.

The Abelian Higgs model discussed here can be regarded as a toy model
of the Higgs mechanism responsible for giving mass to the $W^{\pm}$
and $Z^{0}$ gauge bosons in the Standard Model. In condensed matter
physics the symmetry breaking described by the nonrelativistic version
of the Abelian Higgs model can be used to characterize the onset of a
superconducting phase in the BCS theory, where the complex scalar
field $\Phi$ is associated with the Cooper pairs. In this case the
parameter $\mu^{2}$ depends on the temperature. Above the critical
temperature $T_{c}$, $\mu^{2}(T)>0$ and there is only a symmetric
vacuum $\langle\Phi\rangle=0$. When, on the other hand, $T<T_{c}$ then
$\mu^{2}(T)<0$ and symmetry breaking takes place. The onset of a
nonzero mass of the photon (\ref{eq:mass_photon}) below the critical
temperature explains the Meissner effect:  the
magnetic fields cannot penetrate inside superconductors beyond a
distance of the order ${1\over m_{\gamma}}$.

\section{Anomalies}
\label{sec:anomalies}

So far we did not worry too much about how classical symmetries of a
theory are carried over to the quantum theory. We have implicitly
assumed that classical symmetries are preserved in the process of
quantization, so they are also realized in the quantum theory.

This, however, does not have to be necessarily the case. Quantizing an 
interacting field theory 
is a very involved process that requires regularization and renormalization and
sometimes, it does not matter
how hard we try, there is no way for a classical symmetry to survive quantization. 
When this happens one says that
the theory has an {\em anomaly} (for a review
see \cite{anomalies_review}).
It is important to avoid here the misconception that anomalies appear
due to a bad choice of the way a theory is regularized in the process
of quantization. When we talk about anomalies we mean a classical
symmetry that {\em cannot} be realized in the quantum theory, no matter how
smart we are in choosing the regularization procedure. 

In the following we analyze some examples of anomalies associated with 
global and local symmetries of the classical theory. In Section \ref{sec:renormalization}
we will encounter yet another example of an anomaly, this time associated with the 
breaking of classical scale invariance in the quantum theory.

\subsection{Axial anomaly}
\label{sec:axial_anomaly}

Probably the best known examples of anomalies appear
when we consider axial symmetries.
If we consider a theory of two Weyl spinors $u_{\pm}$
\begin{eqnarray}
\mathcal{L}=i\overline{\psi}\dslash\psi=
iu^{\dagger}_{+}\sigma^{\mu}_{+}\partial_{\mu}u_{+}
+iu^{\dagger}_{-}\sigma^{\mu}_{-}\partial_{\mu}u_{-} \hspace*{0.5cm}
\mbox{with} \hspace*{1cm} \psi=\left(
\begin{array}{c}
u_{+} \\
u_{-}
\end{array}
\right)
\label{eq:weyl_anomaly}
\end{eqnarray}
the Lagrangian is invariant under two types of global U(1)
transformations. In the first one both
helicities transform with the same phase, this is a {\em vector}
transformation:
\begin{eqnarray}
\mbox{U(1)}_{V}:u_{\pm}\longrightarrow e^{i\alpha}u_{\pm},
\label{eq:U(1)V}
\end{eqnarray}
whereas in the second one, the axial $U(1)$, the signs of the phases are
different for the two chiralities
\begin{eqnarray}
\mbox{U(1)}_{A}:u_{\pm}\longrightarrow e^{\pm i\alpha}u_{\pm}.
\label{eq:U(1)A}
\end{eqnarray}
Using Noether's theorem, there are two conserved currents, a vector current
\begin{eqnarray}
J_{V}^{\mu}=\overline{\psi}\gamma^{\mu}\psi=u_{+}^{\dagger}\sigma_{+}^{\mu}
u_{+}+u_{-}^{\dagger}\sigma_{-}^{\mu}u_{-} \hspace*{0.5cm} \Longrightarrow
\hspace*{0.5cm}
\partial_{\mu}J^{\mu}_{V}=0
\end{eqnarray}
and an axial vector current
\begin{eqnarray}
J_{A}^{\mu}=\overline{\psi}\gamma^{\mu}\gamma_{5}\psi
=u_{+}^{\dagger}\sigma_{+}^{\mu}u_{+}
-u_{-}^{\dagger}\sigma_{-}^{\mu}u_{-} \hspace*{0.5cm} \Longrightarrow
\hspace*{0.5cm}
\partial_{\mu}J^{\mu}_{A}=0.
\label{eq:conservation_axial_current}
\end{eqnarray}

The theory described by the Lagrangian (\ref{eq:weyl_anomaly}) can be
coupled to the electromagnetic field. The resulting classical theory
is still invariant under the vector and axial U(1) symmetries
(\ref{eq:U(1)V}) and (\ref{eq:U(1)A}). Surprisingly, upon quantization
it turns out that the conservation of the axial current
(\ref{eq:conservation_axial_current}) is spoiled by quantum effects
\begin{eqnarray}
\partial_{\mu}J^{\mu}_{A}\sim \hbar\, \vec{E}\cdot\vec{B}.
\end{eqnarray}

To understand more clearly how this result comes about we study first
a simple model in two dimensions that captures the relevant physics
involved in the four-dimensional case \cite{jackiw_princeton}. We work
in Minkowski space in two dimensions with coordinates
$(x^{0},x^{1})\equiv (t,x)$ and where the spatial direction is
compactified to a circle $S^{1}$. In this setup we consider a fermion
coupled to the electromagnetic field. Notice that since we are living
in two dimensions the field strength $F_{\mu\nu}$ only has one
independent component that corresponds to the electric field along the
spatial direction, $F^{01}\equiv\mathcal{E}$ (in two dimensions there
are no magnetic fields!).

To write the Lagrangian for the spinor field we need to find a
representation of the algebra of $\gamma$-matrices
\begin{eqnarray}
\{\gamma^{\mu},\gamma^{\nu}\}=2\eta^{\mu\nu} \hspace*{0.5cm} \mbox{with}
\hspace*{0.5cm} \eta=\left(
\begin{array}{rr}
1 & 0 \\
0 & -1
\end{array}
\right).
\end{eqnarray}
In two dimensions the dimension of the representation of the $\gamma$-matrices
is $2^{[{2\over 2}]}=2$. Here take 
\begin{eqnarray}
\gamma^{0}\equiv \sigma^{1}=\left(
\begin{array}{rr}
0 & 1 \\
1 & 0
\end{array}
\right), \hspace*{1cm}
\gamma^{1}\equiv i\sigma^{2}=\left(
\begin{array}{rr}
0 & 1 \\
-1 & 0
\end{array}
\right).
\end{eqnarray}
This is a chiral representation since the matrix $\gamma_{5}$ is 
diagonal\footnote{In any even number of dimensions $\gamma_{5}$ is
defined to satisfy the conditions $\gamma_{5}^{2}=\mathbf{1}$ and
$\{\gamma_{5},\gamma^{\mu}\}=0.$}
\begin{eqnarray}
\gamma_{5}\equiv -\gamma^{0}\gamma^{1}=\left(
\begin{array}{rr}
1 & 0 \\
0 & -1
\end{array}
\right)
\end{eqnarray}
Writing the two-component spinor $\psi$  as
\begin{eqnarray}
\psi=\left(
\begin{array}{c}
u_{+}\\
u_{-}
\end{array}
\right)
\end{eqnarray}
and defining as usual the projectors $P_{\pm}={1\over 2}
(\mathbf{1}\pm\gamma_{5})$ we find that the components $u_{\pm}$ of
$\psi$ are respectively a right- and left-handed Weyl spinor in two
dimensions.

Once we have a representation of the $\gamma$-matrices we can write
the Dirac equation. Expressing it in terms of the components
$u_{\pm}$ of the Dirac spinor we find
\begin{eqnarray}
(\partial_{0}-\partial_{1})u_{+}=0, \hspace*{1cm}
(\partial_{0}+\partial_{1})u_{-}=0.
\label{eq:dirac2d}
\end{eqnarray}
The general solution to these equations can be immediately written as
\begin{eqnarray}
u_{+}=u_{+}(x^{0}+x^{1}), \hspace*{1cm} u_{-}=u_{-}(x^{0}-x^{1}).
\end{eqnarray}
Hence $u_{\pm}$ are two wave packets moving along the spatial
dimension respectively to the left $(u_{+})$ and to the right $(u_{-})$.
Notice that according to our convention the left-moving $u_{+}$
is a right-handed spinor (positive helicity) whereas the right-moving
$u_{-}$ is a left-handed spinor (negative helicity).

\begin{figure}
\centerline{\epsfxsize=6.0truein\epsfbox{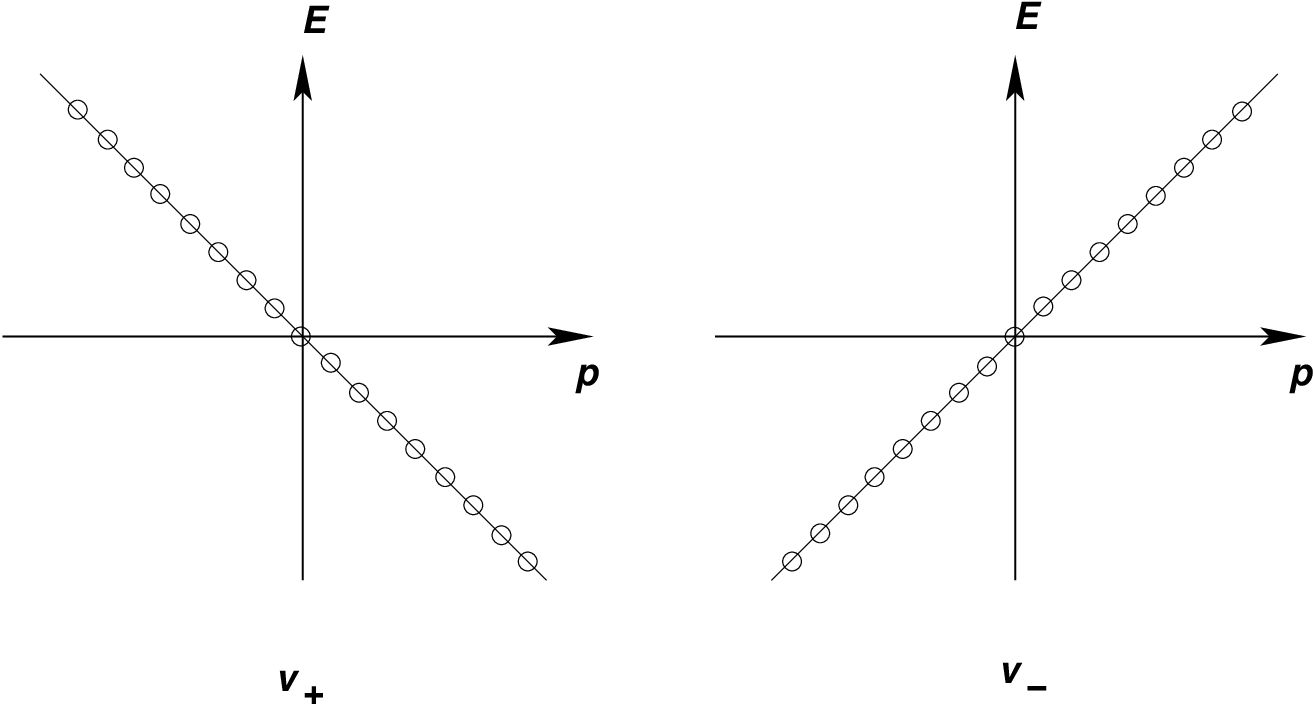}}
\caption[]{Spectrum of the massless two-dimensional Dirac field.}
\label{fig:anomaly1}
\end{figure}

If we want to interpret (\ref{eq:dirac2d}) as the wave equation for
two-dimensional Weyl spinors we have the following wave functions for
free particles with well defined momentum $p^{\mu}=(E,p)$.
\begin{eqnarray}
u_{\pm}^{(E)}(x^{0}\pm x^{1})={1\over \sqrt{L}}e^{-iE(x^{0}\pm x^{1})}
\hspace*{0.5cm}
\mbox{with} \hspace*{0.5cm} p=\mp E.
\label{eq:modes_dirac2d}
\end{eqnarray}
As it is always the case with the Dirac equation we have both positive
and negative energy solutions. For $u_{+}$, since $E=-p$, we
see that the solutions with positive energy are those with negative
momentum $p<0$, whereas the negative energy solutions are plane
waves with $p>0$. For the left-handed spinor $u_{-}$ the situation
is reversed. Besides, since the spatial direction is compact with length
$L$ the momentum $p$ is quantized according to
\begin{eqnarray}
p =  {2\pi n\over L}, \hspace*{1cm} n\in\mathbb{Z}.
\end{eqnarray}
The spectrum of the theory is represented in
Fig. \ref{fig:anomaly1}.

\begin{figure}
\centerline{\epsfxsize=6.0truein\epsfbox{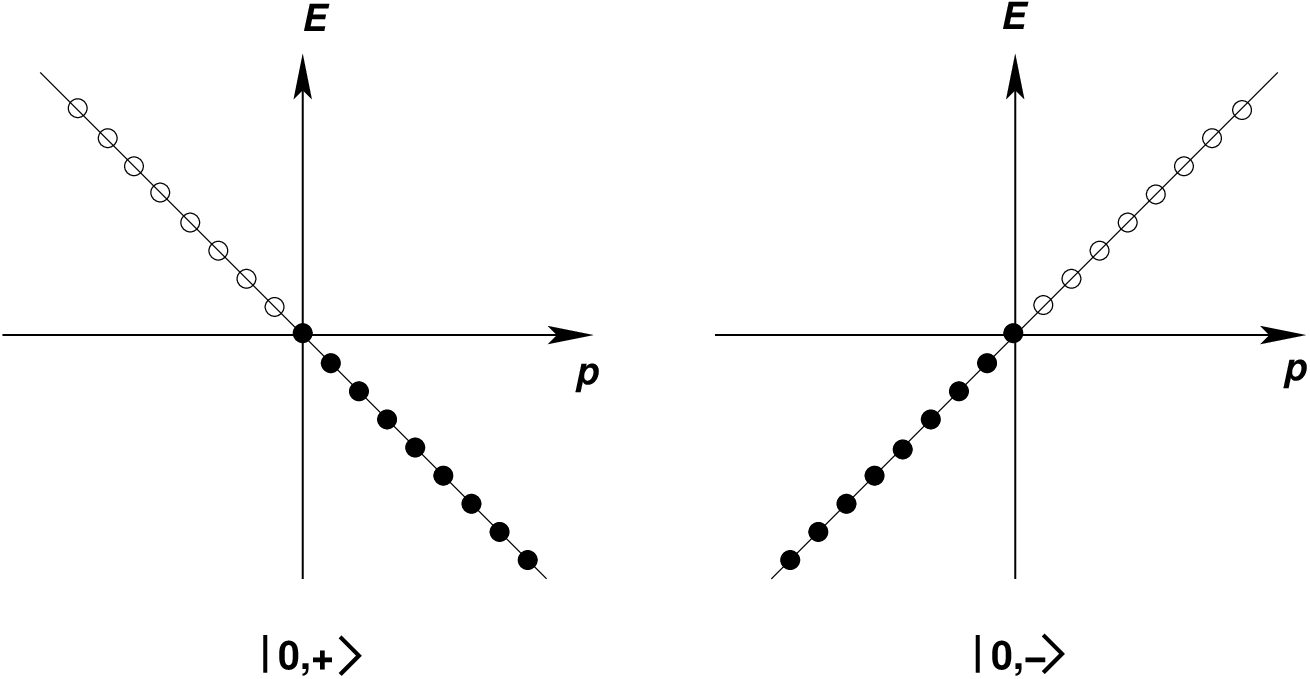}}
\caption[]{Vacuum of the theory.}
\label{fig:anomaly2}
\end{figure}

Once we have the spectrum of the theory the next step is to obtain the vacuum.
As with the Dirac equation in four dimensions  we fill all the states with
$E\leq 0$ (Fig. \ref{fig:anomaly2}). Exciting of a particle in the Dirac
see produces a positive energy fermion plus a hole that is interpreted
as an antiparticle. This gives us the clue 
on how to quantize the theory.
In the expansion of the operator $u_{\pm}$ in terms of the 
modes (\ref{eq:modes_dirac2d}) we associate positive energy states with
annihilation operators whereas the states with negative energy are 
associated with creation operators for the corresponding antiparticle
\begin{eqnarray}
u_{\pm}(x)=\sum_{E>0}\left[
a_{\pm}(E)v_{\pm}^{(E)}(x)
+b_{\pm}^{\dagger}(E)
v_{\pm}^{(E)}(x)^{*}\right].
\label{eq:lagrangian2D}
\end{eqnarray}
The operator $a_{\pm}(E)$ acting on the vacuum $|0,\pm\rangle$
annihilates a particle with positive energy $E$ and momentum $\mp
E$. In the same way $b_{\pm}^{\dagger}(E)$ creates out of the vacuum
an antiparticle with positive energy $E$ and spatial momentum $\mp
E$. In the Dirac sea picture the operator $b_{\pm}(E)^{\dagger}$ is
originally an annihilation operator for a state of the sea with negative
energy $-E$. As in the four-dimensional case the problem of the
negative energy states is solved by interpreting annihilation
operators for negative energy states as creation operators for the
corresponding antiparticle with positive energy (and vice versa).  The
operators appearing in the expansion of $u_{\pm}$ in
Eq. (\ref{eq:lagrangian2D}) satisfy the usual algebra
\begin{eqnarray}
\{a_{\lambda}(E),a_{\lambda'}^{\dagger}(E')\}=
\{b_{\lambda}(E),b_{\lambda'}^{\dagger}(E')\}=\delta_{E,E'}
\delta_{\lambda\lambda'}, \hspace*{0.5cm}
\end{eqnarray}
where we have introduced the label $\lambda,\lambda'=\pm$. Also,
$a_{\lambda}(E)$, $a_{\lambda}^{\dagger}(E)$ anticommute with
$b_{\lambda'}(E')$, $b_{\lambda'}^{\dagger}(E')$.

The Lagrangian of the theory 
\begin{eqnarray}
\mathcal{L}=iu_{+}^{\dagger}(\partial_{0}+\partial_{1})u_{+}
+iu_{-}^{\dagger}(\partial_{0}-\partial_{1})u_{-}
\end{eqnarray} 
is invariant under both U(1)$_{V}$, Eq. (\ref{eq:U(1)V}), and U(1)$_{A}$,
Eq. (\ref{eq:U(1)A}). The associated Noether currents are in this case
\begin{eqnarray}
J^{\mu}_{V}=\left(
\begin{array}{c}
u_{+}^{\dagger}u_{+}+u_{-}^{\dagger}u_{-} \\
-u_{+}^{\dagger}u_{+}+u_{-}^{\dagger}u_{-}
\end{array}
\right), \hspace*{1cm}
J^{\mu}_{A}=\left(
\begin{array}{c}
u_{+}^{\dagger}u_{+}-u_{-}^{\dagger}u_{-} \\
-u_{+}^{\dagger}u_{+}-u_{-}^{\dagger}u_{-}
\end{array}
\right).
\end{eqnarray}
The associated conserved charges are given, for the vector
current by
\begin{eqnarray}
Q_{V}=\int_{0}^{L}dx^{1}\left(u_{+}^{\dagger}u_{+}+u_{-}^{\dagger}u_{-}\right)
\end{eqnarray}
and for the axial current
\begin{eqnarray}
Q_{A}=\int_{0}^{L}dx^{1}\left(u_{+}^{\dagger}u_{+}-u_{-}^{\dagger}u_{-}\right).
\end{eqnarray}
Using the orthonormality relations for the modes $v_{\pm}^{(E)}(x)$
\begin{eqnarray}
\int_{0}^{L}dx^{1}\,v_{\pm}^{(E)}(x)\,v_{\pm}^{(E')}(x)=\delta_{E,E'}
\end{eqnarray}
we find for the conserved charges:
\begin{eqnarray}
Q_{V}&=&\sum_{E>0}\left[a_{+}^{\dagger}(E)a_{+}(E)-b_{+}^{\dagger}(E)
b_{+}(E)+a_{-}^{\dagger}(E)a_{-}(E)-b_{-}^{\dagger}(E)b_{-}(E)\right],
\nonumber \\
Q_{A}&=&\sum_{E>0}\left[a_{+}^{\dagger}(E)a_{+}(E)-b_{+}^{\dagger}(E)
b_{+}(E)-a_{-}^{\dagger}(E)a_{-}(E)+b_{-}^{\dagger}(E)b_{-}(E)\right].
\end{eqnarray}
We see that $Q_{V}$ counts the net number (particles minus
antiparticles) of positive helicity states plus the net number of
states with negative helicity. The axial charge, on the other hand,
counts the net number of positive helicity states minus the number of
negative helicity ones. In the case of the vector current we have
subtracted a formally divergent vacuum contribution to the charge
(the ``charge of the Dirac sea'').

In the free theory there is of course no problem with the conservation
of either $Q_{V}$ or $Q_{A}$, since the occupation numbers do not
change.  What we want to study is the effect of coupling the
theory to electric field $\mathcal{E}$. We work in the gauge
$A_{0}=0$. Instead of solving the problem exactly we are going to
simulate the electric field by adiabatically varying in a long time
$\tau_{0}$ the vector potential $A_{1}$ from zero value to
$-\mathcal{E}\tau_{0}$.  From our discussion in section
\ref{sec:gauge_fields} we know that the effect of the electromagnetic
coupling in the theory is a shift in the momentum according to
\begin{eqnarray}
p\longrightarrow p-eA_{1},
\label{eq:shift}
\end{eqnarray}
where $e$ is the charge of the fermions. Since we assumed that the 
vector potential varies adiabatically, we can assume it to be approximately
constant at each time.

\begin{figure}
\centerline{\epsfxsize=4.0truein\epsfbox{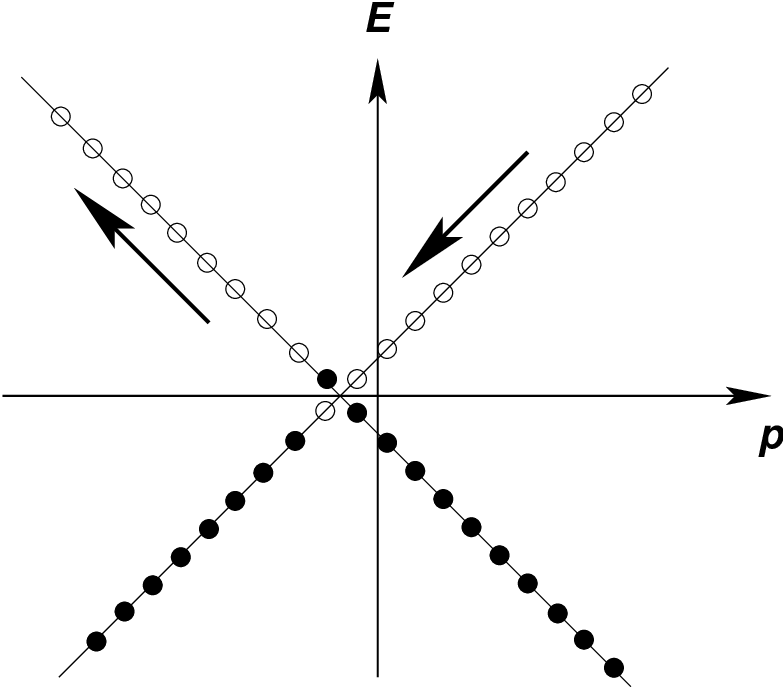}}
\caption[]{Effect of the electric field.}
\label{fig:anomaly3}
\end{figure}

Then, we have to understand what is the effect of (\ref{eq:shift}) on
the vacuum depicted in Fig. (\ref{fig:anomaly2}). What we find is that
the two branches move as shown in Fig. (\ref{fig:anomaly3}) resulting
in some of the negative energy states of the $v_{+}$ branch acquiring
positive energy while the same number of the empty positive energy
states of the other branch $v_{-}$ will become empty negative energy
states. Physically this means that the external electric field
$\mathcal{E}$ creates a number of particle-antiparticle pairs out of
the vacuum. Denoting by $N\sim e\mathcal{E}$ the number of such pairs
created by the electric field per unit time, the final values of the charges
$Q_{V}$ and $Q_{A}$ are
\begin{eqnarray}
Q_{A}(\tau_{0})&=&(N-0)+(0-N)=0, \nonumber \\ 
Q_{V}(\tau_{0})&=&(N-0)-(0-N)=2N.
\end{eqnarray}
Therefore we conclude that the coupling to the electric field produces
a violation in the conservation of the axial charge per unit time
given by $\Delta Q_{A}
\sim e\mathcal{E}$. This implies that
\begin{eqnarray}
\partial_{\mu}J^{\mu}_{A}\sim e\hbar\mathcal{E},
\label{eq:anomaly2d}
\end{eqnarray}
where we have restored $\hbar$ to make clear that the violation in the
conservation of the axial current is a quantum effect. At the
same time $\Delta Q_{V}=0$ guarantees that the vector current remains
conserved also quantum mechanically,  $\partial_{\mu}J_{V}^{\mu}=0$.

We have just studied a two-dimensional example of the Adler-Bell-Jackiw
axial anomaly \cite{ABJ}. The heuristic analysis presented here can be
made more precise by computing the quantity
\begin{eqnarray}
C^{\mu\nu}=\langle 0|T\left[J_{A}^{\mu}(x)J_{V}^{\nu}(0)\right]|0\rangle
=\,\,\,
\parbox{40mm}{
\begin{fmfgraph*}(100,80)
\fmfleft{i}
\fmfright{o}
\fmf{double,label=$J_{A}^{\mu}$}{i,v3}
\fmf{double}{v3,v1}
\fmf{photon}{v4,v2}
\fmf{photon,label=$\gamma$}{v2,o}
\fmf{fermion,left,tension=.2}{v1,v4,v1}
\fmfdotn{v}{1}
\end{fmfgraph*}
}
\end{eqnarray}
The anomaly is given then by $\partial_{\mu}C^{\mu\nu}$. A careful
calculation yields the numerical prefactor missing in Eq. 
(\ref{eq:anomaly2d}) leading to the result
\begin{eqnarray}
\partial_{\mu}J^{\mu}_{A}={e\hbar\over 2\pi}
\varepsilon^{\nu\sigma}F_{\nu\sigma},
\end{eqnarray}
with $\varepsilon^{01}=-\varepsilon^{10}=1$. 

The existence of an anomaly in the axial symmetry that we have
illustrated in two dimensions is present in all even
dimensional of space-times. In particular in four
dimensions the axial anomaly it is given by
\begin{eqnarray}
\partial_{\mu}J^{\mu}_{A}=-{e^{2}\over 16\pi^{2}}
\varepsilon^{\mu\nu\sigma\lambda}F_{\mu\nu}F_{\sigma\lambda}.
\end{eqnarray}
This result has very important consequences in the physics of
strong interactions as we will see in what follows

\subsection{Chiral symmetry in QCD}

Our knowledge of the physics of strong interactions is based on the
theory of Quantum Chromodynamics (QCD) \cite{QCD}. This is a
nonabelian gauge theory with gauge group SU($N_{c}$) coupled to a
number $N_{f}$ of quarks. These are spin-${1\over 2}$ particles
$Q^{i\,f}$ labelled by two quantum numbers: color $i=1,\ldots,N_{c}$
and flavor $f=1,\ldots,N_{f}$. The interaction between them is
mediated by the $N_{c}^{2}-1$ gauge bosons, the gluons $A_{\mu}^{a}$,
$a=1,\ldots,N_{c}^{2}-1$. In the real world $N_{c}=3$ and the number
of flavors is six, corresponding to the number of different quarks: up
($u$), down ($d$), charm ($c$), strange ($s$), top ($t$) and bottom
($b$).

For the time being we are going to study a general theory of QCD with
$N_{c}$ colors and $N_{f}$ flavors. Also, for reasons that will be clear
later we are going to work in the limit of vanishing quark masses,
$m_{f}\rightarrow 0$. In this cases the Lagrangian is given by
\begin{eqnarray}
\mathcal{L}_{\rm QCD}
=-{1\over 4}F_{\mu\nu}^{a}F^{a\,\mu\nu}+\sum_{f=1}^{N_{f}}\left[
i\overline{Q}_{L}^{f}\Dslash Q_{L}^{f}+i\overline{Q}_{R}^{f}\Dslash Q_{R}^{f}
\right],
\label{eq:QCDL}
\end{eqnarray}
where the subscripts $L$ and $R$ indicate respectively left and
right-handed spinors, $Q^{f}_{L,R}\equiv P_{\pm}Q^{f}$, and the field
strength $F_{\mu\nu}^{a}$ and the covariant derivative $D_{\mu}$ are
respectively defined in Eqs. (\ref{eq:fmunuNA}) and
(\ref{eq:covariant_derivative}). Apart from the gauge symmetry, this
Lagrangian is also invariant under a global
U($N_{f}$)$_{L}\times$U($N_{f}$)$_{R}$ acting on the flavor indices
and defined by
\begin{eqnarray}
\mbox{U($N_{f}$)}_{L}:\left\{
\begin{array}{rcl}
Q_{L}^{f}&\rightarrow& \sum_{f'}(U_{L})_{ff'}Q_{L}^{f'} \\
& & \\
Q_{R}^{f}&\rightarrow& Q_{R}^{f}
\end{array}
\right. \hspace*{1cm}
\mbox{U($N_{f}$)}_{R}:\left\{
\begin{array}{rcl}
Q_{L}^{f}&\rightarrow& Q_{L}^{f} \\
& & \\
Q_{R}^{r}&\rightarrow&\sum_{f'}(U_{R})_{ff'}Q_{R}^{f'} 
\end{array}
\right.
\end{eqnarray}
with $U_{L},U_{R}\in\mbox{U}(N_{f})$. Actually, since 
U($N$)=U(1)$\times$SU($N$) this global symmetry group can be 
written as
$\mbox{SU($N_{f}$)}_{L}\times\mbox{SU($N_{f}$)}_{R}\times
\mbox{U(1)}_{L}\times\mbox{U(1)}_{R}$. The
abelian subgroup $\mbox{U(1)}_{L}\times\mbox{U(1)}_{R}$ can be now
decomposed into their vector U(1)$_{B}$ and axial U(1)$_{A}$ subgroups
defined by the transformations
\begin{eqnarray}
\mbox{U(1)}_{B}:\left\{
\begin{array}{rcl}
Q_{L}^{f}&\rightarrow& e^{i\alpha} Q_{L}^{f} \\
& & \\
Q_{R}^{f}&\rightarrow& e^{i\alpha} Q_{R}^{f}
\end{array}
\right. \hspace*{1cm}
\mbox{U(1)}_{A}:\left\{
\begin{array}{rcl}
Q_{L}^{f}&\rightarrow& e^{i\alpha}Q_{L}^{f} \\
& & \\
Q_{R}^{f}&\rightarrow& e^{-i\alpha} Q_{R}^{f} 
\end{array}
\right.
\end{eqnarray}
According to Noether's theorem, associated with these two abelian symmetries
we have two conserved currents:
\begin{eqnarray}
J_{V}^{\mu}=\sum_{f=1}^{N_{f}}\overline{Q}^{f}\gamma^{\mu}\,
Q^{f}, \hspace*{1cm}
J_{A}^{\mu}=\sum_{f=1}^{N_{f}}\overline{Q}^{f}\gamma^{\mu}\gamma_{5}\,
Q^{f}.
\end{eqnarray}
The conserved charge associated with vector charge $J_{V}^{\mu}$ is
actually the baryon number defined as the number of quarks minus
number of antiquarks. 

The nonabelian part of the global symmetry group
SU($N_{f}$)$_{L}\times$SU($N_{f}$)$_{R}$ can also be decomposed into
its vector and axial subgroups,
$\mbox{SU($N_{f}$)}_{V}\times\mbox{SU($N_{f}$)}_{A}$, defined by the
following transformations of the quarks fields
\begin{eqnarray}
\mbox{SU($N_{f}$)}_{V}:\left\{
\begin{array}{rcl}
Q_{L}^{f}&\rightarrow& \sum_{f'}(U_{L})_{ff'}Q_{L}^{f'} \\
& & \\
Q_{R}^{f}&\rightarrow& \sum_{f'}(U_{L})_{ff'}Q_{R}^{f'}
\end{array}
\right. \hspace*{1cm}
\mbox{SU($N_{f}$)}_{A}:\left\{
\begin{array}{rcl}
Q_{L}^{f}&\rightarrow& \sum_{f'}(U_{L})_{ff'}Q_{L}^{f'} \\
& & \\
Q_{R}^{f}&\rightarrow&\sum_{f'}(U_{R}^{-1})_{ff'}Q_{R}^{f'} 
\end{array}
\right.
\end{eqnarray}
Again, the application of Noether's theorem shows the existence of the
following nonabelian conserved charges 
\begin{eqnarray}
J_{V}^{I\,\mu}\equiv \sum_{f,f'=1}^{N_{f}}\overline{Q}^{f}\gamma^{\mu}
(T^{I})_{ff'}Q^{f'}, \hspace*{1cm}
J_{A}^{I\,\mu}\equiv \sum_{f,f'=1}^{N_{f}}\overline{Q}^{f}\gamma^{\mu}
\gamma_{5}(T^{I})_{ff'}Q^{f'}.
\end{eqnarray}
To summarize, we have shown that the initial chiral symmetry of the
QCD Lagrangian (\ref{eq:QCDL}) can be decomposed into its chiral and
vector subgroups according to
\begin{eqnarray}
\mbox{U($N_{f}$)}_{L}\times\mbox{U($N_{f}$)}_{R}=
\mbox{SU($N_{f}$)}_{V}\times\mbox{SU($N_{f}$)}_{A}\times
\mbox{U(1)}_{B}\times\mbox{U(1)}_{A}.
\end{eqnarray}
The question to address now is which part of the classical global
symmetry is preserved by the quantum theory.

As argued in section \ref{sec:axial_anomaly}, the conservation of the
axial currents $J_{A}^{\mu}$ and $J_{A}^{a\,\mu}$ can in principle be 
spoiled due to the presence of an anomaly. In the case of the
abelian axial current $J^{\mu}_{A}$ the relevant quantity is the 
correlation function 
\begin{eqnarray}
C^{\mu\nu\sigma}\equiv
\langle0|T\left[J^{\mu}_{A}(x)j_{\rm gauge}^{a\,\nu}(x')
j_{\rm gauge}^{b\,\sigma}(0)
\right]|0\rangle=\sum_{f=1}^{N_{f}}
\left[\,\,\,\,\parbox{40mm}{
\begin{fmfgraph*}(100,80)
\fmfleft{i}
\fmfright{o1,o2}
\fmf{double,label=$J_{A}^{\mu}$}{i,v4}
\fmf{double}{v4,v1}
\fmf{gluon,label=$g$,tension=0.3}{v2,o1}
\fmf{gluon,label=$g$,tension=0.3}{v3,o2}
\fmfcyclen{fermion,tension=0.05,label=$Q^f$}{v}{3}
\fmfdotn{v}{1}
\end{fmfgraph*}
}
\right]_{\rm symmetric}
\label{eq:triangle1}
\end{eqnarray}
Here $j_{\rm gauge}^{a\,\mu}$ is the nonabelian conserved current 
coupling to the gluon field
\begin{eqnarray}
j_{\rm gauge}^{a\,\mu}\equiv \sum_{f=1}^{N_{f}}\overline{Q}^{f}
\gamma^{\mu}\tau^{a}Q^{f},
\label{eq:gauge_current}
\end{eqnarray} 
where, to avoid confusion with the generators of the global symmetry we have
denoted by $\tau^{a}$ the generators of the gauge group SU($N_{c}$).
The anomaly can be read now from $\partial_{\mu}C^{\mu\nu\sigma}$. If
we impose Bose symmetry with respect to the interchange of the two
outgoing gluons and gauge invariance of the whole expression,
$\partial_{\nu}C^{\mu\nu\sigma}=0=\partial_{\sigma}C^{\mu\nu\sigma}$,
we find that the axial abelian global current has an anomaly given
by\footnote{The normalization of the generators $T^{I}$ of the
global SU($N_{f}$) is given by ${\rm tr\,}(T^{I} T^{J})={1\over
2}\delta^{IJ}$.}
\begin{eqnarray}
\partial_{\mu}J^{\mu}_{A}=-{g^{2}N_{f}\over 32\pi^{2}}\varepsilon^{\mu\nu\sigma
\lambda}F_{\mu\nu}^{a}F^{a\,\mu\nu}.
\label{eq:anomalyabelianaxial}
\end{eqnarray}

In the case of the nonabelian axial global symmetry SU($N_{f}$)$_{A}$
the calculation of the anomaly is made
as above. The result, however, is quite different since in this
case we conclude that the nonabelian axial current $J_{A}^{a\,\mu}$ is
not anomalous.  This can be easily seen by noticing that
associated with the axial current vertex we have a generator $T^{I}$
of SU($N_{f}$), whereas for the two gluon vertices we have the generators
$\tau^{a}$ of the gauge group SU($N_{c}$). Therefore, the triangle diagram
is proportional to the group-theoretic factor 
\begin{eqnarray}
\left[\,\,
\parbox{40mm}{
\begin{fmfgraph*}(100,80)
\fmfleft{i}
\fmfright{o1,o2}
\fmf{double,label=$J_{A}^{I\mu}$}{i,v4}
\fmf{double}{v4,v1}
\fmf{gluon,label=$g$,tension=0.3}{v2,o1}
\fmf{gluon,label=$g$,tension=0.3}{v3,o2}
\fmfcyclen{fermion,tension=0.05,label=$Q^f$}{v}{3}
\fmfdotn{v}{1}
\end{fmfgraph*}
}\right]_{\rm symmetric}\sim
{\rm tr\,}T^{I}\,\,{\rm tr\,}\{\tau^{a},\tau^{b}\}=0
\end{eqnarray}
which vanishes because the generators of SU($N_{f}$) are traceless. 

From here we would conclude that the nonabelian axial symmetry
SU($N_{f}$)$_{A}$ is nonanomalous. However this is not the whole story
since quarks are charged particles that also couple to photons. Hence
there is a second potential source of an anomaly coming from the the one-loop
triangle diagram coupling $J_{A}^{I\,\mu}$ to two photons
\begin{eqnarray}
\langle0|T\left[J^{I\,\mu}_{A}(x)j_{\rm em}^{\nu}(x')j_{\rm em}^{\sigma}(0)
\right]|0\rangle=\sum_{f=1}^{N_{f}}
\left[\,\,\,\,\parbox{40mm}{
\begin{fmfgraph*}(100,80)
\fmfleft{i}
\fmfright{o1,o2}
\fmf{double,label=$J_{A}^{I \mu}$}{i,v4}
\fmf{double}{v4,v1}
\fmf{photon,label=$\gamma$,tension=0.3}{v2,o1}
\fmf{photon,label=$\gamma$,tension=0.3}{v3,o2}
\fmfcyclen{fermion,tension=0.05,label=$Q^{f}$}{v}{3}
\fmfdotn{v}{1}
\end{fmfgraph*}
}
\right]_{\rm symmetric}
\label{eq:triangle3}
\end{eqnarray}
where $j^{\mu}_{\rm em}$ is the electromagnetic current
\begin{eqnarray}
j^{\mu}_{\rm em}=\sum_{f=1}^{N_{f}}q_{f}\,\overline{Q}^{f}\gamma^{\mu}
Q^{f},
\end{eqnarray}
with $q_{f}$ the electric charge of the $f$-th quark flavor. A
calculation of the diagram in (\ref{eq:triangle3}) shows the existence
of an Adler-Bell-Jackiw anomaly given by
\begin{eqnarray}
\partial_{\mu}J^{I\,\mu}_{A}=-{N_{c}\over 16\pi^{2}}\left[\sum_{f=1}^{N_{f}}
(T^{I})_{ff}\,q_{f}^{2}\right]\varepsilon^{\mu\nu\sigma\lambda}F_{\mu\nu}
F_{\sigma\lambda},
\end{eqnarray}
where $F_{\mu\nu}$ is the field strength of the electromagnetic field
coupling to the quarks. The only chance for the anomaly to cancel is
that the factor between brackets in this equation be identically
zero.

Before proceeding let us summarize the results found so far. Because
of the presence of anomalies the axial part of the global chiral
symmetry, SU($N_{f}$)$_{A}$ and U(1)$_{A}$ are not realized quantum
mechanically in general. We found that U(1)$_{A}$ is always affected
by an anomaly. However, because the right-hand side of the anomaly
equation (\ref{eq:anomalyabelianaxial}) is a total derivative, the
anomalous character of $J_{A}^{\mu}$ does not explain the absence of
U(1)$_{A}$ multiplets in the hadron spectrum, since a new current can
be constructed which is conserved. In addition, the nonexistence of
candidates for a Goldstone boson associated with the right quantum
numbers indicates that U(1)$_{A}$ is not spontaneously broken either,
so it has be explicitly broken somehow. This is the so-called U(1)-problem
which was solved by 't Hooft
\cite{thooftU(1)}, who showed how the contribution of quantum transitions
between vacua with topologically nontrivial gauge field configurations
(instantons) results in an explicit breaking of this symmetry.

Due to the dynamics of the SU($N_{c}$) gauge
theory the axial nonabelian symmetry is spontaneously broken due to
the presence at low energies of a vacuum expectation value for the
fermion bilinear $\overline{Q}^{f}Q^{f}$
\begin{eqnarray}
\langle 0|\overline{Q}^{f}Q^{f}|0\rangle\neq 0 \hspace*{2cm}
\mbox{(No summation in $f$!)}.
\label{chisb}
\end{eqnarray}
This nonvanishing vacuum expectation value for the quark bilinear
actually breaks chiral invariance spontaneously to the vector subgroup
SU($N_{f}$)$_{V}$, so the only subgroup of the original global symmetry
that is realized by the full theory at low energy is
\begin{eqnarray}
\mbox{U($N_{f}$)}_{L}\times\mbox{U($N_{f}$)}_{R} \longrightarrow
\mbox{SU($N_{f}$)}_{V}\times\mbox{U(1)}_{B}.
\end{eqnarray}
Associated with this breaking a Goldstone boson should appear with the
quantum numbers of the broken nonabelian current. For example, in the
case of QCD the Goldstone bosons associated with the spontaneously
symmetry breaking induced by the vacuum expectation values $\langle
\overline{u} u\rangle$, $\langle \overline{d}d\rangle$ and $\langle
(\overline{u}d-\overline{d}u)\rangle$ have been identified as the
pions $\pi^{0}$, $\pi^{\pm}$. These bosons are not exactly massless
because of the nonvanishing mass of the $u$ and $d$ quarks. Since the 
global chiral symmetry is already slightly broken by mass terms in 
the Lagrangian, the associated Goldstone bosons also have masses although
they are very light compared to the masses of other hadrons.

In order to have a better physical understanding of the role of
anomalies in the physics of strong interactions we particularize now
our analysis of the case of real QCD. Since the $u$ and $d$ quarks are
much lighter than the other four flavors, QCD at low energies can be
well described by including only these two flavors and ignoring
heavier quarks. In this approximation, from our previous discussion we know
that the low energy global symmetry of the theory is 
SU(2)$_{V}\times$U(1)$_{B}$, where now the vector group SU(2)$_{V}$ is
the well-known isospin symmetry. The axial U(1)$_{A}$ current is 
anomalous due to Eq. (\ref{eq:anomalyabelianaxial}) with $N_{f}=2$.
In the case of the nonabelian axial symmetry SU(2)$_{A}$, taking
into account that $q_{u}={2\over 3}e$ and $q_{d}=-{1\over 3}e$ 
and that the three generators of SU(2) can be written in terms of the
Pauli matrices as $T^{K}={1\over 2}\sigma^{K}$ we find
\begin{eqnarray}
\sum_{f=u,d}(T^{1})_{ff}\,q_{f}^{2}=\sum_{f=u,d}(T^{1})_{ff}\,q_{f}^{2}=0,
\hspace*{1cm}
\sum_{f=u,d}(T^{3})_{ff}\,q_{f}^{2}={e^{2}\over 6}.
\end{eqnarray}
Therefore $J_{A}^{3\,\mu}$ is anomalous. 

Physically, the anomaly in the axial current $J_{A}^{3\,\mu}$ has an
important consequence. In the quark model, the wave function of the
neutral pion $\pi^{0}$ is given in terms of those for the $u$ and $d$ quark
by
\begin{eqnarray}
|\pi^{0}\rangle={1\over \sqrt{2}}\left(|\bar{u}\rangle|u\rangle-
|\bar{d}\rangle|d\rangle\right).
\end{eqnarray}
The isospin quantum numbers of $|\pi^{0}\rangle$ are those of the
generator $T^{3}$. Actually the analogy goes further since 
$\partial_{\mu}J^{3\,\mu}_{A}$ is the operator creating
a pion $\pi^{0}$ out of the vacuum
\begin{eqnarray}
|\pi^{0}\rangle\sim \partial_{\mu}J^{3\,\mu}_{A}|0\rangle.
\end{eqnarray}
This leads to the physical interpretation of the triangle diagram
(\ref{eq:triangle3}) with $J^{3\,\mu}_{A}$ as the one loop
contribution to the decay of a neutral pion into two photons
\begin{eqnarray}
\pi^{0}\longrightarrow 2\gamma\,.
\end{eqnarray}

This is an interesting piece of physics. In 1967 Sutherland and Veltman
\cite{sutherland_veltman} presented a calculation, using current 
algebra techniques, according to which the decay of the pion into two
photons should be suppressed. This however contradicted the
experimental evidence that showed the existence of such a decay. The
way out to this paradox, as pointed out in \cite{ABJ}, is the axial
anomaly. What happens is that the current algebra analysis overlooks
the ambiguities associated with the regularization of divergences in
Quantum Field Theory. A QED evaluation of the triangle diagram leads
to a divergent integral that has to be regularized somehow. It is in
this process that the Adler-Bell-Jackiw axial anomaly appears
resulting in a nonvanishing value for the $\pi^{0}\rightarrow 2\gamma$
amplitude\footnote{An early computation of the triangle diagram for
the electromagnetic decay of the pion was made by Steinberger in
\cite{steinberger}.}.

The existence of anomalies associated with global currents does not
necessarily mean difficulties for the theory. On the contrary, as we
saw in the case of the axial anomaly it is its existence what allows
for a solution of the Sutherland-Veltman paradox and an explanation of
the electromagnetic decay of the pion. The situation, however, is very
different if we deal with local symmetries. A quantum mechanical
violation of gauge symmetry leads to all kinds of problems, from lack
of renormalizability to nondecoupling of negative norm states.  This
is because the presence of an anomaly in the theory implies that the
Gauss' law constraint $\vec{\nabla}\cdot\vec{E}_{a}=\rho_a$ cannot be
consistently implemented in the quantum theory. As a consequence
states that classically are eliminated by the gauge symmetry become
propagating fields in the quantum theory, thus spoiling the consistency
of the theory.

Anomalies in a gauge symmetry can be expected only in chiral theories
where left and right-handed fermions transform in different
representations of the gauge group. Physically, the most interesting
example of such theories is the electroweak sector of the Standard
Model where, for example, left handed fermions transform as doublets
under SU(2) whereas right-handed fermions are singlets. On the other
hand, QCD is free of gauge anomalies since both left- and right-handed
quarks transform in the fundamental representation of SU(3). 

We consider the Lagrangian 
\begin{eqnarray}
\mathcal{L}=-{1\over 4}F^{a\,\mu\nu}F^{a}_{\mu\nu}
+i\sum_{i=1}^{N_{+}}\overline{\psi}^{\,i}_{+}\Dslash^{(+)}\psi^{i}_{+}
+i\sum_{j=1}^{N_{-}}\overline{\psi}^{\,j}_{-}\Dslash^{(-)}\psi^{j}_{-},
\end{eqnarray}
where the chiral fermions $\psi^{i}_{\pm}$ transform according to
the representations $\tau_{i,\pm}^{a}$ of the gauge group $G$
($a=1,\ldots,{\rm dim\,}G$). The covariant derivatives
$D_{\mu}^{(\pm)}$ are then defined by
\begin{eqnarray}
D_{\mu}^{(\pm)}\psi_{\pm}^{i}=\partial_{\mu}\psi_{\pm}^{i}
+igA_{\mu}^{K}\tau^{K}_{i,\pm}\psi_{\pm}^{i}.
\end{eqnarray}
As for global symmetries, anomalies in the gauge symmetry
appear in the triangle diagram with one axial and two vector gauge
current vertices
\begin{eqnarray}
\langle 0|T\left[j^{a\,\mu}_{A}(x)j^{b\,\nu}_{V}(x')
j^{c\,\sigma}_{V}(0)\right]|0\rangle=
\left[\,\,
\parbox{40mm}{
\begin{fmfgraph*}(100,80)
\fmfleft{i}
\fmfright{o1,o2}
\fmf{gluon,label=${j}^{a\mu}_{A}$,tension=0.3}{i,v1}
\fmf{gluon,label=${j}^{b\nu}_{V}$,tension=0.25}{v2,o1}
\fmf{gluon,label=${j}^{c\sigma}_{V}$,tension=0.25}{v3,o2}
\fmfcyclen{fermion,tension=0.05}{v}{3}
\end{fmfgraph*}
}\right]_{\rm symmetric}
\label{eq:triangle4}
\end{eqnarray}
where gauge vector and axial currents $j_{V}^{a\,\mu}$, $j_{A}^{a\,\mu}$ 
are given by 
\begin{eqnarray}
j^{a\mu}_{V}&=&\sum_{i=1}^{N_{+}}\overline{\psi}^{i}_{+}
\tau^{a}_{+}\gamma^{\mu}
\psi^{i}_{+}+\sum_{j=1}^{N_{-}}\overline{\psi}^{j}_{-}\tau^{a}_{-}\gamma^{\mu}
\psi^{j}_{-} ,\nonumber \\
j^{a\mu}_{A}&=&\sum_{i=1}^{N_{+}}\overline{\psi}^{i}_{+}
\tau^{a}_{+}\gamma^{\mu}
\psi^{i}_{+}-\sum_{i=1}^{N_{-}}\overline{\psi}^{j}_{-}\tau^{a}_{-}\gamma^{\mu}
\psi^{j}_{-}.
\end{eqnarray}
Luckily, we do not have to compute the whole diagram in order to find
an anomaly cancellation condition, it is enough if we
calculate the overall group theoretical factor.  In the case of the
diagram in Eq.  (\ref{eq:triangle4}) for every fermion species running
in the loop this factor is equal to
\begin{eqnarray}
{\rm tr\,}\left[\tau^{a}_{i,\pm}\{\tau^{b}_{i,\pm},\tau^{c}_{i,\pm}\}\right],
\end{eqnarray}
where the sign $\pm$ corresponds respectively to the generators of the
representation of the gauge group for the left and right-handed
fermions. Hence the anomaly cancellation condition reads
\begin{eqnarray}
\sum_{i=1}^{N_{+}}
{\rm tr\,}\left[\tau^{a}_{i,+}\{\tau^{b}_{i,+},\tau^{c}_{i,+}\}\right]
-
\sum_{j=1}^{N_{-}}
{\rm tr\,}\left[\tau^{a}_{j,-}\{\tau^{b}_{j,-},\tau^{c}_{j,-}\}\right]
=0.
\end{eqnarray}

Knowing this we can proceed to check the anomaly cancellation in the
Standard Model SU(3)$\times$SU(2)$\times$U(1). Left
handed fermions (both leptons and quarks) transform as doublets with
respect to the SU(2) factor whereas the right-handed components are
singlets.  The charge with respect to the U(1) part, the hypercharge
$Y$, is determined by the Gell-Mann-Nishijima formula
\begin{eqnarray}
Q=T_{3}+Y,
\end{eqnarray}
where $Q$ is the electric charge of the corresponding particle and
$T_{3}$ is the eigenvalue with respect to the third generator of the
SU(2) group in the corresponding representation: $T_{3}={1\over
2}\sigma^{3}$ for the doublets and $T_{3}=0$ for the singlets. For the
first family of quarks ($u$, $d$) and leptons ($e$, $\nu_{e}$) we have
the following field content
\begin{eqnarray}
\mbox{quarks:} & \hspace*{1cm} & \left(
\begin{array}{c}
u^{\alpha}  \\
d^{\alpha}
\end{array}
\right)_{L,{1\over 6}} \hspace*{1cm} u_{R,{2\over 3}}^{\alpha}
\hspace*{1cm} d_{R,{2\over 3}}^{\alpha} \nonumber \\
\mbox{leptons:} & \hspace*{1cm} &\left(
\begin{array}{c}
\nu_{e}  \\
e
\end{array}
\right)_{L,-{1\over 2}} \hspace*{0.8cm} e_{R,-1}
\end{eqnarray}
where $\alpha=1,2,3$ labels the color quantum number and the subscript
indicates the value of the weak hypercharge $Y$. Denoting the
representations of SU(3)$\times$SU(2)$\times$U(1) by
$(n_{c},n_{w})_{Y}$, with $n_c$ and $n_w$ the representations of SU(3)
and SU(2) respectively and $Y$ the hypercharge, the matter content of the
Standard Model consists of a three family replication of the
representations:
\begin{eqnarray}
\mbox{left-handed fermions:} &\hspace*{0.5cm} &
(3,2)^{L}_{1\over 6} \hspace*{1cm} (1,2)^{L}_{-{1\over 2}} \nonumber \\
& & \\
\mbox{right-handed fermions:} &\hspace*{0.5cm} & 
(3,1)^{R}_{2\over 3} \hspace*{1cm} (3,1)^{R}_{-{1\over 3}} \hspace*{1cm}
(1,1)^{R}_{-1}. \nonumber 
\end{eqnarray}
In computing the triangle diagram we have 10 possibilities depending on 
which factor of the gauge group SU(3)$\times$SU(2)$\times$U(1) couples
to each vertex:
\begin{eqnarray*}
\begin{array}{lllll}
\mbox{SU(3)}^{3} & \hspace*{1cm} & \mbox{SU(2)}^{3} &
\hspace*{1cm} &  \mbox{U(1)}^{3} \\
 & & & & \\
\mbox{SU(3)}^{2}\,\mbox{SU(2)} & & \mbox{SU(2)}^{2}\,\mbox{U(1)} & & \\
 & & & & \\
\mbox{SU(3)}^{2}\,\mbox{U(1)} & & \mbox{SU(2)}\,\mbox{U(1)}^{2} & &\\
& & & & \\
\mbox{SU(3)}\,\mbox{SU(2)}^{2} & & &  & \\
& & &  & \\
\mbox{SU(3)}\,\mbox{SU(2)}\,\mbox{U(1)} & & &  & \\
& & & & \\
\mbox{SU(3)}\,\mbox{U(1)}^{2} & &  & & \\
& & & & 
\end{array}
\end{eqnarray*}
It is easy to check that some of them do not give rise
to anomalies. For example the anomaly for the SU(3)$^{3}$ case cancels
because left and right-handed quarks transform in the same
representation. In the case of SU(2)$^{3}$ the cancellation happens term
by term because of the Pauli matrices identity $\sigma^{a}\sigma^{b}=
\delta^{ab}+i\varepsilon^{abc}\sigma^{c}$ that leads to
\begin{eqnarray}
{\rm tr\,}\left[\sigma^{a}\{\sigma^{b},\sigma^{c}\}\right]=2\left(
{\rm tr\,}\sigma^{a}\right)\delta^{bc}=0.
\end{eqnarray}
However the hardest anomaly cancellation condition to satisfy is the
one with three U(1)'s. In this case the absence of anomalies within a
single family is guaranteed by the nontrivial identity
\begin{eqnarray}
\sum_{\rm left}Y_{+}^{3}-\sum_{\rm right}Y_{-}^{3}&=&
{3\times 2\times\left({1\over 6}\right)^{3}}+2\times
\left({-{1\over 2}}\right)^{3} 
-3\times\left({2\over 3}\right)^{3}
-3\times\left(-{1\over 3}\right)^{3}
-(-1)^{3} \nonumber \\
&=&\left(-{3\over 4}\right)+\left({3\over 4}\right)=0.
\end{eqnarray}
It is remarkable that the anomaly exactly cancels between leptons and
quarks. Notice that this result holds even if a right-handed sterile
neutrino is added since such a particle is a singlet under the whole
Standard Model gauge group and therefore does not contribute to the
triangle diagram.  Therefore we see how the matter content of the
Standard Model conspires to yield a consistent quantum field theory.

In all our discussion of anomalies we only  considered
the computation of one-loop diagrams.  It may happen that higher loop
orders impose additional conditions.
Fortunately this is not so:
the Adler-Bardeen theorem \cite{Adler-Bardeen} guarantees that the axial 
anomaly only receives contributions from one loop diagrams. Therefore,
once anomalies are canceled (if possible) at one loop we know that there
will be no new conditions coming from higher-loop diagrams in  
perturbation theory.

The Adler-Bardeen theorem, however, only applies in perturbation
theory.  It is nonetheless possible that nonperturbative effects can
result in the quantum violation of a gauge symmetry. This is precisely
the case pointed out by Witten \cite{Witten-anom} with respect to the
SU(2) gauge symmetry of the Standard Model. In this case the problem
lies in the nontrivial topology of the gauge group SU(2). The
invariance of the theory with respect to gauge transformations which
are not in the connected component  of the identity makes all correlation
functions equal to zero. Only when the number of left-handed
SU(2) fermion doublets is even gauge invariance allows for a nontrivial 
theory. It is again remarkable that the family structure of the Standard
Model makes this anomaly to cancel
\begin{eqnarray}
3\times\left(
\begin{array}{c}
u \\
d
\end{array}
\right)_{L}+1\times
\left(
\begin{array}{c}
\nu_{e} \\
e
\end{array}
\right)_{L}= \mbox{4 SU(2)-doublets},
\end{eqnarray}
where the factor of 3 comes from the number of colors.

\section{Renormalization}
\label{sec:renormalization}

\subsection{Removing infinities}
\label{sec:removinginfinities}
 
From its very early stages, Quantum Field Theory was faced with
infinities. They emerged in the calculation of most physical
quantities, such as the correction to the charge of the electron due
to the interactions with the radiation field. The way these
divergences where handled in the 1940s, starting with Kramers, was
physically very much in the spirit of the Quantum Theory emphasis in
observable quantities: since the observed magnitude of physical
quantities (such as the charge of the electron) is finite, this number
should arise from the addition of a ``bare'' (unobservable) value and
the quantum corrections.  The fact that both of these quantities were
divergent was not a problem physically, since only its finite sum was
an observable quantity.  To make thing mathematically sound, the
handling of infinities requires the introduction of some
regularization procedure which cuts the divergent integrals off at some
momentum scale $\Lambda$. Morally speaking, the physical value of an
observable $\mathcal{O}_{\rm physical}$ is given by
\begin{eqnarray}
\mathcal{O}_{\rm physical}=\lim_{\Lambda\rightarrow\infty}\left[
\mathcal{O}(\Lambda)_{\rm bare}+\Delta\mathcal{O}(\Lambda)_{\hbar}\right],
\end{eqnarray}
where $\Delta\mathcal{O}(\Lambda)_{\hbar}$ represents the regularized quantum
corrections.

To make this qualitative discussion more precise we compute the 
corrections to the electric charge in Quantum Electrodynamics. We consider the
process of annihilation of an electron-positron pair to create a
muon-antimuon pair  $e^{-}e^{+}\rightarrow\mu^{+}\mu^{-}$. To
lowest order in the electric charge $e$ the only diagram contributing is
\begin{eqnarray*}
\parbox{40mm}{
\begin{fmfgraph*}(90,60)
\fmfleft{i1,i2}
\fmfright{o1,o2}
\fmf{fermion,label=$e^-$}{i1,v1}
\fmf{fermion,label=$e^+$}{v1,i2}
\fmf{photon}{v1,v2}
\fmf{photon,label=$\gamma$}{v2,v3}
\fmf{fermion,label=$\mu^+$}{o1,v3}
\fmf{fermion,label=$\mu^-$}{v3,o2}
\end{fmfgraph*}
}
\end{eqnarray*}
However, the corrections at order $e^{4}$ to this result requires the
calculation of seven more diagrams
\begin{eqnarray*}
& &\parbox{40mm}{
\begin{fmfgraph*}(90,60)
\fmfleft{i1,i2}
\fmfright{o1,o2}
\fmf{fermion,label=$e^-$}{i1,v1}
\fmf{fermion,label=$e^+$}{v1,i2}
\fmf{photon}{v1,v2}
\fmf{photon}{v3,v4}
\fmf{fermion,left,tension=.3}{v2,v3,v2}
\fmf{fermion,label=$\mu^+$}{o1,v4}
\fmf{fermion,label=$\mu^-$}{v4,o2}
\end{fmfgraph*}
}+
\parbox{40mm}{
\begin{fmfgraph*}(90,60)
\fmfleft{i1,i2}
\fmfright{o1,o2}
\fmf{fermion,label=$e^-$}{i1,v3}
\fmf{fermion,label=$e^+$}{v4,i2}
\fmf{plain}{v3,v1,v4}
\fmf{photon,tension=0.2}{v3,v4}
\fmf{photon}{v1,v2}
\fmf{photon}{v2,v5}
\fmf{fermion,label=$\mu^+$}{o1,v5}
\fmf{fermion,label=$\mu^-$}{v5,o2}
\end{fmfgraph*}
}+
\parbox{40mm}{
\begin{fmfgraph*}(90,60)
\fmfright{i1,i2}
\fmfleft{o1,o2}
\fmf{fermion,label=$\mu^+$}{i1,v3}
\fmf{fermion,label=$\mu^-$}{v4,i2}
\fmf{plain}{v3,v1,v4}
\fmf{photon,tension=0.2}{v3,v4}
\fmf{photon}{v1,v2}
\fmf{photon}{v2,v5}
\fmf{fermion,label=$e^-$}{o1,v5}
\fmf{fermion,label=$e^+$}{v5,o2}
\end{fmfgraph*}
} \\ & & \\
&+&\parbox{40mm}{
\begin{fmfgraph*}(90,60)
\fmfleft{i1,i2}
\fmfright{o1,o2}
\fmf{fermion,label=$e^-$}{i1,v3}
\fmf{fermion,label=$e^+$}{v1,i2}
\fmf{plain}{v3,v6}
\fmf{plain}{v6,v1}
\fmf{photon,right,tension=0}{v3,v6}
\fmf{photon}{v1,v2}
\fmf{photon}{v2,v5}
\fmf{fermion,label=$\mu^+$}{o1,v5}
\fmf{fermion,label=$\mu^-$}{v5,o2}
\end{fmfgraph*}
}+
\parbox{40mm}{
\begin{fmfgraph*}(90,60)
\fmfleft{i1,i2}
\fmfright{o1,o2}
\fmf{fermion,label=$e^-$}{i1,v1}
\fmf{plain}{v1,v6}
\fmf{plain}{v6,v3}
\fmf{photon,right,tension=0}{v6,v3}
\fmf{fermion,label=$e^+$}{v3,i2}
\fmf{photon}{v1,v2}
\fmf{photon}{v2,v5}
\fmf{fermion,label=$\mu^+$}{o1,v5}
\fmf{fermion,label=$\mu^-$}{v5,o2}
\end{fmfgraph*}
}+
\parbox{40mm}{
\begin{fmfgraph*}(90,60)
\fmfright{i1,i2}
\fmfleft{o1,o2}
\fmf{fermion,label=$\mu^+$}{i1,v1}
\fmf{plain}{v1,v6}
\fmf{plain}{v6,v3}
\fmf{photon,right,tension=0}{v6,v3}
\fmf{fermion,label=$\mu^-$}{v3,i2}
\fmf{photon}{v1,v2}
\fmf{photon}{v2,v5}
\fmf{fermion,label=$e^+$}{o1,v5}
\fmf{fermion,label=$e^-$}{v5,o2}
\end{fmfgraph*}
} \\ & & 
\\
&+&\parbox{40mm}{
\begin{fmfgraph*}(90,60)
\fmfright{i1,i2}
\fmfleft{o1,o2}
\fmf{fermion,label=$\mu^+$}{i1,v3}
\fmf{fermion,label=$\mu^-$}{v1,i2}
\fmf{plain}{v3,v6}
\fmf{plain}{v6,v1}
\fmf{photon,right,tension=0}{v3,v6}
\fmf{photon}{v1,v2}
\fmf{photon}{v2,v5}
\fmf{fermion,label=$e^+$}{o1,v5}
\fmf{fermion,label=$e^-$}{v5,o2}
\end{fmfgraph*}
}
\end{eqnarray*}

In order to compute the renormalization of the charge we consider the
first diagram which takes into account the first correction to the
propagator of the virtual photon interchanged between the pairs due to
vacuum polarization.  We begin by evaluating
\begin{eqnarray}
\parbox{40mm}{
\begin{fmfgraph*}(100,90)
\fmfleft{i}
\fmfright{o}
\fmf{photon}{i,v3}
\fmf{photon}{v3,v1}
\fmf{photon}{v4,v2}
\fmf{photon}{v2,o}
\fmf{fermion,left,tension=.3}{v1,v4,v1}
\end{fmfgraph*}
}={-i\eta^{\mu\alpha}\over q^2+i\epsilon}\left[\hspace*{-0.1cm}
\parbox{30mm}{
\begin{fmfgraph*}(60,60)
\fmfleft{i}
\fmfright{o}
\fmf{phantom}{i,v1}
\fmf{phantom}{v2,o}
\fmf{fermion,left,tension=.3}{v1,v2,v1}
\fmfdotn{v}{2}
\fmflabel{$\alpha$}{v1}
\fmflabel{$\beta$}{v2}
\end{fmfgraph*}
}\hspace*{-1cm}\right]
{-i\eta^{\beta\nu}\over q^2+i\epsilon},
\label{eq:polarization_tensor}
\end{eqnarray}
where the diagram between brackets is given by
\begin{eqnarray}
\parbox{30mm}{
\begin{fmfgraph*}(60,60)
\fmfleft{i}
\fmfright{o}
\fmf{phantom}{i,v1}
\fmf{phantom}{v2,o}
\fmf{fermion,left,tension=.3}{v1,v2,v1}
\fmfdotn{v}{2}
\fmflabel{$\alpha$}{v1}
\fmflabel{$\beta$}{v2}
\end{fmfgraph*}
}\hspace*{-1cm}\equiv \Pi^{\alpha\beta}(q)=
i^{2}(-ie)^{2}(-1)\int {d^{4}k\over (2\pi)^{4}}{
{\rm Tr\,}(\fslash{k}+m_{e})\gamma^{\alpha}(\fslash{k}+\fslash{q}+
m_{e})\gamma^{\beta}\over \left[k^{2}-m_{e}^{2}+i\epsilon\right]
\left[(k+q)^{2}-m_{e}^{2}+i\epsilon\right]}.
\label{eq:Pialphabeta}
\end{eqnarray}
Physically this diagram includes the correction to the propagator due
to the polarization of the vacuum, i.e. the creation of virtual
electron-positron pairs by the propagating photon.  The momentum $q$
is the total momentum of the
electron-positron pair in the intermediate channel.

It is instructive to look at this diagram from the point of view of
perturbation theory in nonrelativistic Quantum Mechanics. In each
vertex the interaction consists of the annihilation (resp. creation)
of a photon and the creation (resp. annihilation) of an
electron-positron pair. This can be implemented by the interaction
Hamiltonian
\begin{eqnarray}
H_{\rm int}=e\int d^{3}x\,\overline{\psi}\gamma^{\mu}\psi A_{\mu}.
\end{eqnarray}
All fields inside the integral can be expressed in terms of
the corresponding creation-annihilation operators for photons, electrons and
positrons. In Quantum Mechanics, the change in the wave function at
first order in the perturbation $H_{\rm int}$ is given by
\begin{eqnarray}
|\gamma,{\rm in}\rangle=|\gamma,{\rm in}\rangle_{0}+\sum_{n}
{\langle n|H_{\rm int}|\gamma,{\rm in}\rangle_{0}\over E_{\rm in}-E_{n}}
|n\rangle
\end{eqnarray}
and similarly for $|\gamma,{\rm out}\rangle$, where we have denoted
symbolically by $|n\rangle$ all the possible states of the
electron-positron pair. Since these states are orthogonal to
$|\gamma,{\rm in}\rangle_{0}$, $|\gamma,{\rm out}\rangle_{0}$, we find
torder $e^{2}$
\begin{eqnarray}
\langle\gamma,{\rm in}|\gamma',{\rm out}\rangle=
{}_{0}\langle\gamma,{\rm in}|\gamma',{\rm out}\rangle_{0}+
\sum_{n}{{}_{0}\langle \gamma,{\rm in}|H_{\rm int}|n\rangle
\,\langle n|H_{\rm int}|\gamma',{\rm out}\rangle_{0} \over 
(E_{\rm in}-E_{n})(E_{\rm out}-E_{n})}+\mathcal{O}(e^{4}).
\end{eqnarray}
Hence, we see that the diagram of Eq.  (\ref{eq:polarization_tensor})
really corresponds to the order-$e^{2}$ correction to the photon
propagator $\langle\gamma,{\rm in}|
\gamma',{\rm out}\rangle$
\begin{eqnarray}
\parbox{40mm}{
\begin{fmfgraph*}(90,60)
\fmfleft{i}
\fmfright{o}
\fmf{photon,label=$\gamma$}{i,v1}
\fmf{photon}{v1,v2}
\fmf{photon,label=$\gamma'$}{v2,o}
\end{fmfgraph*}
}
&\longrightarrow&  \,\,\,\,
{}_{0}\langle\gamma,{\rm in}|\gamma',{\rm out}\rangle_{0} \nonumber \\
\parbox{40mm}{
\begin{fmfgraph*}(90,60)
\fmfleft{i}
\fmfright{o}
\fmf{photon,label=$\gamma$}{i,v3}
\fmf{photon}{v3,v1}
\fmf{photon}{v4,v2}
\fmf{photon,label=$\gamma'$}{v2,o}
\fmf{fermion,left,tension=.3}{v1,v4,v1}
\end{fmfgraph*}
}
&\longrightarrow&  \,\,\,\,
\sum_{n}{\langle \gamma,{\rm in}|H_{\rm int}|n\rangle
\,\langle n|H_{\rm int}|\gamma',{\rm out}\rangle \over 
(E_{\rm in}-E_{n})(E_{\rm out}-E_{n})}.
\end{eqnarray}

Once we understood the physical meaning of the Feynman diagram to be computed
we proceed to its evaluation. In principle there is no problem in
computing the integral in Eq. (\ref{eq:polarization_tensor}) for
nonzero values of the electron mass. However since here we are going
to be mostly interested in seeing how the divergence of the integral
results in a scale-dependent renormalization of the electric charge,
we will set $m_{e}=0$. This is something safe to do, since in the case
of this diagram we are not inducing new infrared divergences in taking
the electron as massless. Implementing gauge invariance and 
using standard techniques in the computation of Feynman diagrams (see
references \cite{ours}-\cite{banks})
the polarization tensor $\Pi_{\mu\nu}(q)$
defined in Eq. (\ref{eq:Pialphabeta}) can be written as
\begin{eqnarray}
\Pi_{\mu\nu}(q)=\left(q^2\eta_{\mu\nu}-q_{\mu}q_{\nu}\right)\Pi(q^{2})
\end{eqnarray}
with
\begin{eqnarray}
\Pi(q)=8e^{2}\int_{0}^{1}dx\int {d^{4}k\over(2\pi)^{4}}
{x(1-x)\over [k^{2}-m^{2}+x(1-x)q^{2}+i\epsilon]^{2}}
\end{eqnarray}
To handle this divergent
integral we have to figure out some procedure to render it finite. This can
be done in several ways, but here we choose to cut the
integrals off at a high energy scale $\Lambda$, where new physics might
be at work, $|p|<\Lambda$. This gives the result
\begin{eqnarray}
\Pi(q^{2})\simeq {e^{2}\over 12\pi^2}\log\left({q^{2}\over \Lambda^{2}}\right)
+\mbox{finite terms}.
\end{eqnarray}
If we would send the cutoff to infinity $\Lambda\rightarrow\infty$ the
divergence blows up and something has to be done about it. 

If we want to make sense out of this, we have to go back to the physical
question that led us to compute Eq. (\ref{eq:polarization_tensor}). 
Our primordial motivation was to compute the corrections to the annihilation
of two electrons into two muons. Including the correction to the 
propagator of the virtual photon we have
\begin{eqnarray}
\parbox{40mm}{
\begin{fmfgraph*}(90,60)
\fmfleft{i1,i2}
\fmfright{o1,o2}
\fmf{fermion}{i1,v1,i2}
\fmf{photon}{v1,v2}
\fmfblob{0.15w}{v2}
\fmf{photon}{v2,v3}
\fmf{fermion}{o1,v3,o2}
\end{fmfgraph*}
}&=&\,\,\,\,\,
\parbox{40mm}{
\begin{fmfgraph*}(90,60)
\fmfleft{i1,i2}
\fmfright{o1,o2}
\fmf{fermion}{i1,v1,i2}
\fmf{photon}{v1,v2}
\fmf{photon}{v2,v3}
\fmf{fermion}{o1,v3,o2}
\end{fmfgraph*}
}+\,\,\,\,\,
\parbox{40mm}{
\begin{fmfgraph*}(90,60)
\fmfleft{i1,i2}
\fmfright{o1,o2}
\fmf{fermion}{i1,v1,i2}
\fmf{photon}{v1,v2}
\fmf{fermion,left,tension=0.3}{v2,v3,v2}
\fmf{photon}{v3,v4}
\fmf{fermion}{o1,v4,o2}
\end{fmfgraph*}
} \nonumber \\
&=&\eta_{\alpha\beta}\left(\overline{v}_{e}\gamma^{\alpha}u_{e}\right)
{e^2\over 4\pi q^2}\left(\overline{v}_{\mu}\gamma^{\beta}u_{\mu}\right)
+\eta_{\alpha\beta}\left(\overline{v}_{e}\gamma^{\alpha}u_{e}\right)
{e^2\over 4\pi q^2}\Pi(q^{2})
\left(\overline{v}_{\mu}\gamma^{\beta}u_{\mu}\right)
\nonumber\\
&=&\eta_{\alpha\beta}\left(\overline{v}_{e}\gamma^{\alpha}u_{e}\right)
\left\{{e^2\over 4\pi q^2}
\left[1+{e^{2}\over 12\pi^{2}}\log\left({q^{2}\over \Lambda^{2}}\right)\right]
\right\}
\left(\overline{v}_{\mu}\gamma^{\beta}u_{\mu}\right).
\label{eq:divergent}
\end{eqnarray}
Now let us imagine that we are performing a $e^{-}\,e^{+}\rightarrow
\mu^{-}\mu^{+}$ with a center of mass energy $\mu$. From the previous result
we can identify the effective charge of the particles at this energy
scale $e(\mu)$ as
\begin{eqnarray}
\parbox{40mm}{
\begin{fmfgraph*}(90,60)
\fmfleft{i1,i2}
\fmfright{o1,o2}
\fmf{fermion}{i1,v1,i2}
\fmf{photon}{v1,v2}
\fmfblob{0.15w}{v2}
\fmf{photon}{v2,v3}
\fmf{fermion}{o1,v3,o2}
\end{fmfgraph*}
}&=&\,\,\,\,\,
\eta_{\alpha\beta}\left(\overline{v}_{e}\gamma^{\alpha}u_{e}\right)
\left[{e(\mu)^2\over 4\pi q^2}\right]
\left(\overline{v}_{\mu}\gamma^{\beta}u_{\mu}\right).
\label{eq:renormalized}
\end{eqnarray}
This charge, $e(\mu)$, is the quantity that is physically measurable
in our experiment.  Now we can make sense of the formally divergent
result (\ref{eq:divergent}) by assuming that the charge appearing in
the classical Lagrangian of QED is just a ``bare'' value that depends
on the scale $\Lambda$ at which we cut off the theory, $e\equiv
e(\Lambda)_{\rm bare}$. In order to reconcile (\ref{eq:divergent})
with the physical results (\ref{eq:renormalized}) we must assume that
the dependence of the bare (unobservable) charge $e(\Lambda)_{\rm bare}$ 
on the cutoff $\Lambda$ is determined by the identity
\begin{eqnarray}
e(\mu)^{2}=e(\Lambda)_{\rm bare} ^{2}\left[1+{e(\Lambda)_{\rm
bare}^{2}\over 12\pi^{2}}\log\left( {\mu^{2}\over\Lambda^{2}
}\right)\right].
\label{eq:runningcoupling}
\end{eqnarray}
If we still insist in removing the cutoff, $\Lambda\rightarrow\infty$
we have to send the bare charge to zero $e(\Lambda)_{\rm bare}
\rightarrow 0$ in such a way that the effective coupling has the finite
value given by the experiment at the energy scale $\mu$. It is not 
a problem, however, that the bare charge is small for large
values of the cutoff, since
the only measurable quantity is the effective charge that remains finite.
Therefore all observable quantities should be expressed
in perturbation theory as a power series in the physical coupling
$e(\mu)^{2}$ and not in the unphysical bare
coupling $e(\Lambda)_{\rm bare}$. 

\subsection{The beta-function and asymptotic freedom}
\label{sec:betafunction}

We can look at the previous discussion, an in particular
Eq. (\ref{eq:runningcoupling}), from a different point of view.  In
order to remove the ambiguities associated with infinities we have
been forced to introduce a dependence of the coupling constant on the
energy scale at which a process takes place. From the expression of
the physical coupling in terms of the bare charge
(\ref{eq:runningcoupling}) we can actually eliminate the cutoff
$\Lambda$, whose value after all should not affect the value of 
physical quantities. Taking into account that we are working in
perturbation theory in $e(\mu)^{2}$, we can express the bare charge
$e(\Lambda)^{2}_{\rm bare}$ in terms of $e(\mu)^{2}$ as
\begin{eqnarray}
e(\Lambda)^{2}=e(\mu)^{2}\left[1+{e(\mu)^{2}\over 12\pi^{2}}\log\left(
{\mu^{2}\over\Lambda^{2} }\right)\right]+\mathcal{O}[e(\mu)^{6}].
\label{eq:inverse}
\end{eqnarray}
This expression allow us to eliminate all dependence in the cutoff in
the expression of the effective charge at a scale $\mu$ by replacing
$e(\Lambda)_{\rm bare}$ in Eq. (\ref{eq:runningcoupling}) by the 
one computed using (\ref{eq:inverse}) at a given reference energy scale 
$\mu_{0}$
\begin{eqnarray}
e(\mu)^{2}=e(\mu_{0})^{2}\left[1+{e(\mu_{0})^{2}\over 12\pi^{2}}\log\left(
{\mu^{2}\over\mu_{0}^{2} }\right)\right].
\label{eq:QEDmu}
\end{eqnarray}

From this equation we can compute, at this order in perturbation theory,
the effective value of the coupling constant at an energy $\mu$, once we know
its value at some reference energy scale $\mu_{0}$. In the case of the 
electron charge we can use as a reference Thompson's scattering at energies
of the order of the electron mass $m_{e}\simeq 0.5 \mbox{ MeV}$, at where
the value of the electron charge is given by the well known value
\begin{eqnarray}
e(m_{e})^{2}\simeq {1\over 137}.
\end{eqnarray}
With this we can compute $e(\mu)^2$ at any other energy scale applying
Eq. (\ref{eq:QEDmu}), for example at the electron mass
$\mu=m_{e}\simeq 0.5\,$MeV.  However, in computing the electromagnetic
coupling constant at any other scale we must take into account the
fact that other charged particles can run in the loop in
Eq. (\ref{eq:divergent}).  Suppose, for example, that we want to
calculate the fine structure constant at the mass of the $Z^{0}$-boson
$\mu=M_{Z}\equiv 92$ GeV. Then we should include in
Eq. (\ref{eq:QEDmu}) the effect of other fermionic Standard Model
fields with masses below $M_{Z}$.  Doing this, we find\footnote{In the
first version of these notes the argument used to show the growing of
the electromagnetic coupling constant could have led to confusion to
some readers. To avoid this potential problem we include in the
equation for the running coupling $e(\mu)^{2}$ the contribution of all
fermions with masses below $M_{Z}$. We thank Lubos Motl for bringing
this issue to our attention.}
\begin{eqnarray}
e(M_{Z})^{2}
=e(m_{e})^{2}\left[1+{e(m_{e})^{2}\over 12\pi^{2}}\left(\sum_{i}q_{i}^{2}
\right)
\log\left(
{M_{Z}^{2}\over m_e^{2} }\right)\right],
\label{eq:coupling_general}
\end{eqnarray}
where $q_{i}$ is the charge in units of the electron charge of the
$i$-th fermionic species running in the loop and we sum over all
fermions with masses below the mass of the $Z^{0}$ boson.  This
expression shows how the electromagnetic coupling grows with
energy. However, in order to compare with the experimental value of
$e(M_{Z})^{2}$ it is not enough with including the effect of fermionic
fields, since also the $W^{\pm}$ bosons can run in the loop
($M_{W}<M_{Z}$). Taking this into account, as well as threshold
effects, the value of the electron charge at the scale $M_{Z}$ is
found to be
\cite{pdg}
\begin{eqnarray}
e(M_{Z})^{2}\simeq {1\over 128.9}\,\,.  
\end{eqnarray}

This growing of the effective fine structure constant with energy can
be understood heuristically by remembering that the effect of the
polarization of the vacuum shown in the diagram of
Eq. (\ref{eq:polarization_tensor}) amounts to the creation of a
plethora of electron-positron pairs around the location of the charge.
These virtual pairs behave as dipoles that, as in a dielectric medium,
tend to screen this charge and decreasing its value at long distances
(i.e. lower energies).

The variation of the coupling constant with energy is usually encoded in
Quantum Field Theory in the {\em beta function} defined by
\begin{eqnarray}
\beta(g)=\mu{dg\over d\mu}.
\label{eq:beta_function_def}
\end{eqnarray}
In the case of QED the beta function can be computed from Eq. (\ref{eq:QEDmu})
with the result
\begin{eqnarray}
\beta(e)_{\rm QED}={e^{3}\over 12\pi^{2}}.
\label{eq:betaQED}
\end{eqnarray}
The fact that the coefficient of the leading term in the beta-function
is positive $\beta_{0}\equiv {1\over 6\pi}>0$ gives us the overall
behavior of the coupling as we change the
scale. Eq. (\ref{eq:betaQED}) means that, if we start at an energy
where the electric coupling is small enough for our perturbative
treatment to be valid, the effective charge grows with the energy
scale. This growing of the effective coupling constant with energy
means that QED is infrared safe, since the perturbative approximation
gives better and better results as we go to lower energies. Actually,
because the electron is the lighter electrically charged particle and
has a finite nonvanishing mass the running of the fine structure
constant stops at the scale $m_{e}$ in the well-known value ${1\over
137}$. Would other charged fermions with masses below $m_{e}$ be
present in Nature, the effective value of the fine structure constant
in the interaction between these particles would run further to lower
values at energies below the electron mass.

On the other hand if we increase the energy scale $e(\mu)^2$ grows
until at some scale the coupling is of order one and the perturbative
approximation breaks down. In QED this is known as the problem of the
Landau pole but in fact it does not pose any serious threat to the
reliability of QED perturbation theory: a simple calculation shows
that the energy scale at which the theory would become strongly
coupled is $\Lambda_{\rm Landau}\simeq 10^{277}
\mbox{ GeV}$. However, we know that QED does not live that long! 
At much lower scales we expect electromagnetism to be unified with
other interactions, and even  if  this is not the case we will enter 
the uncharted territory of quantum gravity at energies of the order
of $10^{19}$ GeV. 

So much for QED. The next question that one may ask at this stage is
whether it is possible to find quantum field theories with a behavior
opposite to that of QED, i.e. such that they become weakly coupled at
high energies. This is not a purely academic question. In the late
1960s a series of deep-inelastic scattering experiments carried out at
SLAC showed that the quarks behave essentially as free particles
inside hadrons.  The apparent problem was that no theory was known
at that time that would become free at very short distances: the
example set by QED seem to be followed by all the theories that were
studied. This posed a very serious problem for Quantum Field Theory as
a way to describe subnuclear physics, since it seemed that its
predictive power was restricted to electrodynamics but failed
miserably when applied to describe strong interactions.

Nevertheless, this critical time for Quantum Field Theory turned out to be
its finest hour. In 1973 David Gross and Frank Wilczek
\cite{gross-wilczek} and David Politzer \cite{politzer} showed that
nonabelian gauge theories can actually display the required
behavior. For the QCD Lagrangian in Eq. (\ref{eq:QCDL}) the beta
function is given by\footnote{The expression of the beta function of
QCD was also known to 't Hooft \cite{thooftAF}.  There are even
earlier computations in the russian literature \cite{russian-beta}.}
\begin{eqnarray}
\beta(g)=-{g^{3}\over 16\pi^{2}}\left[{11\over 3}N_{c}-
{2\over 3}N_{f}\right].
\end{eqnarray}
In particular, for real QCD ($N_{C}=3$, $N_{f}=6$) we have that
$\beta(g)=-{7g^{3}\over 16\pi^{2}}<0$. This means that for a theory
that is weakly coupled at an energy scale $\mu_{0}$ the coupling constant
decreases as the energy increases $\mu\rightarrow\infty$. This explain the
apparent freedom of quarks inside the hadrons: when the quarks are very 
close together their effective color charge tend to zero.
This phenomenon is called {\em asymptotic freedom}.

Asymptotic free theories display a behavior that is opposite to that 
found above in QED. At high energies their coupling constant approaches 
zero whereas at low energies they become strongly coupled (infrared slavery).
This features are at the heart of the success of QCD as a theory of 
strong interactions, since this is exactly the type of behavior found in 
quarks: they are quasi-free particles inside the hadrons but the interaction
potential potential between them increases at large distances.

Although asymptotic free theories can be handled in the
ultraviolet, they become extremely complicated in the infrared. In the
case of QCD it is still to be understood (at least analytically) how
the theory confines color charges and generates the spectrum of 
hadrons, as well as the breaking of the chiral symmetry (\ref{chisb}).

\begin{figure}
\centerline{\epsfxsize=3.0truein\epsfbox{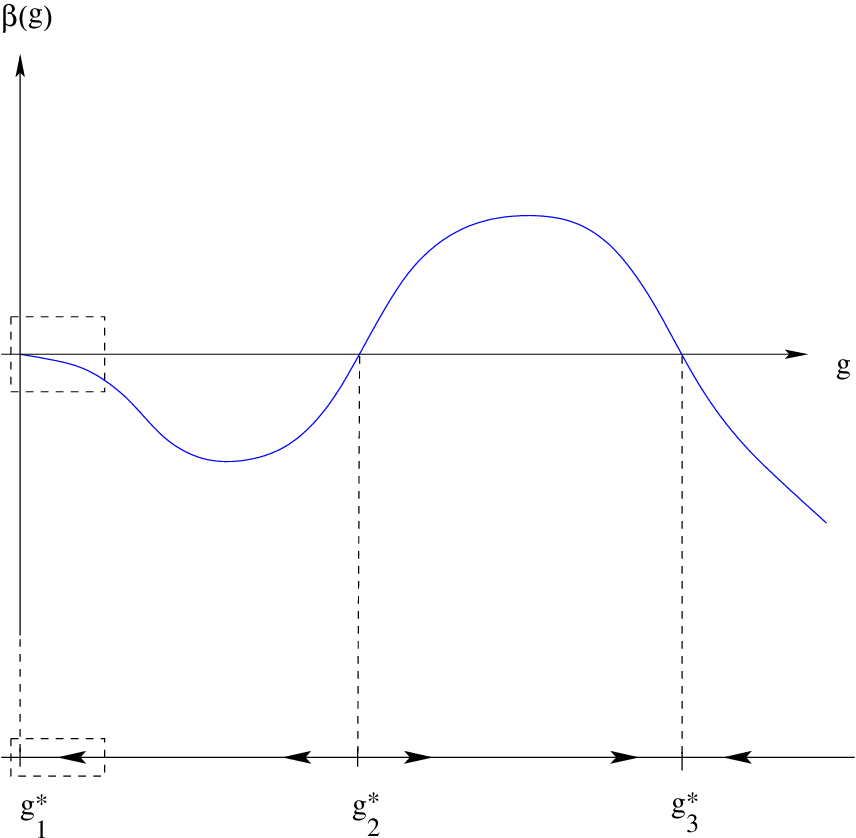}}
\caption[]{Beta function for a hypothetical theory with three fixed points $g_{1}^{*}$, $g_{2}^{*}$ and $g_{3}^{*}$. A perturbative
analysis would capture only the regions shown in the boxes.}
\label{fig:beta_function}
\end{figure}

In general, the ultraviolet and infrared properties of a theory are controlled
by the fixed points of the beta function, i.e. those values of the coupling constant
$g$ for which it vanishes
\begin{eqnarray}
\beta(g^{*})=0.
\end{eqnarray}
Using perturbation theory we have seen that for both QED and QCD one of such fixed points occurs at 
zero coupling, $g^{*}=0$. However, our analysis also showed that the two theories present radically 
different behavior at high and low energies. From the point of view of the beta function, the difference lies
in the energy regime at which the coupling constant approaches its critical value. This is in fact governed by the sign of 
the beta function around the critical coupling. 

We have seen above that when the beta function 
is negative close to the fixed point (the case of QCD) the coupling tends to its critical value, $g^{*}=0$, as the energy is increased. 
This means that the critical point is {\em ultraviolet stable}, i.e. it is an attractor as we evolve
towards higher energies. If, on the contrary, the beta function is positive (as it happens in QED) the 
coupling constant approaches the critical value as the energy decreases. This  is the case of 
an {\em infrared stable} fixed point. 

This analysis that we have motivated with the examples of QED and QCD is completely general and can 
be carried out for any
quantum field theory. In Fig. \ref{fig:beta_function} we have represented the beta function for a hypothetical theory
with three fixed points located at couplings $g_{1}^{*}$, $g_{2}^{*}$ and $g_{3}^{*}$. The arrows in the line
below the plot represent the evolution of the coupling constant as the energy increases. 
From the analysis presented above we see that $g_{1}^{*}=0$ and $g_{3}^{*}$ are  ultraviolet stable fixed points, 
while the fixed point $g_{2}^{*}$ is infrared stable. 

In order to understand the high and low energy behavior of a quantum field theory it is then crucial to know the
structure of the beta functions associated with its couplings.  
This can be a very difficult task, since perturbation theory only allows the study of the
theory around ``trivial" fixed points, i.e. those that occur at zero coupling like the case of $g_{1}^{*}$ in Fig. 
\ref{fig:beta_function}. On the other hand, any ``nontrivial'' fixed point occurring in a theory 
(like $g_{2}^{*}$ and $g_{3}^{*}$) cannot be captured in perturbation theory and requires a full nonperturbative analysis.

The moral to be learned from our discussion above is that dealing with the ultraviolet divergences in a quantum field 
theory has the consequence, among others, of introducing an energy dependence in the measured value of the 
coupling constants of the theory (for example the electric charge in QED). This happens even in the case of
renormalizable theories without mass terms. These theories are scale invariant at the classical level because the
action does not contain any dimensionful parameter. In this case 
the running of the coupling constants can be seen as resulting from a quantum breaking of classical scale invariance:
different energy scales in the theory are distinguished by different values of the coupling constants. Remembering what we learned
in Section \ref{sec:anomalies}, we conclude that classical scale invariance is an anomalous symmetry. One 
heuristic way to
see how the conformal anomaly comes about is to notice that the regularization of an otherwise scale invariant field theory requires
the introduction of an energy scale (e.g. a cutoff). This breaking of scale invariance cannot be restored after 
renormalization.

Nevertheless, scale invariance is not lost forever in the quantum theory.  It is recovered at the fixed points of the 
beta function where, by definition, the coupling does not run. To understand how this happens we go back to a scale
invariant classical field theory whose field $\phi(x)$ transform under coordinate rescalings as
\begin{eqnarray}
x^{\mu}\longrightarrow \lambda x^{\mu}, \hspace*{1cm}
\phi(x)\longrightarrow \lambda^{-\Delta}\phi(\lambda^{-1}x),
\label{eq:rescaling_conf}
\end{eqnarray}
where $\Delta$ is called the canonical scaling dimension of the field. An example of such a theory is
a massless $\phi^{4}$ theory
in four dimensions
\begin{eqnarray}
\mathcal{L}={1\over 2}\partial_{\mu}\phi\,\partial^{\mu}\phi
-{g\over 4!}\phi^{4},
\label{eq:lambda_phi_4}
\end{eqnarray}
where the scalar field has canonical scaling dimension $\Delta=1$.  The Lagrangian
density transforms as
\begin{eqnarray}
\mathcal{L}\longrightarrow \lambda^{-4}\mathcal{L}[\phi]
\end{eqnarray}
and the classical action remains invariant\footnote{In a $D$-dimensional theory the canonical scaling dimensions
of the fields coincide with its engineering dimension: $\Delta={D-2\over 2}$ for bosonic fields and $\Delta={D-1\over 2}$ for fermionic ones.
For a Lagrangian with no dimensionful parameters classical scale invariance follows then from dimensional analysis.}.

If scale invariance is preserved under quantization, the Green's functions transform as
\begin{eqnarray}
\langle\Omega|T[\phi'(x_{1})\ldots\phi'(x_{n})]|\Omega\rangle=
\lambda^{n\Lambda}\langle\Omega|T[\phi(\lambda^{-1}x_{1})\ldots\phi(\lambda^{-1}x_{n})]|\Omega\rangle.
\label{eq:transf_greenfunc_freetheor}
\end{eqnarray}
This is precisely what happens in a free theory. In an interacting theory the running of the coupling constant
destroys classical scale invariance at the quantum level. Despite of this, at the fixed points of the beta function
the Green's functions transform again according to \eqref{eq:transf_greenfunc_freetheor} where $\Delta$ is replaced
by
\begin{eqnarray}
\Delta_{\rm anom}=\Delta+\gamma^{*}.
\end{eqnarray}
The canonical scaling dimension of the fields are corrected by $\gamma^{*}$, which is called the anomalous
dimension. They carry 
the dynamical information about the high-energy behavior of the theory.

\subsection{The renormalization group}

In spite of its successes, the renormalization procedure presented above
can be seen as some kind of prescription or recipe to get rid of the
divergences in an ordered way. This discomfort about renormalization
was expressed in occasions by comparing it with ``sweeping the
infinities under the rug''. However thanks to Ken Wilson to a large extent
\cite{wilson} the process of renormalization is now understood in
a very profound way as a procedure to incorporate the effects of
physics at high energies by modifying the value of the parameters that
appear in the Lagrangian.

\begin{figure}
\centerline{\epsfxsize=4.5truein\epsfbox{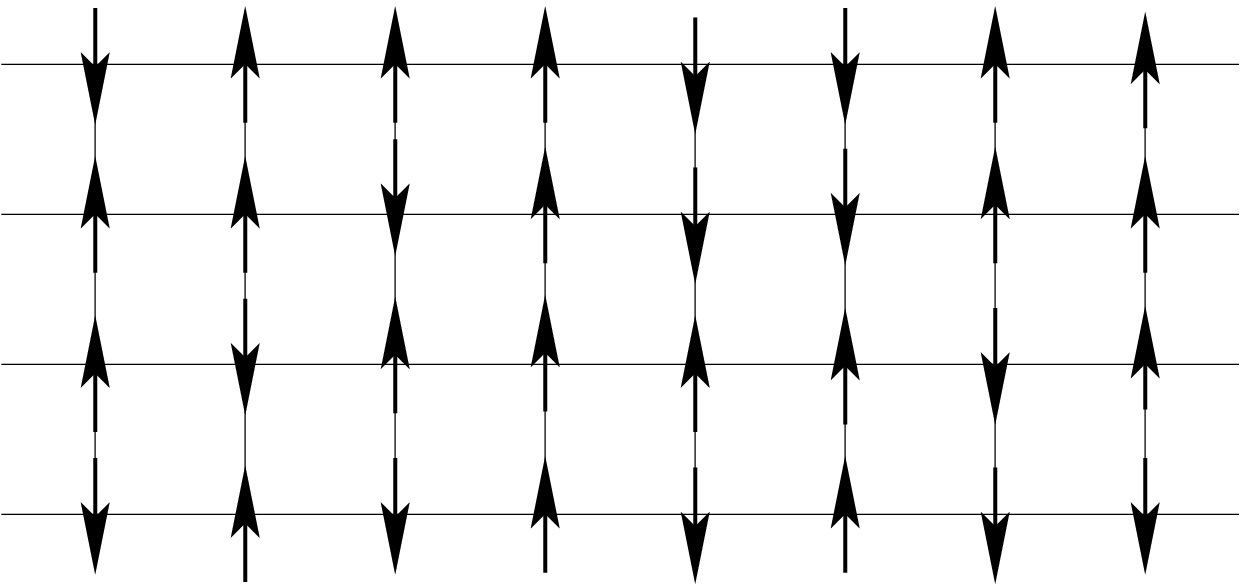}}
\caption[]{Systems of spins in a two-dimensional square lattice.}
\label{fig:lattice1}
\end{figure}

{\bf Statistical mechanics.}
Wilson's ideas are both simple and profound and consist in thinking 
about Quantum Field Theory as the analog of a thermodynamical description 
of a statistical system. To be more precise, let us consider an Ising spin
system in a two-dimensional square lattice as the one depicted in Fig 
\ref{fig:lattice1}. In terms of the spin variables ${s}_{i}=\pm{1\over 2}$,  
where $i$ labels the lattice site, the Hamiltonian of the system is given by
\begin{eqnarray}
H=-J\sum_{\langle i,j\rangle}{s}_{i}\, {s}_{j},
\end{eqnarray}
where $\langle i,j\rangle$ indicates that the sum extends over 
nearest neighbors and $J$ is the coupling constant between neighboring spins 
(here we consider that there is no external magnetic field).
The starting point to study the statistical mechanics of this system
is the partition function defined as
\begin{eqnarray}
\mathcal{Z}=\sum_{\{{s_{i}}\}}e^{-\beta H},
\label{eq:partition_function}
\end{eqnarray}
where the sum is over all possible configurations of the spins and
$\beta={1\over T}$ is the inverse temperature. For $J>0$ the Ising
model presents spontaneous magnetization below a critical temperature
$T_{c}$, in any dimension higher than one. Away from this temperature
correlations between spins decay exponentially at large distances
\begin{eqnarray}
\langle {s}_{i}{s}_{j}\rangle \sim e^{-{|x_{ij}|\over \xi}},
\end{eqnarray}
with $|x_{ij}|$ the distance between the spins located in the $i$-th
and $j$-th sites of the lattice. This expression serves as a
definition of the correlation length $\xi$ which sets the
characteristic length scale at which spins can influence each other by
their interaction through their nearest neighbors.

\begin{figure}
\centerline{\epsfxsize=3.5truein\epsfbox{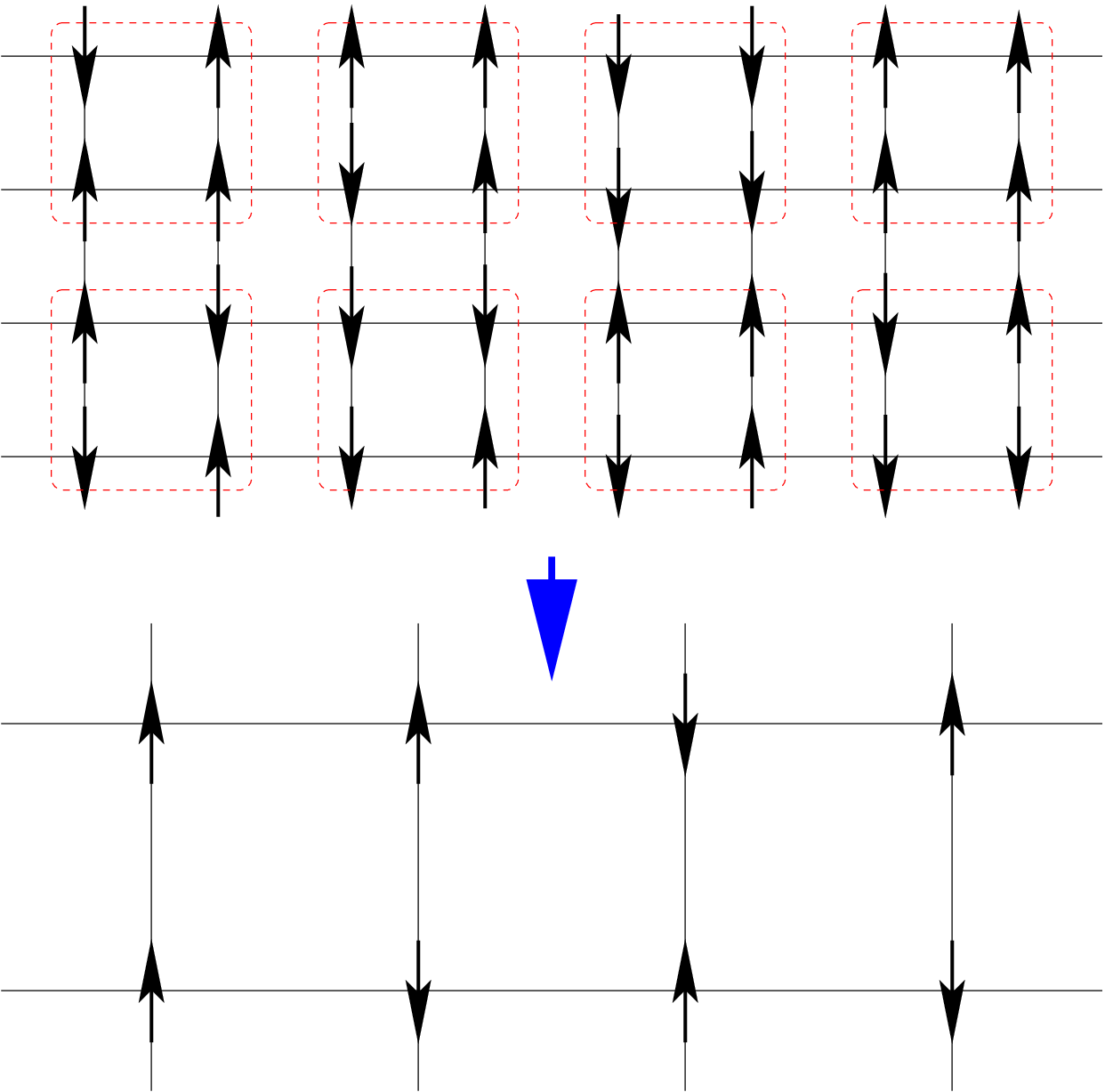}}
\caption[]{Decimation of the spin lattice. Each block in the upper 
lattice is replaced by an effective spin computed according to the 
rule (\ref{eq:constraint_lattice}). Notice also that the size of the
lattice spacing is doubled in the process.}
\label{fig:lattice2}
\end{figure}

Suppose now that we are interested in a macroscopic description of
this spin system. We can capture the relevant physics by integrating
out somehow the physics at short scales. A way in which this can be
done was proposed by Leo Kadanoff \cite{kadanoff} and consists in
dividing our spin system in spin-blocks like the ones showed in Fig
\ref{fig:lattice2}.  Now we can construct another spin system where
each spin-block of the original lattice is replaced by an effective
spin calculated according to some rule from the spins contained in
each block $B_{a}$
\begin{eqnarray}
\{{s}_{i}:i\in B_{a}\}\hspace*{0.5cm}
\longrightarrow \hspace*{0.5cm}
{s}_{a}^{\,\,\, (1)}.
\end{eqnarray}
For example we can define the effective spin associated with the block 
$B_{a}$ by taking the majority rule with an additional prescription in
case of a draw
\begin{eqnarray}
{s}_{a}^{\,\,\,(1)}={1\over 2}{\rm sgn\,}\left(
\sum_{i\in B_{a}}s_{i}\right),
\label{eq:constraint_lattice}
\end{eqnarray}
where we have used the sign function, ${\rm sign}(x)\equiv {x\over
|x|}$, with the additional definition ${\rm sgn}(0)=1$. This procedure
is called decimation and leads to a new spin system with a doubled
lattice space.

The idea now is to rewrite the partition function
(\ref{eq:partition_function}) only in terms of the new effective spins
${s}_{a}^{\,\,\,(1)}$. Then we start by splitting the sum over 
spin configurations into two nested sums, one over the spin blocks and 
a second one over the spins within each block
\begin{eqnarray}
\mathcal{Z}=\sum_{\{\vec{s}\}}e^{-\beta H[{s}_{i}]}=
\sum_{\{\vec{s}^{\,(1)}\}}\sum_{\{\vec{s}\in B_a\}}\delta
\left[{s}^{\,\,\,
(1)}_{a}-{\rm sign}\left(\sum_{i\in B_{a}}{s}_{i}\right)\right]
e^{-\beta H[{s}_{i}]}.
\end{eqnarray}
The interesting point now is that the sum over spins inside each
block can be written as the exponential of 
a new effective Hamiltonian depending only on the effective spins,
$H^{(1)}[{s}^{\,\,(1)}_{a}]$
\begin{eqnarray}
\sum_{\{{s\in B_a}\}}\delta\left[{s}^{\,\,\,
(1)}_{a}-{\rm sign}\left(\sum_{i\in B_{a}}{s}_{i}\right)\right]
e^{-\beta H[{s}_{i}]}
=e^{-\beta H^{(1)}[{s}^{\,(1)}_{a}]}.
\end{eqnarray}
The new Hamiltonian is of course more complicated
\begin{eqnarray}
H^{(1)}=-J^{(1)}\sum_{\langle i,j\rangle}s_{i}^{(1)}s_{j}^{(1)}
+\ldots
\end{eqnarray}
where the dots stand for other interaction terms between the effective
block spins. This new terms appear because in the process of
integrating out short distance physics we induce interactions between
the new effective degrees of freedom. For example the interaction
between the spin block variables $s^{(1)}_{i}$ will in general not be
restricted to nearest neighbors in the new lattice.  The important
point is that we have managed to rewrite the partition function solely
in terms of this new (renormalized) spin variables ${s}^{\,\,(1)}$
interacting through a new Hamiltonian $H^{(1)}$
\begin{eqnarray}
\mathcal{Z}=\sum_{\{{s}^{\,(1)}\}}e^{-\beta H^{(1)}[{s}^{\,(1)}_{a}]}.
\end{eqnarray}

Let us now think about the space of all possible Hamiltonians for our
statistical system including all kinds of possible couplings between
the individual spins compatible with the symmetries of the system.  If
denote by $\mathcal{R}$ the decimation operation, our previous
analysis shows that $\mathcal{R}$ defines a map in this space of
Hamiltonians
\begin{eqnarray}
\mathcal{R}:H\rightarrow H^{(1)}.
\end{eqnarray}
At the same time the operation $\mathcal{R}$ replaces a lattice with
spacing $a$ by another one with double spacing $2a$. As a consequence
the correlation length in the new lattice measured in units of the
lattice spacing is divided by two, $\mathcal{R}:\xi\rightarrow
{\xi\over 2}$.  

Now we can iterate the operation $\mathcal{R}$ an indefinite number of
times. Eventually we might reach a Hamiltonian $H_{\star}$ that is not
further modified by the operation $\mathcal{R}$
\begin{eqnarray}
H\stackrel{\mathcal{R}}{\longrightarrow} H^{(1)}
\stackrel{\mathcal{R}}{\longrightarrow} H^{(2)}
\stackrel{\mathcal{R}}{\longrightarrow} \ldots 
\stackrel{\mathcal{R}}{\longrightarrow} H_{\star}.
\end{eqnarray}
The fixed point Hamiltonian $H_{\star}$ is
{\em scale invariant} because it does not change as $\mathcal{R}$ is
performed. Notice that because of this invariance the correlation
length of the system at the fixed point do not change under
$\mathcal{R}$.  This fact is compatible with the transformation
$\xi\rightarrow {\xi\over 2}$ only if $\xi=0$ or $\xi=\infty$. Here we
will focus in the case of nontrivial fixed points with infinite
correlation length. 

The space of Hamiltonians can be parametrized by specifying the values
of the coupling constants associated with all possible interaction
terms between individual spins of the lattice. If we denote by 
$\mathcal{O}_{a}[s_{i}]$ these (possibly infinite) interaction terms, the
most general Hamiltonian for the spin system under study can be written as
\begin{eqnarray}
H[s_{i}]=\sum_{a=1}^{\infty}\lambda_{a}\mathcal{O}_{a}[s_{i}],
\end{eqnarray}
where $\lambda_{a}\in\mathbb{R}$ are the coupling constants for the
corresponding operators. These constants can be thought of as
coordinates in the space of all Hamiltonians. Therefore the operation
$\mathcal{R}$ defines a transformation in the set of coupling
constants
\begin{eqnarray}
\mathcal{R}:\lambda_{a}\longrightarrow \lambda_{a}^{(1)}.
\end{eqnarray}
For example, in our case we started with a Hamiltonian in which only
one of the coupling constants is different from zero (say
$\lambda_{1}=-J$). As a result of the decimation $\lambda_{1}\equiv
-J\rightarrow -J^{(1)}$ while some of the originally vanishing
coupling constants will take a nonzero value. Of course, for the fixed
point Hamiltonian the coupling constants do not change under the scale
transformation $\mathcal{R}$.

Physically the transformation $\mathcal{R}$ integrates out short
distance physics. The consequence for physics at long distances is
that we have to replace our Hamiltonian by a new one with different values
for the coupling constants. That is, our ignorance of the details of
the physics going on at short distances result in a {\em
renormalization} of the coupling constants of the Hamiltonian that
describes the long range physical processes.  It is important to
stress that although $\mathcal{R}$ is sometimes called a
renormalization group transformation in fact this is a
misnomer. Transformations between Hamiltonians defined by
$\mathcal{R}$ do not form a group:
since these transformations proceed by integrating out degrees of
freedom at short scales they cannot be inverted.

In statistical mechanics fixed points under renormalization group
transformations with $\xi=\infty$ are associated with phase
transitions.  From our previous discussion we can conclude that the
space of Hamiltonians is divided in regions corresponding to the
basins of attraction of the different fixed points.  We can ask
ourselves now about the stability of those fixed points.  Suppose we
have a statistical system described by a fixed-point Hamiltonian
$H_{\star}$ and we perturb it by changing the coupling constant
associated with an interaction term $\mathcal{O}$. This is equivalent
to replace $H_{\star}$ by the perturbed Hamiltonian
\begin{eqnarray}
H=H_{\star}+\delta\lambda\,\mathcal{O},
\end{eqnarray}
where $\delta\lambda$ is the perturbation of the coupling constant
corresponding to $\mathcal{O}$ (we can also consider perturbations in
more than one coupling constant). At the same time thinking of the 
$\lambda_{a}$'s as coordinates in the space of all Hamiltonians this
corresponds to moving slightly away from the position of the fixed
point.

The question to decide now is in which direction the renormalization group 
flow will take the perturbed system. Working at first order in 
$\delta\lambda$ there are three possibilities:
\begin{itemize}
\item
The renormalization group flow takes the system back to the fixed point.
In this case the corresponding interaction $\mathcal{O}$ is called 
{\em irrelevant}.

\item
$\mathcal{R}$ takes the system away from the fixed point. If this is 
what happens the interaction is called {\em relevant}.

\item
It is possible that the perturbation actually does not take
the system away from the fixed point at first order in $\delta\lambda$. 
In this case the interaction is said to be {\em marginal} and
it is necessary to go to higher orders in $\delta\lambda$ 
in order to decide whether the system moves to or away the fixed point,
or whether  we have a family of  fixed points.

\end{itemize}

\begin{figure}
\centerline{\epsfxsize=4.5truein\epsfbox{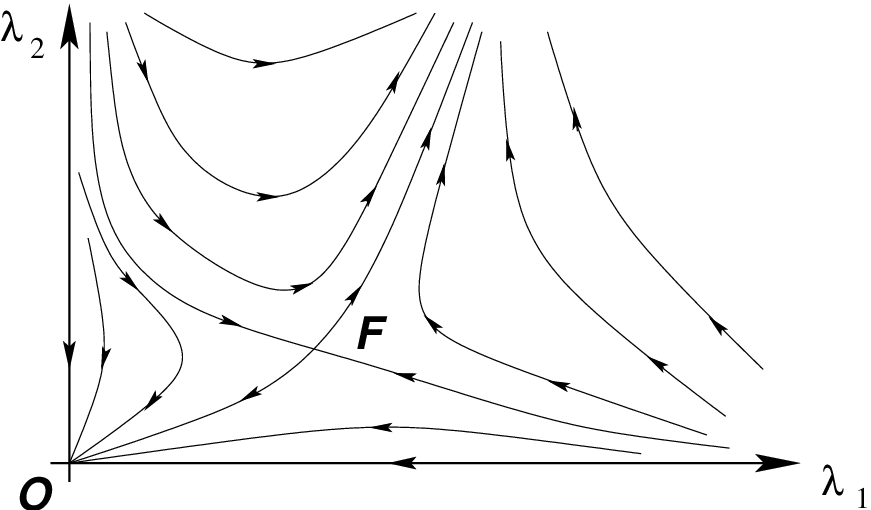}}
\caption[]{Example of a renormalization group flow.}
\label{fig:flows}
\end{figure}

Therefore we can picture the action of the renormalization group
transformation as a flow in the space of coupling constants. In
Fig. \ref{fig:flows} we have depicted an example of such a flow in the
case of a system with two coupling constants $\lambda_{1}$ and
$\lambda_{2}$. In this example we find two fixed points, one at the
origin $O$ and another at $F$ for a finite value of the couplings. The
arrows indicate the direction in which the renormalization group flow
acts. The free theory at $\lambda_{1}=\lambda_{2}=0$ is a stable fix
point since any perturbation $\delta\lambda_{1},\delta\lambda_{2}>0$
makes the theory flow back to the free theory at long distances. On
the other hand, the fixed point $F$ is stable with respect to certain
type of perturbations (along the line with incoming arrows) whereas
for any other perturbations the system flows either to the free theory
at the origin or to a theory with infinite values for the couplings.

{\bf Quantum field theory.}
Let us see now how these ideas of the renormalization group apply to
Field Theory. Let us begin with a quantum field theory defined by the
Lagrangian
\begin{eqnarray}
\mathcal{L}[\phi_{a}]=\mathcal{L}_{0}[\phi_{a}]+\sum_{i}g_{i}
\mathcal{O}_{i}[\phi_{a}],
\end{eqnarray}
where $\mathcal{L}_{0}[\phi_{a}]$ is the kinetic part of the
Lagrangian and $g_{i}$ are the coupling constants associated with the
operators $\mathcal{O}_{i}[\phi_{a}]$. In order to make sense of the
quantum theory we introduce a cutoff in momenta $\Lambda$. In principle
we include all operators $\mathcal{O}_{i}$ compatible with the symmetries
of the theory.

In section \ref{sec:betafunction} we saw how in the cases of QED and QCD,
the value of the coupling constant changed with the scale from its
value at the scale $\Lambda$. We can understand now this behavior
along the lines of the analysis presented above for the Ising
model. If we would like to compute the effective dynamics of the
theory at an energy scale $\mu<\Lambda$ we only have to integrate out
all physical models with energies between the cutoff $\Lambda$ and the
scale of interest $\mu$.  This is analogous to what we did in the
Ising model by replacing the original spins by the block spins. In the
case of field theory the effective action $S[\phi_{a},\mu]$ at scale
$\mu$ can be written in the language of functional integration as
\begin{eqnarray}
e^{iS[\phi_{a}',\mu]}=\int_{\mu<p<\Lambda} \prod_{a}\mathcal{D}\phi_{a}\,e^{iS[\phi_{a},
\Lambda]}.
\label{eq:pathintegral}
\end{eqnarray}
Here $S[\phi_{a},\Lambda]$ is the action at the cutoff scale 
\begin{eqnarray}
S[\phi_{a},\Lambda]=\int d^{4}x\,\left\{\mathcal{L}_{0}[\phi_{a}]
+\sum_{i}g_{i}(\Lambda)\mathcal{O}_{i}[\phi_{a}]\right\}
\end{eqnarray}
and the functional integral in Eq. (\ref{eq:pathintegral}) is carried
out only over the field modes with momenta in the range
$\mu<p<\Lambda$. The action resulting from integrating out the physics
at the intermediate scales between $\Lambda$ and $\mu$ depends not on 
the original field variable $\phi_{a}$ but on some renormalized field
$\phi_{a}'$. At the same time the couplings $g_{i}(\mu)$ differ from
their values at the cutoff scale $g_{i}(\Lambda)$. This is analogous
to what we learned in the Ising model: by integrating out short distance physics
we ended up with a new Hamiltonian depending on renormalized effective
spin variables and with renormalized values for the coupling constants.
Therefore the resulting effective action at scale $\mu$ can be written as
\begin{eqnarray}
S[\phi_{a}',\mu]=\int d^{4}x\,\left\{\mathcal{L}_{0}[\phi_{a}']
+\sum_{i}g_{i}(\mu)\mathcal{O}_{i}[\phi_{a}']\right\}.
\label{eq:renormalizedaction}
\end{eqnarray}
This Wilsonian interpretation of renormalization sheds light to what in 
section \ref{sec:removinginfinities} might have looked just a smart way to 
get rid of the infinities. The running of the coupling constant with 
the energy scale can be understood now as a way of incorporating
into an effective action at scale $\mu$ the effects of field excitations
at higher energies $E>\mu$. 

As in statistical mechanics there are also quantum field theories
that are fixed points of the renormalization group flow, i.e.  whose
coupling constants do not change with the scale. We have encountered 
them already in Section \ref{sec:betafunction} when studying the
properties of the beta function. The most trivial
example of such theories are massless free quantum field theories, but
there are also examples of four-dimensional interacting quantum field
theories which are scale invariant. Again we can ask the question of
what happens when a scale invariant theory is perturbed with some
operator.  In general the perturbed theory is not scale invariant
anymore but we may wonder whether the perturbed theory flows at low
energies towards or away the theory at the fixed point.

In quantum field theory this can be decided by looking at the
canonical dimension $d[\mathcal{O}]$ of the operator
$\mathcal{O}[\phi_{a}]$ used to perturb the theory at the fixed point.
In four dimensions the three possibilities are defined by:

\begin{itemize}

\item
$d[\mathcal{O}]>4$: irrelevant perturbation. The running of
the coupling constants takes the theory back to the fixed point.

\item
$d[\mathcal{O}]<4$: relevant perturbation. 
At low energies the theory flows away from the scale-invariant theory.

\item
$d[\mathcal{O}]=4$: marginal deformation. The direction of the flow
cannot be decided only on dimensional grounds.

\end{itemize}

As an example, let us consider first a massless fermion theory perturbed
by a four-fermion interaction term
\begin{eqnarray}
\mathcal{L}=i\overline{\psi}\dslash\psi
-{1\over M^{2}}(\overline{\psi}\psi)^{2}.
\label{eq:four-fermion}
\end{eqnarray}
This is indeed a perturbation by an irrelevant operator, since in
four-dimensions $[\psi]={3\over 2}$. Interactions generated by the
extra term are suppressed at low energies since typically their
effects are weighted by the dimensionless factor ${E^{2}\over M^2}$,
where $E$ is the energy scale of the process.  This means that as we
try to capture the relevant physics at lower and lower energies the
effect of the perturbation is weaker and weaker rendering in the
infrared limit $E\rightarrow 0$ again a free theory.  Hence, the
irrelevant perturbation in (\ref{eq:four-fermion}) makes the theory
flow back to the fixed point.

On the other hand relevant operators dominate the physics at low
energies. This is the case, for example, of a mass term. As we lower
the energy the mass becomes more important and once the
energy goes below the mass of the field its dynamics is completely
dominated by the mass term. This is, for example, how Fermi's theory
of weak interactions emerges from the Standard Model at energies below
the mass of the $W^{\pm}$ boson
\begin{eqnarray*}
\parbox{40mm}{
\begin{fmfgraph*}(90,70)
\fmfleft{i1}
\fmfright{o1,o2,o3}
\fmf{fermion,label=$u$}{i1,v1}
\fmf{fermion,label=$d$}{v1,o1}
\fmf{photon,label=$W^{+}$,tension=0.5}{v1,v2}
\fmf{fermion,label=${\nu}_e$,tension=0.5}{v2,o2}
\fmf{fermion,label=$e^{+}$,tension=0.5}{o3,v2}
\end{fmfgraph*}
} \Longrightarrow
\hspace*{0.5cm}\parbox{40mm}{
\begin{fmfgraph*}(90,70)
\fmfleft{i1}
\fmfright{o1,o2,o3}
\fmf{fermion,label=$u$}{i1,v1}
\fmf{fermion,label=$d$,tension=0.4}{v1,o1}
\fmf{fermion,label=${\nu}_e$,tension=0.3}{v1,o2}
\fmf{fermion,label=$e^+$}{o3,v1}
\end{fmfgraph*}
} 
\end{eqnarray*}
At energies below $M_{W}=80.4$ GeV the dynamics of the $W^{+}$ boson
is dominated by its mass term and therefore becomes nonpropagating, giving
rise to the effective four-fermion Fermi theory.

To summarize our discussion so far, we found that while relevant
operators dominate the dynamics in the infrared, taking the theory
away from the fixed point, irrelevant perturbations become suppressed
in the same limit. Finally we consider the effect of marginal
operators. As an example we take the interaction term in
massless QED,
$\mathcal{O}=\overline{\psi}\gamma^{\mu}\psi\,A_{\mu}$. Taking into
account that in $d=4$ the dimension of the electromagnetic potential
is $[A_{\mu}]=1$ the operator $\mathcal{O}$ is a marginal perturbation.
In order to decide whether the fixed point theory
\begin{eqnarray}
\mathcal{L}_{0}=-{1\over 4}F_{\mu\nu}F^{\mu\nu}+i\overline{\psi}
\Dslash\psi
\end{eqnarray}
is restored at low energies or not we need to study the perturbed
theory in more detail. This we have done in section
\ref{sec:removinginfinities} where we learned that the effective coupling in
QED decreases at low energies.  Then we conclude that the perturbed
theory flows towards the fixed point in the infrared.

As an example of a marginal operator with the opposite behavior we can
write the Lagrangian for a SU($N_{c}$) gauge theory, 
$\mathcal{L}=-{1\over 4}F_{\mu\nu}^{a}F^{a\,\mu\nu}$, as
\begin{eqnarray}
\mathcal{L}&=&-{1\over 4}\left(\partial_{\mu}A_{\nu}^{a}-
\partial_{\nu}A_{\mu}^{a}\right)\left(\partial^{\mu}A^{a\,\nu}-
\partial^{\nu}A^{a\,\mu}\right)-4gf^{abc}A_{\mu}^{a}A_{\nu}^{b}\,
\partial^{\mu}A^{c\,\nu} \nonumber 
\\
&+& g^{2}f^{abc}f^{ade}A_{\mu}^{b}A_{\nu}^{c}A^{d\,\mu}A^{e\,\nu}
\equiv \mathcal{L}_{0}+\mathcal{O}_{g},
\end{eqnarray}
i.e. a marginal perturbation of the free theory described by
$\mathcal{L}_{0}$, which is obviously a fixed point under
renormalization group transformations. Unlike the case of QED
we know that the full theory is asymptotically free, so the
coupling constant grows at low energies. This implies that the 
operator $\mathcal{O}_{g}$ becomes more and more important in the infrared
and therefore the theory flows away the fixed point in this limit.

It is very important to notice here that in the Wilsonian view the
cutoff is not necessarily regarded as just some artifact to remove
infinities but actually has a physical origin. For example in the case
of Fermi's theory of $\beta$-decay there is a natural cutoff
$\Lambda=M_{W}$ at which the theory has to be replaced by the Standard
Model. In the case of the Standard Model itself the cutoff can be
taken at Planck scale $\Lambda\simeq 10^{19}$ GeV or the Grand
Unification scale $\Lambda\simeq 10^{16}$ GeV, where new degrees of 
freedom are expected to become relevant. The cutoff serves the
purpose of cloaking the range of energies at which new physics has to
be taken into account.

Provided that in the Wilsonian approach the quantum theory is always
defined with a physical cutoff, there is no fundamental difference
between renormalizable and nonrenormalizable theories. Actually, a 
renormalizable field theory, like the Standard Model, can generate 
nonrenormalizable operators at low energies such as the effective 
four-fermion interaction of Fermi's theory. They are not sources of any
trouble if we are interested in the physics at scales much below the
cutoff, $E\ll\Lambda$, since their contribution to the amplitudes will be
suppressed by powers of ${E\over \Lambda}$.

\section{Special topics}

\subsection{Creation of particles by classical fields}
\label{sec:schwinger}

{\bf Particle creation by a classical source.} 
In a free quantum field theory the total number of particles contained
in a given state of the field is a conserved quantity. For example,
in the case of the quantum scalar field studied in section  
\ref{sec:class2quant} we have that the number operator commutes with 
the Hamiltonian
\begin{eqnarray}
\widehat{n}\equiv \int {d^{3}k\over (2\pi)^{3}}{1\over 2\omega_{k}}\alpha^{\dagger}(\vec{k})
\alpha(\vec{k}), \hspace*{1cm}
[\widehat{H},\widehat{n}]=0.
\end{eqnarray}
This means that any states with a well-defined number of particle 
excitations will preserve this number at all times. The situation, however,
changes as soon as interactions are introduced, since in this case
particles can be created and/or destroyed as a result of the dynamics.

Another case in which the number of particles might change is if the 
quantum theory is coupled to a classical source. The archetypical example
of such a situation is the Schwinger effect, in which a classical 
strong electric field produces the creation of electron-positron pairs
out of the vacuum. However, before plunging into this more involved situation
we can illustrate the relevant physics involved in the creation of 
particles by classical sources with the help of the simplest example:
a free scalar field theory coupled to a classical
external source $J(x)$. The action for such a theory can be written as
\begin{eqnarray}
S=\int d^{4}x\left[{1\over 2}\partial_{\mu}\phi(x)\partial^{\mu}\phi(x)
-{m^2\over 2}\phi(x)^2+J(x)\phi(x)\right],
\end{eqnarray}
where $J(x)$ is a real function of the coordinates.  Its
identification with a classical source is obvious once we calculate
the equations of motion
\begin{eqnarray}
\left(\nabla^2+m^2\right)\phi(x)=J(x).
\label{eq:KGJ}
\end{eqnarray}
Our plan is to quantize this theory but, unlike the case analyzed in
section \ref{sec:class2quant}, now the presence of the source $J(x)$
makes the situation a bit more involved. The general solution to the
equations of motion can be written in terms of the retarded Green
function for the Klein-Gordon equation as
\begin{eqnarray}
\phi(x)=\phi_{0}(x)+i\int d^{4}x'\,G_{R}(x-x')J(x'),
\label{eq:phi_general_J}
\end{eqnarray}
where $\phi_{0}(x)$ is a general solution to the homogeneous
equation and
\begin{eqnarray}
G_{R}(t,\vec{x})&=&\int {d^{4}k\over (2\pi)^4}{i\over k^2-m^2+i\epsilon\, {\rm sign}(k^{0})}e^{-ik\cdot x}
\nonumber \\
&=&
i\,\theta(t)\int {d^{3}k\over (2\pi)^{3}}{1\over 2\omega_{k}}
\left(e^{-i\omega_{k}t+\vec{k}\cdot\vec{x}}-
e^{i\omega_{k}t-i\vec{p}\cdot\vec{x}}\right),
\label{eq:greenfunc}
\end{eqnarray}
with $\theta(x)$ the Heaviside step function. The integration contour
to evaluate the integral over $p^{0}$ surrounds the poles at
$p^{0}=\pm \omega_{k}$ from above. Since
$G_{R}(t,\vec{x})=0$ for $t<0$, the function $\phi_{0}(x)$ corresponds
to the solution of the field equation at $t\rightarrow -\infty$,
before the interaction with the external source\footnote{We could have
taken instead the advanced propagator $G_{A}(x)$ in which case
$\phi_{0}(x)$ would correspond to the solution to the equation at
large times, after the interaction with $J(x)$.}

To make the argument simpler we assume that $J(x)$ 
is switched on at $t=0$, and only last for a time $\tau$, that is 
\begin{eqnarray}
J(t,\vec{x})=0 \hspace*{1cm} \mbox{if $t<0$ or $t>\tau$}.
\end{eqnarray} 
We are interested in a solution of (\ref{eq:KGJ}) for times after the
external source has been switched off, $t>\tau$. In this case 
the expression (\ref{eq:greenfunc}) can be written in terms of the
Fourier modes $\widetilde{J}(\omega,\vec{k})$ of the source as
\begin{eqnarray}
\phi(t,\vec{x})=\phi_{0}(x)
+i\int{d^{3}k\over (2\pi)^{3}}{1\over 2\omega_{k}}\left[
\widetilde{J}(\omega_{k},\vec{k})e^{-i\omega_{k}t+i\vec{k}\cdot\vec{x}}-
\widetilde{J}(\omega_{k},\vec{k})^{*}
e^{i\omega_{k}t-i\vec{k}\cdot\vec{x}}\right].
\label{eq:J}
\end{eqnarray}
On the other hand, the general solution $\phi_{0}(x)$ has been already
computed in Eq.  (\ref{eq:general_sol_phi}). Combining this result
with Eq. (\ref{eq:J}) we find the following expression for the late
time general solution to the Klein-Gordon equation in the presence of
the source
\begin{eqnarray}
\phi(t,x)&=&\int{d^{3}k\over (2\pi)^{3}}{1\over \sqrt{2\omega_{k}}}\left\{
\left[\alpha(\vec{k})+{i\over \sqrt{2\omega_{k}}}
\widetilde{J}(\omega_{k},\vec{k})\right]
e^{-i\omega_{k}t+i\vec{k}\cdot\vec{x}} \right. \nonumber \\
&+& \left.\left[\alpha^{*}(\vec{k})-{i\over \sqrt{2\omega_{k}}}
\widetilde{J}(\omega_{k},\vec{k})^{*}\right]
e^{i\omega_{k}t-i\vec{k}\cdot\vec{x}} 
\right\}.
\label{eq:afterJ}
\end{eqnarray}
We should not forget that this is a solution valid for times $t>\tau$, i.e.
once the external source has been disconnected. On the other hand, 
for $t<0$ we find from Eqs. (\ref{eq:phi_general_J}) and 
(\ref{eq:greenfunc}) that the general solution is given by 
Eq. (\ref{eq:general_sol_phi}). 

Now we can proceed to quantize the theory. The conjugate momentum
$\pi(x)=\partial_{0}\phi(x)$ can be computed from
Eqs. (\ref{eq:general_sol_phi}) and (\ref{eq:afterJ}). Imposing the
canonical equal time commutation relations (\ref{eq:etccr}) we find
that $\alpha(\vec{k})$, $\alpha^{\dagger}(\vec{k})$ satisfy the
creation-annihilation algebra (\ref{eq:Lorentz_inv_cr}). From our
previous calculation we find that for $t>\tau$ the expansion of the
operator $\phi(x)$ in terms of the creation-annihilation operators
$\alpha(\vec{k})$, $\alpha^{\dagger}(\vec{k})$ can be obtained from the one
for $t<0$ by the replacement
\begin{eqnarray}
\alpha(\vec{k}) &\longrightarrow& \beta(\vec{k})\equiv 
\alpha(\vec{k})+{i\over \sqrt{2\omega_{k}}}
\widetilde{J}(\omega_{k},\vec{k}), \nonumber \\
\alpha^{\dagger}(\vec{k}) &\longrightarrow&  
\beta^{\dagger}(\vec{k}) \equiv \alpha^{\dagger}(\vec{k})
-{i\over \sqrt{2\omega_{k}}}\widetilde{J}(\omega_{k},\vec{k})^{*}.
\end{eqnarray}
Actually, since $\widetilde{J}(\omega_{k},\vec{k})$ is a c-number, the
operators $\beta(\vec{k})$, $\beta^{\dagger}(\vec{k})$ satisfy the
same algebra as $\alpha(\vec{k})$, $\alpha^{\dagger}(\vec{k})$ and
therefore can be interpreted as well as a set of creation-annihilation
operators. This means that we can define two vacuum states,
$|0_{-}\rangle$, $|0_{+}\rangle$ associated with both sets of
operators
\begin{eqnarray}
\left.
\begin{array}{c}
\alpha(\vec{k})|0_{-}\rangle =0  \\
\\
\beta(\vec{k})|0_{+}\rangle =0
\end{array}
\right\}
\hspace*{1cm} \forall \,\,\vec{k}.
\end{eqnarray}

For an observer at $t<0$, $\alpha(\vec{k})$ and $\alpha(\vec{k})$ are the
natural set of creation-annihilation operators in terms of which to 
expand the field operator $\phi(x)$. After the usual zero-point
energy subtraction the Hamiltonian is given by
\begin{eqnarray}
\widehat{H}^{(-)}={1\over 2}\int {d^{3}k\over (2\pi)^{3}}\,\alpha^{\dagger}(\vec{k})
\alpha(\vec{k})
\end{eqnarray}
and the ground state of the spectrum for this observer is the vacuum
$|0_{-}\rangle$. At the same time, a second observer at $t>\tau$ will also
see a free scalar quantum field (the source has been switched off
at $t=\tau$) and consequently will expand $\phi$ in terms of the
second set of creation-annihilation operators $\beta(\vec{k})$,
$\beta^{\dagger}(\vec{k})$. In terms of this operators the Hamiltonian 
is written as
\begin{eqnarray}
\widehat{H}^{(+)}={1\over 2}\int {d^{3}k\over (2\pi)^{3}}\,\beta^{\dagger}(\vec{k})
\beta(\vec{k}).
\end{eqnarray}
Then for this late-time observer the ground state of the Hamiltonian 
is the second vacuum state $|0_{+}\rangle$.

In our analysis we have been working in the Heisenberg picture,
where states are time-independent and the time dependence comes in the
operators. Therefore the states of the theory are globally
defined. Suppose now that the system is in the ``in'' ground state
$|0_{-}\rangle$. An observer at $t<0$ will find that there are no
particles
\begin{eqnarray}
\widehat{n}^{(-)}|0_{-}\rangle =0.
\end{eqnarray}
However the late-time observer will find that the state $|0_{-}\rangle$
contains an average number of particles given by
\begin{eqnarray}
\langle 0_{-}|\widehat{n}^{(+)}|0_{-}\rangle =
\int {d^{3}k\over (2\pi)^{3}}{1\over 2\omega_{k}}\left|
\widetilde{J}(\omega_{k},\vec{k})\right|^{2}.
\end{eqnarray}
Moreover, $|0_{-}\rangle$ is no longer the ground state for the ``out''
observer. On the contrary, this state have a vacuum expectation value for 
$\widehat{H}^{(+)}$
\begin{eqnarray}
\langle 0_{-}|\widehat{H}^{(+)}|0_{-}\rangle 
= {1\over 2}\int {d^{3}k\over (2\pi)^{3}}\left|\tilde{J}(\omega_{k},\vec{k})
\right|^2.
\label{eq:0H0}
\end{eqnarray}

The key to understand what is going on here lies in the fact that the
external source breaks the invariance of the theory under space-time
translations. In the particular case we have studied here where
$J(x)$ has support over a finite time interval $0<t<\tau$, this
implies that the vacuum is not invariant under time translations, so
observers at different times will make different choices of vacuum
that will not necessarily agree with each other. This is clear in our
example.  An observer in $t<\tau$ will choose the vacuum to be the
lowest energy state of her Hamiltonian, $|0_{-}\rangle$. On the other
hand, the second observer at late times $t>\tau$ will naturally choose
$|{0}_{+}\rangle$ as the vacuum. However, for this second observer,
the state $|0_{-}\rangle$ is not the vacuum of his Hamiltonian, but
actually an excited state that is a superposition of states with
well-defined number of particles. In this sense it can be said that
the external source has the effect of creating particles out of the
``in'' vacuum.  Besides, this breaking of time translation invariance 
produces a violation in the energy conservation as we see
from Eq. (\ref{eq:0H0}). Particles are actually created from the
energy pumped into the system by the external source. 

{\bf The Schwinger effect.}
A classical example of creation of particles by a external field was 
pointed out by Schwinger \cite{schwinger} and consists of the creation
of electron-positron pairs by a strong electric field. In order to 
illustrate this effect we are going to follow a heuristic argument
based on the Dirac sea picture and the WKB approximation.

\begin{figure}
\centerline{\epsfxsize=4.5truein\epsfbox{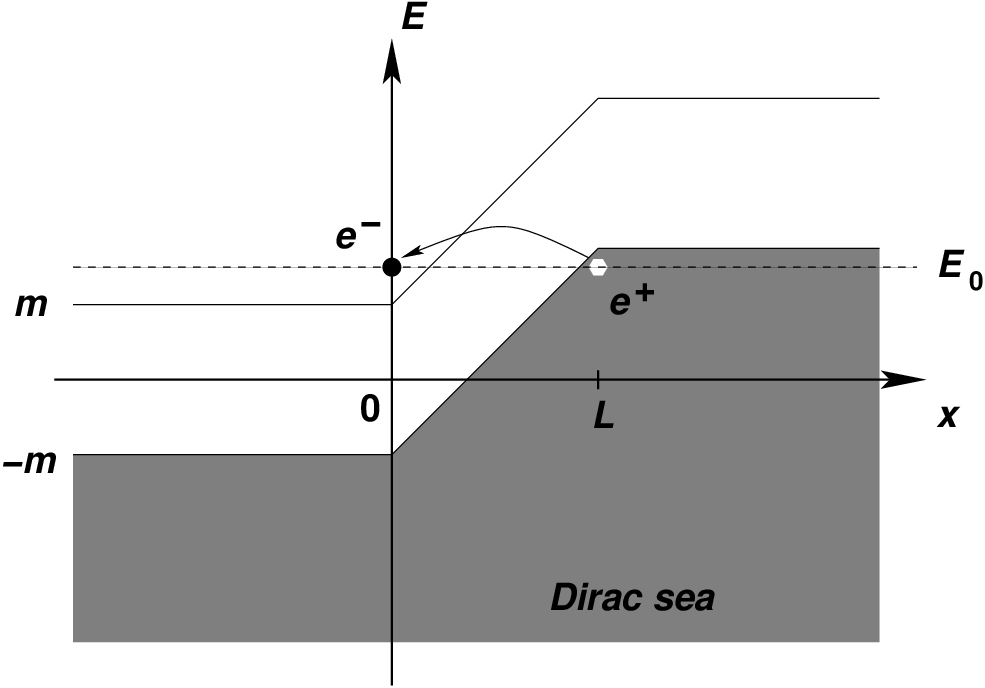}}
\caption[]{Pair creation by a electric field in the Dirac sea picture.}
\label{fig:schwinger}
\end{figure}

In the absence of an electric field the vacuum state of a
spin-${1\over 2}$ field is constructed by filling all the negative
energy states as depicted in Fig. \ref{fig:dirac_sea}. Let us now
connect a constant electric field
$\vec{\mathcal{E}}=\mathcal{E}\vec{u}_{x}$ in the range $0<x<L$
created by a electrostatic potential
\begin{eqnarray}
V(\vec{r})=\left\{
\begin{array}{ccc}
0 & \,\,\,\,\, & \,\,x<0 \\
-\mathcal{E}x & & 0<x<L \\
-\mathcal{E}L  & & \hspace*{0.9cm} x> L
\end{array}
\right.
\end{eqnarray}
After the field has been switched on, the Dirac sea looks like
in Fig. \ref{fig:schwinger}. In particular we find that if
$e\mathcal{E}L>2m$ there are negative energy states at $x>L$ with
the same energy as the positive energy states in the region
$x<0$. Therefore it is possible for an electron filling a negative
energy state with energy close to $-2m$ to tunnel through the
forbidden region into a positive energy state. The interpretation of
such a process is the production of an electron-positron pair out of
the electric field.

We can compute the rate at which such pairs are produced by using 
the WKB approximation. Focusing for simplicity on an electron on top
of the Fermi surface near $x=L$ with energy $E_{0}$,
the transmission coefficient in this approximation is given by\footnote{
Notice that the electron satisfy the relativistic dispersion relation
$E=\sqrt{\vec{p}^{\,2}+m^{2}}+V$ and therefore $-p_{x}^{2}=m^{2}-(E-V)^{2}
+\vec{p}_{T}^{\,\,2}$. The integration limits are set by those values of
$x$ at which $p_{x}=0$.}
\begin{eqnarray}
T_{\rm WKB}&=&\exp\left[-2
\int_{{1\over e\mathcal{E}}\left({E_{0}-\sqrt{m^{2}+\vec{p}_{T}^{\,\,2}}}
\right)}^{
{1\over e\mathcal{E}}\left({E_{0}+\sqrt{m^{2}+\vec{p}_{T}^{\,\,2}}}\right)}
dx\,\sqrt{m^2-\left[
E_{0}-e\mathcal{E}(x-x_{0})\right]^{2}+\vec{p}_{T}^{\,\,2}}\right] 
\nonumber \\
&=& \exp\left[-{\pi\over e\mathcal{E}}\left(\vec{p}_{T}^{\,\,2}+m^2
\right)\right],
\end{eqnarray}
where $p^{2}_{T}\equiv p_{y}^{2}+p_{z}^{2}$. This gives the transition
probability per unit time and per unit cross section $dydz$ for
an electron in the Dirac sea with transverse momentum $\vec{p}_{T}$
and energy $E_{0}$. To get the total probability per unit time and per
unit volume we have to integrate over all possible values of
$\vec{p}_{T}$ and $E_{0}$. Actually, in the case of the energy,
because of the relation between $E_{0}$ and the coordinate $x$ at
which the particle penetrates into the barrier we can write ${dE_{0}\over
2\pi}={e\mathcal{E}\over 2\pi}dx$ and the total probability per unit
time and per unit volume for the creation of a pair is given by
\begin{eqnarray}
W=2\left({e\mathcal{E}\over 2\pi}\right)\int {d^{2}p_{T}\over (2\pi)^{2}}
e^{-{\pi\over e\mathcal{E}}\left(\vec{p}_{T}^{\,\,2}+m^2\right)}
={e^2\mathcal{E}^{2}\over 4\pi^{3}}e^{-{\pi\,m^{2}\over e\mathcal{E}}},
\label{eq:pair_prod}
\end{eqnarray}
where the factor of $2$ accounts for the two polarizations of the
electron. 

Then production of electron-positron pairs is
exponentially suppressed and it is only sizeable for strong electric fields.
To estimate its order of magnitude it is useful to restore the powers
of $c$ and $\hbar$ in (\ref{eq:pair_prod})
\begin{eqnarray}
W={e^2\mathcal{E}^{2}\over 4\pi^{3}c\hbar^{2}}e^{-{\pi\,m^{2}c^{3}\over 
\hbar e\mathcal{E}}}
\end{eqnarray}
The exponential suppression of the pair production disappears when the
electric field reaches the critical value $\mathcal{E}_{\rm crit}$ at
which the exponent is of order one
\begin{eqnarray}
\mathcal{E}_{\rm crit}={ m^{2}c^{3}\over \hbar e}\simeq 1.3\times 10^{16}\, 
{\rm V}\,{\rm cm}^{-1}.
\end{eqnarray}
This is indeed a very strong field which is extremely difficult to 
produce. A similar effect, however, takes place also in a time-varying 
electric field \cite{BY} and there is the hope that pair production could
be observed in the presence of the alternating electric field produced by a 
laser.

The heuristic derivation that we followed here can be made more
precise in QED. There the decay of the vacuum into electron-positron
pairs can be computed from the imaginary part of the effective action
$\Gamma[A_{\mu}]$ in the presence of a classical gauge potential
$A_{\mu}$
\begin{eqnarray}
i\Gamma[A_{\mu}]&\equiv& 
\parbox{40mm}{
\begin{fmfgraph*}(90,70)
\fmfleft{i1}
\fmfright{o1}
\fmf{photon,tension=3}{i1,v1}
\fmf{phantom,tension=2}{v2,o1}
\fmf{fermion,right,tension=1}{v1,v2,v1}
\end{fmfgraph*}
}\hspace*{-2cm}+\parbox{40mm}{
\begin{fmfgraph*}(90,70)
\fmfleft{i1}
\fmfright{o1}
\fmf{photon,tension=3}{i1,v1}
\fmf{photon,tension=3}{v2,o1}
\fmf{fermion,right,tension=1.2}{v1,v2,v1}
\end{fmfgraph*}
}\hspace*{-0.8cm}+\hspace*{-.5cm}
\parbox{40mm}{
\begin{fmfgraph*}(90,70)
\fmfkeep{fermion}
\fmfleft{i1,i2}
\fmfright{o1}
\fmf{photon,tension=4}{i1,v1}
\fmf{photon,tension=4}{i2,v3}
\fmf{photon,tension=4}{v2,o1}
\fmf{fermion,tension=2,left=1.4/3}{v1,v3}
\fmf{fermion,tension=2,left=2/3}{v3,v2}
\fmf{fermion,tension=2,left=2/3}{v2,v1}
\end{fmfgraph*}
}\hspace*{-0.8cm}+\ldots \nonumber \\
&=& \log\,\,\det\left[1-ie\Fslash{A}{1\over i\dslash - m}
\right].
\end{eqnarray}
This determinant can be computed using the standard heat kernel techniques.
The probability of pair production is proportional to the imaginary part
of $i\Gamma[A_{\mu}]$ and gives 
\begin{eqnarray}
W={e^2\mathcal{E}^{2}\over 4\pi^{3}}\sum_{n=1}^{\infty}{1\over n^{2}}
e^{-n{\pi\,m^{2}\over e\mathcal{E}}}.
\label{eq:schwinger_total}
\end{eqnarray}
Our simple argument based on tunneling in the Dirac sea gave only the leading
term of Schwinger's result (\ref{eq:schwinger_total}). The remaining terms
can be also captured in the WKB approximation by taking into account the 
probability of production of several pairs, i.e. the tunneling of more than
one electron through the barrier.

Here we have illustrated the creation of particles by semiclassical
sources in Quantum Field Theory using simple examples. Nevertheless,
what we learned has important applications to the study of
quantum fields in curved backgrounds.  In Quantum Field Theory in
Minkowski space-time the vacuum state is invariant under the
Poincar\'e group and this, together with the covariance of the theory
under Lorentz transformations, implies that all inertial observers
agree on the number of particles contained in a quantum state.  The
breaking of such invariance, as happened in the case of coupling to a
time-varying source analyzed above, implies that it is not possible
anymore to define a state which would be recognized as the vacuum by
all observers.

This is precisely the situation when fields are quantized on curved
backgrounds.  In particular, if the background is time-dependent (as
it happens in a cosmological setup or for a collapsing star) different
observers will identify different vacuum states. As a consequence what
one observer call the vacuum will be full of particles for a different
observer. This is precisely what is behind the phenomenon of Hawking
radiation \cite{hawking}. The emission of particles by a physical black 
hole formed from gravitational collapse of a star is the consequence of
the fact that the vacuum state in the asymptotic past contain
particles for an observer in the asymptotic future. As a consequence, 
a detector located far away from the black hole detects a stream of 
thermal radiation with temperature
\begin{eqnarray}
T_{\rm Hawking}={\hbar c^{3}\over 8\pi G_{N}\,k\,M}
\end{eqnarray}
where $M$ is the mass of the black hole, $G_{N}$ is Newton's constant
and $k$ is Boltzmann's constant.  There are several ways in which this
results can be obtained. A more heuristic way is perhaps to think of
this particle creation as resulting from quantum tunneling of
particles across the potential barrier posed by gravity
\cite{parikh_wilczek}.

\subsection{Supersymmetry}
\label{sec:supersymmetry}

One of the things that we have learned in our journey around the
landscape of Quantum Field Theory is that our knowledge of the
fundamental interactions in Nature is based on the idea of symmetry,
and in particular gauge symmetry. The Lagrangian of the Standard Model
can be written just including all possible renormalizable
terms (i.e.  with canonical dimension smaller o equal to 4) compatible
with the gauge symmetry SU(3)$\times$SU(2)$\times$U(1) and Poincar\'e
invariance.  All attempts to go beyond start
with the question of how to extend the symmetries of the Standard Model.

As explained in Section \ref{sec:smatrix}, in a quantum field theoretical description of the interaction of 
elementary particles the basic observable quantity to compute is
the scattering or $S$-matrix giving the probability amplitude for 
the scattering of a number of incoming particles with a certain momentum
into some final products
\begin{eqnarray}
\mathcal{A}(\mbox{in}\longrightarrow\mbox{out})=
\langle \vec{p}_{1}{}',\ldots;\mbox{out}
|\vec{p}_{1},\ldots;\mbox{in}
\rangle.
\end{eqnarray}
An explicit symmetry of the theory has to be necessarily a symmetry of the 
$S$-matrix. Hence it is fair to ask what is the largest symmetry of the $S$-matrix.

Let us ask this question in the simple case of the scattering of two particles
with four-momenta $p_{1}$ and $p_{2}$ in the $t$-channel 
\begin{eqnarray*}
\parbox{40mm}{
\begin{fmfgraph*}(90,70)
\fmfleft{i1,i2}
\fmfright{o1,o2}
\fmf{fermion,tension=0.5,label=$p_1$}{i1,v1}
\fmf{fermion,tension=0.5,label=$p_1'$}{v1,o1}
\fmf{fermion,tension=0.5,label=$p_2$}{i2,v1}
\fmf{fermion,tension=0.5,label=$p_2'$}{v1,o2}
\fmfblob{0.5w}{v1}
\end{fmfgraph*}
}
\end{eqnarray*}
We will make the usual assumptions regarding positivity of the energy
and analyticity.  Invariance of the theory under the Poincar\'e
group implies that the amplitude can only depend on the scattering
angle $\vartheta$ through
\begin{eqnarray}
t=(p_1'-p_1)^2=2\left(m_{1}^{2}-p_{1}\cdot p_{1}'\right)
=2\left(
m_{1}^{2}-E_{1}E_{1}'+|\vec{p}_{1}||\vec{p}_{1}{}'|\cos\vartheta\right).
\end{eqnarray}
If there would be any extra bosonic symmetry of the theory 
it would restrict the scattering angle to a set of discrete values. 
In this case the $S$-matrix cannot be analytic 
since it would vanish everywhere except for the discrete values selected
by the extra symmetry. 

Actually, the only way to extend the symmetry of the theory without
renouncing to the analyticity of the scattering amplitudes is to
introduce ``fermionic'' symmetries, i.e. symmetries whose generators
are anticommuting objects \cite{susy}. This means that in addition to
the generators of the Poincar\'e group\footnote{The generators
$M^{\mu\nu}$ are related with the ones for boost and rotations
introduced in section \ref{sec:lorentz} by $J^{i}\equiv M^{0i}$,
$M^{i}={1\over 2} \varepsilon^{ijk}M^{jk}$. In this section we also
use the ``dotted spinor'' notation, in which spinors in the
$(\mathbf{1\over 2},\mathbf{0})$ and $(\mathbf{0},\mathbf{1\over 2})$
representations of the Lorentz group are indicated respectively by
undotted ($a,b,\ldots$) and dotted ($\dot{a},\dot{b},\ldots$)
indices.} $P^{\mu}$, $M^{\mu\nu}$ and the ones for the internal gauge
symmetries ${G}$, we can introduce a number of fermionic generators
$Q^{I}_{a}$, $\overline{Q}_{\dot{a}\,I}$ ($I=1,\ldots,\mathcal{N}$),
where $\overline{Q}_{\dot{a}\,I}=(Q^{I}_{a})^{\dagger}$. The most
general algebra that these generators satisfy is the
$\mathcal{N}$-extended supersymmetry algebra \cite{hls}
\begin{eqnarray}
\{Q^{I}_{a},\overline{Q}_{\dot{b}\,J}\}&=&
2\sigma^{\mu}_{a\dot{b}}P_{\mu}\delta^{I}_{\,\,\,J}, \nonumber \\
\{Q^{I}_{a},Q^{J}_{b}\}&=&2\varepsilon_{ab}\mathcal{Z}^{IJ}, \\
\{\overline{Q}^{I}_{\dot{a}},\overline{Q}^{J}_{\dot{b}}\}&=&
2\varepsilon_{\dot{a}\dot{b}}\overline{\mathcal{Z}}^{IJ},
\end{eqnarray}
where $\mathcal{Z}^{IJ}\in\mathbb{C}$ commute with any other generator
and satisfies $\mathcal{Z}^{IJ}=-\mathcal{Z}^{JI}$. Besides we have
the commutators that determine the Poincar\'e transformations of the
fermionic generators $Q^{I}_{a}$, $Q_{\dot{a}\,J}$
\begin{eqnarray}
[Q^{I}_{a},P^{\mu}]&=&[\overline{Q}_{\dot{a}\,I},P^{\mu}]=0, \nonumber \\
{}[Q^{I}_{a},M^{\mu\nu}]&=&{1\over 2} 
(\sigma^{\mu\nu})_{a}^{\,\,\,b}Q^{I}_{b} ,
\label{eq:transfQ}\\
{}[\overline{Q}_{a\,I},M^{\mu\nu}]&=&-{1\over 2} 
(\overline{\sigma}^{\mu\nu})_{\dot{a}}^{\,\,\,\dot{b}}
\,\overline{Q}_{\dot{b}\,I},
\nonumber 
\end{eqnarray}
where $\sigma^{0i}=-i\sigma^{i}$,
$\sigma^{ij}=\varepsilon^{ijk}\sigma^{k}$ and
$\overline{\sigma}^{\mu\nu}=(\sigma^{\mu\nu})^{\dagger}$.  These
identities simply mean that $Q^{I}_{a}$, $\overline{Q}_{\dot{a}\,J}$
transform respectively in the $(\mathbf{1\over 2},\mathbf{0})$ and
$(\mathbf{0},\mathbf{1\over 2})$ representations of the Lorentz
group.

We know that the presence of a global symmetry in a theory implies that the
spectrum can be classified in multiplets with respect to that
symmetry. In the case of supersymmetry start with the case
case $\mathcal{N}=1$ in which there is a single pair of supercharges
$Q_{a}$, $\overline{Q}_{\dot{a}}$ satisfying the algebra
\begin{eqnarray}
\{Q_{a},\overline{Q}_{\dot{b}}\}=2\sigma^{\mu}_{a\dot{b}}P_{\mu}, 
\hspace*{1cm} \{Q_{a},Q_{b}\}=\{\overline{Q}_{\dot{a}},\overline{Q}_{\dot{b}}
\}=0.
\label{eq:susyN=1}
\end{eqnarray}
Notice that in the $\mathcal{N}=1$ case there is no possibility of having
central charges. 

We study now the representations of the supersymmetry algebra
(\ref{eq:susyN=1}), starting with the massless case. Given a state
$|k\rangle$ satisfying $k^{2}=0$, we can always find a reference frame
where the four-vector $k^{\mu}$ takes the form
$k^{\mu}=(E,0,0,E)$. Since the theory is Lorentz covariant we can obtain the
representation of the supersymmetry algebra in this frame where the
expressions are simpler. In particular, the right-hand side of the
first anticommutator in Eq. (\ref{eq:susyN=1}) is given by
\begin{eqnarray}
2\sigma^{\mu}_{a\dot{b}}P_{\mu}=2(P^{0}-\sigma^{3}P^{3})=
\left(
\begin{array}{cc}
0 & 0 \\
0 & 4E
\end{array}
\right).
\end{eqnarray}
Therefore the algebra of supercharges in the massless case reduces to
\begin{eqnarray}
\{Q_{1},Q^{\dagger}_{1}\}&=&\{Q_{1},Q_{2}^{\dagger}\}=0, \nonumber \\
\{Q_{2},Q_{2}^{\dagger}\}&=&4E .
\end{eqnarray}
The commutator $\{Q_{1},Q^{\dagger}_{1}\}=0$ implies that the action
of $Q_{1}$ on any state gives a zero-norm state of the Hilbert space
$|\!|Q_{1}|\Psi\rangle|\!|=0$. If we want the theory to preserve unitarity
we must eliminate these null states from the spectrum. This is equivalent
to setting $Q_{1}\equiv 0$. On the other hand, in terms of the second 
generator $Q_{2}$ we can define the operators
\begin{eqnarray}
a={1\over 2\sqrt{E}}Q_{2}, \hspace*{1cm} 
a^{\dagger}={1\over 2\sqrt{E}}Q_{2}^{\dagger},
\end{eqnarray}
which satisfy the algebra of a pair of fermionic creation-annihilation
operators, $\{a,a^{\dagger}\}=1$, $a^{2}=(a^{\dagger})^{2}=0$.  Starting with
a vacuum state $a|\lambda\rangle=0$ with
helicity $\lambda$ we can build the massless multiplet
\begin{eqnarray}
|\lambda\rangle, \hspace*{1cm} |\lambda+\mbox{${1\over 2}$}\rangle\equiv 
a^{\dagger}|\lambda\rangle.
\end{eqnarray}
Here we consider two important cases:
\begin{itemize}
\item
Scalar multiplet: we take the vacuum state to have zero helicity
$|0^{+}\rangle$ so the multiplet consists of a scalar and a
helicity-${1\over 2}$ state
\begin{eqnarray}
|0^{+}\rangle, \hspace*{1cm} |\,\mbox{${1\over 2}$}\rangle\equiv
a^{\dagger}|0^{+}\rangle.
\end{eqnarray}
However, this multiplet is not invariant under the CPT transformation which
reverses the sign of the helicity of the states. In order to have
a CPT-invariant theory we have to add to this multiplet its CPT-conjugate
which can be obtain from a vacuum state with helicity $\lambda=-{1\over 2}$
\begin{eqnarray}
|0^{-}\rangle, \hspace*{1cm} |\,\mbox{$-{1\over 2}$}\rangle.
\end{eqnarray}
Putting them together we can combine the two zero helicity states
with the two fermionic ones into the degrees of freedom of a 
complex scalar field and a Weyl (or Majorana) spinor.

\item
Vector multiplet: now we take the vacuum state to have helicity 
$\lambda={1\over 2}$,
so the multiplet contains also a massless state with helicity $\lambda=1$
\begin{eqnarray}
|\,\mbox{${1\over 2}$}\rangle, \hspace*{1cm} |1\rangle\equiv a^{\dagger}
|\,\mbox{${1\over 2}$}\rangle.
\label{eq:vector}
\end{eqnarray}
As with the scalar multiplet we add the CPT conjugated obtained from a 
vacuum state with helicity $\lambda=-1$
\begin{eqnarray}
|\,-\mbox{${1\over 2}$}\rangle, \hspace*{1cm} |-1\rangle,
\end{eqnarray}
which together with (\ref{eq:vector}) give the propagating
states of a gauge field and a spin-${1\over 2}$ gaugino.

\end{itemize}
In both cases we see the trademark of supersymmetric theories: the
number of bosonic and fermionic states within a multiplet are the same.

In the case of extended supersymmetry we have to repeat the previous
analysis for each supersymmetry charge. At the end, we have
$\mathcal{N}$ sets of fermionic creation-annihilation operators
$\{a^{I},a_{I}^{\dagger}\}=\delta^{I}_{\,\,\,J}$, $(a_{I})^{2}
=(a_{I}^{\dagger})^{2}=0$. Let us work out the case of $\mathcal{N}=8$
supersymmetry. Since for several reasons we do not want to have states
with helicity larger than $2$, we start with a vacuum state
$|-2\rangle$ of helicity $\lambda=-2$.  The rest of the states of the
supermultiplet are obtained by applying the eight different creation
operators $a_{I}^{\dagger}$ to the vacuum:
\begin{eqnarray}
\lambda=2:& &a_{1}^{\dagger}\ldots a_{8}^{\dagger}|-2\rangle \hspace*{2.3cm}
\binom{8}{8}=\mbox{1 state},  \nonumber \\
\lambda={3\over 2}: & &a_{I_{1}}^{\dagger}\ldots 
a_{I_7}^{\dagger}|-2\rangle
\hspace*{2cm} \binom{8}{7}=\mbox{8 states}, \nonumber \\
\lambda=1: & &a_{I_{1}}^{\dagger}\ldots a_{I_6}^{\dagger}|-2\rangle 
\hspace*{2cm} \binom{8}{6}=\mbox{28 states}, \nonumber \\
\lambda={1\over 2}:& &a_{I_{1}}^{\dagger}\ldots 
a_{I_5}^{\dagger}|-2\rangle 
\hspace*{2cm} \binom{8}{5}=\mbox{56 states}, \nonumber \\
\lambda=0: & &a_{I_{1}}^{\dagger}\ldots a_{I_4}^{\dagger}|-2\rangle
\hspace*{2cm} \binom{8}{4}=\mbox{70 states},  \\
\lambda=-{1\over 2}: & &a_{I_{1}}^{\dagger}a_{I_2}^{\dagger}
a_{I_3}^{\dagger}|-2\rangle 
\hspace*{2.1cm} \binom{8}{3}=\mbox{56 states}, \nonumber \\
\lambda=-1: & &a_{I_{1}}^{\dagger}a_{I_2}^{\dagger}
|-2\rangle 
\hspace*{2.6cm} \binom{8}{2}=\mbox{28 states}, \nonumber \\
\lambda=-{3\over 2}: & &a_{I_{1}}^{\dagger}
|-2\rangle 
\hspace*{3.1cm} \binom{8}{1}=\mbox{8 states}, \nonumber \\
\lambda=-2: & &|-2\rangle
\hspace*{5cm} \mbox{1 state}. \nonumber 
\end{eqnarray}
Putting together the states with opposite helicity we find that the
theory contains:
\begin{itemize}
\item
1 spin-2 field $g_{\mu\nu}$ (a graviton),
\item
8 spin-${3\over 2}$ gravitino fields $\psi_{\mu}^{I}$,
\item
28 gauge fields $A_{\mu}^{[IJ]}$,
\item
56 spin-${1\over 2}$ fermions $\psi^{[IJK]}$,
\item
70 scalars $\phi^{[IJKL]}$,
\end{itemize}
where by $[IJ...]$ we have denoted that the indices are
antisymmetrized.  We see that, unlike the massless multiplets of
$\mathcal{N}=1$ supersymmetry studied above, this multiplet is CPT
invariant by itself. As in the case of the massless $\mathcal{N}=1$
multiplet, here we also find as many bosonic as fermionic states:
\begin{eqnarray*}
\begin{array}{lrr}
\mbox{bosons:} &  1+28+70+28+1=128 & \mbox{states}, \\
\mbox{fermions:} & 8+56+56+8= 128 & \mbox{states}.
\end{array}
\end{eqnarray*}

Now we study briefly the case of massive representations $|k\rangle$,
$k^{2}=M^{2}$. Things become simpler if we work in the rest frame
where $P^{0}=M$ and the spatial components of the momentum
vanish. Then, the supersymmetry algebra becomes:
\begin{eqnarray}
\{Q_{a}^{I},\overline{Q}_{\dot{b}\,J}\}=2M\delta_{a\dot{b}}
\delta^{I}_{\,\,\,J}.
\end{eqnarray}
We proceed now in a similar way to the massless case by defining the
operators
\begin{eqnarray}
a_{a}^{I}\equiv {1\over \sqrt{2M}}Q^{I}_{a}, \hspace*{1cm}
a_{\dot{a}\,I}^{{\dagger}}\equiv {1\over
\sqrt{2M}}\overline{Q}_{\dot{a}
\,I}.
\end{eqnarray}
The multiplets are found by choosing a vacuum state with a definite
spin.  For example, for $\mathcal{N}=1$ and taking a spin-0 vacuum
$|0\rangle$ we find three states in the multiplet transforming irreducibly 
with respect to the Lorentz group:
\begin{eqnarray}
|0\rangle, \hspace*{1cm} a_{\dot{a}}^{\dagger}|0\rangle, 
\hspace*{1cm} \varepsilon^{\dot{a}\dot{b}}a_{\dot{a}}^{\dagger}
a_{\dot{b}}^{\dagger}|0\rangle,
\end{eqnarray}
which, once transformed back from the rest frame, correspond to the
physical states of two spin-0 bosons and one spin-${1\over 2}$
fermion.  For $\mathcal{N}$-extended supersymmetry the corresponding
multiplets can be worked out in a similar way.

The equality between bosonic and fermionic degrees of freedom is at
the root of many of the interesting properties of supersymmetric
theories.  For example, in section \ref{sec:theories&lagrangians} we
computed the divergent vacuum energy contributions for each real bosonic or
fermionic propagating degree of freedom is\footnote{For a boson, this
can be read off Eq. (\ref{eq:hamiltonianKG}). In the case of fermions,
the result of Eq. (\ref{eq:fermion_hamiltonian}) gives the vacuum
energy contribution of the four real propagating degrees of freedom of
a Dirac spinor.}
\begin{eqnarray}
E_{\rm vac}=\pm{1\over 2}\delta(\vec{0})\int d^{3}p\, \omega_{p},
\end{eqnarray}
where the sign $\pm$ corresponds respectively to bosons and fermions.
Hence, for a supersymmetric theory the vacuum energy contribution
exactly cancels between bosons and fermions.  This boson-fermion
degeneracy is also responsible for supersymmetric quantum field
theories being less divergent than nonsupersymmetric ones.


\section*{Appendix: A crash course in Group Theory}
\label{sec:appendix}
\renewcommand{\thesection}{A}

In this Appendix we summarize some basic facts about
Group Theory. Given a group ${G}$ a representation of
${G}$ is a correspondence between the elements of $G$ and 
the set of linear operators acting on a vector space $V$, such that
for each element of
the group $g\in{G}$ there is a linear operator $D(g)$
\begin{eqnarray}
D(g):V\longrightarrow V
\end{eqnarray}
satisfying the group operations
\begin{eqnarray}
D(g_{1})D(g_{2})=D(g_{1}g_{2}), \hspace*{1cm}
D(g_{1}^{-1})=D(g_{1})^{-1}, \hspace*{1cm} g_{1},g_{2}\in\mathcal{G}.
\end{eqnarray}
The representation $D(g)$ is irreducible if and only if the only 
operators $A:V\rightarrow V$ commuting with all the elements
of the representation $D(g)$ are the ones proportional to the identity
\begin{eqnarray}
[D(g),A]=0, \,\,\,\forall g\hspace*{1cm}\Longleftrightarrow 
\hspace*{1cm} A=\lambda\mathbf{1},\hspace*{0.5cm} \lambda\in\mathbb{C}
\end{eqnarray}
More intuitively, we can say that a representation is irreducible if
there is no proper subspace $U\subset V$ (i.e. $U\neq V$ and $U\neq
\emptyset$) such that $D(g)U\subset U$ for every element 
$g\in G$.

Here we are specially interested in Lie groups whose elements are
labelled by a number of continuous parameters. In mathematical terms
this means that a Lie group is a manifold $\mathcal{M}$ together with
an operation $\mathcal{M}\times\mathcal{M}\longrightarrow \mathcal{M}$
that we will call multiplication that satisfies the associativity
property $g_{1}\cdot(g_{2}\cdot g_{3}) =(g_{1}\cdot g_{2})\cdot g_{3}$
together with the existence of unity $g\mathbf{1}=
\mathbf{1}g=g$,for every $g\in\mathcal{M}$ and inverse $gg^{-1}=g^{-1}g=
\mathbf{1}$.

The simplest example of a Lie group is SO(2), the group of rotations
in the plane. Each element $R(\theta)$ is labelled by the rotation
angle $\theta$, with the multiplication acting as
$R(\theta_{1})R(\theta_{2})=R(\theta_{1}+\theta_{2})$. Because the angle
$\theta$ is defined only modulo $2\pi$, the manifold of SO(2) is a
circumference $S^{1}$.

One of the interesting properties of Lie groups is that in a neighborhood
of the identity element they can be expressed in terms of a set of
generators $T^{a}$ ($a=1,\ldots,{\rm dim\,}G$) as
\begin{eqnarray}
D(g)=\exp(-i\alpha_{a}T^{a})\equiv \sum_{n=0}^{\infty}{(-i)^{n}\over n!}
\alpha_{a_1}\ldots\alpha_{a_n}T^{a_1}\ldots T^{a_n},
\end{eqnarray}
where $\alpha_{a}\in\mathbb{C}$ are a set of coordinates of
$\mathcal{M}$ in a neighborhood of $\mathbf{1}$.  Because of the
general Baker-Campbell-Haussdorf formula, the multiplication of two
group elements is encoded in the value of the commutator of two
generators, that in general has the form
\begin{eqnarray}
[T^{a},T^{b}]=if^{abc}T^{c},
\end{eqnarray}
where $f^{abc}\in\mathbb{C}$ are called the structure constants. 
The set of generators with the commutator operation form the Lie algebra 
associated with the Lie group. Hence, given a representation of the 
Lie algebra of generators we can construct a representation of the group
by exponentiation (at least locally near the identity).

We illustrate these concept with some particular examples. 
For SU(2) each group element is labelled by
three real number $\alpha_{i}$, $i=1,2,3$. We have two basic
representations: one is the fundamental representation (or spin
${1\over 2}$) defined by
\begin{eqnarray}
D_{1\over 2}(\alpha_{i})=e^{-{i\over 2}\alpha_{i}\sigma^{i}},
\end{eqnarray}
with $\sigma^{i}$ the Pauli matrices. The second one is the 
adjoint (or spin 1) representation which can be written as
\begin{eqnarray}
D_{1}(\alpha_{i})=e^{-i\alpha_{i}J^{i}},
\end{eqnarray}
where 
\begin{eqnarray}
J^{1}=\left(
\begin{array}{ccc}
0 & 0 & 0 \\
0 & 0 & 1 \\
0 & -1 & 0
\end{array}
\right), \hspace*{1cm}
J^{2}=\left(   
\begin{array}{ccc}
0 & 0 & -1 \\
0 & 0 & 0 \\
1 & 0 & 0
\end{array}
\right), \hspace*{1cm}
J^{3}=\left(
\begin{array}{ccc}
0 & 1 & 0 \\
-1 & 0 & 0 \\
0 & 0 & 0
\end{array}
\right).
\label{eq:j=1}
\end{eqnarray}
Actually, $J^{i}$ ($i=1,2,3$) generate rotations around the $x$, $y$
and $z$ axis respectively. Representations of spin $j\in
\mathbb{N}+{1\over 2}$ can be also constructed with dimension
\begin{eqnarray}
{\rm dim\,}D_{j}(g)=2j+1.
\end{eqnarray}

As a second example we consider SU(3). This group has two basic
three-dimensional representations denoted by $\mathbf{3}$ and
$\mathbf{\overline{3}}$ which in QCD are associated with the transformation
of quarks and antiquarks under the color gauge symmetry
SU(3). The elements of these representations can be written as
\begin{eqnarray}
D_{\mathbf{3}}(\alpha^{a})=e^{{i\over 2}\alpha^{a}\lambda_{a}},
\hspace*{1cm} D_{\mathbf{\overline{3}}}(\alpha^{a})
=e^{-{i\over 2}\alpha^{a}{\lambda}_{a}^{T}} \hspace*{1cm}
(a=1,\ldots,8),
\end{eqnarray}
where $\lambda_{a}$ are the eight hermitian Gell-Mann matrices
\begin{eqnarray}
\lambda_{1}&=&\left(
\begin{array}{ccc}
0 & 1 & 0 \\
1 & 0 & 0 \\
0 & 0 & 0
\end{array}
\right), \hspace*{1cm}
\lambda_{2}=\left(
\begin{array}{ccc}
0 & -i & 0 \\
i & 0 & 0 \\
0 & 0 & 0
\end{array}
\right), \hspace*{1cm}
\lambda_{3}=\left(
\begin{array}{ccc}
1 & 0 & 0 \\
0 & -1 & 0 \\
0 & 0 & 0
\end{array}
\right), \nonumber \\
\lambda_{4}&=&\left(
\begin{array}{ccc}
0 & 0 & 1 \\
0 & 0 & 0 \\
1 & 0 & 0
\end{array}
\right), \hspace*{1cm}
\lambda_{5}=\left(
\begin{array}{ccc}
0 & 0 & -i \\
0 & 0 & 0 \\
i & 0 & 0
\end{array}
\right), \hspace*{1cm}
\lambda_{6}=\left(
\begin{array}{ccc}
0 & 0 & 0 \\
0 & 0 & 1 \\
0 & 1 & 0
\end{array}
\right), \\
\lambda_{7}&=&\left(
\begin{array}{ccc}
0 & 0 & 0 \\
0 & 0 & -i \\
0 & i & 0
\end{array}
\right), \hspace*{1cm}
\lambda_{8}=\left(
\begin{array}{ccc}
{1\over \sqrt{3}} & 0 & 0 \\
0 & {1\over \sqrt{3}} & 0 \\
0 & 0 & -{2\over\sqrt{3}}
\end{array}
\right). \nonumber 
\end{eqnarray}
Hence the generators of the representations $\mathbf{3}$ and $\overline{
\mathbf{3}}$ are given by
\begin{eqnarray}
T^{a}(\mathbf{3})={1\over 2}\lambda_{a}, \hspace*{1cm}
T^{a}(\overline{\mathbf{3}})=-{1\over 2}\lambda_{a}^{T}.
\end{eqnarray}

Irreducible representations can be classified in three groups: real, 
complex and pseudoreal. 

\begin{itemize}

\item
Real representations: a representation is said to be real 
if there is a {\em symmetric matrix} $S$ which acts as intertwiner 
between the generators and their complex conjugates
\begin{eqnarray}
\overline{T}^{a}=-ST^{a}S^{-1}, \hspace*{1cm} S^{T}=S.
\label{eq:STS1}
\end{eqnarray}
This is for example the case of the adjoint representation of SU(2) generated
by the matrices (\ref{eq:j=1})

\item
Pseudoreal representations: are the ones for which 
an {\em antisymmetric matrix} $S$ exists with the property
\begin{eqnarray}
\overline{T}^{a}=-ST^{a}S^{-1}, \hspace*{1cm} S^{T}=-S.
\label{eq:STS2}
\end{eqnarray}
As an example we can mention the spin-${1\over 2}$ representation of 
SU(2) generated by ${1\over 2}\sigma^{i}$. 

\item
Complex representations: finally, a representation is complex if the
generators and their complex conjugate are not related by a similarity
transformation. This is for instance the case of the two three-dimensional
representations $\mathbf{3}$ and $\mathbf{\overline{3}}$ of SU(3).

\end{itemize}

There are a number of invariants that can be constructed associated
with an irreducible representation $R$ of a Lie group ${G}$ and that
can be used to label such a representation. If $T^{a}_{R}$ are the
generators in a certain representation $R$ of the Lie algebra, it is
easy to see that the matrix $\sum_{a=1}^{{\rm
dim\,}{G}}T^{a}_{R}T^{a}_{R}$ commutes with every generator
$T^{a}_{R}$. Therefore, because of Schur's lemma, it has to be
proportional to the identity\footnote{Schur's lemma states that
if there is a matrix $A$ that commutes with all elements of an irreducible 
representation of a Lie algebra, then $A=\lambda\mathbf{1}$, for some $\lambda\in\mathbb{C}$.}. 
This defines the Casimir invariant
$C_{2}(R)$ as
\begin{eqnarray}
\sum_{a=1}^{{\rm dim\,}{G}}T^{a}_{R}T^{a}_{R}
=C_{2}(R)\mathbf{1}.
\label{eq:c2}
\end{eqnarray}
A second invariant $T_{2}(R)$ associated with a representation $R$ can
also be defined by the identity
\begin{eqnarray}
{\rm Tr\,}T^{a}_{R}T^{b}_{R}=T_{2}(R)\delta^{ab}.
\label{eq:t2}
\end{eqnarray}
Actually, taking the trace in Eq. (\ref{eq:c2}) and combining the result with
(\ref{eq:t2}) we find that both invariants are related by the identity
\begin{eqnarray}
C_{2}(R)\,{\rm dim\,}R=T_{2}(R)\,{\rm dim\,}G,
\end{eqnarray}
with ${\rm dim\,}R$ the dimension of the representation $R$.

These two invariants appear frequently in quantum field theory
calculations with nonabelian gauge fields. For example $T_{2}(R)$
comes about as the coefficient of the one-loop calculation of the
beta-function for a Yang-Mills theory with gauge group ${G}$. 
In the case of SU(N), for the fundamental representation,
we find the values
\begin{eqnarray}
C_{2}(\mbox{fund})={N^{2}-1\over 2N}, \hspace*{1cm}
T_{2}(\mbox{fund})={1\over 2},
\end{eqnarray}
whereas for the adjoint representation the results are
\begin{eqnarray}
C_{2}(\mbox{adj})=N, \hspace*{1cm} T_{2}(\mbox{adj})=N.
\end{eqnarray}

A third invariant $A(R)$ is specially important in the calculation of
anomalies.  As discussed in section (\ref{sec:anomalies}), the 
chiral anomaly in gauge theories is proportional to the group-theoretical
factor ${\rm Tr\,}\left[T^{a}_{R}\{T^{b}_{R},T^{c}_{R}\}\right]$.
This leads us to define $A(R)$ as
\begin{eqnarray}
{\rm Tr\,}\left[T^{a}_{R}\{T^{b}_{R},T^{c}_{R}\}\right]=A(R)
d^{abc},
\end{eqnarray}
where $d^{abc}$ is symmetric in its three indices and does not depend
on the representation. Therefore, the cancellation of anomalies in 
a gauge theory with fermions transformed in the representation $R$ of
the gauge group is guaranteed if the corresponding invariant $A(R)$
vanishes.

It is not difficult to prove that $A(R)=0$ if the
representation $R$ is either real or pseudoreal. Indeed, if this is
the case, then there is a matrix $S$ (symmetric or antisymmetric) that
intertwins the generators $T^{a}_{R}$ and their complex
conjugates $\overline{T}^{a}_{R}=-ST^{a}_{R}S^{-1}$. Then, using the
hermiticity of the generators we can write
\begin{eqnarray}
{\rm Tr\,}\left[T^{a}_{R}\{T^{b}_{R},T^{c}_{R}\}\right]=
{\rm Tr\,}\left[T^{a}_{R}\{T^{b}_{R},T^{c}_{R}\}\right]^{T}=
{\rm Tr\,}\left[\overline{T}^{a}_{R}\{
\overline{T}^{b}_{R},\overline{T}^{c}_{R}\}\right].
\end{eqnarray}
Now, using (\ref{eq:STS1}) or (\ref{eq:STS2}) we have
\begin{eqnarray}
{\rm Tr\,}\left[\overline{T}^{a}_{R}\{
\overline{T}^{b}_{R},\overline{T}^{c}_{R}\}\right]=
-{\rm Tr\,}\left[ST^{a}_{R}S^{-1}\{ST^{b}_{R}S^{-1},
ST^{c}_{R}S^{-1}\}\right]=
-{\rm Tr\,}\left[T^{a}_{R}\{T^{b}_{R},T^{c}_{R}\}\right],
\end{eqnarray}
which proves that ${\rm Tr\,}\left[T^{a}_{R}\{T^{b}_{R},T^{c}_{R}\}\right]$
and therefore $A(R)=0$ whenever the representation is real or 
pseudoreal. Since the gauge anomaly in four dimensions 
is proportional to $A(R)$ this means that
anomalies appear only when the fermions transform in a complex representation
of the gauge group.

\end{fmffile}


\begin{thebibliography}{99}

\bibitem{ours}
L.~\'Alvarez-Gaum\'e and M.~A.~V\'azquez-Mozo, {\it An Invitation to 
Quantum Field Theory,} Springer 2011.

\bibitem{bjorken}
J.~D.~Bjorken and S.~D.~Drell, {\it Relativistic Quantum Fields},
McGraw-Hill 1965.

\bibitem{itzykson}
C.~Itzykson and J.-B.~Zuber, {\it Quantum Field Theory}, McGraw-Hill
1980.

\bibitem{ramond}
P.~Ramond, {\it Field Theory: A Modern Primer}, Addison-Wesley 1990.

\bibitem{peskin}
M.~E.~Peskin and D.~V.~Schroeder, {\it An Introduction to Quantum 
Field Theory}, Addison Wesley 1995.

\bibitem{weinberg}
S.~ Weinberg, {\it The Quantum Theory of Fields}, Vols. 1-3, Cambridge
1995

\bibitem{deligne}
P. Deligne et al. (editors), {\it Quantum Fields and Strings: a Course
for Mathematicians}, American Mathematical Society 1999.

\bibitem{zee}
A.~ Zee, {\it Quantum Field Theory in a Nutshell}, Princeton 2003.

\bibitem{dewitt}
B.~S.~DeWitt, {\it The Global Approach to Quantum Field Theory},
Vols. 1 \& 2, Oxford 2003.

\bibitem{nair}
V.~P.~Nair, {\it Quantum Field Theory. A Modern Perspective}, 
Springer 2005.

\bibitem{banks}
T.~Banks, {\it Modern Quantum Field Theory}, Cambridge 2008.

\bibitem{klein}
O.~Klein, {\it Die Reflexion von Elektronen an einem Potentialsprung nach der 
Relativischen Dynamik von Dirac}, Z. Phys. {\bf 53} (1929) 157.

\bibitem{klein-review1}
B.~R.~Holstein, {\it Klein's paradox}, Am. J. Phys. {\bf 66} (1998) 507. 

\bibitem{klein-review2}
N.~Dombey and A.~Calogeracos, {\it Seventy years of the Klein paradox}, 
Phys. Rept. {\bf 315} (1999) 41.
\\
N.~Dombey and A.~Calogeracos, {\it History and Physics of the Klein Paradox}, 
Contemp. Phys. {\bf 40} (1999) 313 ({\tt quant-ph/9905076}).

\bibitem{Sauter}
F. Sauter, {\it Zum Kleinschen Paradoxon}, Z. Phys. {\bf 73} (1932) 547.

\bibitem{casimir}
H.~B.~G.~Casimir, {\it On the attraction between two perfectly conducting plates}, 
Proc. Kon. Ned. Akad. Wet. {\bf 60} (1948) 793.

\bibitem{casimir-reviews}
G.~Plunien, B.~M\"uller and W.~Greiner, {\it The Casimir Effect},  
Phys. Rept. {\bf 134} (1986) 87.
\\
K.~A.~Milton, {\it The Casimir Effect: Physical Manifestation of Zero-Point 
Energy}, {\tt (hep-th/9901011)}.
\\
K.~A.~Milton, {\it The Casimir effect: recent controversies and progress},  
J. Phys. {\bf A37} (2004) R209 {\tt (hep-th/0406024)}.
\\
S.~K.~Lamoreaux, {\it The Casimir force: background, experiments, and
applications}, Rep. Prog. Phys. {\bf 68} (2005) 201.

\bibitem{sparnaay}
M.~J.~Sparnaay, {\it Measurement of attractive forces between flat 
plates}, Physica {\bf 24} (1958) 751.

\bibitem{abramowitz}
M.~Abramowitz and I.~A.~Stegun, {\it Handbook of Mathematical Functions}, 
Dover 1972.

\bibitem{aharonov_bohm}
Y.~Aharonov and D.~Bohm, {\it Significance of the electromagnetic potentials
in the quantum theory}, Phys. Rev. {\bf 115} (1955) 485.

\bibitem{dirac-monopole}
P.~A.~M.~Dirac, {\it Quantised Singularities in the Electromagnetic Field}, 
Proc. Roy. Soc. {\bf 133} (1931) 60.

\bibitem{dodelson}
S.~Dodelson, {\it Modern Cosmology}, Academic Press 2003.

\bibitem{dirac_book} P.~A.~M.~Dirac, {\it Lectures on Quantum Mechanics}, 
Dover 2001.

\bibitem{henneaux-teitelboim}
M.~Henneaux and C.~Teitelboim, {\it Quantization of Gauge Systems}, Princeton
1992.

\bibitem{jackiw_rmp} 
R.~Jackiw, {\it Quantum meaning of classical field theory}, 
Rev. Mod. Phys. {\bf 49} (1977) 681
\\
R.~Jackiw, {\it Introduction to the Yang-Mills
quantum theory}, Rev. Mod. Phys. {\bf 52} (1980) 661.


\bibitem{beyond}
P.~Ramond, {\it Journeys Beyond the Standard Model}, Perseus Books 1999.
\\
R.~.N~. Mohapatra, {\it Unification and Supersymmetry. The 
Frontiers of Quark-Lepton Physics}, Springer 2003.

\bibitem{solid_state}
C.~P.~Burguess, {\it Goldstone and pseudogoldstone bosons in nuclear,
particle and condensed matter physics}, Phys. Rept. {\bf 330} (2000) 193
{\tt (hep-th/9808176)}.


\bibitem{anomalies_review}
L.~\'Alvarez-Gaum\'e, {\it An introduction to anomalies}, in: ``Fundamental
problems of gauge field theory'', eds. G.~Velo and A.~S.~Wightman, 
Plenum Press 1986.

\bibitem{jackiw_princeton} 
R.~Jackiw, {\it Topological investigations of quantized gauge theories}, 
in: ``Current Algebra and Anomalies'', eds. S.~B.~Treiman, R.~Jackiw, 
B.~Zumino and E.~Witten, Princeton 1985.

\bibitem{ABJ}
S.~Adler, {\it Axial-Vector Vertex in Spinor Electrodynamics,}
Phys. Rev. {\bf 177} (1969) 2426. 
\\ 
J.~S.~Bell and R.~Jackiw, {\it A
PCAC puzzle: $\pi^{0}\rightarrow 2\gamma$ in the sigma model}, Nuovo
Cimento {\bf A60} (1969) 47.

\bibitem{steinberger}
J.~Steinberger, {\it On the Use of Substraction Fiels and the Lifetimes
of Some Types of Meson Decay}, Phys. Rev. {\bf 76} (1949) 1180.

\bibitem{QCD}
F.~J.~Yndur\'ain, {\it The Theory of Quark and Gluon Interactions}, Springer
1999.

\bibitem{thooftU(1)}
G.~'t Hooft, {\it How the instantons solve the U(1) problem}, Phys. Rept. 
{\bf 142} (1986) 357.

\bibitem{sutherland_veltman}
D.~G.~Sutherland, {\it Current Algebra and Some Nonstrong Mesonic Decays}, 
Nucl. Phys. {\bf B2} (1967) 433.
\\
M.~J.~G.~Veltman, {\it Theoretical aspects of high-energy 
neutrino interactions}, Proc. R. Soc. {\bf A301} (1967) 107.


\bibitem{Adler-Bardeen}
S.~L.~Adler and W.~A.~Bardeen, {\it Absence of higher order corrections 
in the anomalous axial vector divergence equation}, Phys. Rev. {\bf 182}
(1969) 1517.

\bibitem{Witten-anom}
E.~Witten, {\it An SU(2) anomaly}, Phys. Lett. {\bf B117} (1982) 324.

\bibitem{pdg}
S.~Eidelman et al. {\it Review of Particle Phhysics}, Phys. Lett. {\bf B592}
(2004) 1 ({\tt http://pdg.lbl.gov}). 

\bibitem{gross-wilczek}
D.~J.~Gross and F.~Wilczek, {\it Ultraviolet behavior of nonabelian gauge 
theories}, 
Phys. Rev. Lett. {\bf 30} (1973) 1343.

\bibitem{politzer}
H.~D.~Politzer, {\it Reliable perturbative results for strong interations?}, 
Phys. Rev. Lett. {\bf 30} (1973) 1346.

\bibitem{thooftAF}
G.~'t Hooft, remarks at the {\it Colloquium on Renormalization of Yang-Mills
fields and applications to particle physics}, Marseille 1972.

\bibitem{russian-beta}
I.~B.~Khriplovich, {\it Green's functions in theories with a non-abelian
gauge group}, Yad. Fiz. {\bf 10} (1969) 409 [Sov. J. Nucl. Phys. {\bf 10}
(1970) 235].
\\
M.~V.~Terentiev and V.~S.~Vanyashin, {\it The vacuum polarization of a
charged vector field}, Zh. Eskp. Teor. Fiz. {\bf 48} (1965) 565
[Sov. Phys. JETP {\bf 21} (1965) 375].

\bibitem{wilson}
K.~G.~Wilson, {\it Renormalization group and critical phenomena
1. Renormalization group and the Kadanoff scaling picture,}
Phys. Rev. {\bf B4} (1971) 3174.
\\
K.~G.~Wilson, {\it Renormalization group and critical phenomena 2. Phase space cell
analysis of critical behavior}, Phys. Rev. {\bf B4} (1971) 3184
\\
K.~G.~Wilson, {\it The renormalization group and critical phenomena}, Rev. Mod.
Phys. {\bf 55} (1983) 583.

\bibitem{kadanoff}
L.~P.~Kadanoff, {\it Scaling Laws for Ising Models Near $T_c$}, Physics 
{\bf 2} (1966) 263.

\bibitem{schwinger}
J.~Schwinger, {\it On Gauge Invariance and Vacuum Polarization}, 
Phys. Rev. {\bf 82} (1951) 664.

\bibitem{BY} 
E.~Brezin and C.~Itzykson, {\it Pair Production in Vacuum by 
an Alternating Field}, Phys. Rev. {\bf D2} (1970) 1191.

\bibitem{hawking}
S.~W.~Hawking, {\it Particle Creation by Black Holes}, Commun. Math. Phys.
{\bf 43} (1975) 199.

\bibitem{parikh_wilczek}
M.~K.~Parikh and F.~Wilczek, {\it Hawking Radiation as Tunneling}, Phys. Rev. 
Lett. {\bf 85} (2000) 5042 ({\tt hep-th/9907001})

\bibitem{susy}
Yu.~A.~Golfand and E.~P.~Likhtman, {\it Extension of the Algebra of Poincar\'e 
group generators and violations of P-invariance}, JETP Lett. {\bf 13}
(1971) 323. 
\\
D.~V.~Volkov and V.~P.~Akulov, {\it Is the Neutrino a Goldstone Particle}, 
Phys. Lett. {\bf B46} (1973) 109.
\\
J.~Wess and B.~Zumino, {\it A Lagrangian Model Invariant under Supergauge
Transformations}, Phys. Lett. {\bf B49} (1974) 52.

\bibitem{hls} R.~Haag, J.~ \L opusza\'nski and M.~Sohnius, 
{\it All possible generators of supersymmetries of the S-matrix}, Nucl. Phys.
{\bf B88} (1975) 257.

\end{thebibliography}
\end{document}